\documentclass{article}
\usepackage{graphicx} 
\usepackage{matlab-prettifier}
\usepackage{JackPackage}
\usepackage{multirow}
\usepackage{graphicx}
\usepackage{subcaption}
\usepackage[normalem]{ulem}
\usepackage{authblk}
\usepackage{hyperref}
\usepackage{cleveref}
\usepackage[title]{appendix}
\usepackage{overpic}
\usepackage{float}
\usepackage{xcolor}

\usepackage[backend=biber, bibstyle=ieee, citestyle=ieee, doi=false, url=false]{biblatex}

\AtEveryBibitem{
  \ifentrytype{online}{
  }{
    \clearfield{url}
    \clearfield{urldate}
  }

  \ifboolexpr{ test {\iffieldundef{journal}} and test {\iffieldundef{journaltitle}} }{
  }{
    \clearfield{eprint}
    \clearfield{archivePrefix}
    \clearfield{primaryClass}
  }

}
  
\Crefname{subfigure}{Figure}{Figures}

\renewcommand{\thesubsection}{\thesection.\arabic{subsection}}
\newtheorem{assumption}{Assumption}

\date{} 

\begin{document}
\renewcommand{\labelenumi}{(\alph{enumi})}
\begin{center}
    {\Large \bfseries A Frequency-Learning Interpolation-Based Coordinate Descent Method to Optimize Parameterized Quantum Circuits on Quantum Hardware} \\[1.5em]
    Jack Mandell$^1$, Adam Leicester$^2$, Zixu Wang$^2$, Jian Shi$^{2,3}$,  Haimeng Zhang$^4$, and Yangyang Xu$^{1,*}$ \\[1.5em]

    \footnotesize
    $^1$Department of Mathematical Sciences, Rensselaer Polytechnic Institute, Troy, New York, USA \\
    $^2$Department of Materials Science and Engineering, Rensselaer Polytechnic Institute, Troy, New York, USA \\
    $^3$Department of Physics, Rensselaer Polytechnic Institute, Troy, New York, USA \\
    $^4$IBM Quantum, IBM T.J. Watson Research Center, Yorktown Heights, NY, USA. \\
    $^*$Corresponding author: \href{mailto:xuy21@rpi.edu}{xuy21@rpi.edu}
\end{center}

\begin{abstract}
Structure-aware optimizers have been the focus of recent study for the optimization of parameters in variational quantum circuits. Many of these optimizers follow a coordinate-based paradigm, making single parameter updates by using classical approximations to the cost function generated from quantum sampling. However, the performance of structure-aware optimizers on real quantum hardware at large system size has yet to be demonstrated. Recently, Interpolation-based Coordinate Descent (ICD) has been proposed to unify many of these optimizers into one framework. ICD requires a preset frequency set for interpolation, which can be theoretically derived, but for which the cost function is generally sparse over and may require manual tuning. To address these issues, we propose Frequency-learning Interpolation-based Coordinate Descent (FLICD) to adaptively learn frequencies online, allowing easier scaling up to larger systems. We demonstrate that our algorithm is more accurate than an existing zeroth-order method COBYLA and requires fewer quantum resources than ICD on simulator. Using IBM quantum hardware, we execute FLICD on systems up to 100 qubits and observe robust performance with an increasing number of layers.
\end{abstract}


\section{Introduction}

Variational Quantum Algorithms (VQAs) are a paradigm of hybrid quantum-classical algorithms designed for noisy intermediate-scale quantum (NISQ) era quantum computers. These classes of algorithms are intended to balance the quantum mechanical nature and noise accumulation of current hardware computer. Some examples of these algorithms include Variational Quantum Eigensolver (VQE) \cite{peruzzoVariationalEigenvalueSolver2014} and Quantum Approximate Optimization Algorithm (QAOA) \cite{farhiQuantumApproximateOptimization2014}. 

At their core, these algorithms seek to find system ground state energy, specified by the system's Hamiltonian $M$, from a variational form. These variational forms are encoded into parameterized quantum circuits (PQCs) denoted by $U(\boldsymbol{\theta})$ with classical parameters $\boldsymbol{\theta}$. For a $q$-qubit quantum system, the task is to solve
\begin{align} \label{def:problem_def}
    \min _ {\boldsymbol{\theta} \in \mathbb{R}^m} f(\boldsymbol{\theta}) = \langle \psi_0 | U(\boldsymbol{\theta})^\dagger M U(\boldsymbol{\theta})|\psi_0\rangle.
\end{align}
The parameters $\boldsymbol{\theta}$ are optimized via a classical optimizer. These types of algorithms involve repeated communication between the quantum and classical computers: the quantum computer computes the (noisy) function value at a given parameter assignment and the classical optimizer proposes new parameter assignments based on information from the previous and current function values. This process is repeated until some stopping condition is met. It has been shown that the functional landscape generated by parameterized quantum circuits can be highly non-convex, have many local minima, and can generally be an NP-hard problem~\cite{bittelTrainingVariationalQuantum2021,anschuetzBarrenPlateausQuantum2022}.

The optimizers for these problems fall into three main categories: zeroth-order, gradient-based, and structure-aware. \textit{Zeroth-order optimizers}, such as COBYLA and SPSA, have been used for optimizing parameters in QAOA on quantum hardware \cite{wangDepthIndependentLinearChain2025,cogburnHardwareAwareQAOAHoneypot2026a,azfarImprovingQAOAFeasibility2026}. \textit{Gradient-based optimizers} numerically estimate the gradient to update the parameters. These methods have been benchmarked in a variety of studies such as in \cite{pellow-jarmanEffectClassicalOptimizers2024,bonet-monroigPerformanceComparisonOptimization2023}.  \textit{Structure-aware optimizers} utilize the specific structure of parameterized quantum circuits to make parameter updates. For many optimizers in this class, the parameter updates are made coordinate-wise by taking advantage of the shape of the cost function when restricted to a single parameter $\theta_j$. Lai et al. unify these methods into a single algorithmic framework, known as Interpolation-based Coordinate Descent (ICD) and show that other structure-aware optimizers are specific instances of ICD \cite{laiInterpolationbasedCoordinateDescent2026}.

We will explore structure-aware optimizers in this work. The main motivation behind structure-aware optimizer design has been for decreasing the number of function calls needed to find minima \cite{laiInterpolationbasedCoordinateDescent2026,parrishJacobiDiagonalizationAnderson2019,liEfficientRobustParameter2024,nakanishiSequentialMinimalOptimization2020,jagerFastGradientfreeOptimization2025}. While theoretical convergence analysis is sparse for structure-aware optimizers, some research has begun to show convergence, such as for Rotosolve \cite{pramanikOneCoordinateTime2026}. However, many commonly-used ansatz have more complicated underlying structure, such as having  multiple gates use the same angle parameter, making it hard for these ansatz to be studied from a theoretical standpoint. Empirical research shows that structure-aware optimizers may have increased performance compared to off-the-shelf optimizers in finding better minima  \cite{laiInterpolationbasedCoordinateDescent2026,ostaszewskiStructureOptimizationParameterized2021}.   

Though a variety of structure-aware optimizers have been proposed in the literature, the level of testing of the algorithms has been limited. This limitation in the experimentation can be quantified based on two metrics: closeness to true quantum computing conditions and system size. In Algorithm 1 of \cite{vidalCalculusParameterizedQuantum2018}, the Fourier transform is used to recover Fourier coefficients, and no experimentation is given. For the studies that do have experimentation, conditions and set-up vary. For the Jacobi+Anderson algorithm \cite{parrishJacobiDiagonalizationAnderson2019}, function values are computed from exact wavefunction expectations on 4-qubit systems. In the Sequential Optimization with Approximate Parabola (SOAP) algorithm \cite{liEfficientRobustParameter2024}, which is based on parabolic approximation for the unitary couple-cluster ansatz, systems of up to 20 qubits are tested. However, the statistical nature of function calls are modeled by Gaussian noise, as opposed to shot-wise averages \cite{liEfficientRobustParameter2024}. Some studies use noiseless quantum simulators, such as Sequential Minimal Optimization (SMO) of \cite{nakanishiSequentialMinimalOptimization2020} and ICD of \cite{laiInterpolationbasedCoordinateDescent2026}, which both investigate algorithm performance in different shot regimes, with system sizes between 5 and 12 qubits. Other proposed algorithms include testing on quantum simulator and real quantum hardware. RotoSolve \cite{ostaszewskiStructureOptimizationParameterized2021} and ExcitationSolve \cite{jagerFastGradientfreeOptimization2025} are  tested on IBM quantum hardware, using computers \textit{ibm\_melbourne} and \textit{ibm\_quebec}, respectively. However, the systems deployed are no larger than 6 qubits. Few studies have utilized these proposed methods to do benchmarking on larger systems. There is a study where researchers at IonQ apply the Jacobi Iteration method \cite{parrishJacobiDiagonalizationAnderson2019} as a subroutine in combinatorial optimization using trapped-ion quantum computers up to 32 qubits. This makes the performance of structure-aware optimizers at scale on real quantum hardware an open question. 

It is unclear from previous literature how structure-aware optimizers actually perform at scale when hardware noise is present. Some of the studies claim that structure-aware optimizers will be robust on NISQ era devices \cite{nakanishiSequentialMinimalOptimization2020,jagerFastGradientfreeOptimization2025}. One argument is that if noise is systematic, such as a  consistent stretching or shrinking of the function values, position of minimizers remain unchanged \cite{jagerFastGradientfreeOptimization2025}. However, if noise is not systematic, where function values are biased in various directions, the location of the minimizers may change structurally. It is uncertain if uni-variate function recovery can be robust to such noise conditions. Experimentation is needed to determine the validity of these arguments.

Lastly, current structure-aware optimizers require certain niche knowledge about a problem instance structure to utilize the algorithm in practice. For this point, we highlight ICD as it generalizes many of the structure-aware optimizers into one unified framework \cite{laiInterpolationbasedCoordinateDescent2026}. ICD requires a specified frequency set for interpolation of each $\theta_j$. While a theoretical set of frequencies can be known in theory, in practice, this set may be unnecessarily large: contribution from higher-order frequencies decays rapidly or is sparse, and inherent function value noise may render using higher-order frequencies futile if the noise floor dominates. When deploying ICD, the frequency set is generally chosen to be a much smaller set, preserving algorithm performance while saving function evaluations. However, this frequency subset must be manually chosen. For small system sizes on a simulator, it may be easy to check which frequencies work the best. However, this same frequency set may not work for a larger system size and manually adjusting the frequency set may become impracticable at a large-size system. 

\textbf{Contributions.} We address these open questions in this work and summarize our contributions below:
\begin{itemize}
    \item We design \textit{Frequency-Learning Interpolation-Based Coordinate Descent} to solve problem \eqref{def:problem_def}. FLICD uses the same principles as ICD to make parameter updates, but makes changes to increase usability and scalability.  The main difference between ICD and FLICD is that FLICD seeks to \textit{learn} the necessary frequencies for interpolation during the algorithm execution, as opposed to specifying a fixed frequency set. FLICD is also controlled by a relative error hyper-parameter, as opposed to a preset shot count. This allows the algorithm to be more easily adapted to different problems and at larger problem scale. 
    \item We explore a more general method to recover the uni-variate cost function, namely via least-squares approximation with possible oversampling of data. Theoretically, we extend the work of \cite{laiInterpolationbasedCoordinateDescent2026} and show that 
    for the case of $r$ equidistant frequencies and $P\geq 2r+1$ sample points, choosing points evenly-spaced by $\frac{2\pi}{P}$ is optimal, assuming constant variance. We show in this scenario the allocation between the shots per function evaluation and number of sample points does not matter in theory for a fixed total number of shots. In practice, when trying to recover a very sparse representation of the cost function, increasing the number of sample points and decreasing the number of shots yields lower error compared to solving a square system, i.e. with $P = 2r+1$. However, when enough frequencies are used, the redistribution does not yield any difference in accuracy. Empirically, we also explore least-squares approximation in the case of non-equidistant frequencies with evenly-spaced sample points, and show that the accuracy and stability of the recovered cost function can be greatly improved at higher-order frequency interpolation by changing the allocation between number of shots and sample points.   
     
    \item On simulator, we benchmark FLICD against a zeroth-order optimizer COBYLA and the ICD method on a 20-qubit problem. FLICD achieves lower objective value than COBYLA, regardless of the number of layers. Likewise, 
    FLICD is more efficient than ICD in the high shot regime  and more accurate than ICD in the low shot regime when the number of function parameters is increased. 

    \item We confirm that hardware noise is systematic on IBM quantum hardware. From interpolating the cost function generated from a 16-qubit system, there is a global shrinking of the function values toward zero: the overall function structure is preserved as well as the location of minimizers. This supports why structure-aware optimizers may still be deployable on hardware.
    
    \item On IBM quantum hardware, we deploy FLICD on increasing system sizes of up to 100 qubits. The same hyper-parameters work at all system sizes, and FLICD is able to learn important frequencies. We observe that the ansatz structure greatly determines what frequencies are learned and the required number of shots needed to achieve a desired approximation error. Some learned frequency sets remain the same with system size, while the algorithm learns higher-order frequencies to achieve the same relative error with increasing system size. 
\end{itemize}

\textbf{Outline of the Work.} The rest of this paper is organized as follows. In section \ref{sec:Problem setting}, we define the problem setting in more detail and investigate properties of the cost function.  In section  \ref{sec:Proposed methods}, we give the FLICD algorithm and its various subroutines. Lastly, in section \ref{sec:Numerical Experiments}, we provide experiments to demonstrate the performance of FLICD on simulator and hardware and compare it with 
existing methods. 

\textbf{Notation.}
The superscript  $^\dagger$ denotes the Hermitian conjugate, or the transpose if real-valued. 
For two sets $A,B$ in $\mathbb{R}$ and $c \in \mathbb{R}$, we define the set addition and scalar-set multiplication by 
\begin{align*}
    A+B &:= \{a+b:a\in A,b \in B \}, \text{ and } 
    c\cdot A := \{ca: a \in A \}.
\end{align*}
$\mathcal{N}(\mu,\sigma^2)$ denotes the Gaussian distribution with mean $\mu$ and variance $\sigma^2$. Given an integer $N>0$, $[N]$ denotes the set $\{0,1,2,\dots,N-1 \}$. A vector $\mathbf{x}$ is put into a column form as $\mathbf{x}=[x_1,\ldots,x_t]^\dagger$ or simply denoted as $\mathbf{x}=(x_k)_{k=1}^t$. If $A\in  \mathbb{R}^{n \times n}$ is a matrix, then the condition number is denoted by $\kappa(A)$ and the set of eigenvalues (spectrum) is denoted by $\Lambda(A)$. 


\section{Problem setting}  \label{sec:Problem setting}

For a $q$-qubit quantum system with $N:=2^q$, our task is to solve problem \eqref{def:problem_def}
where $M \in \mathbb{C}^{N \times N}$ is a Hermitian quantum observable, $|\psi_0\rangle \in \mathbb{C}^N $ is an initial state, and $U(\boldsymbol{\theta}) \in \mathbb{C}^{N \times N}$ is a PQC of the form
\begin{align*}
    U(\boldsymbol{\theta}) = V_m U_m(\theta_m) \cdots  V_1 U_1(\theta_1).
\end{align*}
In the above for $j = 1,2,\dots,m$, $V_j$ is a fixed gate and $U_j(\theta_j) = \exp(i H_j\theta_j)$ is a rotation-like gate for some Hermitian generator matrix $H_j \in \mathbb{C}^{N \times N} $.  For notational convenience, we will also notate $U(\boldsymbol{\theta})|\psi_0\rangle $ as $| \psi (\boldsymbol{\theta})\rangle   $. 

It is common to describe an ansatz $U({\boldsymbol{\theta}})$ with $m$ parameters as having a ``layered" structure. Suppose that a positive integer $p$ divides $m$ and define $L := \frac{m}{p}$. It is said that $U({\boldsymbol{\theta}})$ is a \textit{$p$-layer ansatz} if for all $\ell \in  \{1,2\dots,L \}$, the following two strings of equalities hold:
\begin{align}
    H_\ell = H_{\ell+L} =  H_{\ell+2L} =\cdots = H_{\ell+(p-1)L}\label{eq:H-cycle} \\
    V_\ell = V_{\ell+L} =  V_{\ell+2L} =\cdots = V_{\ell+(p-1)L}.
\end{align}
From the definition, $L$ denotes the number of parameters per layer. Importantly, the individual parameters $\{\theta_j\}_{j=1}^m$ are still different each layer: only the Hermitian generators are repeated.

\subsection{Statistics of quantum measurement } \label{subsec:stats_quant_mech}


The quantum computer gives estimates of $f(\boldsymbol{\theta}) = \langle \psi(\boldsymbol{\theta })| M | \psi(\boldsymbol{\theta }) \rangle$ via a random variable $\Lambda$ drawn from a discrete random variable distribution $p_{\boldsymbol{\theta}}$. If $M = \sum_{l = 1}^{N} \lambda_l P_l$, where $P_l$ is the projection onto the eigenstate corresponding to eigenvalue $\lambda_l$, then $p_{\boldsymbol{\theta}}$ is given by
\begin{align*}
    \mathrm{Prob}\{\Lambda = \lambda_l \} = \langle \psi(\boldsymbol{\theta })| P_l | \psi(\boldsymbol{\theta }) \rangle \text{ for } l = 1,2,\dots,N
\end{align*}
in accordance with the quantum measurement postulate. The expectation of $\Lambda$ is exactly
\begin{align} \label{random variable}
    \text{E}_{\Lambda \sim p_{\boldsymbol{\theta }}} [\Lambda] &= f(\boldsymbol{\theta }),
\end{align}
and the variance
\begin{align*}
    \text{Var}_{\Lambda \sim p_{\boldsymbol{\theta }}} [\Lambda] &=  g(\boldsymbol{\theta}) - [f(\boldsymbol{\theta })]^2 = :\sigma^2(\boldsymbol{\theta }),
\end{align*}
where 
\begin{align*}
    g(\boldsymbol{\theta}) := \langle \psi (\boldsymbol{\theta }) | M^2 |\psi(\boldsymbol{\theta }) \rangle.
\end{align*} 

For more details on $\Lambda$, refer to \cite{laiInterpolationbasedCoordinateDescent2026}. 
The function estimate is resampled independently $n$ times, where $n$ is canonically referred to as the number of shots, to obtain i.i.d samples $\Lambda_1,\dots,\Lambda_n$. The sample mean $\bar{\Lambda}=\frac{1}{n}(\Lambda_1 + \cdots + \Lambda_n)$ is used as an estimator of $f(\boldsymbol{\theta})$ with $\bar{\Lambda} $ approximately following $ \mathcal{N}(f(\boldsymbol{\theta}),\frac{\sigma^2 (\theta)} { n})$ when $n$ is large by the central limit theorem.

\subsection{Structure of the cost function}

Define the map $f_j:\theta_j \mapsto f(\boldsymbol{\theta})$ by keeping $\theta_k$ for $k \neq j$ fixed. $f_j(\theta_j)$ is determined by the values of the other parameters $\theta_k$ for $k \neq j$, even though this dependence is not explicitly written. By changing these other $\theta_k$, a different uni-variate restriction arises. By separating out $\theta_j$, $f_j(\theta_j)$ can be rewritten as
\begin{align} \label{def:split_notation}
  f_j(\theta_j) =  f(\theta_{<j},\theta_j,\theta_{>j}) = \langle \psi_{<j} | U_j(\theta_j) ^\dagger M_{>j} U_j(\theta_j) |\psi_{<j} \rangle 
\end{align}
where 
\begin{align} \label{def:split_notation_psi}
    |\psi_{<j} \rangle   := V_{j-1}U_{j-1}(\theta_{j-1})\cdots V_1 U_1(\theta_1) |\psi_0\rangle
\end{align}
and 
\begin{align} \label{def:split_notation_M}
    M_{>j} :=V_{j}^\dagger  \cdots U_m(\theta_m) ^\dagger  V_m ^\dagger M V_{m}U_{m}(\theta_{m})\cdots V_j .
\end{align}
Since the restriction is to $\theta_j$,  $|\psi_{<j}\rangle $ and $M_{>j}$ are some fixed vector and matrix in terms of $\theta_k$ for $k\neq j$. 

In \cite{laiInterpolationbasedCoordinateDescent2026}, it is shown that $f_j(\theta_j)$ takes on the form of a finite-term Fourier series with the Fourier frequencies determined exclusively by the Hermitian generator $H_j$. If the eigenvalues of $H_j$ are denoted by $\{ \lambda_l^j\}_{l=1}^{N}$, then the frequency set $\boldsymbol{\Omega}^j$ is given by
\begin{align} \label{def:Omega^j}
    \boldsymbol{\Omega}^j :=  \left\{|\lambda_l^j - \lambda_{l'}^j| > 0: \forall l,l' = 1,2,\dots,N\right\},\hspace{4mm}  \forall j = 1,2,\dots,m.
\end{align}
Let $r_j := | \boldsymbol{\Omega}^j|$. When accessing frequencies in $\boldsymbol{\Omega}^j$, they are indexed in an increasing order and referred to as $ \{ \Omega_k^j\}_{k=1}^{r_j} $. 
The uni-variate restricted function is of the form
\begin{align} \label{ab_rep}
    f_j(\theta_j) = \frac{1}{\sqrt{2}}a_0 + \sum_{k=1}^{r_j} \left[  a_k \cos(\Omega_k^j \theta_j) + b_k \sin(\Omega_k^j \theta_j)\right],
\end{align}
where $a_0,\{a_k\}_{k=1}^{r_j}$ and $\{b_k\}_{k=1}^{r_j}$ are real-valued coefficients. The function can also be written in the standard Fourier representation form as 
\begin{align}  \label{c_rep}
    f_j(\theta_j) =\sum_{k=-r_j}^{r_j} c_k e^{i\Omega_k^j \theta_j}
\end{align}
with $\Omega_0^j = 0$ for all $ j = 1,2,\dots,m$. The linear transformation of the complex-valued coefficients $\{c_k\}_{k=1}^{r_j}$ is given by
\begin{align} \label{ab_to_c_coeff_change}
    \begin{cases}
        c_0 \hspace{3mm}:= \frac{1}{\sqrt{2}}a_0 \\
        c_k \hspace{3mm}:= \frac{a_k}{2} - \frac{b_k}{2} i, \hspace{4mm} \forall k=1,\dots,r_j, \\
        c_{-k} \hspace{1mm}:= \frac{a_k}{2} + \frac{b_k}{2}i,\hspace{4mm} \forall k=1,\dots,r_j.
    \end{cases}
\end{align}

\subsection{Properties of the uni-variate restricted function}

In this subsection, we describe the theoretical frequency set in more detail and demonstrate how the frequencies contribute in practice. The goal is to give a high-level idea of how sparse the function spectrum can be over the theoretical frequency set. 

\subsubsection{Theoretical analysis}
\label{subsubsec:theoretical_analysis}
It was seen earlier that the frequency set for $f_j(\theta_j)$, given by $\boldsymbol{\Omega}^j$, consists of all positive differences in eigenvalues of $H_j$. For a general $H_j$, obtaining this set directly is impractical as it involves eigenvalues of exponentially large matrices. But, for certain classes of generators, there are guarantees of the structure of $\boldsymbol{\Omega}^j$. In many cases, $H_j$ is a sum of much simpler matrices. In \cite{laiInterpolationbasedCoordinateDescent2026}, it is noted that if $H_j$ is of the form $H_j = \sum_{k = 1}^K \pm P_k$
where $\{P_k\}_{k=1}^K$ are commuting Pauli terms, then $\boldsymbol{\Omega}^j$ contains evenly-spaced positive integers. A generalization can be made as follows:

\begin{lemma} \label{lemma:theory_freqs}
If $  H_j = \sum_{k = 1}^K (\Omega c_k) P_k$ where $\{c_k\}_{k=1}^K$ are integers, $\Omega$ is a positive real number, and $\{P_k\}_{k=1}^K$ are commuting non-identity Pauli terms, then $\boldsymbol{\Omega}^j$ contains only positive integer multiples of $\Omega$. If the Pauli terms are also algebraically independent, then $\Omega^j_{max} := \max\{ \boldsymbol{\Omega}^j \} = 2 \Omega \sum_{k=1}^K |c_k| $.
\end{lemma} 
\begin{proof}

The Pauli terms commute and all non-identity Pauli terms have spectrum $\{ +1,-1\}$. Hence, $\Lambda(H_j)$ can be estimated as follows:
  \begin{align} 
      \Lambda(H_j) &= \Lambda \left( \sum_{k = 1}^K \Omega c_k P_k \right) \\
      &\subseteq  \label{proof:alg_indep} \sum_{k = 1}^K \Lambda \left( \Omega c_k P_k \right) 
      =  \sum_{k = 1}^K \Omega c_k \cdot \Lambda \left(  P_k \right)  
      =  \Omega \sum_{k = 1}^K  c_k \cdot \{ \pm 1\} 
      =  \Omega \sum_{k = 1}^K  \{ \pm c_k\}.
  \end{align} 
Since the $\{c_k\}_{k=1}^K$ are integers, the set addition over $k$ must also give integers. Hence, $\Lambda(H_j)$ contains only integer multiples of $\Omega$. Recall that the frequencies are positive differences in the eigenvalues, therefore the frequencies in $\boldsymbol{\Omega}^j$ must also be positive integer multiples of $\Omega$. 

Now we prove the second claim. The subset sign in \eqref{proof:alg_indep}  can be promoted to equality under the assumption of algebraic independence. Then, note that if $\lambda \in \Lambda(H_j)$, then $-\lambda \in \Lambda(H_j)$. If $\lambda_{max}:=\max\{\Lambda(H_j)\}$, then  $\Omega_{max}^j = \lambda_{max} - ( - \lambda_{max}) = 2\lambda_{max}$.  Since $\lambda \in \Lambda(H_j)$ can be written as a sum of $K$ terms each chosen from $\{ \pm c_k\}$, pick out the positive values in each $\{ \pm c_k\}$. Hence,   $\lambda_{max}  = \Omega \sum_{k=1}^K |c_k|$ and $\Omega_{max}^j = 2\Omega \sum_{k=1}^K |c_k|$. 
\end{proof}
In summary, the frequencies of $f_j(\theta_j)$ are contained in $ \{ \Omega,2\Omega,3\Omega,\dots, \Omega^j_{max}\}$. While the frequencies stay inside this range, the contribution to each frequency may vary based on the values of $\theta_k$ for $k \neq j$. To understand the dependence more explicitly, 
we view the cost function representation in 
\eqref{def:split_notation}. By Appendix A in \cite{laiInterpolationbasedCoordinateDescent2026}, if $\{ |\xi_u\rangle \}_{u=1}^N$ is the eigenbasis of $H_j$, then the Fourier coefficients can be written in terms of projections $\langle \xi_u|\psi_{<j} \rangle$ and $\langle \xi_u |M_{>j}| \xi_v \rangle$ for $u,v = 1,2,\dots,N$. 
When the values of other $\theta_k$ for $k\neq j$ change, these overlap and projection terms change as well, and the contribution to each frequency may be quite different. 


\subsubsection{Empirical analysis } \label{section:theoretical param dep}

While we save most of the numerical results to section \ref{sec:Numerical Experiments}, the takeaway from empirical results is central to motivate our algorithm design. Before giving our algorithm, we provide some numerical results to demonstrate certain empirical properties of these uni-variate restricted functions. 

From section \ref{subsubsec:theoretical_analysis}, different $\theta_k$ for $k \neq j$ may give different contribution from each frequency in theory. Recall $f_j(\theta_j)$ has a standard-form finite Fourier expansion \eqref{c_rep} with coefficients $\{c_k\}_{k=-r_j}^{r_j}$. Notice $c_{-k} = \overline{c_k}$ and $|c_{-k}|^2 = |c_{k}|^2$ for $k = 1,2,\dots,r_j$ as $f_j(\theta_j)$ is real-valued. Thus we ignore coefficients of negative frequencies and let $\boldsymbol{c} := (c_1,\dots,c_2)$. To quantify the contribution from each frequency, define the \textit{Fourier coefficient magnitude distribution }
\begin{align} \label{def:d}
    \bold{d} = (d_1,d_2,\dots,d_{r}) := \left( \frac{|c_1|^2}{||\boldsymbol{c}||^2},\frac{|c_2|^2}{||\boldsymbol{c}||^2},\dots,\frac{|c_{r}|^2}{||\boldsymbol{c}||^2} \right),
\end{align}
where each component $d_k$ measures the contribution that $\Omega^j_k$ makes in $f_j(\theta_j)$. 

We now give an example of what $f_j$ can look like in practice for various choices of $\{\theta_k\}_{k\neq j}$. We define $f(\boldsymbol{\theta})$ as the cost function generated from a Max-cut problem of a 16-vertex random regular degree 5 graph with weights chosen uniformly at random from $\{1,2 \}$ using a $3$-layer Linear Chain QAOA ansatz \cite{wangDepthIndependentLinearChain2025}. In this example, $\boldsymbol{\theta} \in \mathbb{R}^6$ and we restrict to $\theta_1$ and vary the remaining parameters. For a more detailed description of this problem and ansatz setup, refer to section \ref{subsec:problem_setup}. The goal here is simply to give a quick visualization.



\begin{figure}[h]
    \centering
    \begin{subfigure}{0.32\textwidth}
        
        \centering
        \captionsetup{justification=centering}
        \includegraphics[width=\linewidth]{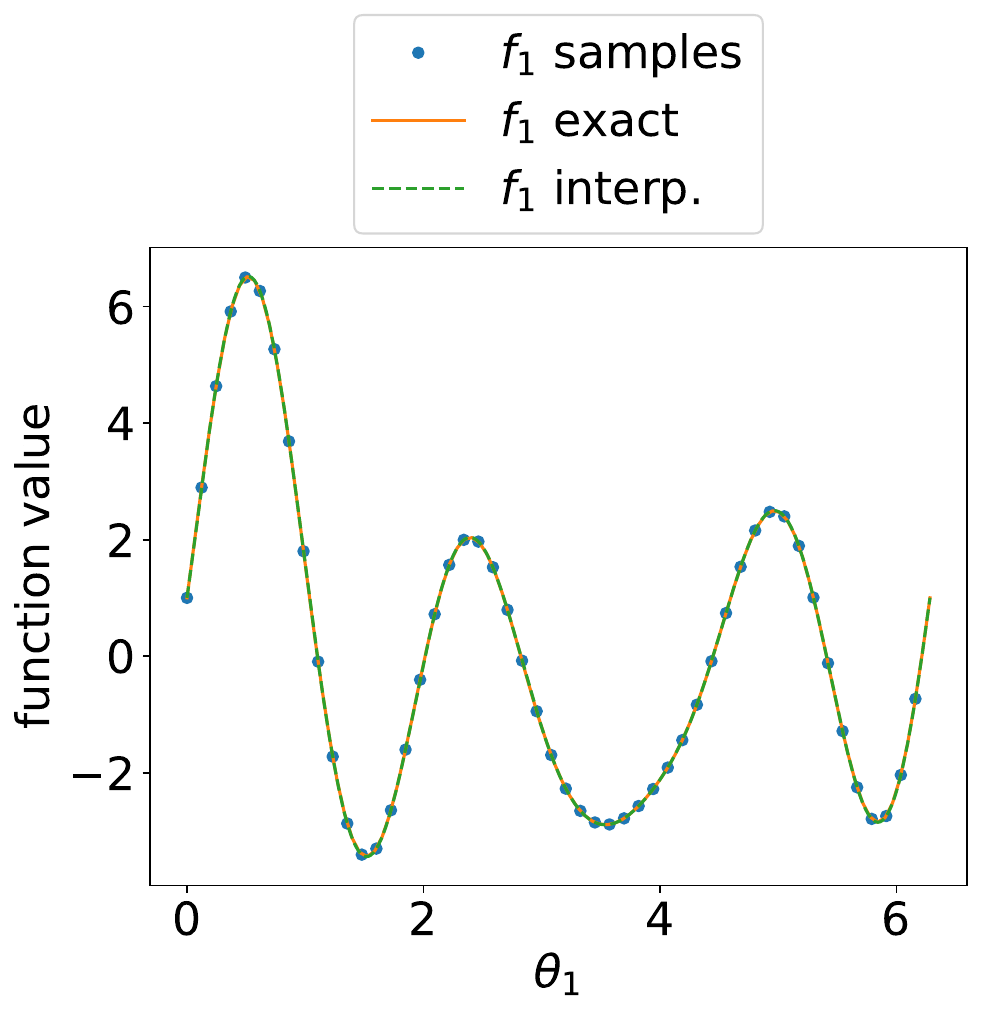}
        \caption{$\theta_3 = \theta_5 = \frac{\pi}{3}, $ \\ \hspace{1.18cm}$\theta_2 = \theta_4 = \theta_6 = \frac{\pi}{7}$}
        \label{subfig:pi_2}
    \end{subfigure}
    \hfill
    \begin{subfigure}{0.335\textwidth}
        \centering
        \captionsetup{justification=centering}

        \includegraphics[width=\linewidth]{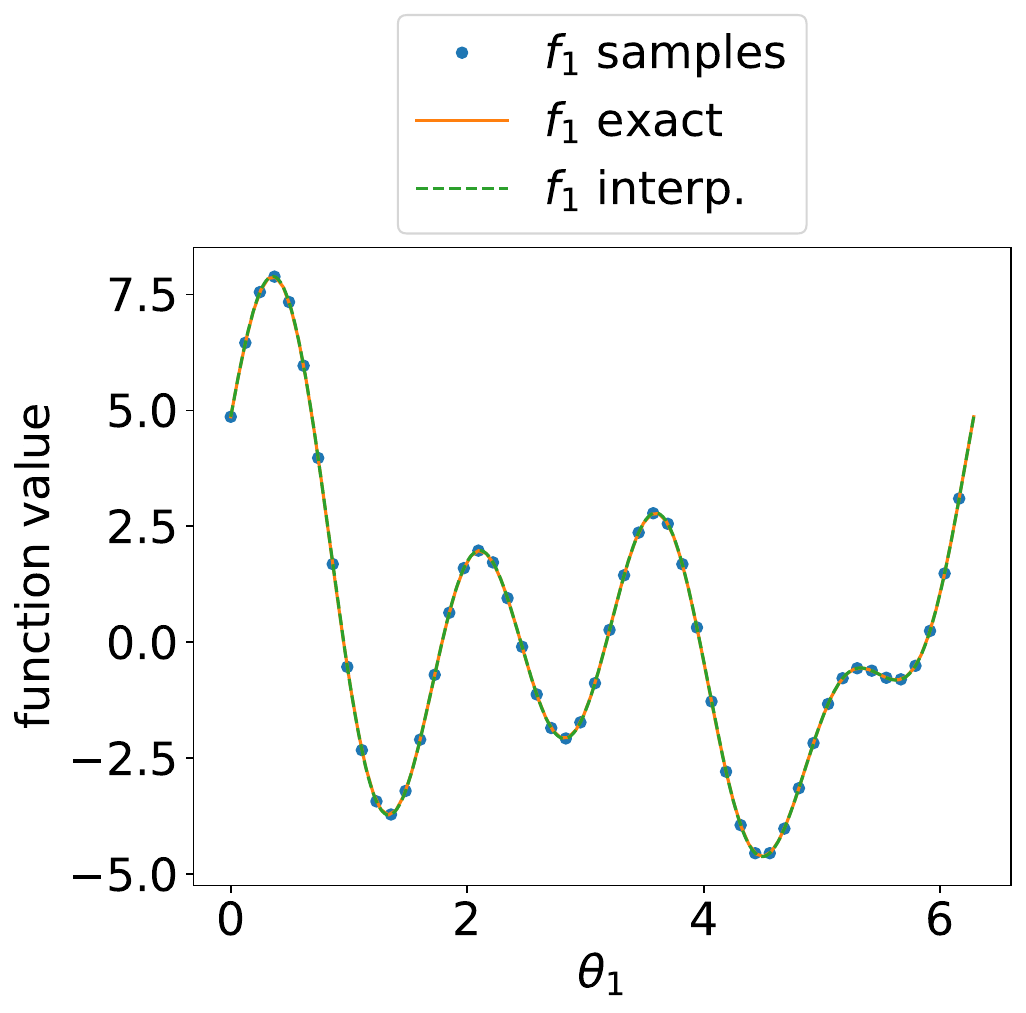}
        \caption{$\theta_3 = \theta_5 = \frac{\pi}{5}, $\\ \hspace{1.18cm} $\theta_2 = \theta_4 = \theta_6 = \frac{\pi}{8}$}
        \label{subfig:seed=55}
    \end{subfigure}
    \hfill
    \begin{subfigure}{0.32\textwidth}
        \centering
        \captionsetup{justification=centering}

        \includegraphics[width=\linewidth]{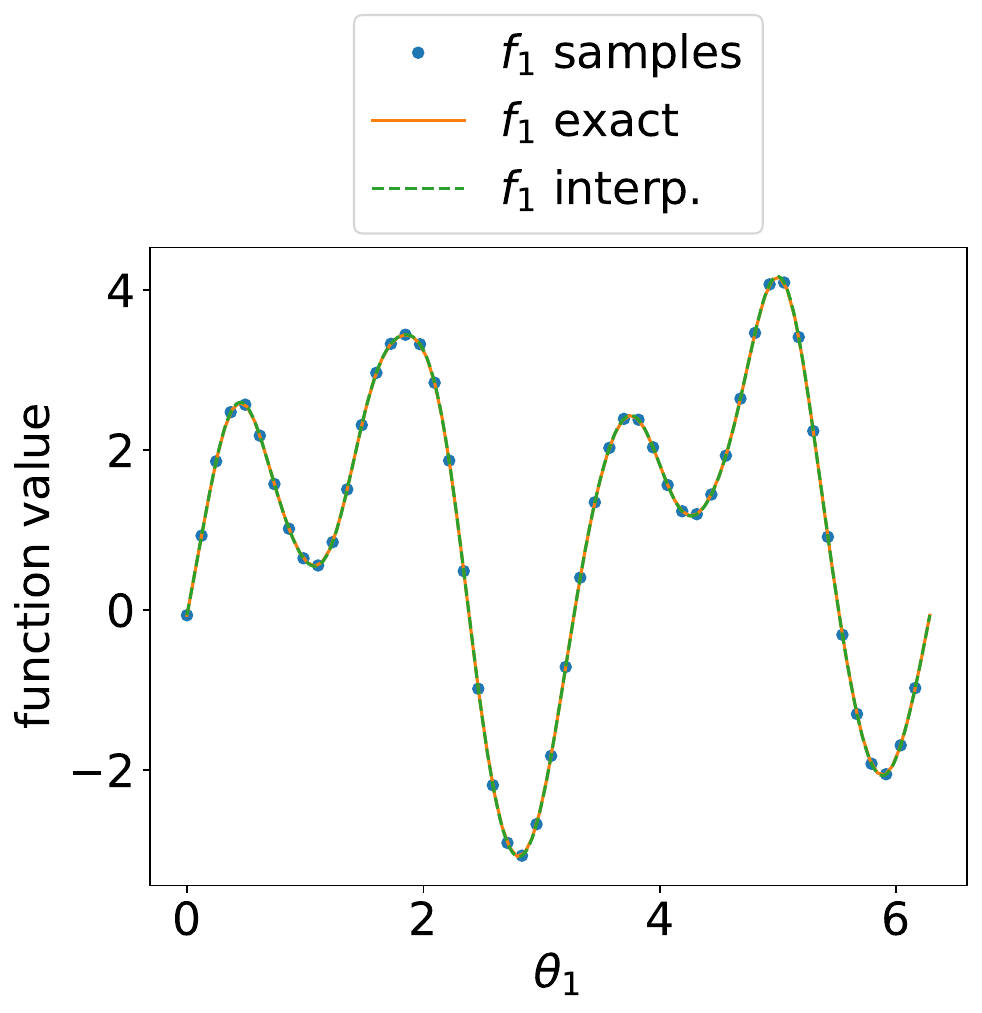}
        \caption{$\theta_k \sim \mathrm{Unif}[0,2 \pi)$\\ \hspace{0cm} i.i.d. $\forall k\geq 2$}
        \label{subfig:random}
    \end{subfigure}

    \vspace{0.5em}

    \begin{subfigure}{0.32\textwidth}
        \centering
        \captionsetup{justification=centering}

        \includegraphics[width=\linewidth]{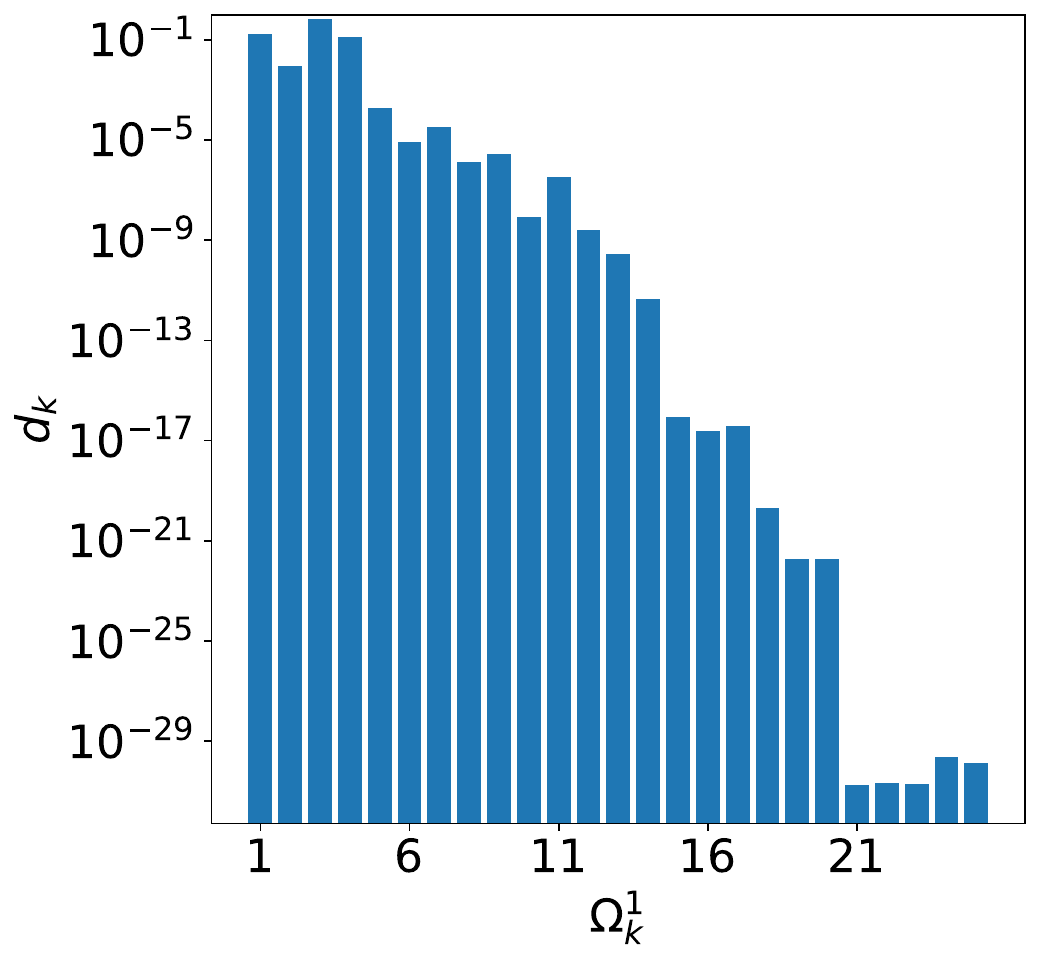}
        \caption{$\theta_3 = \theta_5 = \frac{\pi}{3}, $ \\ \hspace{1.18cm}$\theta_2 = \theta_4 = \theta_6 = \frac{\pi}{7}$}
    \end{subfigure}
    \hfill
    \begin{subfigure}{0.32\textwidth}
        \centering
        \captionsetup{justification=centering}

        \includegraphics[width=\linewidth]{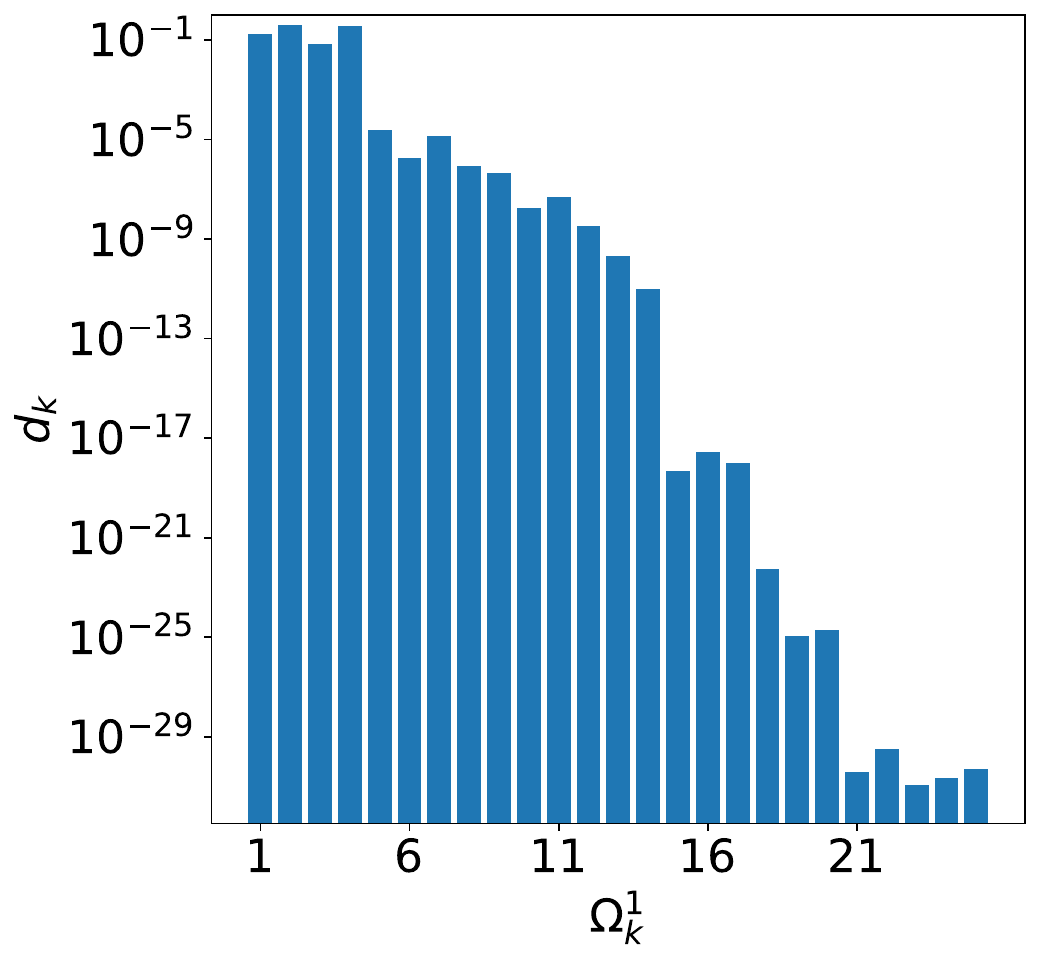}
        \caption{$\theta_3 = \theta_5 = \frac{\pi}{5}, $\\ \hspace{1.18cm} $\theta_2 = \theta_4 = \theta_6 = \frac{\pi}{8}$}
    \end{subfigure}
    \hfill
    \begin{subfigure}{0.32\textwidth}
        \centering
        \captionsetup{justification=centering}

        \includegraphics[width=\linewidth]{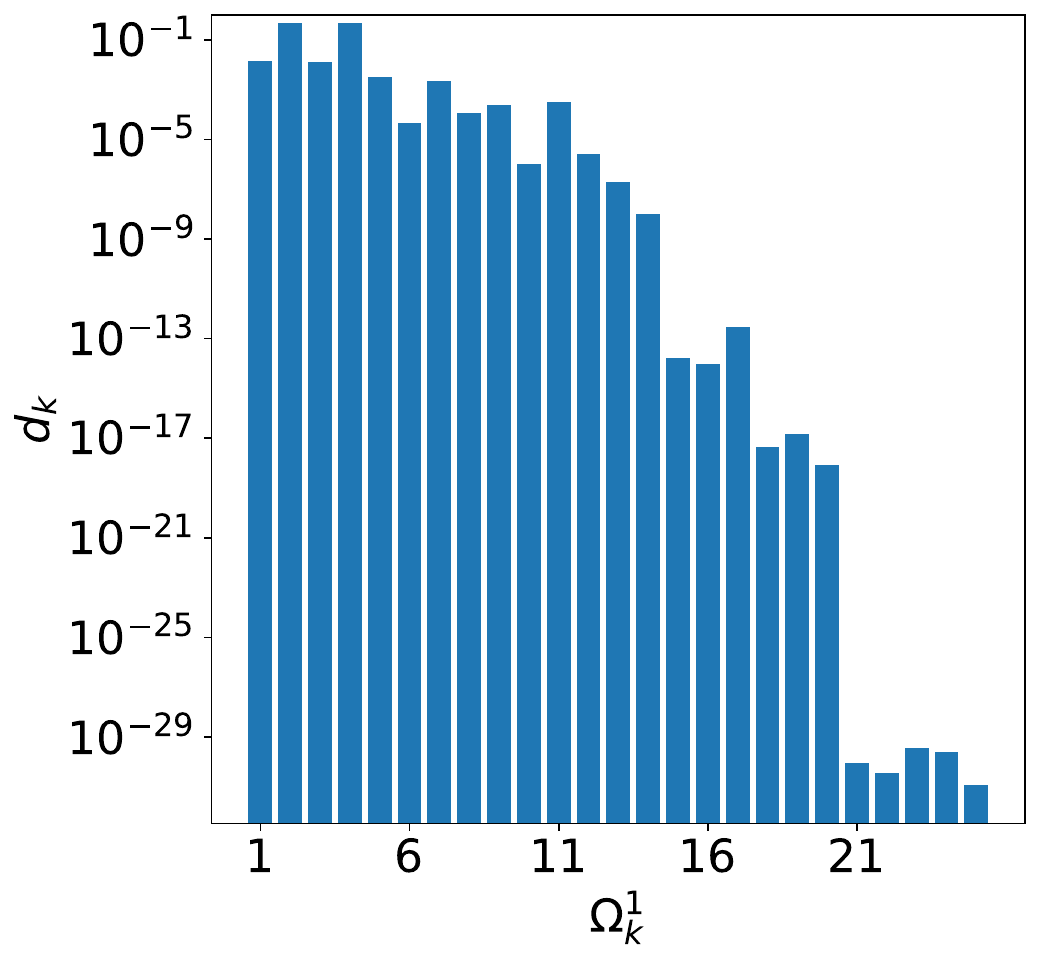}
        \caption{$\theta_k \sim \mathrm{Unif}[0,2 \pi)$ \\ \hspace{0cm}i.i.d. $\forall k\geq2$}
    \end{subfigure}

    \caption{$f_1(\theta_1)$ for different choices of $\{ \theta_k\}_{k = 2}^6$. Orange line denotes $f_1$ plotted using exact Statevector calculation. Green dashed line represents interpolation of $f_1$ using exact interpolation at points in blue. The distribution $\mathbf{d}$ is obtained from the interpolant of $f_1$.}
    \label{fig:function diff parameter values}
\end{figure}

Across all of the parameter regimes of $\{ \theta_k \}_{k \neq 1}$ in Figure \ref{fig:function diff parameter values}, higher-order frequencies tend to have lower contributions, with a consistent rate of decay. The exact contribution can depend on the setting of the other parameters and it is not necessarily monotonic.  After around frequency 15, the contribution becomes on the order of machine precision. The exact contribution cannot be inferred below this line, but most likely the contribution continues to approach $0$. 

While the type of behavior we observe cannot be proven to appear in every problem class, we believe this sparsity is evident in enough problem classes to form the basis for Frequency-learning Interpolation-based Coordinate Descent (FLICD), as was noted in the discussion section of \cite{laiInterpolationbasedCoordinateDescent2026}. 

\section{Proposed methods} \label{sec:Proposed methods}

In this section, we provide technical details to construct our algorithm. Let $f_j$ be the uni-variate function restricted on $\theta_j$ as defined in \eqref{def:split_notation}. For simplicity, we will use $x$ in replace of $\theta_j$ for the generic single variable. Due to quantum hardware noise, we do not have access to an exact evaluation of $f_j$. Following Assumption 1 of \cite{laiInterpolationbasedCoordinateDescent2026}, we assume the availability of a noisy estimate of $f_j$,  where the deviation follows a zero-mean Gaussian distribution. This assumption is made only for theoretical analysis; Experiments on quantum simulator and hardware follow the statistics of quantum measurement in subsection \ref{subsec:stats_quant_mech}. 

\begin{assumption}[$n$-shot noisy estimate of function value]\label{assump:noisy-est}
For any $j=1,\ldots,m$, let $f_j$ be the uni-variate function given in \eqref{def:split_notation}. We are able to obtain its noisy estimate $\tilde{f}_j^{(n)} (x)$ at any $x \in \mathbb{R}$ by $n$ shots such that 
\begin{align}\label{eq:x-var}
    \epsilon_x = \tilde{f_j}^{(n)} (x) - f_j(x) \sim \mathcal{N}\left(0,\frac{\sigma^2}{n}\right),
\end{align}
where the variance $\sigma^2$ is independent of $x$.  
\end{assumption}

Let $\bold{\Omega}$ be a frequency set of cardinality $t=|\bold{\Omega}|$ and its entries be ordered increasingly as $\Omega_1 < \Omega_2 < \cdots < \Omega_t$. Also let $\mathbf{x} = [x_0, x_1,\ldots,x_{P-1}]^\dagger\in \mathbb{R}^P$ be the vector $P$ sample values. We define the \textit{interpolation matrix} $A_{\bold{\Omega},\mathbf{x}} \in \mathbb{R}^{P\times (2t+1)}$ as
\begin{align}\label{eq:def-A-Omage-x}
    A_{\bold{\Omega},\mathbf{x}} := \begin{bmatrix}
        1/\sqrt{2}&  \cos (\Omega_1x_0) & \sin (\Omega_1x_0) & \cdots &  \cos (\Omega_t x_0) & \sin (\Omega_t x_0) \\
        1/\sqrt{2}&  \cos (\Omega_1x_1) & \sin (\Omega_1x_1) & \cdots  & \cos (\Omega_t x_1)  & \sin (\Omega_t x_1)\\
        \vdots & \vdots & \vdots & \ddots & \vdots & \vdots  \\
        1/\sqrt{2}&  \cos (\Omega_1x_{P-1}) & \sin (\Omega_1x_{P-1}) & \cdots  & \cos (\Omega_t x_{P-1}) & \sin (\Omega_t x_{P-1})\\
    \end{bmatrix} .
\end{align}
We will use $A_{\bold{\Omega},\mathbf{x}}$ to formulate least-squares problems to fit $\{f_j\}_{j=1}^m$.

\subsection{Overview}

Our algorithm is inspired by Interpolation-based Coordinate Descent (ICD), proposed in  \cite{laiInterpolationbasedCoordinateDescent2026}. Before giving our algorithm, we first summarize ICD and its main results. 

At each iteration, ICD selects an index $j$ and updates $\theta_j$. In a typical coordinate-based method, the following update is performed:
\begin{align*}
    \theta_j \leftarrow \mathrm{argmin}_{x \in \mathbb{R}}f_j(x).
\end{align*}
However, a full description of $f_j$ is not accessible on a quantum computer: only noisy estimates are accessible via the distribution $\Lambda$, according to statistics of quantum measurement. Instead, a classical approximation $\hat{f}_j$ to $f_j$ is learned; in this case, $\hat{f}_j$ is obtained via sampling function values on the quantum computer and using interpolation on a classical machine. In ICD, the update rule is thus
\begin{align*}
    \theta_j \leftarrow \mathrm{argmin}_{x \in \mathbb{R}}\hat{f}_j(x).
\end{align*}
The key computation of ICD is the construction of $\hat{f}_j$. Recall that $f_j$ has a finite Fourier expansion of the form \eqref{ab_rep}. To estimate the coefficient vector $\bold{z}:= [a_0,a_1,b_1,\dots,a_r,b_r]^\dagger$, where for simplicity of notation we denote $r = r_j$, the following procedure is used:
\begin{enumerate}
    \item Choose a set of sample points and form a vector $\bold{x} := [x_0,x_1,\ldots,x_{2r}]^\dagger$;
    \item With $n$ shots per function evaluation, obtain noisy estimates  $\bold{y_{obs}} = [ \tilde{f}^{(n)}_j(x_0),\dots,\tilde{f}^{(n)}_j(x_{2r})]^\dagger$;
    \item Solve the system $A_{\bold{\Omega}^j, \bold{x}} \hat{\bold{z}} = \bold{y_{obs}}  $ to obtain $\hat{\bold{z}}=[\hat{a}_0,\hat{a}_1,\hat{b}_1,\dots,\hat{a}_r,\hat{b}_r]^\dagger$, where $\bold{\Omega}^j$ is given in \eqref{def:Omega^j} and $A_{\bold{\Omega}^j, \bold{x}} \in \mathbb{R}^{(2r+1) \times (2r+1)}$ is defined in \eqref{eq:def-A-Omage-x}. 
 \item Form $\hat{f}_j$ by using expression \eqref{ab_rep} with coefficients contained in $\hat{\bold{z}}$. 
\end{enumerate}
In \cite{laiInterpolationbasedCoordinateDescent2026}, much of the analysis is proving the right choice of interpolation nodes. It is shown that if $\boldsymbol{\Omega}^j$ contains equidistant frequencies, then the evenly-spaced points $\bold{x} = \{ \frac{2\pi}{2r+1} k\}_{k=0}^{2r}$ are optimal, under Assumption \ref{assump:noisy-est}, according to various metrics. In numerical experiments of \cite{laiInterpolationbasedCoordinateDescent2026}, the full frequency set $\boldsymbol{\Omega}^j$ is not used, but a much smaller set $\boldsymbol{{\Omega}}^{j,*} \subseteq \boldsymbol{\Omega}^j$ is used to interpolate. It is claimed that since $f_j$ is generally sparse over $\boldsymbol{\Omega}^j$, the performance of ICD can be maintained even with $\boldsymbol{{\Omega}}^{j,*}$, while also drastically reducing the number of required total shots in the process. 

For ICD to perform well, $\boldsymbol{{\Omega}}^{j,*}$ must be chosen correctly by the user. However, it is not obvious how to choose $\boldsymbol{{\Omega}}^{j,*}$ without performing some prior computation. If $\boldsymbol{{\Omega}}^{j,*}$ does not have enough frequencies, then $\hat{f}_j$ is not a good approximation of $f_j$ and the updates to $\theta_j$ may be harmful. If $\boldsymbol{{\Omega}}^{j,*}$ has many extra unnecessary frequencies, then this slows down the algorithm and leads to extra total shots. In addition, using extra frequencies can harm the fitting due to noise and worse conditioning. 

To address the issue in ICD, we propose FLICD. Instead of fixing the frequencies used for interpolation at the start,  FLICD learns online the required frequency set. In this way, a required level of accuracy can be reached without manually fixing frequencies a priori. The success of FLICD relies in how well the algorithm is able to learn these frequencies and how efficiently and robustly the procedure can work in the presence of noisy function values and hardware noise.



\subsection{Least-squares approximation}\label{sec:least-square}

In ICD, interpolation with a square system is used to recover the uni-variate restricted function. We now extend its analysis to a more general method for recovering the Fourier coefficients.  

If there are $P = 2r+1$ sample points and the function evaluation at each point requires $n$ shots, the total number of shots to obtain $\hat{f}_j$ is $Q := nP$. With this setup, the interpolation matrix is square. However, a more general way to recover the Fourier coefficients is through least-squares approximation, by possibly oversampling data, i.e. $P \geq 2r+1$. By allowing $P$ to vary, there is freedom in how to distribute the total shot budget $Q$ between  $n$ and $P$. The question then becomes whether $\hat{f}_j$ can be improved purely by setting a different allocation between $n$ and $P$.  In the following sections, we analyze recovering the Fourier coefficients using least-squares approximation. The theoretical analysis is similar to that of ICD, but we will highlight important differences. 

For convenience, we abuse notation and drop the index $j$ in the rest of this subsection.  The frequency set $\bold{\Omega}^j$ is simply denoted as $\bold{\Omega}=\{ \Omega_k\}_{k=1}^{r}$, while the uni-variate function $f_j$ is denoted as $f$.
%
Given the integer $P \geq 2r+1$ and the vector of sampled points $\bold{x} := [x_0,x_1,\dots,x_{P-1}]^ \dagger \in \mathbb{R}^{P}$, let 
the interpolation matrix $ A_\bold{x} = A_{\bold{\Omega}, \bold{x}} \in \mathbb{R}^{P \times (2r+1)}$ be as in definition \eqref{eq:def-A-Omage-x}. 
With a vector of (possibly noisy) function evaluations $\mathbf{y}$ at the sampled points, we can recover  
the Fourier coefficient vector $\bold{z} := [ a_0,a_1,b_1,\dots,a_r,b_r ]^\dagger \in \mathbb{R}^{2r+1}$ 
via solving the following least-squares problem:
\begin{align} \label{def:least_squares_problem}
    \bold{z}^* := \underset{\bold{z} \in \mathbb{R}^{2r+1}}{\mathrm{argmin}} || A_\bold{x}  \bold{z} - \bold{y}||^2.
\end{align}
Suppose that $A_\bold{x}$ has full rank. 
Then
\begin{align}\label{eq:recover-z}
    \bold{z}^* = ( A_\bold{x} ^\dagger A_\bold{x}  )^{-1} A_\bold{x} ^{\dagger} \bold{y}.
\end{align}
Below, consider $\mathbf{y}=\bold{y_x} := [f(x_0),\dots,f(x_{P-1})]^\dagger$ and $\mathbf{y}=\bold{y_{obs}} := [ \tilde{f}^{(n)}(x_0),\dots,\tilde{f}^{(n)}(x_{P-1})]^\dagger$ in the recovery of Fourier coefficients in \eqref{eq:recover-z}, where $\tilde{f}^{(n)}$ satisfies the condition in Assumption~\ref{assump:noisy-est}. We denote their corresponding recovered $\mathbf{z}$-vector as $\mathbf{z}^*_{\mathbf{x}}$ and $\hat{\mathbf{z}}^*_{\mathbf{x}}$ respectively. 

Define the error vector $\bold{e} = [\epsilon_0,\epsilon_1,\dots,\epsilon_{P-1}]^\dagger \sim \mathcal{N}(0,\frac{\sigma^2}{n}I)$ and thus 
$$\bold{y_{obs}}= \bold{y_{x}} + \bold{e}, \quad \text{ and }\quad \bold{\hat{z}_x^*} = \bold{z}^*_{\mathbf{x}}  + ( A_\bold{x}^\dagger A_\bold{x} )^{-1} A_\bold{x}^{\dagger} \bold{e}.$$ 
Let the \textit{estimation error} and the \textit{squared error} (SE) respectively be
\begin{align*}
    \Delta \bold{\hat{z}_x^*} := \bold{\hat{z}_x^*} - \bold{{z}^*_x} = ( A_\bold{x}^\dagger A_\bold{x} )^{-1} A_\bold{x}^{\dagger} \bold{e},\quad\text{ and }\quad \mathrm{SE}(\bold{\hat{z}_x^*}) := ||\Delta \bold{\hat{z}_x^*}||^2.
\end{align*}
Then $\mathbb{E}[\Delta \bold{\hat{z}_x^*}] = 0$ and  
\begin{align*}
    \text{Var} [\bold{\hat{z}_x^*}] & = \mathbb{E}[\Delta \bold{\hat{z}_x^*} (\Delta \bold{\hat{z}_x^*})^\dagger]  = ( A_\bold{x}^\dagger A_\bold{x} )^{-1} A_\bold{x}^{\dagger}  \cdot \mathbb{E}[\bold{e} \bold{e}^\dagger]  \cdot A_\bold{x} [( A_\bold{x}^\dagger A_\bold{x} )^{-1}]^\dagger \\
    & = \frac{\sigma^2}{n} [( A_\bold{x}^\dagger A_\bold{x} )^{-1}]^\dagger  = \frac{\sigma^2}{n} ( A_\bold{x}^\dagger A_\bold{x} )^{-1}.
\end{align*}
Hence,
\begin{align*} \label{eq:E[SE] expression}
    \mathbb{E}\left[\mathrm{SE}(\bold{\hat{z}_x^*})\right] 
     &=\mathbb{E} \left[\mathrm{tr} \left( (\Delta \bold{\hat{z}_x^*})^\dagger  \Delta \bold{\hat{z}_x^*}  \right) \right]\\[.6em] 
    & =\mathbb{E} \left[\mathrm{tr} \left(\Delta \bold{\hat{z}_x^*} (\Delta \bold{\hat{z}_x^*})^\dagger  \right) \right] = \mathrm{tr} \left( \mathbb{E}[\Delta \bold{\hat{z}_x^*} (\Delta \bold{\hat{z}_x^*})^\dagger] \right) = \frac{\sigma^2}{n}  \text{tr} \left(( A_\bold{x}^\dagger A_\bold{x} )^{-1} \right).
\end{align*}
To obtain the best interpolation, one can minimize the expected SE by optimizing the choice of sampled points $\mathbf{x}$. In section 4.3 of the Supplementary Information of \cite{laiInterpolationbasedCoordinateDescent2026}, the optimal sample points are found for $P = 2r+1$ by assuming equidistant frequencies. Below, we extend the result to the more general case of $P\ge 2r+1$ under the same assumption. While we focus on the case of equidistant frequencies in the theoretical analysis, we will explore how least-squares approximation improves the fitting accuracy in the case of non-equidistant frequencies in numerical experiments.

\begin{assumption}[equidistant frequencies] \label{assump:eq-dist-freqs} 
The frequencies $\bold{\Omega}$ satisfy $\Omega_k = k, \forall\, k=1,2,\dots,r$.
\end{assumption}

The above assumption can be weakened to $\Omega_k = k\Omega, \forall\, k=1,2,\dots,r$ for some $\Omega > 0$. 
Assuming $\Omega = 1$ will make our analysis easier but does not lose the generality, since we can simply adjust the domain of the function. Suppose we want to compute the Fourier coefficient vector $\bold{z}$ of $f(x)$ in the form of \eqref{ab_rep} with frequencies $\Omega_k = k\Omega, \forall\, k = 1,2,\dots,r$. Define $f_\Omega(x):= f(\frac{x}{\Omega})$. Then, $f_\Omega(x)$ has frequencies $\Omega_k = k$ for $k = 1,2,\dots,r$. Hence, Assumption~\ref{assump:eq-dist-freqs} holds for $f_\Omega(x)$, and our method can be applied to fit $f_\Omega(x)$. 

\begin{theorem} \label{theorem:MSE_optimality}
Under Assumptions \ref{assump:noisy-est} and \ref{assump:eq-dist-freqs}, let $P$ be an integer such that $P \geq 2r+1$. Define $\bold{x}^*\in\mathbb{R}^P$ with $x_k^* = \frac{2\pi}{P}k, \forall\, k=0,1,\ldots,P-1$ and denote $ A_\bold{x} = A_{\bold{\Omega}, \bold{x}}$ by the definition in \eqref{eq:def-A-Omage-x}. 
Then, 
    $$ \bold{x}^*=\underset{ \substack{\bold{x} \in [0,2\pi]^P  \\ x_k \text{ distinct}} } {\mathrm{argmin}}  \mathbb{E}[\mathrm{SE}(\bold{\hat{z}_x^*})] = \frac{\sigma^2}{n} \mathrm{tr}( (A_{\bold{x}}^\dagger A_{\bold{x}}) ^{-1}),$$ \\
and 
$\mathbb{E}[\mathrm{SE}(\bold{\hat{z}_{\bold{x}^*}^*})] = \frac{2\sigma^2(2r+1)}{nP}$. 
For any $\varepsilon > 0$, if $nP =  \left\lceil\frac{2\sigma^2 (2r+1)}{\varepsilon}\right\rceil$,  then $\mathbb{E}[\mathrm{SE}(\bold{\hat{z}_{x^*}^*})] \le \varepsilon$. 

\end{theorem} 


According to Theorem \ref{theorem:MSE_optimality}, the expected SE can be controlled solely based on the total number of shots, disregarding the individual contribution of $n$ and $P$. This holds in an ideal scenario, i.e., when Assumptions \ref{assump:noisy-est} and \ref{assump:eq-dist-freqs} hold. It will provide a useful guidance for us on how to choose $n$ and $P$ to achieve a prescribed error tolerance in practice. The real situation may deviate from this ideal scenario, 
motivating the need for experimentation to support this claim. Firstly, 
if $P$ is made very large while $nP$ is fixed, forcing $n$ to become very small, 
then the central limit theorem will not apply, i.e., the error in the observed function value will deviate significantly from Gaussian distribution. Secondly, the constant variance assumption will simplify the analysis; however, practically it is expected that $\sigma^2$ in \eqref{eq:x-var} is correlated with $f_j(x)$. 
Thirdly, Theorem \ref{theorem:MSE_optimality} does not reveal how the error changes across $n$ and $P$ when performing function approximation over a more sparse frequency set, which will be a key subroutine of our proposed FLICD method. 
These will be demonstrated by numerical experiments.

\subsection{Empirical function variance and norm calculation}

To learn important frequencies, Theorem \ref{theorem:MSE_optimality} can be used as a guide. Since $\sigma^2$ is an unknown quantity, we use an empirical value to estimate it, in order to set $n$ and $P$ to achieve a desired error tolerance $\varepsilon$. 

Under Assumption \ref{assump:eq-dist-freqs}, $f_j$ is periodic with period $2\pi$, and thus one approach is to use the \emph{average variance} across the domain $[0,2\pi]$. Recall that the quantum computer gives estimates of $f(\boldsymbol{\theta})$ via a random variable $\Lambda$ drawn from a discrete distribution $p_{\boldsymbol{\theta}}$. 
To estimate $f_j(\theta_j)$, let $p_{\theta_j}$ be the discrete distribution of $p_{\boldsymbol{\theta}}$ by fixing $\theta_k$ for all $k \neq j$. Then 
\begin{align}
    \text{E}_{\Lambda \sim p_{\theta_j}} [\Lambda] &= f_j(\theta_j) \\ 
    \text{Var}_{\Lambda \sim p_{\theta_j}} [\Lambda] &= \langle \psi ({\theta_j}) | M^2 |\psi({\theta_j}) \rangle - [f_j({\theta_j})]^2 =:\sigma^2({\theta_j}) \label{eq:variance_calc}
\end{align}
where $\psi(\theta_j) = U(\theta_{<j},\theta_j,\theta_{> j})|\psi_0\rangle $. For notational convenience, we define 
\begin{align*}
    g_j(\theta_j) := \langle \psi ({\theta_j}) | M^2 |\psi({\theta_j}) \rangle.
\end{align*} 

To generate an estimate $\tilde{\sigma}^2({\theta_j})$ for $\sigma(\theta_j)$, 
we obtain estimates $\tilde{f}^{(n_0)}_j(\theta_j)$ and $\tilde{g}^{(n_0)}_j(\theta_j)$ on the quantum computer by using $n_0$ shots, and combine with expression \eqref{eq:variance_calc}. This $n_0$ is different from the number of shots $n$ used when performing least-squares approximation with problem \eqref{def:least_squares_problem}. At any given $\theta_j$, we let 
\begin{align} \label{def:point-wise_variance_est}
    \tilde{\sigma}^2({\theta_j}) := \tilde{g}^{(n_0)}_j(\theta_j) - \tilde{f}^{(n_0)}_j(\theta_j)^2.
\end{align}
Then taking $K$ sample points $\{ x_k\}_{k=1}^K$ in $[0,2\pi)$, we use the \textit{empirical average variance}
\begin{align} \label{def:avg sample variance}
     \bar{\sigma}_{K}^2 := \frac{1}{K} \sum_{k=1}^{K} \tilde{\sigma}^2({x_k})
\end{align}
to replace $\sigma^2$ for setting $n$ and $P$ in our algorithm. 


Notice that the minimizer of $f_j(\theta_j)$ does not change if the function is scaled positively by a constant factor. This indicates that controlling the relative error in estimating the function values aligns well with our goal. We estimate the $L_2$-norm of $f_j$, namely, $||f_j||_2:=\sqrt{  \int_{0}^{2\pi} | f_j(\theta_j)| ^2 d{\theta_j} }$, by the empirical value: 
%
%
%
\begin{align} \label{def:L2_norm_n_K}
    ||f_j ||_{2,K} := \sqrt{ \frac{2\pi}{K} \sum_{k=1}^K     |\tilde{f}^{(n_0)}_j(x_k)|^2 }.   
\end{align}
Now given a target (relative) error tolerance $\varepsilon$, we can set $n$ and $P$ such that
$$nP = \left\lceil \frac{2\bar{\sigma}_{K}^2 (2r+1)}{\varepsilon ||f_j ||_{2,K}}\right\rceil.$$

\textbf{Jittered grids.} The computations of $\bar{\sigma}_{K}^2$ and $ ||f_j ||_{2,K}$ in \eqref{def:avg sample variance} and \eqref{def:L2_norm_n_K} both require choosing sample points $\{ x_k\}_{k=1}^K$. 
We use jittered grids \cite{cookStochasticSamplingComputer1986}, which are a common sample set that greatly removes chances of aliasing when under-sampling from data with possibly higher-order frequencies. More precisely, we take $x_k = \frac{2 \pi}{K}(k+\frac{1}{2}) +\beta_k$, 
where $\beta_k \sim \mathrm{Unif}([-\frac{\pi}{K},\frac{\pi}{K}])$ independently for $k=0,1,2,\dots,K-1$. 

\subsection{Adaptive frequency update}

One key technique of our proposed FLICD method is the adaptive frequency update, 
whose pseudocode is given in Algorithm \ref{alg:one}. 
Our idea is to gradually add higher-order frequencies one-by-one and monitor how the fitting changes. 
Below we describe our procedure in a formal way. 

In the rest of this subsection, we again abuse notation for convenience and denote $f(x)$ as a generic uni-variate function, i.e., the restriction on a generic $j$-th variable $\theta_j$ in \eqref{def:split_notation}. Suppose that the theoretical frequency set is $\bold{\Omega} = \{ \Omega_1,\Omega_2,\dots,\Omega_r\}$ in an increasing order. 
For each $1\leq t\leq r$, let $\bold{\Omega}^{(t)} := \{ \Omega_1,\Omega_2,\dots,\Omega_t \}$. 
Following the discussions in Section~\ref{sec:least-square}, we form a least-squares problem by using each $\bold{\Omega}^{(t)}$ and matrix $A_{\bold{x},\bold{\Omega}} \in \mathbb{R}^{{P(t)}\times (2t+1)}$ to obtain  
a set of Fourier coefficients $\bold{\hat{w}}^{(t)} \in \mathbb{R}^{2t+1}$. 
Using expression \eqref{ab_rep} with coefficients $\bold{\hat{w}}^{(t)}$, this defines a function approximation $\hat{f}^{(t)}$. 

Given $\varepsilon_{rel} > 0 $, ideally, we aim at finding the minimum $t$ by the adaptive update such that the fitting error is in the order of $\varepsilon_{abs}:= ||f||_2\cdot \varepsilon_{rel}$. However, practically we do not have access to $f$ and thus will not be able to compute its $L_2$-norm or measure the difference between an approximation and the ground truth $f$. Hence, we track the change of two consecutive approximations $\hat{f}^{(t-1)}$ and $\hat{f}^{(t)}$ defined as
\begin{align} \label{def:MSE}
    \mathrm{MSE}(\hat{f}^{(t)},\hat{f}^{(t-1)}) := \frac{1}{2\pi} \int_{0}^{2\pi } \left| \hat{f}^{(t)}(x) - \hat{f}^{(t-1)}(x)\right|^2 dx.
\end{align}
We accept the first $t\ge2$ such that 
$$\mathrm{MSE}(\hat{f}^{(t)},\hat{f}^{(t-1)}) \leq \gamma \cdot ||f||_{2,K}  \cdot \varepsilon_{rel},$$
where $||f||_{2,K}$ is a numerical approximation of $||f||_2$ as in \eqref{def:L2_norm_n_K} and $\gamma \geq 1$ is a hyper-parameter that increases the threshold to take into account random fluctuations of the error about the expected value.

\SetKwComment{Comment}{/* }{ */}
\RestyleAlgo{ruled}
\begin{algorithm} 
\setstretch{1.2} 
\caption{ $\hat{f} = $ \textbf{least\_squares\_fit}$(f,n,\bold{x},\boldsymbol{\Omega}) $} \label{alg:zero}
\KwData{Uni-variate restricted function $f$, number of shots $n$, vector of $P$ sample points $\bold{x}= \{ x_k\}_{k=1}^P$, and  frequency list $\boldsymbol{\Omega} = \{ \Omega_1,\dots,\Omega_t \}$.}
\KwResult{ $\hat{f}$ as an approximation of $f$}
    Compute interpolation matrix $A_{\bold{x},\bold{\Omega}} \in \mathbb{R}^{{P}\times (2t+1)}$ using \eqref{eq:def-A-Omage-x}.\\
    Obtain noisy function estimates of $f$ using $n$ shots per sample point and form $\bold{y_{obs}} = [\tilde{f}^{(n)}(x_1),\tilde{f}^{(n)}(x_2),\dots,\tilde{f}^{(n)}(x_P)]^\dagger \in \mathbb{R}^P$.\\ 
    Solve least-squares system for $\bold{\hat{z}_x} \in \mathbb{R}^{2t+1}$ by
    $$\bold{\hat{z}_x^*} := \underset{\bold{\hat{z}_x} \in \mathbb{R}^{2t+1}}{\mathrm{argmin}} || A_{\bold{x},\bold{\Omega}} \bold{\hat{z}_x} - \bold{y_{obs}}||^2.$$
      \\  
    Recover approximation $\hat{f}$ using expression \eqref{ab_rep} with coefficients $\bold{\hat{z}_x^*}$ and frequencies $\boldsymbol{\Omega}$. 
\end{algorithm}

\SetKwComment{Comment}{/* }{ */}
\RestyleAlgo{ruled}
\begin{algorithm} 
\setstretch{1.2} 
\caption{$(\hat{f}^{(t-1)},\boldsymbol{\Omega}^{(t-1)},P^{(t-1)},\tilde{n},t-1) = \textbf{adaptive\_update}(f,t_0,\varepsilon_{rel},\gamma,n,K,\boldsymbol{\Omega}) $}\label{alg:one}
\KwData{Uni-variate restricted function $f(x)$, starting frequency index $t_0$, relative error bound $\varepsilon_{rel}$, threshold error factor $\gamma \geq 1$, sample size factor $\alpha \geq 1 $, number of shots $n_0$ and grid size $K$ for variance and norm estimation, 
and frequency list $\boldsymbol{\Omega} = \{ \Omega_1,\dots,\Omega_r \}$.}
\KwResult{$\hat{f}^{(t-1)}$ as an approximation of $f$, learned frequencies $\boldsymbol{\Omega}^{(t-1)}$, number of sample points $P^{(t-1)}$, shots per function evaluation $\tilde{n}$, index $t-1$ for highest learned frequency.}
Compute 
$\bar{\sigma}_{K}^2$ and $||f||_{2,K}$ 
using \eqref{def:avg sample variance} and \eqref{def:L2_norm_n_K} with jittered grid points\\
Set $\varepsilon_{abs} = ||f||_{2,K} \cdot \varepsilon_{rel} $ and 
$\tilde{n} = \lceil \frac{2 \bar{\sigma}_{K}^2 }{\alpha \varepsilon_{abs} }\rceil $ \\
\For{$t = t_0,t_0+1,\dots,r$}{
    $\boldsymbol{\Omega}^{(t)} = \{ \Omega_1.\dots,\Omega_t\}$; \hspace{8mm}
    $P^{(t)} = \lceil\alpha (2t+1)\rceil$; \hspace{8mm}
    $\bold{x}^{(t)} = \left( \frac{2 \pi k} {P^{(t)}  }\right)_{k=0}^{P^{(t)}-1}$; \\
    Call Algorithm~\ref{alg:zero} to have $\hat{f}^{(t)} \leftarrow $ \textbf{least\_squares\_fit}$(f,\tilde{n},\bold{x}^{(t)},\boldsymbol{\Omega}^{(t)}) $ \\
    \If{$t > t_0$}{
        \If{$\mathrm{MSE}(\hat{f}^{(t)},\hat{f}^{(t-1)}) < \gamma \cdot \varepsilon_{abs}$ }{
        \textbf{return} $(\hat{f}^{(t-1)},\boldsymbol{\Omega}^{(t-1)},P^{(t-1)},\tilde{n},t-1)$
    }
    }

}
\end{algorithm}


\subsection{Frequency-learning interpolation-based coordinate descent}

We can now formally introduce the FLICD, given in Algorithm \ref{alg:two}. To explain the high-level structure of the algorithm, we provide some methodology here.

FLICD takes advantage of the ansatz structure to learn frequency sets efficiently. Since $U(\boldsymbol{\theta})$ has a $p$-layer structure with $L$ parameters per layer, the hermitian generators obey the equalities in \eqref{eq:H-cycle}. 
Hence, the frequency sets also obey
\begin{align*}
    \bold{\Omega}^\ell = \bold{\Omega}^{\ell+L} =  \bold{\Omega}^{\ell+2L} =\cdots = \bold{\Omega}^{\ell+(p-1)L} \text{ for } \ell = 1,2,\dots,L.
\end{align*}
We say that two indices $j,k \in \{ 1,2,\dots,m\}$ have the same \textit{layer position} if $(j - k) \mod L = 0$. The above equalities indicate that all indices with the same layer position give rise to the same theoretical frequency set. If all the uni-variate restrictions have sparse Fourier coefficients, then there exists a small frequency set that would work for all indices in the same layer position. This idea inspires our strategy of frequency learning. Instead of learning $m$ different independent smaller frequency sets, we learn only $L$ frequency sets and use them 
for parameter updates with the same layer position. 

FLICD adopts two alternating ways of updating parameters. Each $\theta_j$ in a given layer $k \in \{1,2,\dots,p \}$ is updated by either of the following schemes: 
\begin{itemize}
    \item (\textit{Adaptive frequency update}) Use Algorithm \ref{alg:one} to learn or update frequency set for the update;
    \item (\textit{Fixed frequency update}) Use the frequency set that has been previously learned within the same layer position for the update.
\end{itemize}
Both schemes use the least-squares problem subroutine given by Algorithm \ref{alg:zero}. To specify how often the adaptive versus fixed frequency update occurs, we introduce a positive integer $S$ such that an adaptive frequency update is performed every $S$ layers, with fixed frequency updates applied to the following $S-1$ layers. For example, when $p=4$ and $S=3$, an adaptive frequency update is first performed on layer $1$, followed by fixed frequency updates on layers $2$ and $3$. The next adaptive frequency update is then performed on layer $4$, followed by fixed frequency updates on layers $1$ and $2$. The layers are updated cyclically, returning to layer $1$ after layer $4$.


\SetKwFor{While}{while}{}{end}
\SetKwFor{For}{for}{}{end}
\SetKwFor{ForEach}{foreach}{}{end foreach}
\SetKwComment{Comment}{/* }{ */}
\RestyleAlgo{ruled}
    
\begin{algorithm}
\setstretch{1.2}
\caption{Frequency learning interpolation-based coordinate descent (FLICD)}\label{alg:two}
\KwData{Initial Parameters $\boldsymbol{\theta}^0=[\theta_1^0,\dots,\theta_m^0]$, adaptive update every $S$ layer updates, total number of layer updates $T\ge S$, relative error bound $\varepsilon_{rel}$, threshold error factor $\gamma\geq 1$,  number of shots $n_0$ and grid size $K$ for variance and norm estimation $n_0$, 
and frequency lists $\boldsymbol{\Omega}^{\ell}$ for $\ell = 1,2,\dots,L$}
\KwResult{Optimized parameters $\boldsymbol{\theta}^{LT}$ }

Set $i = 0$ and
$t^{(\ell)} = 1$ for $\ell = 1,2,\dots,L$ \tcp*[f]{start from lowest frequency}\\
\For(\tcp*[f]{count layer updates}){$k=0$ \KwTo $T-1$} {
    \For(\tcp*[f]{current layer position}){$\ell=1$ \KwTo $L$} {
        $j$ $=$  $(i \mod m) + 1$ \tcp*[f]{cyclically update parameters}\\
        Define  $f_j(\theta_j) := f(\boldsymbol{\theta}^i_{<j},\theta_j,\boldsymbol{\theta}^i_{>j})$ \\
        \uIf(\tcp*[f]{adaptive frequency update}){$k \mod S = 0$ 
        }{$(\hat{f}_j,\boldsymbol{\Omega}^{\ell}_{\mathrm{adp}},P^{(\ell)},\tilde{n}^{(\ell)},t^{(\ell)})  \leftarrow\textbf{ adaptive\_update}(f_j,t^{(\ell)},\varepsilon_{rel},\gamma,n_0,K,\boldsymbol{\Omega}^{\ell} )$ \\
            
        }  
        \Else(\tcp*[f]{fixed frequency update}){
            Let $\bold{x} = \left( \frac{2 \pi k} {P^{(\ell)}  }\right)_{k=0}^{P^{(\ell)}-1}$ and 
            $\hat{f}_j \leftarrow $ \textbf{least\_squares\_fit}$(f_j,\tilde{n}^{\ell},\bold{x},\boldsymbol{\Omega}^{\ell}_{\mathrm{adp}}) $ \\
        }
        Find $\theta_j^* = \mathrm{argmin}_{x \in  \mathbb{R}} \hat{f}_j(x)$ \\
        Set $\theta_j^{i+1} = \theta_j^*$ and 
        $\theta_u^{i+1} = \theta_u^i$ for all $u \neq j$ \\
        Increase $i \leftarrow i + 1$
    }

}
\end{algorithm}


\section{Numerical experiments} \label{sec:Numerical Experiments}

We demonstrate the performance of FLICD and its subroutines on quantum simulator and hardware by solving Max-cut problem instances using a QAOA-type ansatz. We also benchmark FLICD against existing optimizers. 

The circuits are constructed using Qiskit \cite{qiskit2024}. For simulator experiments, the built-in AerSimulator is used to perform noiseless simulation for systems up to 20 qubits. For hardware experiments, two different IBM superconducting qubit QPUs are utilized: r2 Heron processors (\textit{ibm\_marrakesh} and \textit{ibm\_fez}) and r1 Nighthawk processor (\textit{ibm\_rensselaer}). For more information on circuit construction and transpilation, see Appendix \ref{appendix:circuit_exection}. We solve the argmin subproblem (line 11 of Algorithm \ref{alg:two}) via grid search on a uniform grid of size 2000 over $[0,2\pi]$, since the uni-variate restricted functions are $2\pi$-periodic in the experiments. 

\subsection{Problem setup} \label{subsec:problem_setup}

We refer to Pauli-x and Pauli-z operators by $X$ and $Z$, respectively. If $P$ is a Pauli operator, we denote $P_i$ as $P$ acting on the $i^{th}$ qubit. 

\subsubsection{Max-cut problem} 
Given an undirected graph $G = (V,E)$ with vertex set $V = [q]$ and weight function $w:E  \rightarrow \mathbb{R}$, the goal of Max-cut is to determine a cut $(S,V \setminus S)$ of the vertices where $S \subset V$ such that sum of edge weights across the cut is maximized. This problem can be written as a quadratic unconstrained binary optimization problem (QUBO) in the form
\begin{align*}
    \max_{\bold{x} \in \{ 0,1\}^q } \sum_{(i,j) \in E} w_{ij} (x_i + x_j - 2x_i x_j).
\end{align*}
It can be canonically transformed to a problem that a quantum computer can solve, and the cost Hamiltonian is
\begin{align*}
     H_C = \sum_{(i,j) \in E} \frac{1}{2} w_{ij} Z_iZ_j.
\end{align*}

For our experiments, we restrict to random regular degree-$d$ weighted graphs. This means that each vertex is connected to exactly $d$ other vertices chosen uniformly at random. Given a set of non-negative integers $\{ b_k \}_{k=0}^{K-1}$ and a probability distribution $\{ p_k\}_{k=0}^{K-1}$ such that $p_k \geq 0$ for all $k \in [K]$ and $\sum_{k=0}^{K-1} p_k = 1$, the weight $w_{ij}$ on an edge $(i,j) \in E$ is chosen i.i.d. at random according to $\textrm{Prob}\{ w_{ij} = b_k\} = p_k$ for $k \in [K]$. 

\subsubsection{Linear chain QAOA ansatz}
The Linear Chain QAOA (LC-QAOA) ansatz \cite{wangDepthIndependentLinearChain2025}, is created by restricting entanglement between vertices along a path through the vertices in the instance graph. This ansatz has layer depth independent of problem size, which makes it amenable to run on quantum hardware. 
 
To construct the ansatz, first find a path $(k_0,k_1,\dots,k_{q-1})$ through the vertices of $G$, which can be found via depth first search (DFS). To make the circuit construction simpler, relabel all the vertices in the graph via the map $k_j \mapsto j$ for $j \in [q]$. This makes the path through the vertices simply $(0,1,2,\dots,q-1)$, and induces a new edge set $E' := \{(u,v) \text{ for } (k_u,k_v) \in E  \}$ and weight function $w':E' \rightarrow \mathbb{R} $ where $w'_{uv} :=w_{k_uk_v} $. The relabeled graph is referred to as $G'=(V,E')$. The $p$-layer LC-QAOA ansatz, using $G'$, is defined as 
\begin{align*}
    U_{\text{LC}}(\boldsymbol{\theta}) := \prod_{k=1}^{p} \exp\left(-i \frac{\theta_{2k-1}}{2} H^{\mathrm{LC}}_{zz}\right)\cdot \exp \left({-i \frac{\theta_{2k}}{2}  H_x} \right)
\end{align*}
where $H^{\mathrm{LC}}_{zz} := \sum_{j=0}^{q-2} w'_{j,j+1} Z_{j} Z_{j+1}$ and $H_{x} = \sum_{j = 0}^{q-1} X_j$. It is necessary to use the relabeled graph to generate the cost function. The new cost Hamiltonian is given by
\begin{align*}
    H_C' :=  \sum_{(j,k) \in E'} \frac{1}{2} w_{jk} Z_j Z_k.
\end{align*}

\subsubsection{Example construction of graph and ansatz}

\begin{figure}[p]
    \centering

    \begin{subfigure}[t]{0.4\textwidth}
        \centering
        \includegraphics[width=\linewidth]{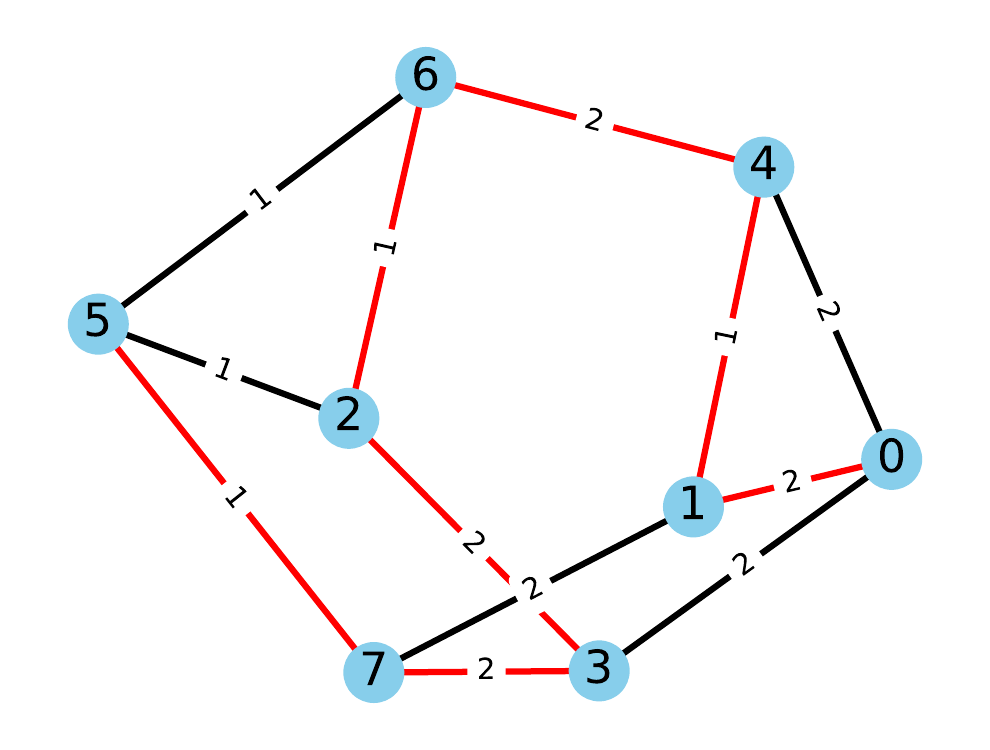} 
        \caption{Graph $G$, with highlighted path through the vertices}
        \label{fig:graph_G}
    \end{subfigure}
    \hfill
    \begin{subfigure}[t]{0.43\textwidth}
        \centering
        \includegraphics[width=\linewidth]{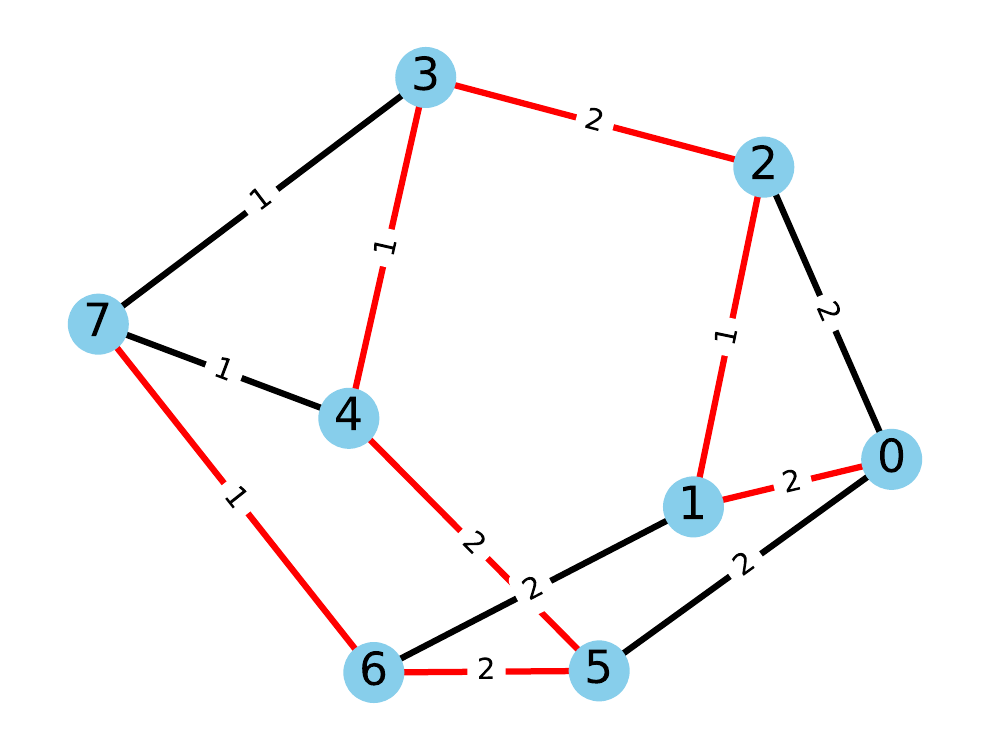}
        \caption{Graph $G'$ after relabeling, with highlighted path through the vertices.}
        \label{fig:graph_G'}
    \end{subfigure}

    \vspace{0.3em}
       \centering
    \begin{subfigure}[t]{0.5\textwidth}
        \centering
        \includegraphics[width=\linewidth]{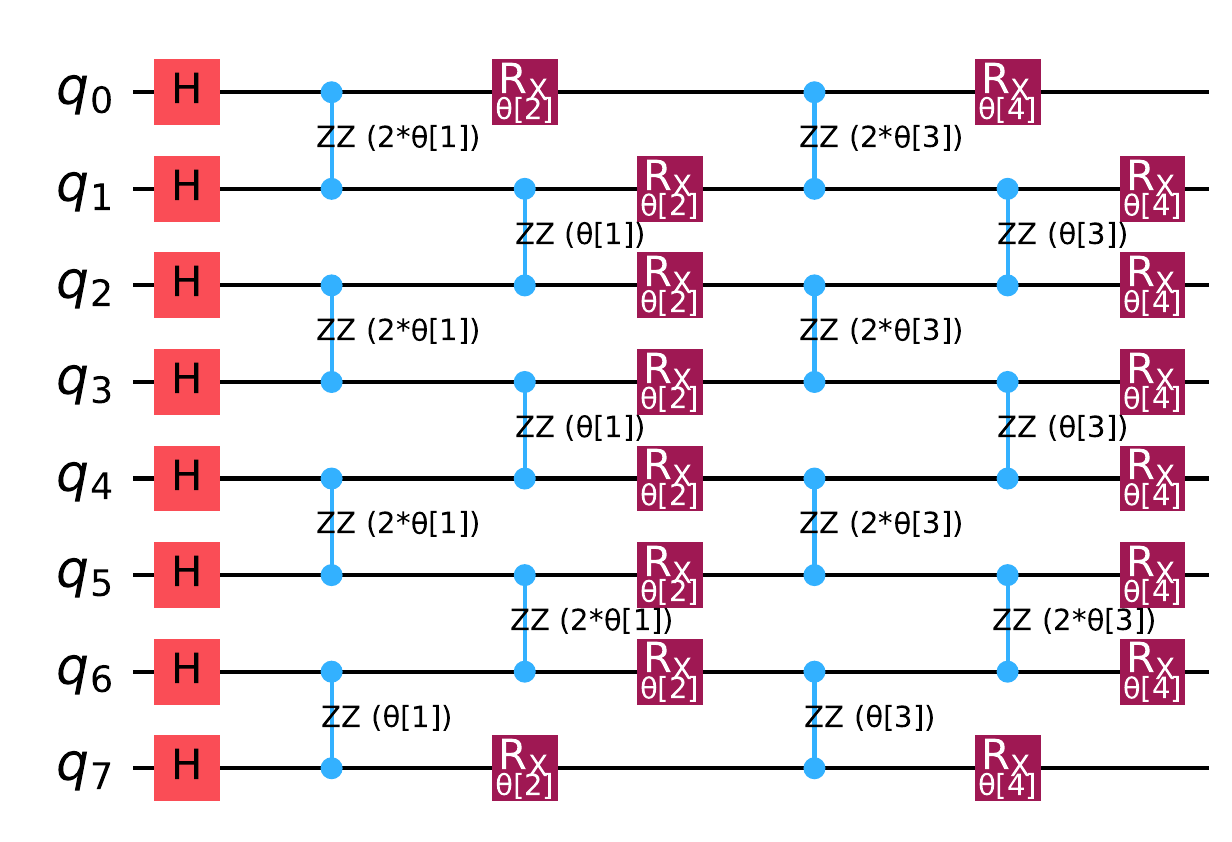}
        \caption{Picture of 2-layer LC-QAOA ansatz for graph $G'$ }
        \label{fig:ansatz for example instance graph}
    \end{subfigure}
    \

    \label{fig:example instance graph}
    \caption{Construction of LC-QAOA ansatz from a random regular degree-3 graph with 8 vertices. Weights are chosen at random from $\{1,2 \}$ with equal probability.The pre-factors in front of $\theta_1$ and $\theta_3$ in the RZZ layers come from the corresponding weights in the graph.}

    \vspace{0.1em}

    \centering
    \includegraphics[width=.50\linewidth]{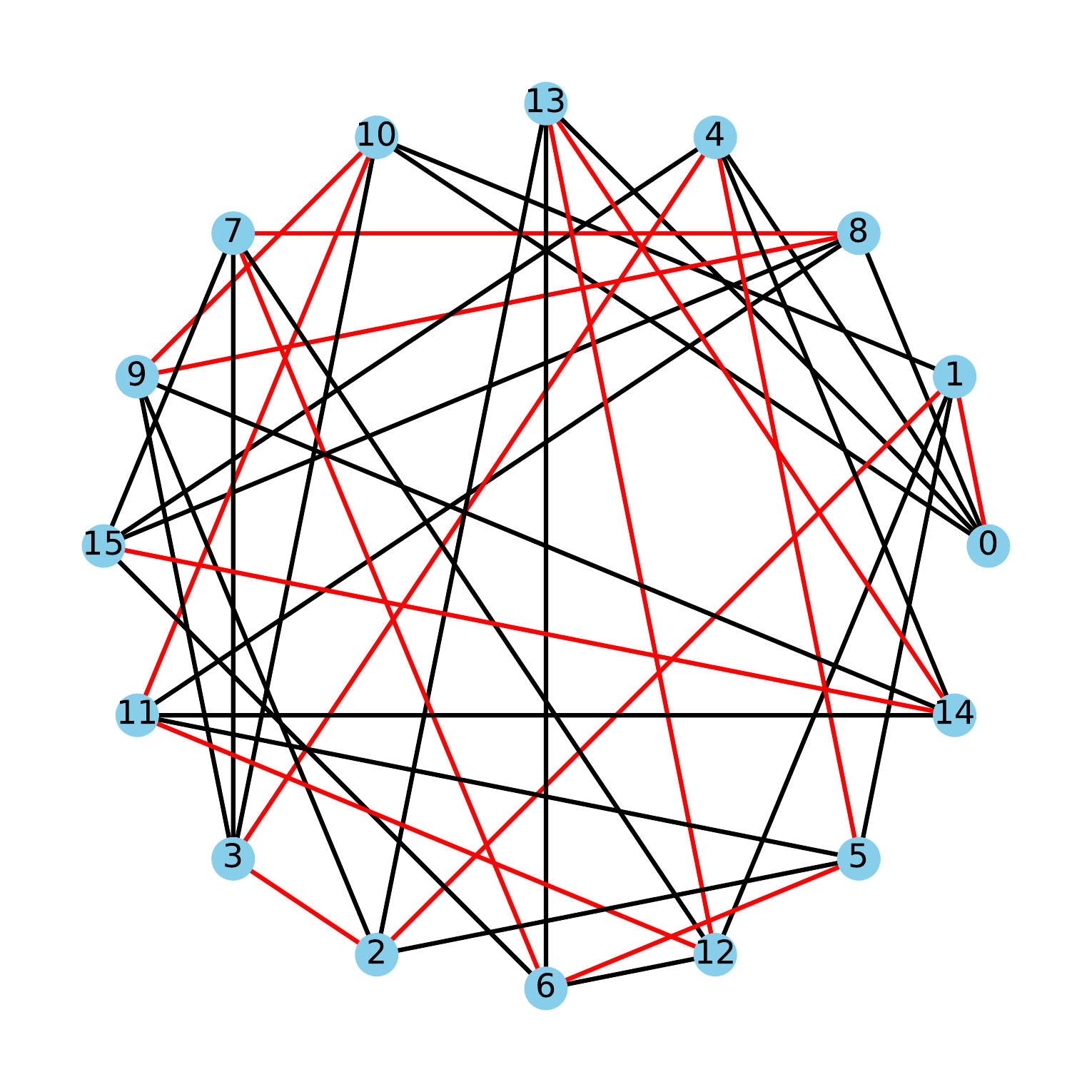}
    \caption{Random regular degree-5 graph after relabeling of vertices. Linear chain highlighted with vertices in increasing order. Two different weight distribution are used, which are specified in the the Appendix. }
    \label{fig:16 vertex graph}
    
\end{figure}

To better understand the problem setup, we give a simple example. Figure \ref{fig:graph_G} shows the problem graph $G$ with a highlighted DFS path through the vertices. Using this path, the graph is relabeled to form the graph $G'$, shown in Figure  \ref{fig:graph_G'}. Figure \ref{fig:ansatz for example instance graph} shows the ansatz for the graph $G'$ with $p=2$ layers. 

\subsection{Empirical variance and norm experiments} \label{sec:Empirical variance and norm experiments}

In this subsection, we demonstrate computation of the empirical average variance and $L_2$-norm using simulator backend, which is used to set the shot count during the adaptive update and abosulte error bound.

We use a 16-vertex random regular degree-5 graph with weights chosen in $\{1,2 \}$ with equal probability. This graph is shown in Figure \ref{fig:16 vertex graph}, with weight matrix given by \eqref{matrix:weights_1,2} in Appendix \ref{subappendix:weights}. The LC-QAOA ansatz is set with $p = 3$ layers, so the variable $\boldsymbol{\theta} \in \mathbb{R}^6$. To obtain a restriction $f_1(\theta_1)$ to the first variable, we pick $\theta_k  \sim \mathrm{Unif}[0,2 \pi)$ i.i.d. for $k \in \{2,3,4,5,6 \}$. This $f_1(\theta_1)$ is exactly the same function as in Figure \ref{subfig:random}.  For each $K \in \{4,8 \}$, we fix a jittered grid of size $K$ and set a shot count $n_0 \in \{50,500,5000 \}$.

\begin{figure}[p]
    \centering
    \begin{minipage}{5.5in}
        \centering
        \begin{subfigure}[b]{0.31\textwidth}
            \centering
            \begin{overpic}[width=\linewidth]{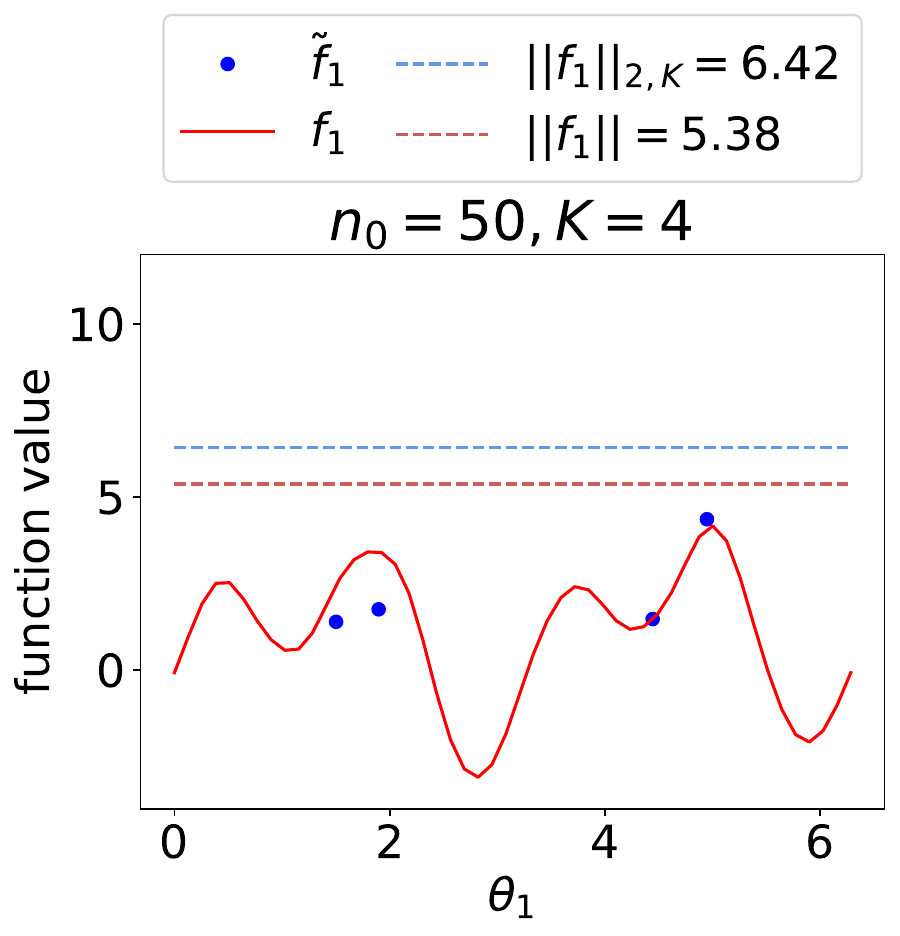}
                \put(6, 0){(a)}
            \end{overpic}
            \label{subfig:f_K4_n50}
        \end{subfigure}\hfill
        \begin{subfigure}[b]{0.31\textwidth}
            \centering
            \begin{overpic}[width=\linewidth]{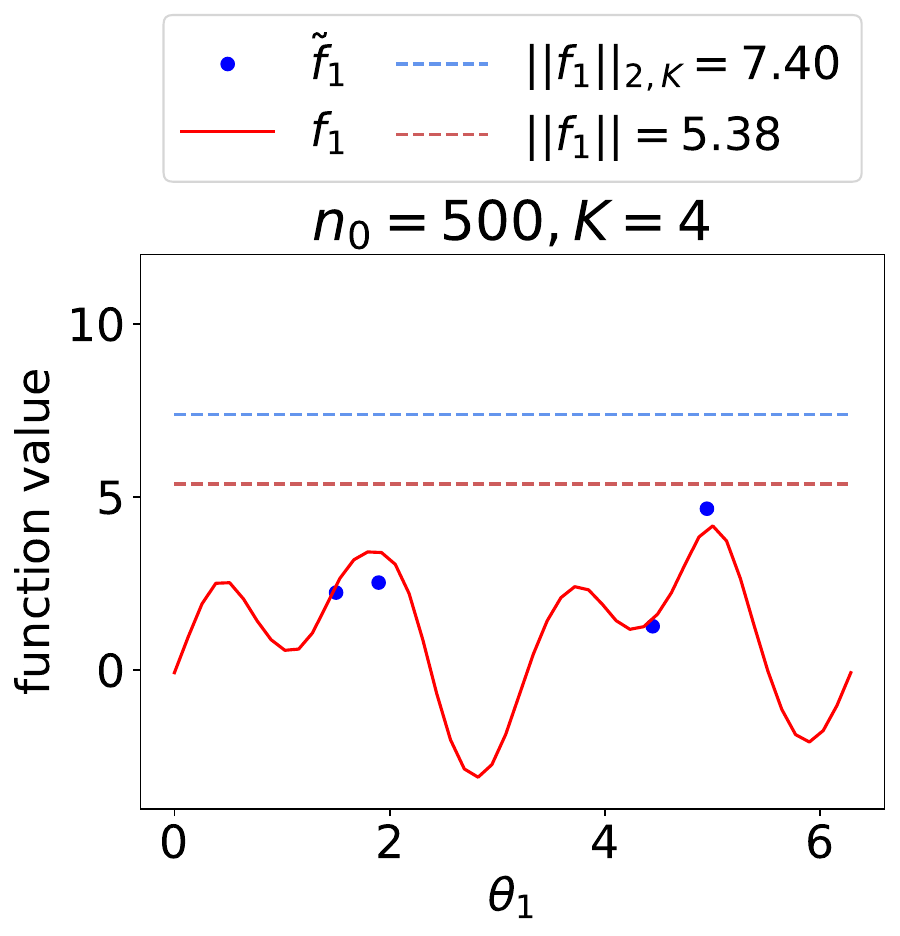}
                \put(6, 0){(b)}
            \end{overpic}
            \label{subfig:f_K4_n500}
        \end{subfigure}\hfill
        \begin{subfigure}[b]{0.31\textwidth}
            \centering
            \begin{overpic}[width=\linewidth]{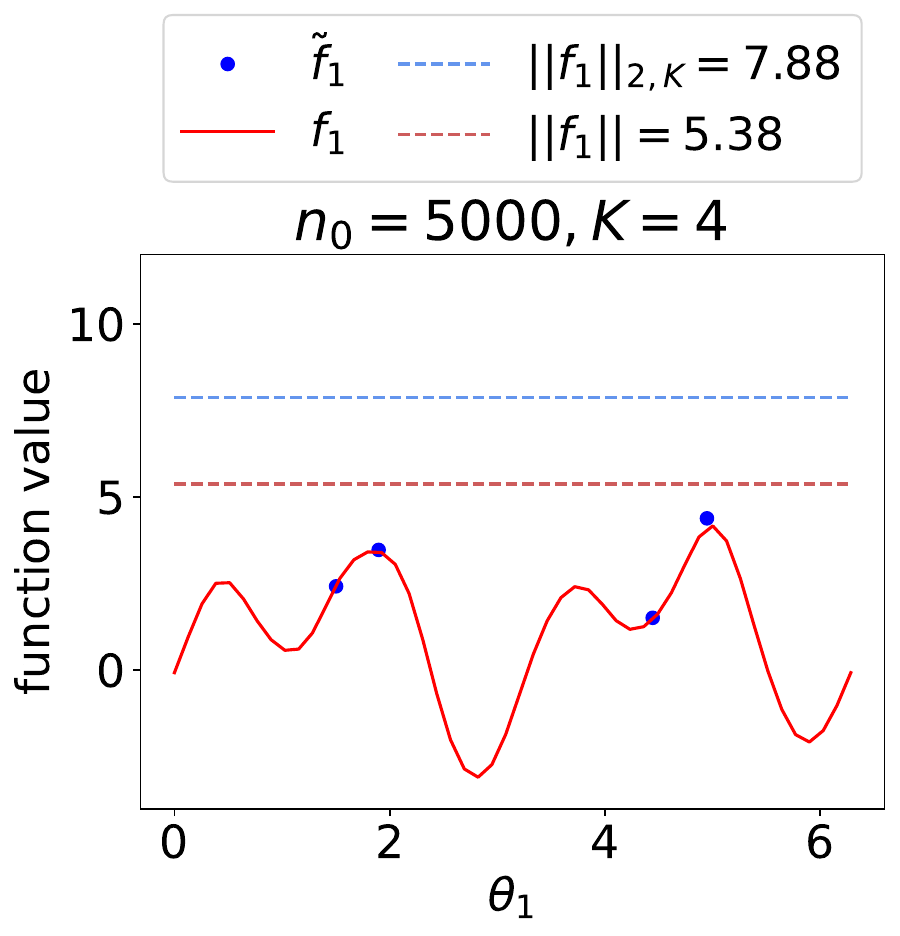}
                \put(6, 0){(c)}
            \end{overpic}
            \label{subfig:f_K4_n5000}
        \end{subfigure}
        
        \vspace{0.5em}

        \begin{subfigure}[b]{0.31\textwidth}
            \centering
            \begin{overpic}[width=\linewidth]{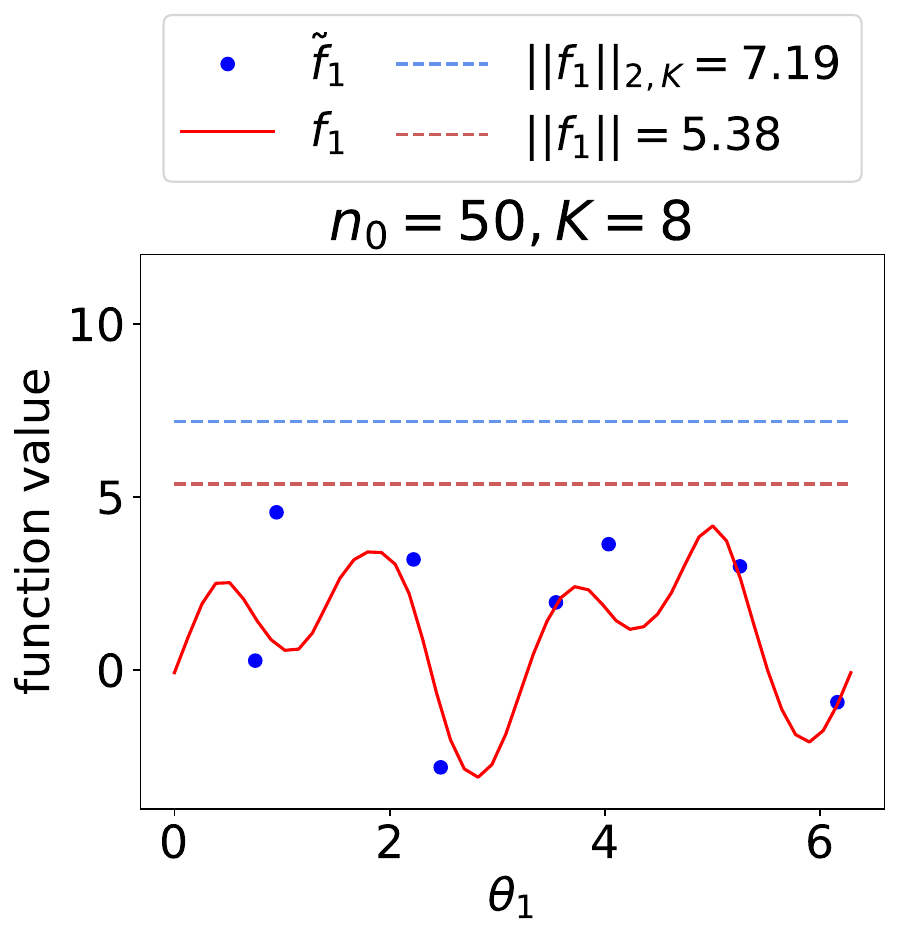}
                \put(6, 0){(d)}
            \end{overpic}
            \label{subfig:f_K8_n50}
        \end{subfigure}\hfill
        \begin{subfigure}[b]{0.31\textwidth}
            \centering
            \begin{overpic}[width=\linewidth]{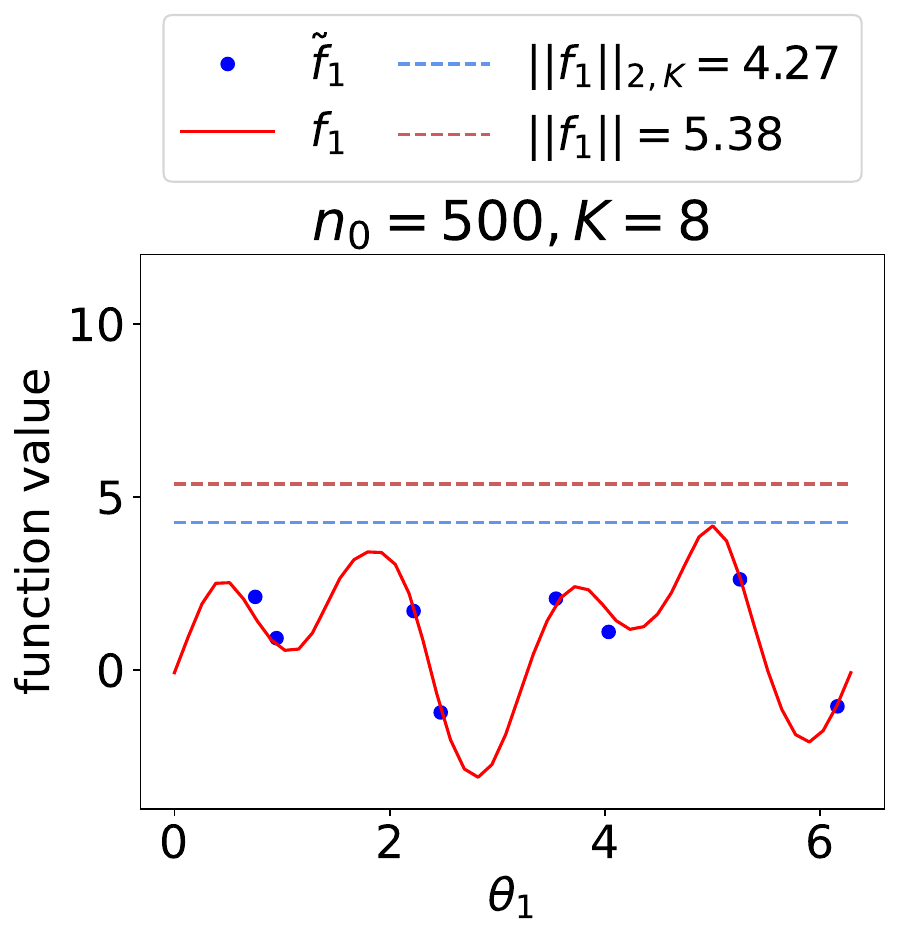}
                \put(6, 0){(e)}
            \end{overpic}
            \label{subfig:f_K8_n500}
        \end{subfigure}\hfill
        \begin{subfigure}[b]{0.31\textwidth}
            \centering
            \begin{overpic}[width=\linewidth]{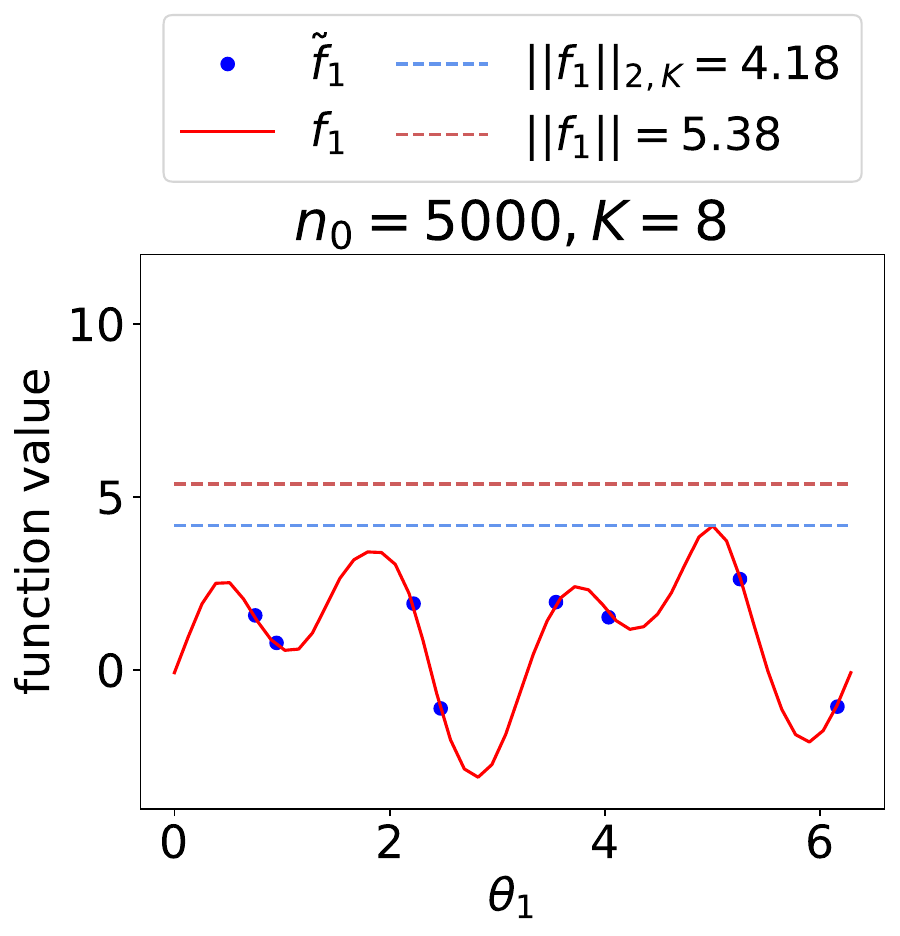}
                \put(6, 0){(f)}
            \end{overpic}
            \label{subfig:f_K8_n5000}
        \end{subfigure}

    \end{minipage}
    \caption{$f_1(\theta_1)$ and estimates of $\|f_1\|_2$ on $[0,2\pi]$ for different $K$ and $n_0$. $f_1(\theta_1)$ is generated from a 3-layer LC-QAOA ansatz of a 16-vertex random regular degree-5 graph with weights chosen uniformly at random in $\{1,2\}$. Parameters $\theta_k \sim \textrm{Unif}([0,2\pi])$ $\forall k \geq 2$.}
    \label{fig:f_L2_norm}
\end{figure}

\begin{figure}[p]
    \centering
    \begin{minipage}{5.5in}
        \centering
        \begin{subfigure}[b]{0.31\textwidth}
            \centering
            \begin{overpic}[width=\linewidth]{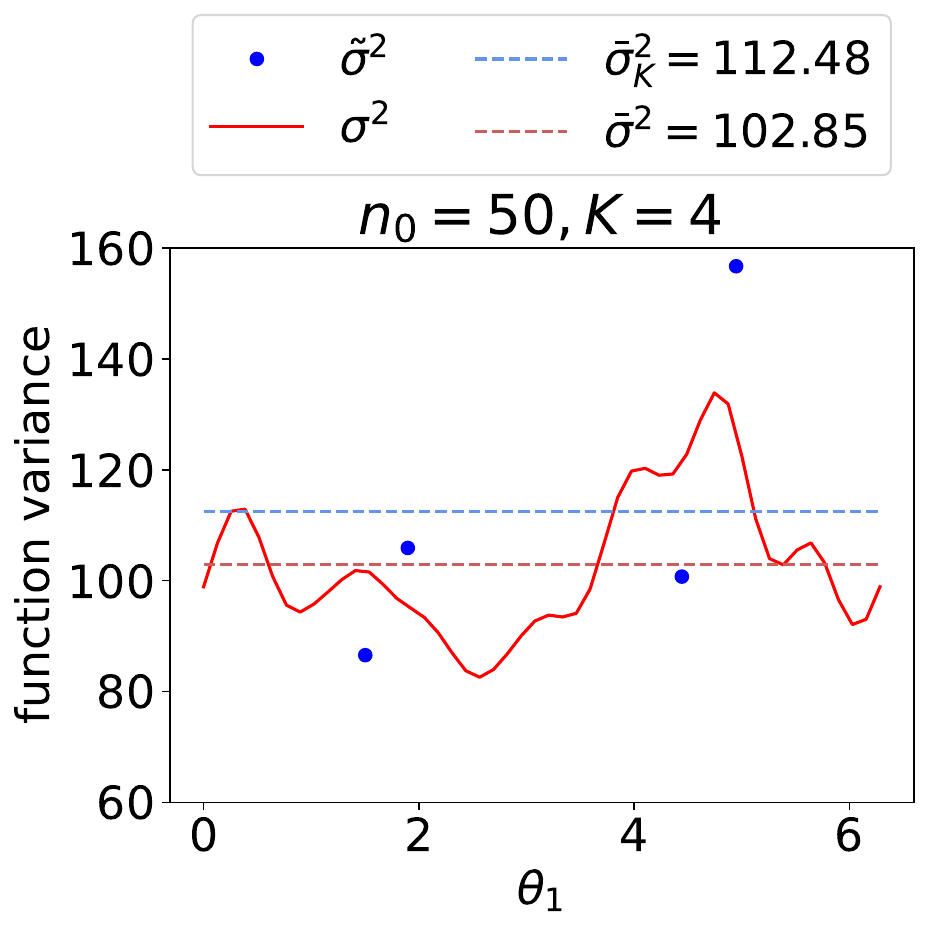}
                \put(6, 0){(a)}
            \end{overpic}
            \label{subfig:sigma_K4_n50}
        \end{subfigure}\hfill
        \begin{subfigure}[b]{0.31\textwidth}
            \centering
            \begin{overpic}[width=\linewidth]{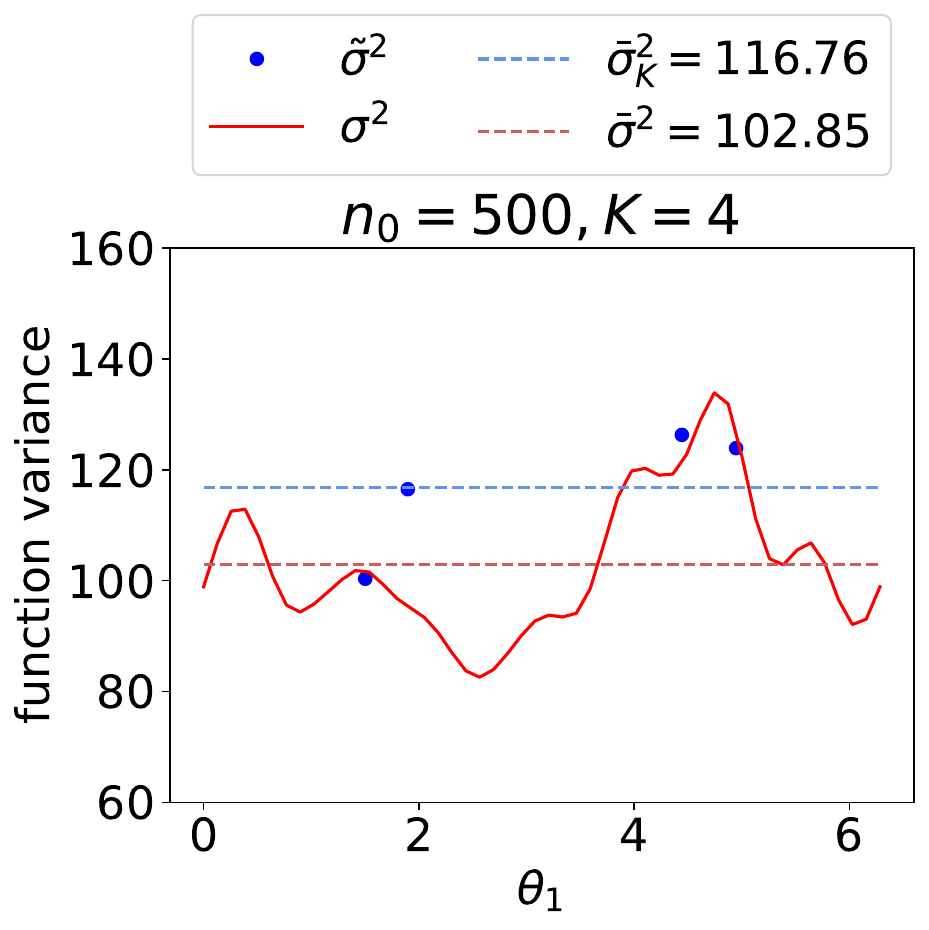}
                \put(6, 0){(b)}
            \end{overpic}
            \label{subfig:sigma_K4_n500}
        \end{subfigure}\hfill
        \begin{subfigure}[b]{0.31\textwidth}
            \centering
            \begin{overpic}[width=\linewidth]{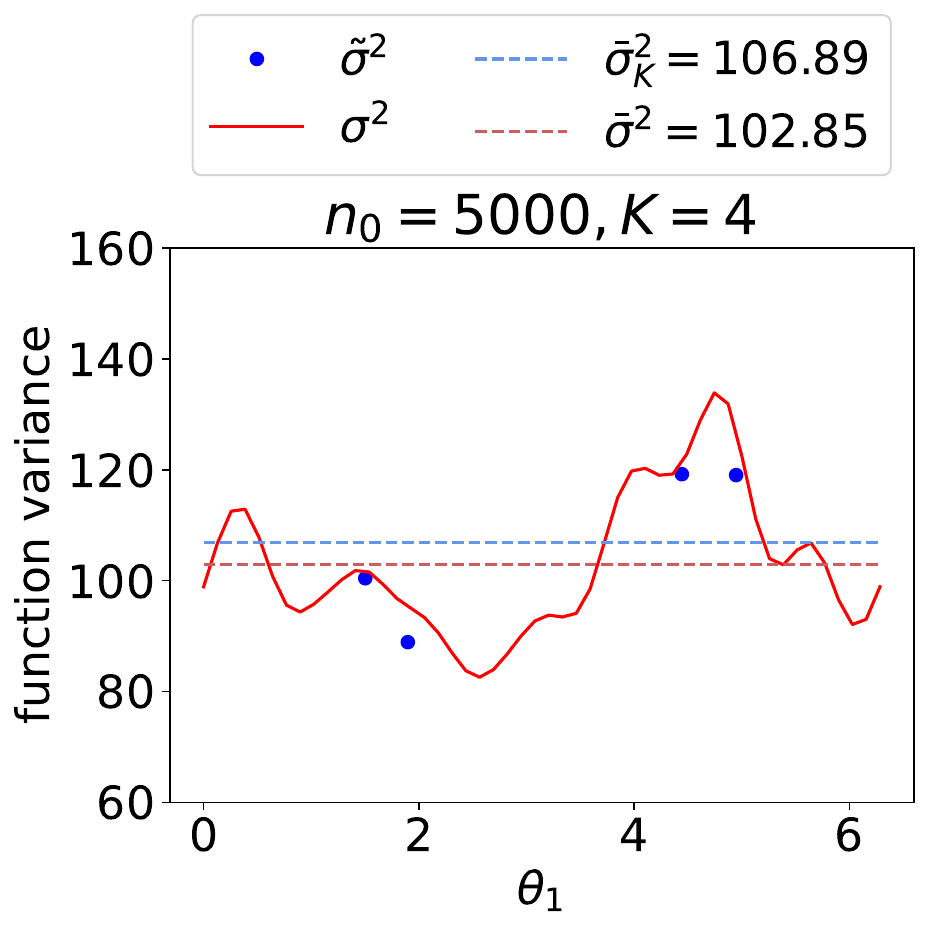}
                \put(6, 0){(c)}
            \end{overpic}
            \label{subfig:sigma_K4_n5000}
        \end{subfigure}
        
        \vspace{0.5em}

        \begin{subfigure}[b]{0.31\textwidth}
            \centering
            \begin{overpic}[width=\linewidth]{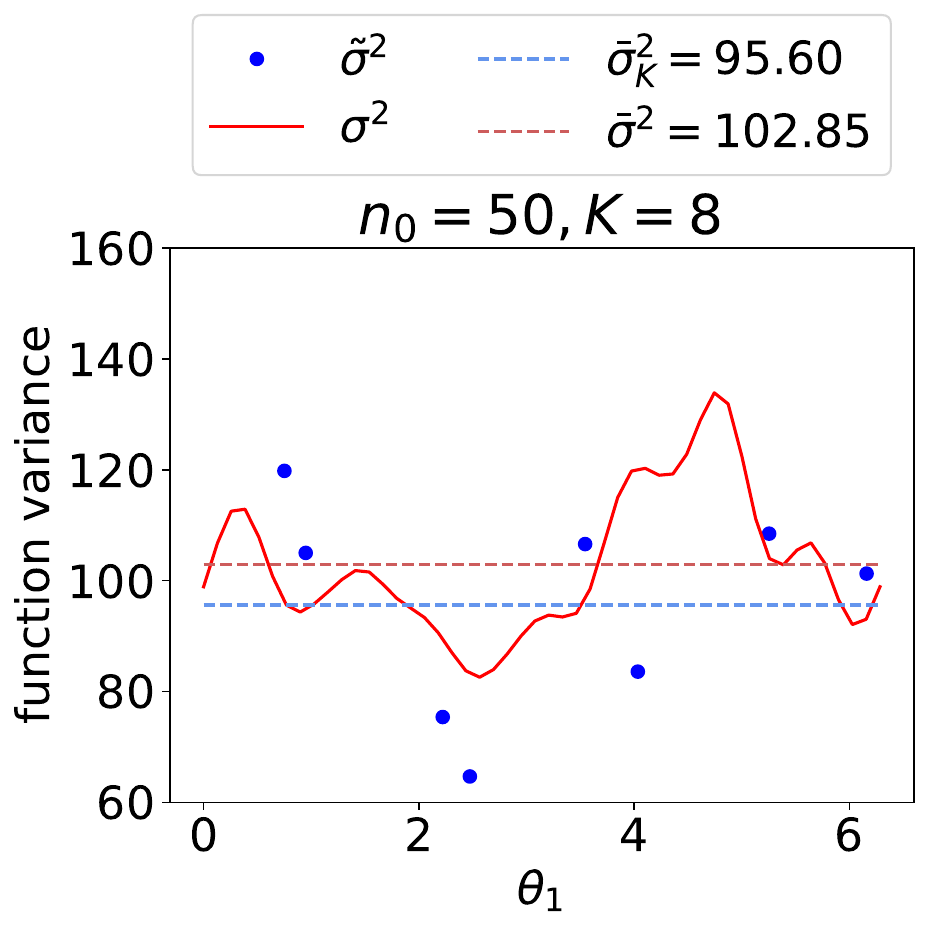}
                \put(6, 0){(d)}
            \end{overpic}
            \label{subfig:sigma_K8_n50}
        \end{subfigure}\hfill
        \begin{subfigure}[b]{0.31\textwidth}
            \centering
            \begin{overpic}[width=\linewidth]{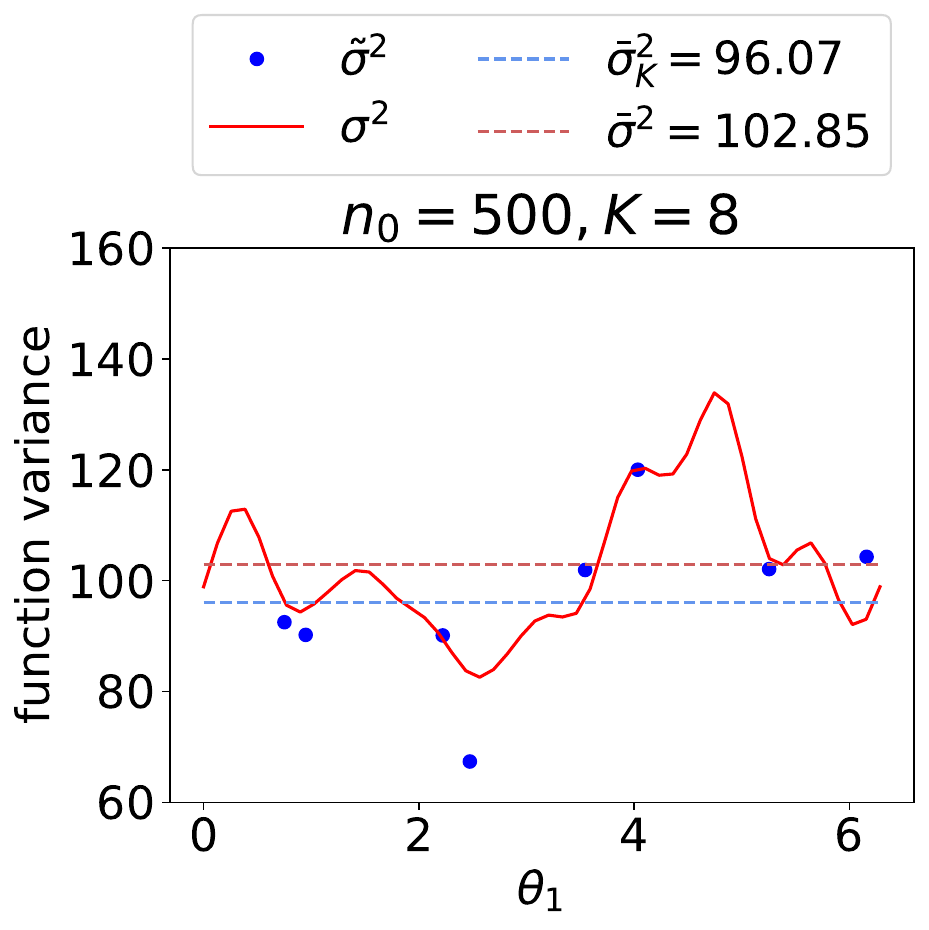}
                \put(6, 0){(e)}
            \end{overpic}
            \label{subfig:sigma_K8_n500}
        \end{subfigure}\hfill
        \begin{subfigure}[b]{0.31\textwidth}
            \centering
            \begin{overpic}[width=\linewidth]{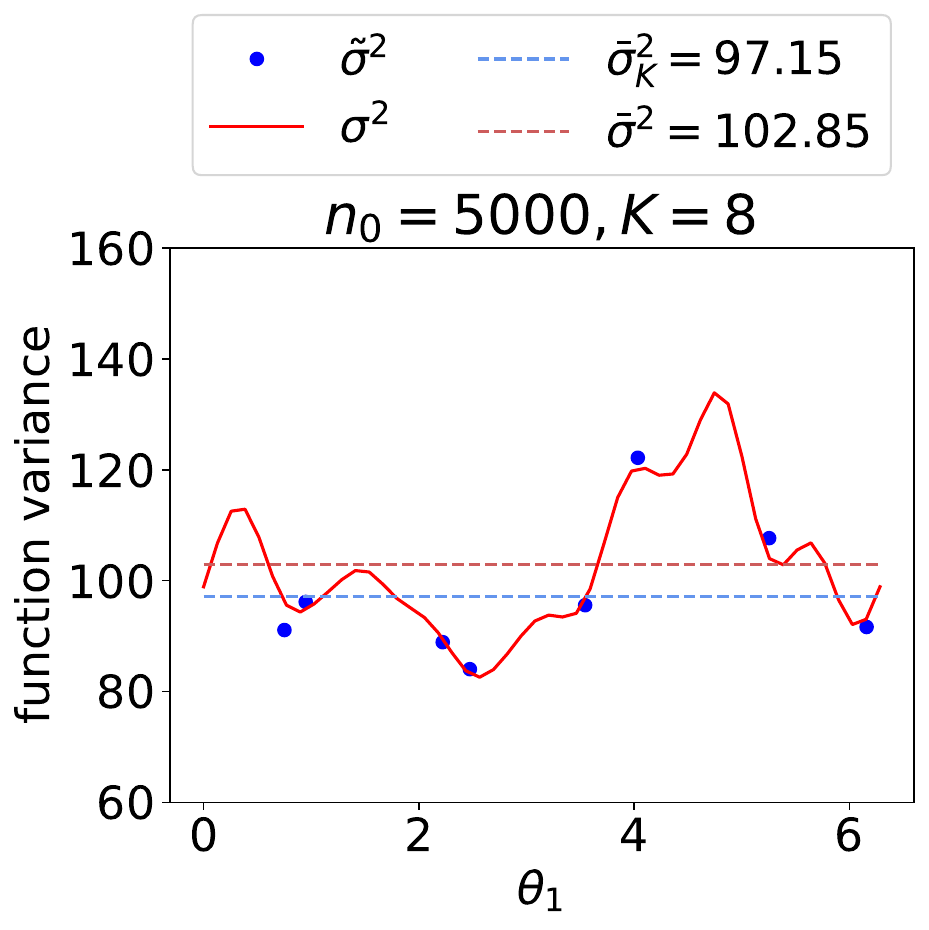}
                \put(6, 0){(f)}
            \end{overpic}
            \label{subfig:sigma_K8_n5000}
        \end{subfigure}

    \end{minipage}
    \caption{$\sigma^2(\theta_1)$ and estimates of $\bar{\sigma}^2$ on $[0,2\pi]$ for different $K$ and $n_0$. $\sigma^2(\theta_1)$ is the variance at $\theta_1$ of $f_1(\theta_1)$ as in Figure \ref{fig:f_L2_norm}.}
    \label{fig:variance}
\end{figure}

Figure \ref{fig:f_L2_norm} shows $f_1(\theta_1)$ with the true and empirical $L_2$-norm, for the various $K$ and $n_0$. The estimates $\tilde{f}^{(n_0)}_1(x_k)$ at the jittered-grid points are denoted by the blue dots. Similarly, Figure~\ref{fig:variance} shows the point-wise $\sigma^2(\theta_1)$ with the true and empirical average variance, for various $K$ and $n_0$. The estimates $\tilde{\sigma}^2(x_k)$ at the jittered-grid points are denoted by the blue dots.


From the figures, we observe that even with a grid of size $K = 4$, a decent estimate for the average variance and $L_2$-norm can be computed. When $K = 4$, increasing $n_0$ generally makes the estimate for both $\sigma^2_{K}$ and $||f_1||_{2,K}$ more accurate. However, when $K = 8$, even with a small $n_0$ leads to a very accurate value of the estimate. In fact, the breakdown of $n_0$ and $K$ can be quite important. For example, using $n_0 = 50$ and $K =8$ leads to a more accurate estimate than $n_0 = 500$ and $K = 4$, even though the latter uses more overall shots. As a note, the constant variance assumption does not hold in practice: the variance can fluctuate quite large. However, the variance is concentrated around a similar order of magnitude. 

\subsection{Adaptive frequency update experiments} \label{sec:Adaptive update experiments}

In this subsection, we demonstrate Algorithm \ref{alg:one} on both simulator and hardware backends in the case of equidistant and non-equidistant frequencies. 

For the purposes of testing, two slight changes are made to Algorithm 1. The first change is that the true values of $\bar{\sigma}^2$ and $||f_j||_2$ are used to set the shot count $\tilde{n}$. This is to isolate error due to function fitting and not to uncertainty in the shot count estimate. Secondly, we use a highest frequency stopping condition for termination of the algorithm, to analyze the behavior of Algorithm \ref{alg:one} well past where 
a relative error based termination would be. Algorithm \ref{alg:one} computes the metric $\textrm{MSE}(f_j^{(t)},f_j^{(t-1)})$, but for comparison, the MSE against the true function $\textrm{MSE}(f_j^{(t)},f_j)$ is computed using Statevector evaluations of $f_j$, at each step $t$. On simulator backend, all cases are repeated 3 times, and we report mean and standard deviation; on hardware backend, each experiment is only performed once.  

\subsubsection{Adaptive frequency update on equidistant frequencies}

We utilize the same random regular degree-5 graph in Figure \ref{fig:16 vertex graph}, but with weights chosen in $\{1,2,4 \}$ with uniform probability and a 3-layer LC-QAOA ansatz. The weight matrix is given by \eqref{weights_1,2,4} in Appendix \ref{subappendix:weights}. The parameters will be the same randomly chosen $\theta_k$ for $k > 1$ as in section \ref{sec:Empirical variance and norm experiments}. The goal here is to recover $f_1$ using Algorithm \ref{alg:one}. 
We choose $\varepsilon_{rel}$ in $\{ 0.01,0.05,0.1,0.2\}$ to test various desired accuracy levels. To analyze the effect of least squares, $\alpha$ is chosen in $\{1,2,4\}$, with $\alpha = 1$ corresponding to a square system.

\begin{figure}[p]
    \centering
    \begin{minipage}{6.5in}
        \centering
        \begin{subfigure}[b]{0.495\textwidth}
            \centering
            \begin{overpic}[width=\textwidth]{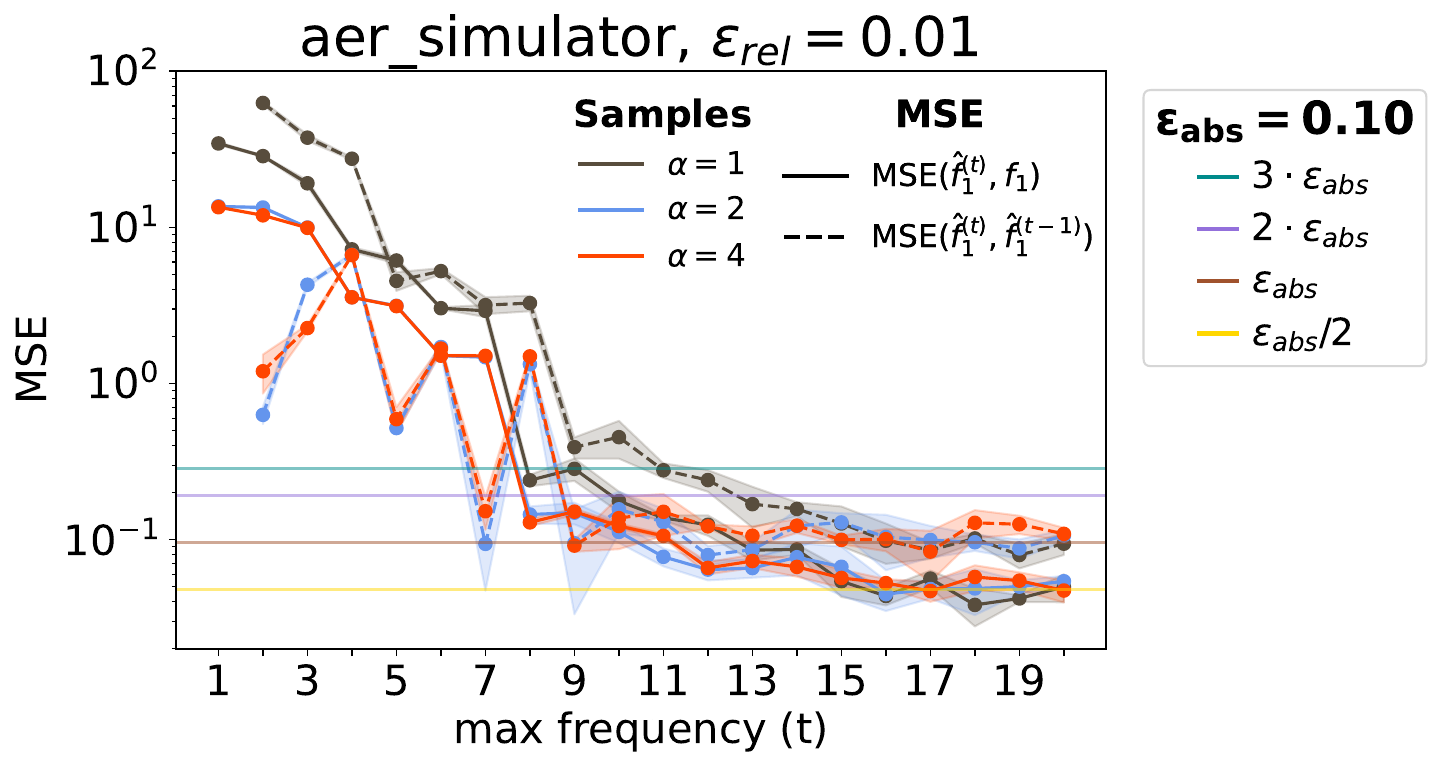}
                \put(6, 0){\text{(a)}}
            \end{overpic}
            \label{fig:sim-eps-001}
        \end{subfigure}
        \hfill
        \begin{subfigure}[b]{0.495\textwidth}
            \centering
            \begin{overpic}[width=\textwidth]{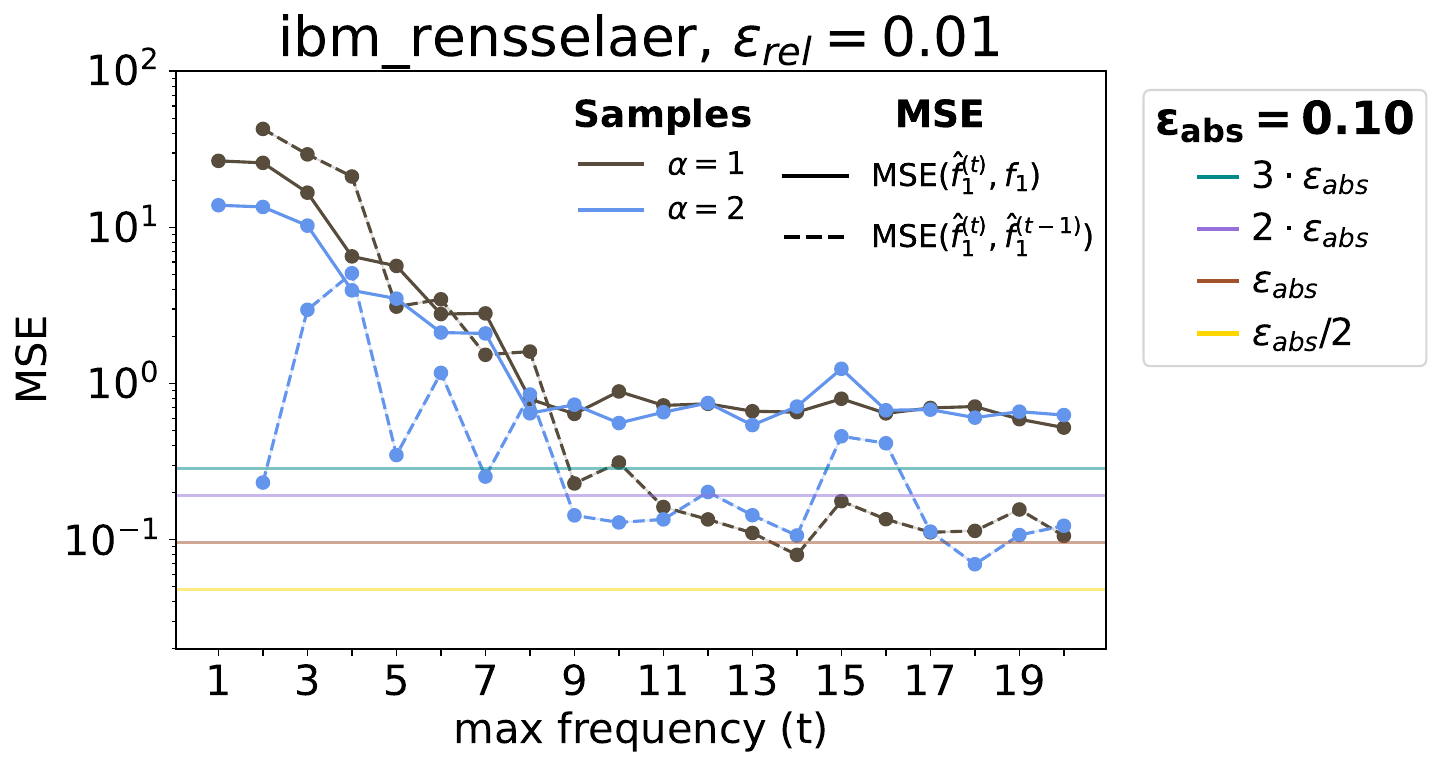}
                \put(6, 0){\text{(b)}}
            \end{overpic}
            \label{fig:hw-eps-001}
        \end{subfigure}

        \vspace{1mm}

        \begin{subfigure}[b]{0.495\textwidth}
            \centering
            \begin{overpic}[width=\textwidth]{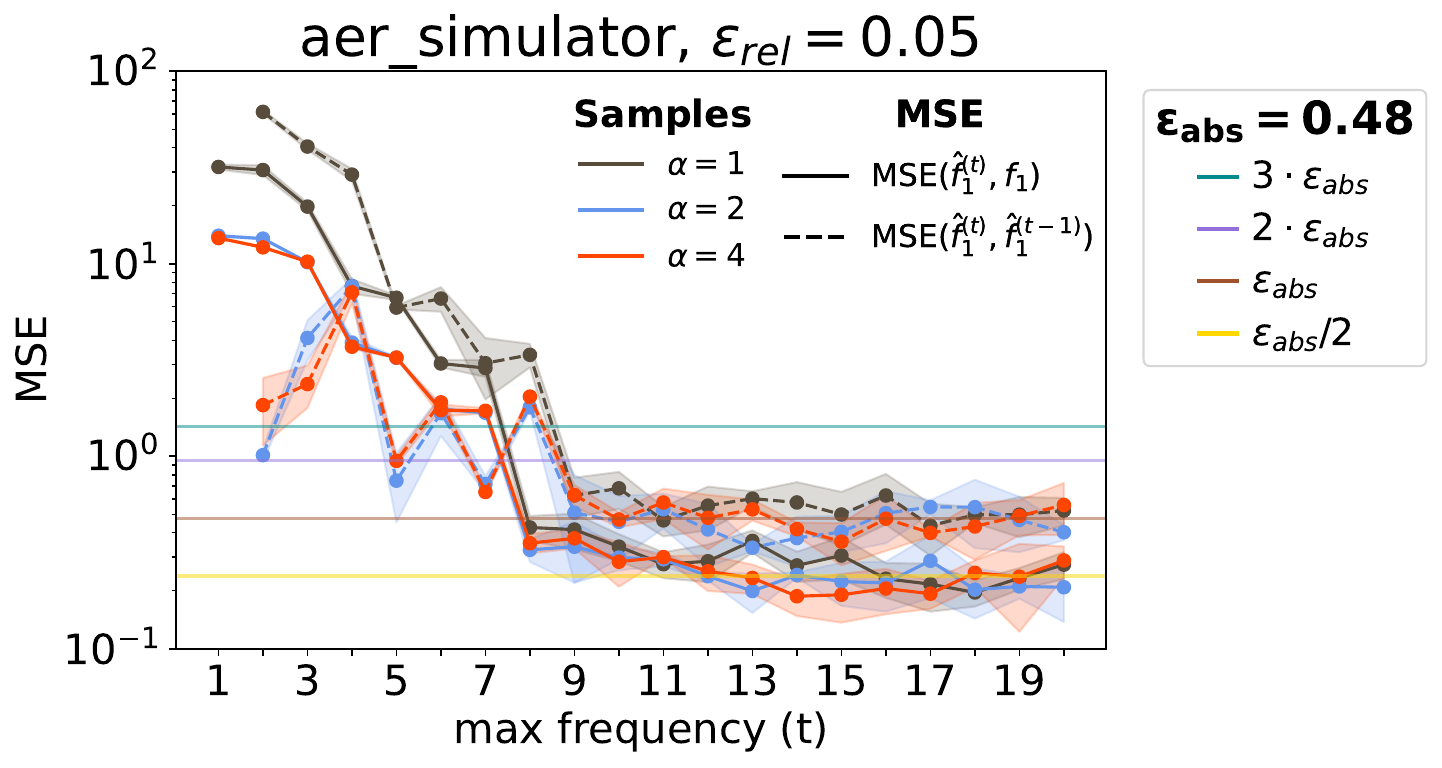}
                \put(6, 0){\text{(c)}}
            \end{overpic}
            \label{fig:sim-eps-005}
        \end{subfigure}
        \hfill
        \begin{subfigure}[b]{0.495\textwidth}
            \centering
            \begin{overpic}[width=\textwidth]{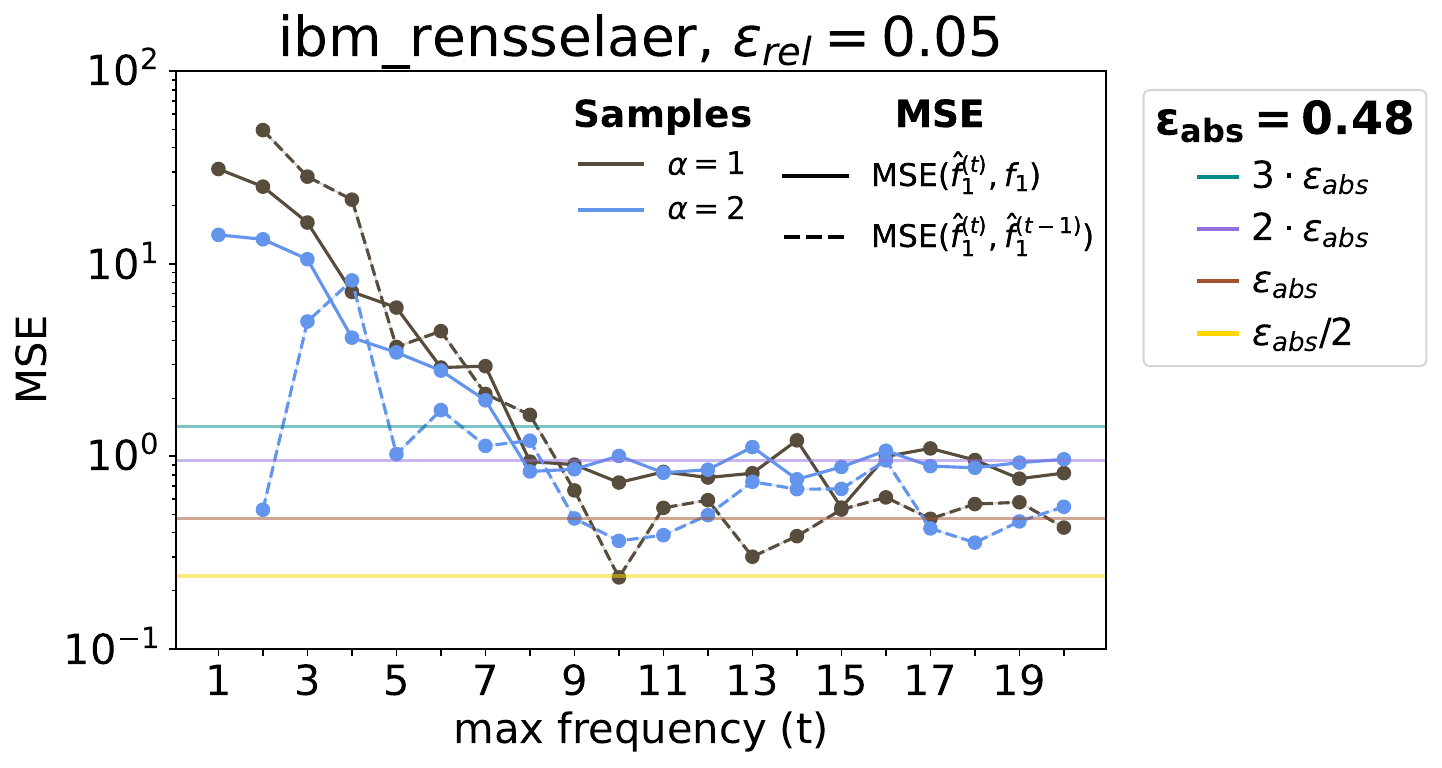}
                \put(6, 0){\text{(d)}}
            \end{overpic}
            \label{fig:hw-eps-005}
        \end{subfigure}

        \vspace{1mm}

        \begin{subfigure}[b]{0.495\textwidth}
            \centering
            \begin{overpic}[width=\textwidth]{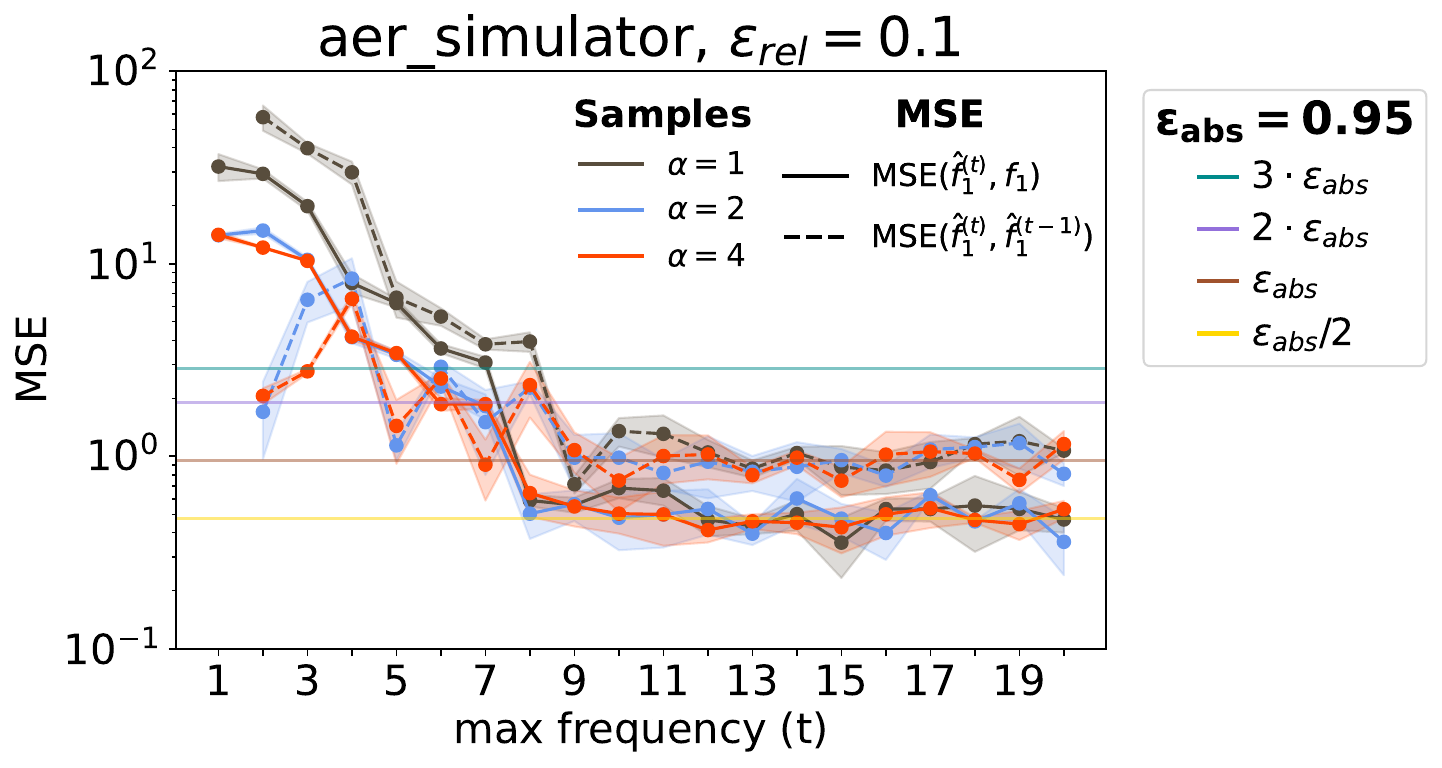}
                \put(6, 0){\text{(e)}}
            \end{overpic}
            \label{fig:sim-eps-01}
        \end{subfigure}
        \hfill
        \begin{subfigure}[b]{0.495\textwidth}
            \centering
            \begin{overpic}[width=\textwidth]{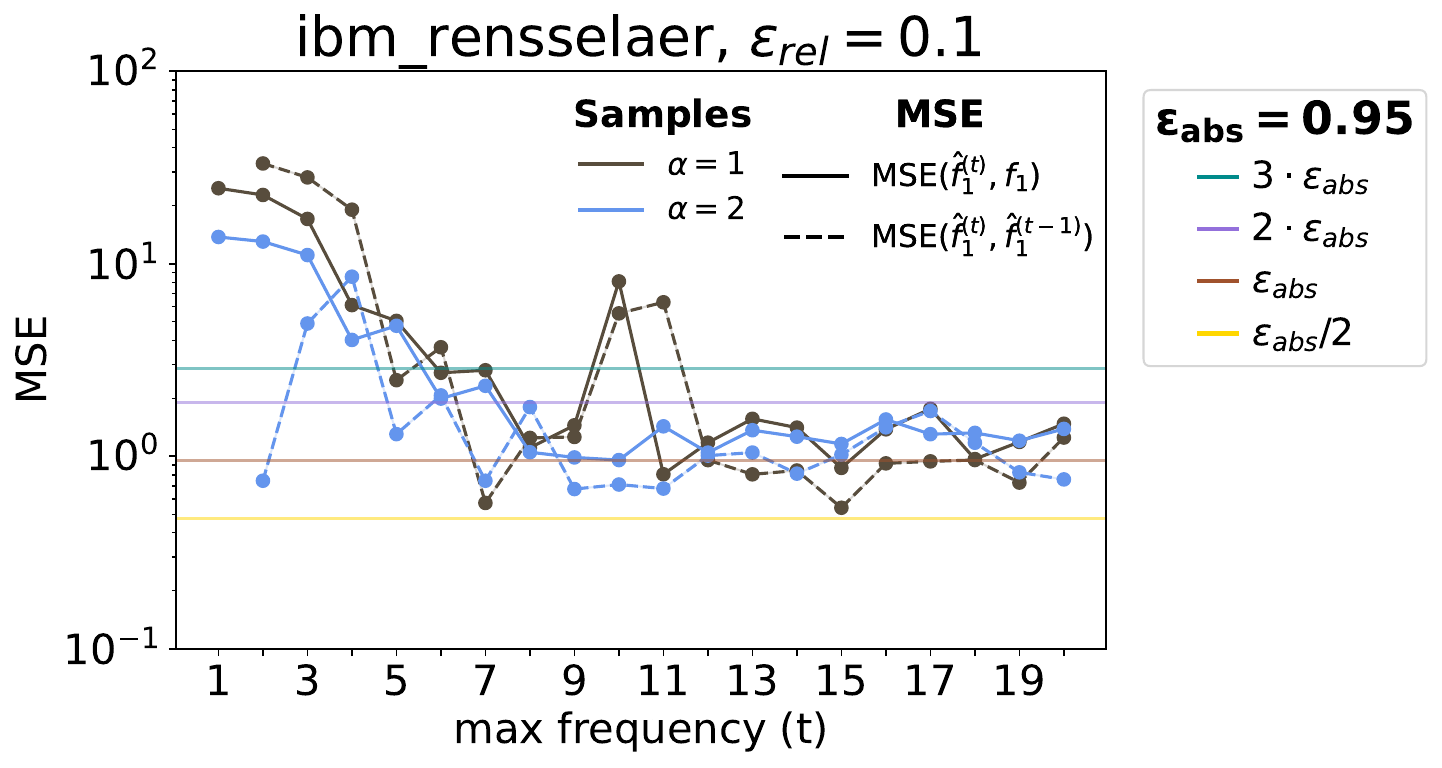}
                \put(6, 0){\text{(f)}}
            \end{overpic}
            \label{fig:hw-eps-01}
        \end{subfigure}

        \vspace{1mm}

        \begin{subfigure}[b]{0.495\textwidth}
            \centering
            \begin{overpic}[width=\textwidth]{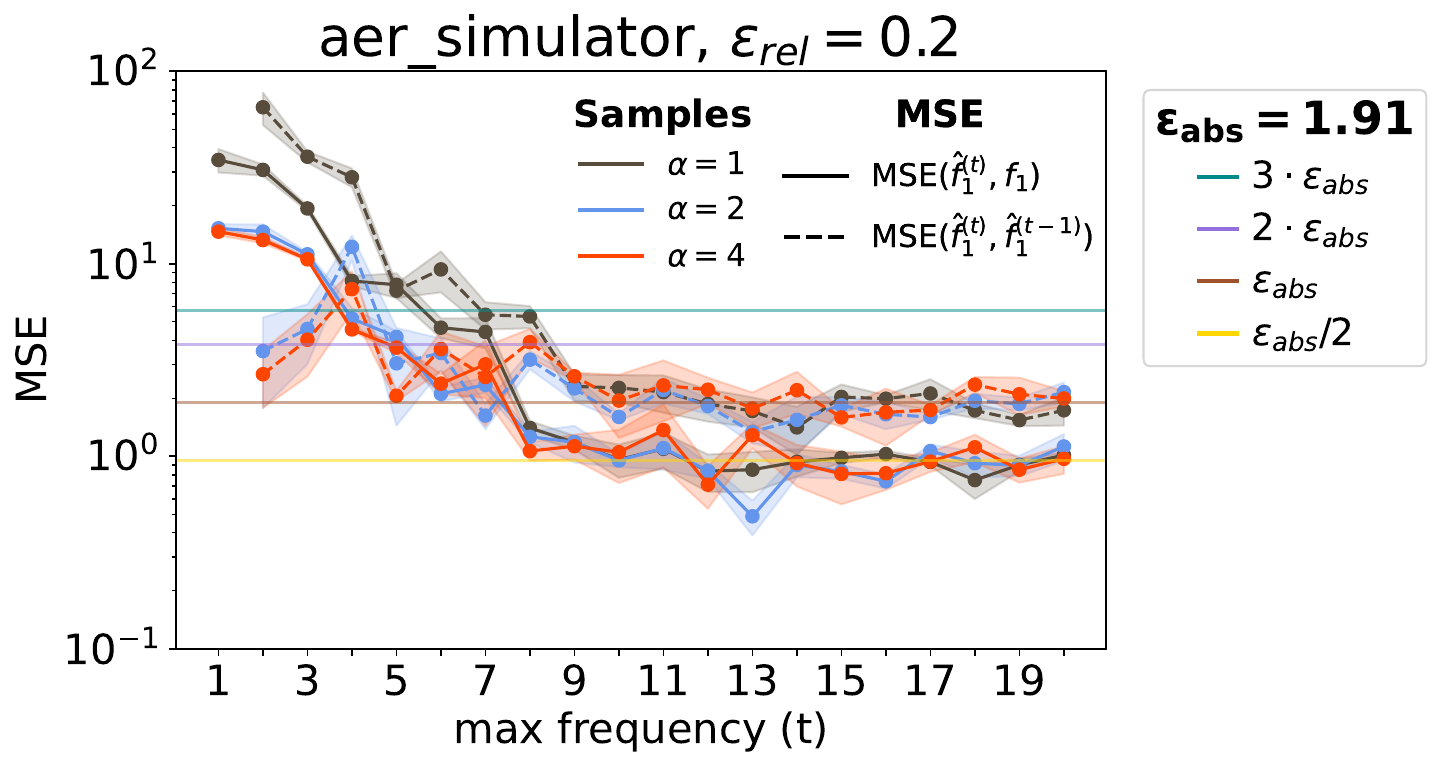}
                \put(6, 0){\text{(g)}}
            \end{overpic}
            \label{fig:sim-eps-02}
        \end{subfigure}
        \hfill
        \begin{subfigure}[b]{0.495\textwidth}
            \centering
            \begin{overpic}[width=\textwidth]{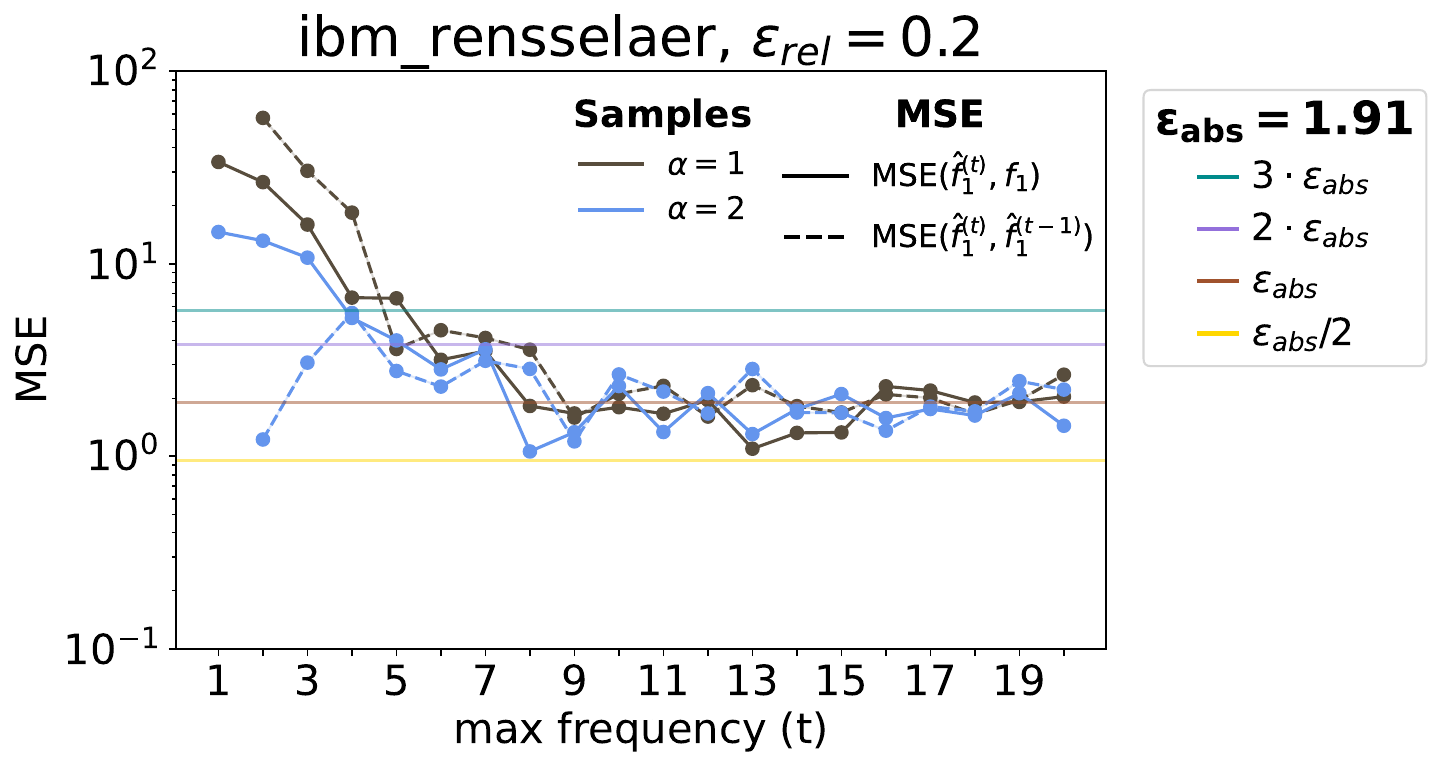}
                \put(6, 0){\text{(h)}}
            \end{overpic}
            \label{fig:hw-eps-02}
        \end{subfigure}
    \end{minipage}
    
    \captionsetup{}
    \caption{Adaptive frequency update performance on $f_1(\theta_1)$ from a 3-layer LC-QAOA ansatz of a 16-vertex random regular degree-5 graph with weights chosen uniformly at random in $\{1,2,4\}$. Left column (a, c, e, g) with AerSimulator and right column (b, d, f, h) with \textit{ibm\_rensselaer} hardware for $\epsilon_{rel} \in \{0.01, 0.05, 0.1, 0.2\}$.}
    \label{fig:adaptive_update}
\end{figure}

\begin{figure}[p] 
    \centering
    \begin{subfigure}[b]{0.48\textwidth}
        
        \centering
        \begin{overpic}[width=\textwidth]{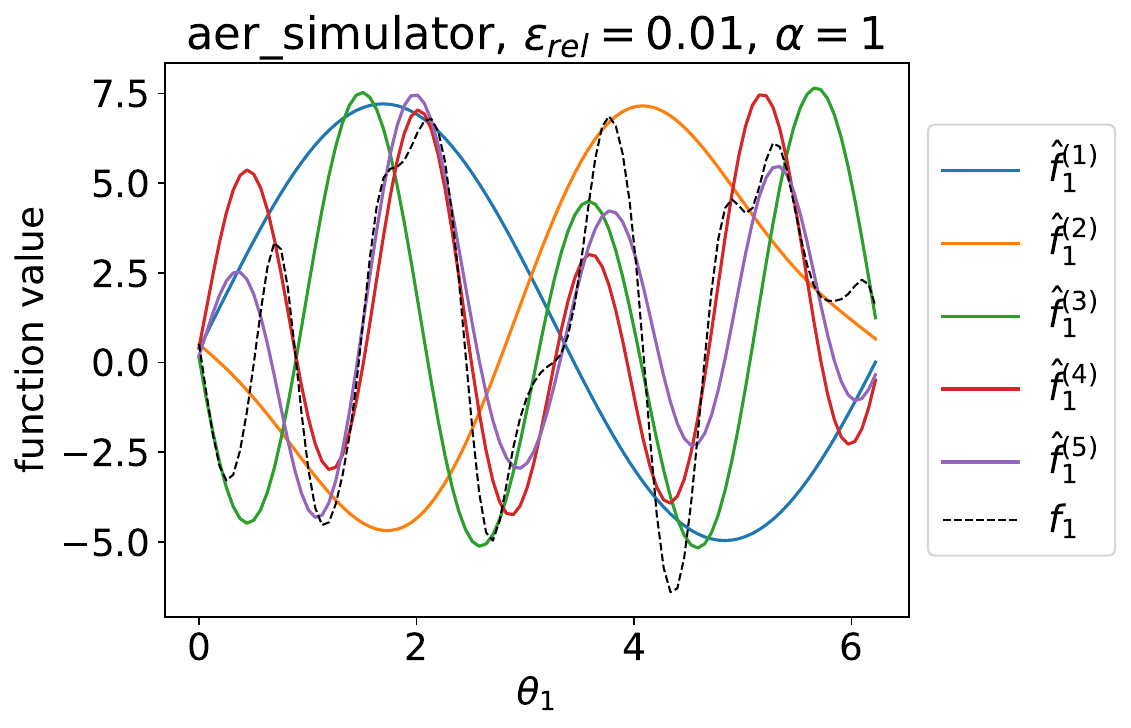}
            \put(6, 0){\text{(a)}}
        \end{overpic}
        \phantomcaption
        \label{fig:eps=0.01_alpha=1_low_sim}
    \end{subfigure}
    \hfill
    \begin{subfigure}[b]{0.48\textwidth}
        \centering
        \begin{overpic}[width=\textwidth]{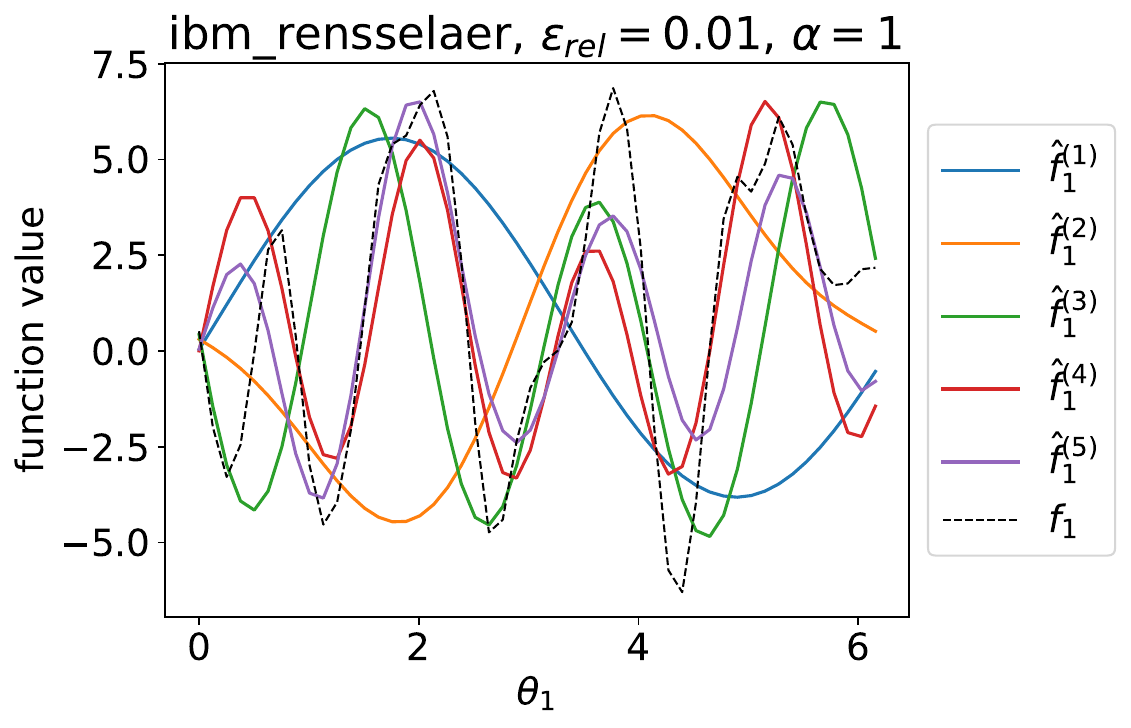}
            \put(6, 0){\text{(b)}}
        \end{overpic}
        \phantomcaption
        \label{fig:eps=0.01_alpha=1_low_hardware}
    \end{subfigure}

    \vspace{0.05em} 

    \begin{subfigure}[b]{0.48\textwidth}
        \centering
        \begin{overpic}[width=\textwidth]{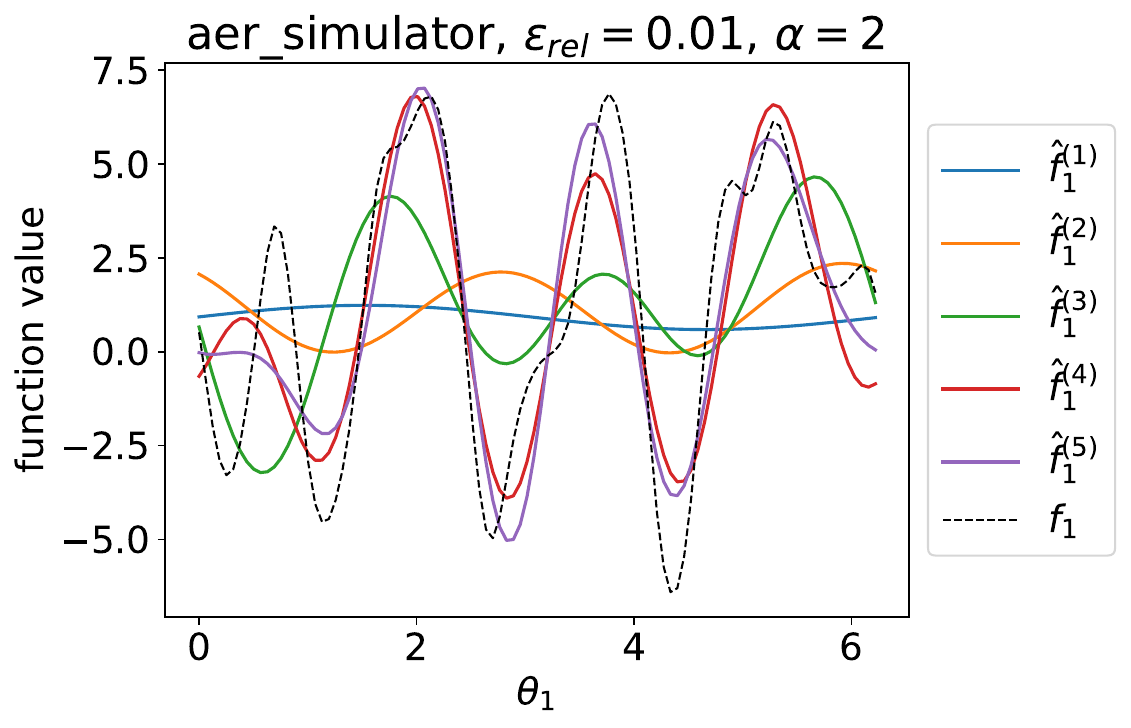}
            \put(6, 0){\text{(c)}}
        \end{overpic}
        \phantomcaption
        \label{fig:eps=0.01_alpha=2_low_sim}
    \end{subfigure}
    \hfill
    \begin{subfigure}[b]{0.48\textwidth}
        \centering
        \begin{overpic}[width=\textwidth]{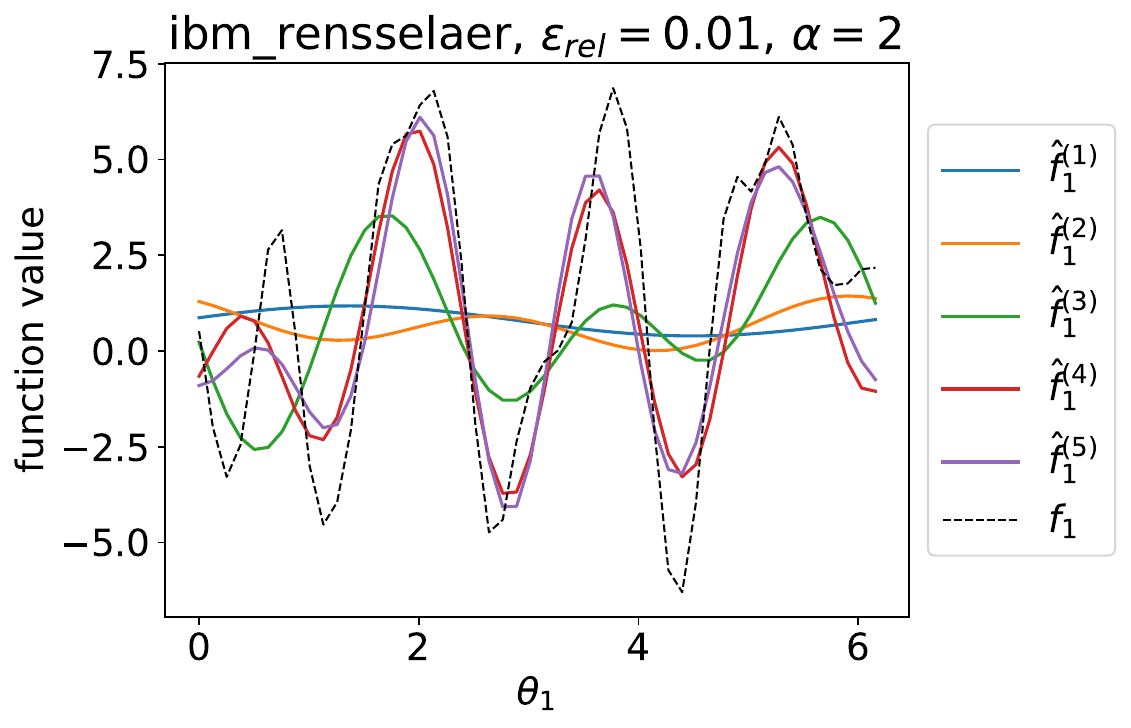}
            \put(6, 0){\text{(d)}}
        \end{overpic}
        \phantomcaption
        \label{fig:eps=0.01_alpha=2_low_hardware}
    \end{subfigure}

    \vspace{0.05em}

    \begin{subfigure}[b]{0.48\textwidth}
        \centering
        \begin{overpic}[width=\textwidth]{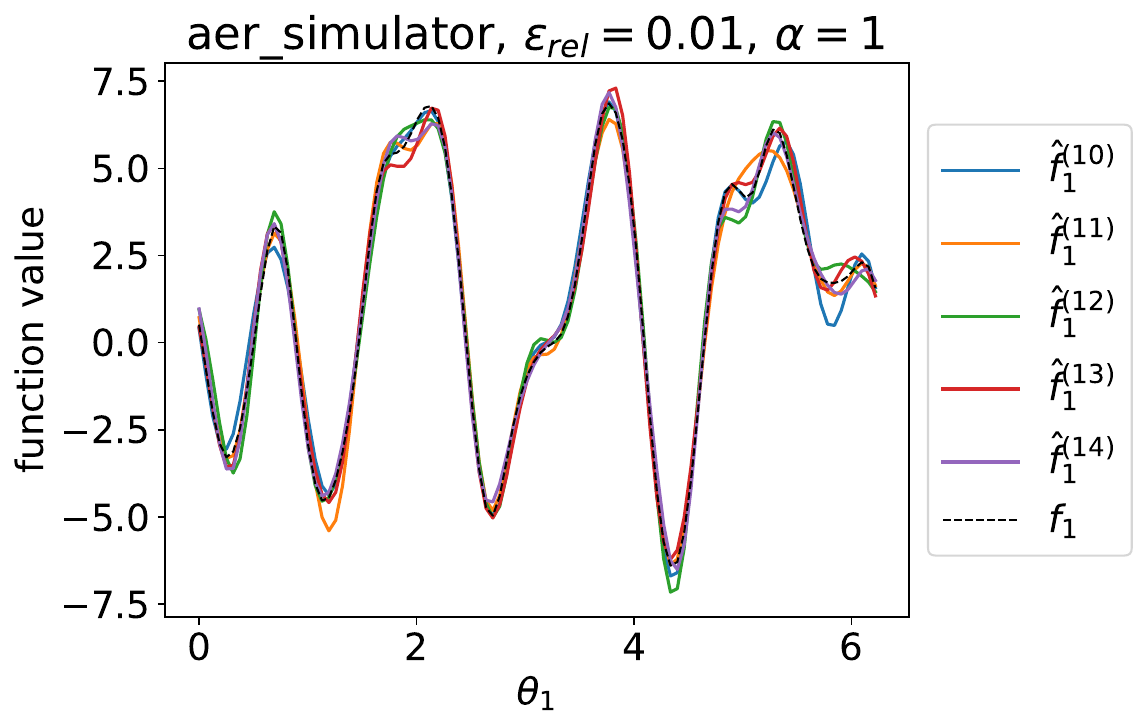}
            \put(6, 0){\text{(e)}}
        \end{overpic}
        \phantomcaption
        \label{fig:eps=0.01_alpha=1_high_sim}
    \end{subfigure}
    \hfill
    \begin{subfigure}[b]{0.48\textwidth}
        \centering
        \begin{overpic}[width=\textwidth]{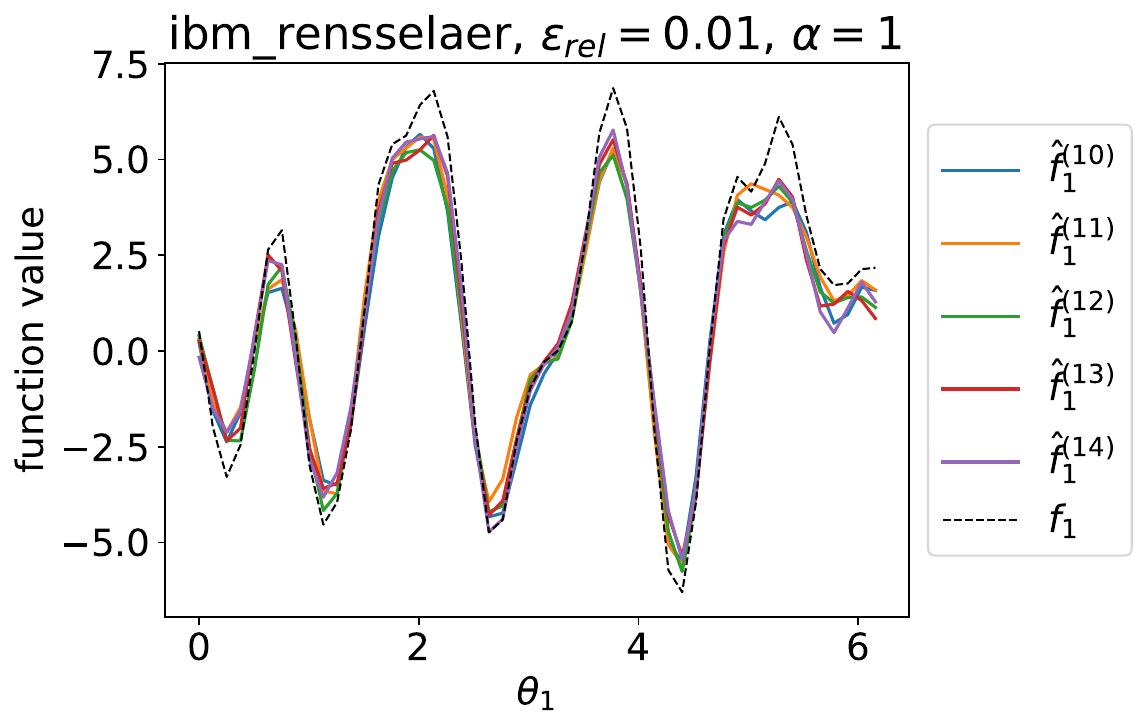}
            \put(6, 0){\text{(f)}}
        \end{overpic}
        \phantomcaption
        \label{fig:eps=0.01_alpha=1_high_hardware}
    \end{subfigure}

    \vspace{0.05em}

    \begin{subfigure}[b]{0.48\textwidth}
        \centering
        \begin{overpic}[width=\textwidth]{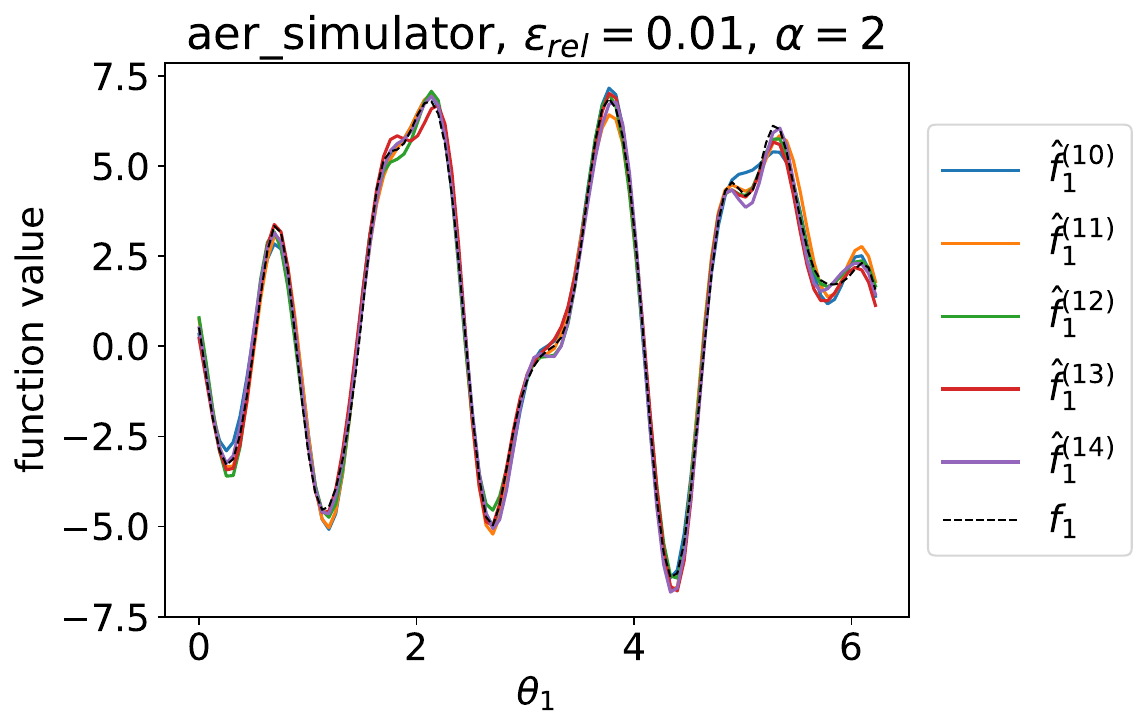}
            \put(6, 0){\text{(g)}}
        \end{overpic}
        \phantomcaption
        
        \label{fig:sub7}
    \end{subfigure}
    \hfill
    \begin{subfigure}[b]{0.48\textwidth}
        \centering
        \begin{overpic}[width=\textwidth]{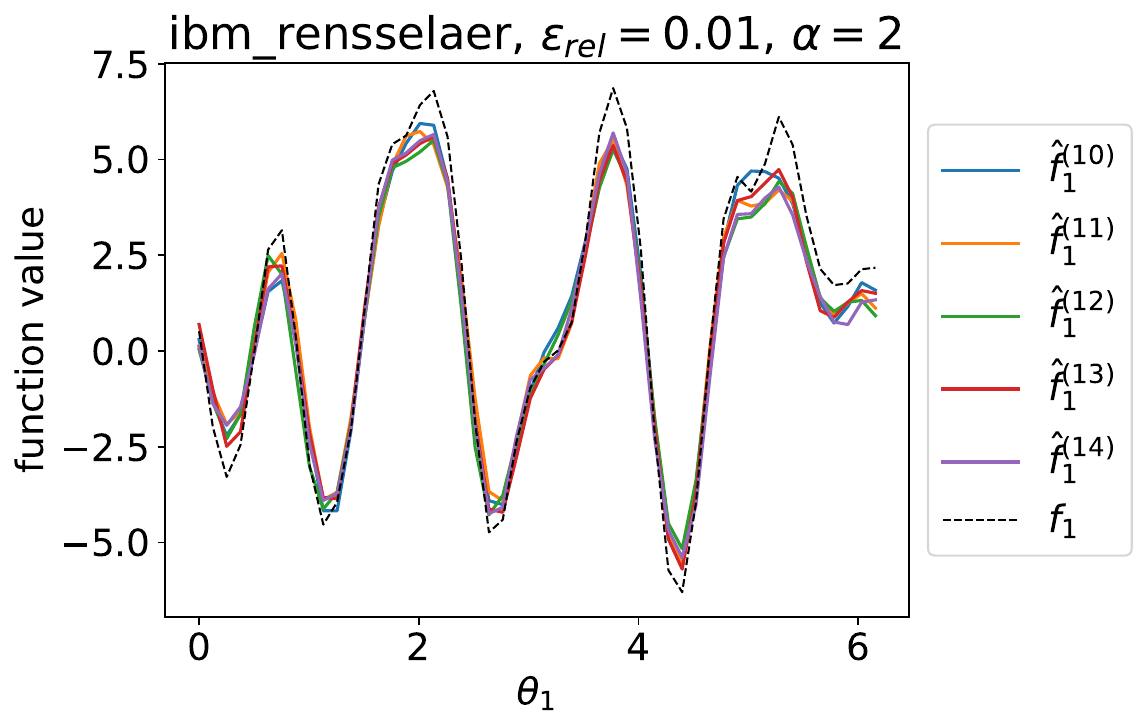}
            \put(6, 0){\text{(h)}}
        \end{overpic}
        \phantomcaption
        
        \label{fig:sub8}
    \end{subfigure}

    \caption{Plots of $f_1$ and $\hat{f}_1^{(t)}$ for $t$ in range 1 to 5 and range 10 to 14. Left column (a, c, e, g) with AerSimulator and right column (b, d, f, h) with \textit{ibm\_rensselaer} for $\varepsilon_{rel} = 0.01$. For simulator result, function plots come from first execution of Algorithm \ref{alg:one}. }
    \label{fig:adaptive_f_vals}
\end{figure}

The left column of Figure \ref{fig:adaptive_update} shows the results of Algorithm \ref{alg:one} using simulator backend. It can be seen that the results for $\alpha=2$ and $\alpha = 4$ are nearly identical across the entire range of $\varepsilon_{rel}$, though the results for $\alpha = 1$ are markedly different. The difference in lower-order frequency fitting can be seen by comparing Figures \ref{fig:eps=0.01_alpha=1_low_sim} and \ref{fig:eps=0.01_alpha=2_low_sim}. When $\alpha = 2$, the best fitting functions have lower amplitude and thus are closer to each other, while for $\alpha = 1$, the recovered functions have larger amplitude and are much further from the next iterate. The true MSE is consistently lower when $\alpha > 1$ when using lower-order frequency, meaning solving an overdetermined least-squares approximation is better than solving a square system in this regime. However, when the true MSE error becomes in the order of $\varepsilon_{abs}$, the true MSE is identical across all $\alpha$. By analyzing the consecutive MSE, we notice that convergence is sublinear and nearly monotonic decreasing in $t$ when $\alpha =1$, while large oscillations are present when $\alpha > 1$.

The right column of Figure \ref{fig:adaptive_update} shows the results of Algorithm \ref{alg:one} using hardware on \textit{ibm\_rensselaer}. A similar trend is found for the values of $\alpha$: the consecutive MSE with $\alpha = 1$ follows a more stable decrease compared to $\alpha > 1$. The most notable difference between hardware and simulator is in the true MSE as $\varepsilon_{abs}$ becomes smaller. At higher frequency fitting, the true MSE begins to lag behind its corresponding values on simulator. This is most obvious when $\varepsilon_{abs} = 0.01$ and the true MSE becomes higher than the consecutive MSE. The error is not only higher than expected, but flattens out even as more frequencies are added, 
indicating some structural difference in $f_1$ on hardware. This idea is supported by comparing Figure \ref{fig:eps=0.01_alpha=1_high_sim} and \ref{fig:eps=0.01_alpha=1_high_hardware}. On hardware in Figure \ref{fig:eps=0.01_alpha=1_high_hardware}, the iterates have very small difference between each other, but large common deviation from the true function. Hardware noise is systemically changing the function landscape. Based on the recovered functions, it appears that this change is mostly a shrinking in function value toward zero. 

These results have important impacts for FLICD. Experiments show that more frequencies do not necessary improve the fitting when interpolating with noisy function values; After a certain frequency is reached, the MSE flattens out confirming that the noise floor has been reached. Lastly, the consecutive MSE is a robust metric that can be used to determine stopping condition. Even though hardware may structurally alter $f_1$, the overall structure of the recovered function is well preserved and any recovered argmin is still valid. Lastly, we will need to use $\alpha = 1$ for FLICD in order to have stable descent in consecutive MSE.

\subsubsection{Adaptive frequency update on non-equidistant frequencies}

In this subsection, we investigate the performance of the adaptive frequency update in the case of non-equidistant frequencies. We utilize a 8-vertex random regular degree-3 graph with weights chosen from $\{6,7 \}$ with uniform probability and a 3-layer LC-QAOA ansatz. The parameters chosen are the same as those 
in section \ref{sec:Empirical variance and norm experiments}. Algorithm \ref{alg:one} is used to recover $f_1$, focusing on $\varepsilon_{rel} = 0.05$. The full set of frequencies are computed by using Lemma \ref{lemma:theory_freqs}, and the size of the least-square system is controlled by $\alpha$ in $\{1,1.5,2\}$.

\begin{figure}[htbp] 
    \centering
    \begin{subfigure}[b]{0.6\textwidth}
        \centering
        \includegraphics[width=\textwidth]{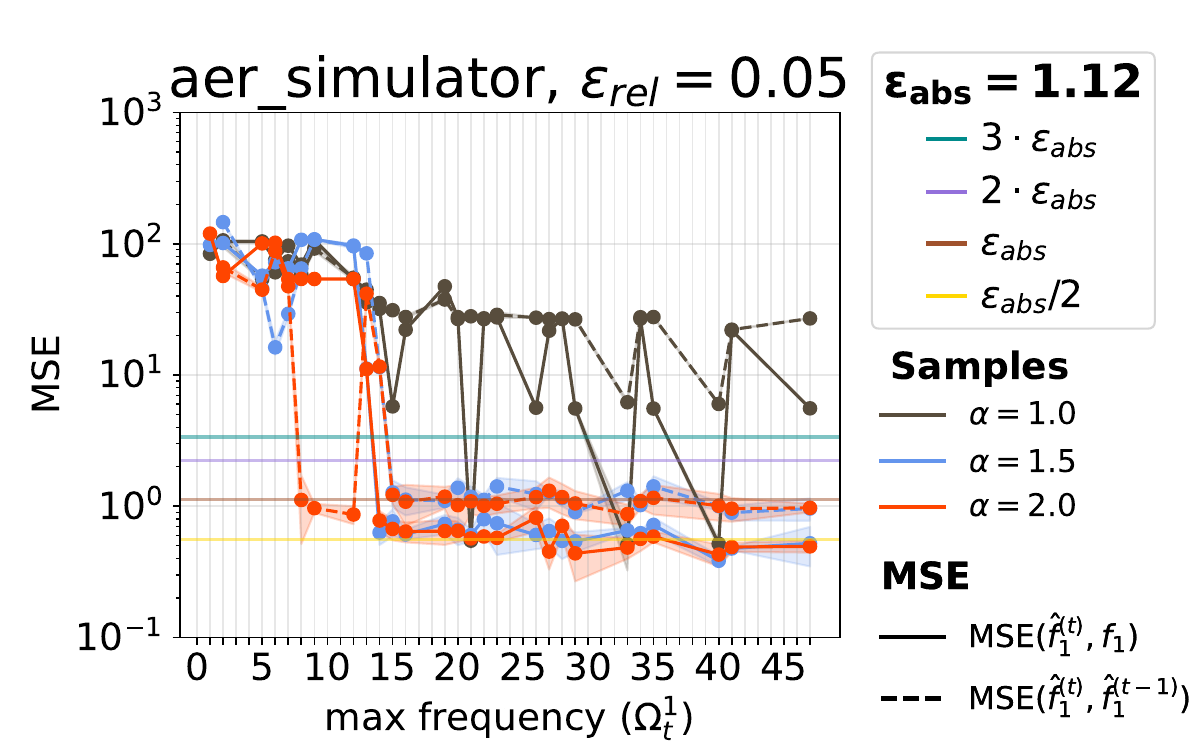}
        \caption{}
        \label{fig:ill_cond_sub1}
    \end{subfigure}
    \hfill
    \begin{subfigure}[b]{0.39\textwidth}
        \centering
        \includegraphics[width=\textwidth]{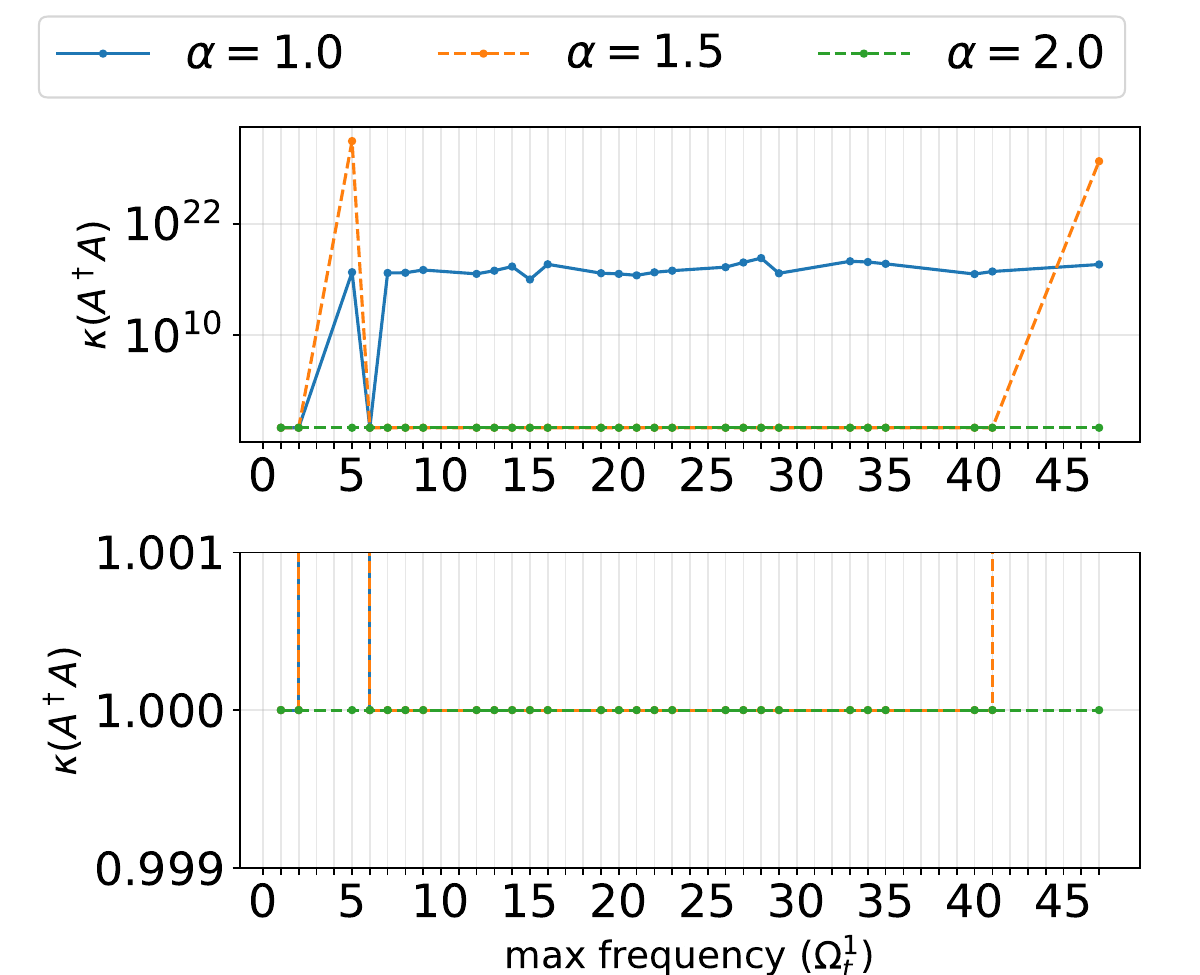}
        \caption{}
        \label{fig:ill_cond_sub2}
    \end{subfigure}
    \caption{(left) Adaptive frequency update performance on $f_1(\theta_1)$ from a 3-layer LC-QAOA ansatz of a 8-vertex random regular degree-3 graph with weights chosen uniformly at random in $\{6,7\}$. (right) Condition number of the interpolation matrix at each step $t$ of the adaptive frequency update.}
    \label{fig:ill_cond_adaptive_update}
\end{figure}

Figure \ref{fig:ill_cond_sub1} shows the MSE at each step $t$. Note that dots are only on certain integer frequencies, making up the non-equidistant frequency set. At low frequency interpolation, the true MSE is similar across all values of $\alpha$. Contrary to the case of equidistant frequencies, there is a clear distinction in behavior when higher-order frequencies are added. When $\alpha = 1$, the true MSE can often spike well above the expected order of magnitude. However, when $\alpha > 1$, the true MSE levels off at order $\varepsilon_{abs}$, with the true MSE roughly half the consecutive MSE. Even though the expression used to compute the number of shots was derived assuming equidistant frequencies, this number of shots still yields the desired error threshold in the case of non-equidistant frequencies when $\alpha>1$. The consecutive MSE for $\alpha=2$ dips below the threshold prematurely. This is remedied by choosing $\alpha$ between $1$ and $2$, such as $\alpha = 1.5$ in this case.

To 
explain why interpolation with a square system leads to unstable fitting, in Figure \ref{fig:ill_cond_sub2} we show the conditioning of the least-squares problem at each step of the algorithm. Since the Fourier coefficients are derived from the normal equations of the least-squares problem, the stability of the solution is governed by $\kappa(A^\dagger A)$, where $A \in \mathbb{R}^{P \times 2t+1}$ is the interpolation matrix at step $t$. For $\alpha=1$, the condition number is excessively high as $t$ increases. By taking evenly-spaced sample points with higher $P$, the condition number can be brought down to $1$.

For a general set of frequencies, it is shown in \cite{laiInterpolationbasedCoordinateDescent2026} that interpolation with a square system can be made more stable by finding the optimally-spaced sample points.  This requires minimizing $||A^{-1}_{\bold{x}}||^2_{F}$ over $\bold{x}$ (Algorithm 2 of \cite{laiInterpolationbasedCoordinateDescent2026}) or equivalently $\kappa(A_{\bold{x}})$, which may be very difficult and computationally  expensive. 
In the case of non-equidistant frequencies, choosing a least squares approach with evenly-spaced sample points may provide another option to avoid accurately optimizing the interpolation matrix. 

\subsection{FLICD experiments}

In this subsection, we run FLICD on simulator and hardware backends. Random regular degree-3 graphs with 20 vertices are used on simulator and with $20,40,$ and $100$ vertices on hardware. The edge weights are assigned from $\{ 1,2,5\}$ with respective probabilities $\{0.4,0.4,0.2\}$. The LC-QAOA ansatz is used with the number of layers $p \in \{1,3,5 \}$.

\subsubsection{Comparison of FLICD to  COBYLA and ICD} \label{subsubsec:Comparison of FLICD to  COBYLA and ICD}

We first compare FLICD against the zeroth-order optimizer COBYLA and the ICD method. Recall that our goal is to preserve the performance of ICD, while increasing its efficiency. Hence, benchmarking FLICD against ICD is essential. ICD was compared against other methods such as randomized coordinate descent and stochastic gradient descent in \cite{laiInterpolationbasedCoordinateDescent2026}. While there are various zeroth-order optimizers, COBYLA is commonly used in the QAOA literature, so it is the most apparent choice.  In the following experiments, all algorithms are run from the same five initial parameter vectors. We average the metrics of the algorithms across these five runs, and the results can be seen in Figure \ref{fig:q=20_avg_convergence_combined}. For information about individual runs, see Appendix \ref{appendix:Comparison of algorithm performance over each run}.

\textbf{FLICD setup}. FLICD requires several hyper-parameters, and  the most critical parameter that requires some investigation is $\varepsilon_{rel}$. We summarize the parameter setting below: 

\begin{itemize}
    
    \item $S$: Set to $S = p$, meaning the adaptive frequency update occurs always on the first layer of the ansatz. 
    \item $\varepsilon_{rel}$: The relative error is chosen to be $\varepsilon_{rel} \in \{ 0.01,0.05,0.10\}$. 
    \item $\gamma$: The threshold error factor is set to $\gamma = 3$. 
    \item $\alpha$: The sample size factor is set to $\alpha = 1$.
    \item $n_0$: number of shots for variance and norm estimation is set to $n_0 = 500$.
    \item $K$: grid size for variance and norm estimation is set to $K = 8$.
\end{itemize}

\textbf{COBYLA setup}. COBYLA itself is a hyper-parameter free method, but querying the function value from the quantum computer requires a specified number of shots. To compare COBYLA and FLICD in a fair way, we treat this number of shots as a hyper-parameter of COBYLA. It is given by $n \in \{ 1000,5000,10000\}$ per function evaluation. This range covers both low and high shot regimes to establish performance of COBYLA. 

\textbf{ICD setup}. ICD requires a fixed frequency set for each $\theta_j$ and a number of shots per function evaluation. We utilize the full frequency sets. The shot per function evaluation is given by $n \in \{20,100,500,2500\}$. This is to compare ICD to FLICD across the full range of shot regimes.

\textbf{Stopping conditions}. \label{section:stop_cond} Since COBYLA and FLICD have different notions of iterations, we keep track of the accumulated shot total during each algorithm. For each $p$, we set the shot total $n_{tot}$ as follows: $n_{tot} = 10^6$ when $p = 1$, $n_{tot} = 1.5\times 10^6$ when $ p =3$, and $n_{tot} = 2.0 \times 10^6$ when $p = 5$. For COBYLA, we also specify a tolerance of $10^{-30}$ and stop when either the tolerance has been reached or shot total has been reached. Note that a tolerance of 0 was initially tried for COBYLA, but this led the iterates to eventually diverge. Hence, we choose the tiny but positive tolerance of $10^{-30}$. 
For each of FLICD and ICD, the algorithm stops only when the shot total has been reached. 

\textbf{Initial Parameters}. The initial parameters $\boldsymbol{\theta}^0$ are chosen  $\boldsymbol{\theta}^0 \sim \textrm{Unif}([0,2\pi]^{2p})$, with the randomness controlled by a specified seed value. 





\textbf{FLICD outperforms COBYLA.} \Cref{fig:q=20_avg_convergence_p=1_COBYLA,fig:q=20_avg_convergence_p=3_COBYLA,fig:q=20_avg_convergence_p=5_COBYLA} show that there is a clear increase in performance using FLICD when compared to COBYLA for each  $p$. Most notably, the fewer problem parameters, the larger the gap between FLICD and COBYLA. This gap is especially stark when $p = 1$ and there are only two problem parameters: the objective value for FLICD is more than three times lower. Meanwhile, FLICD can efficiently navigate to much lower objective values using similar number of total shots, when $\varepsilon_{rel} = 0.05$ and $0.10$. As the number of layers increases, this performance gap is lessened. 


However, the difference is not only that FLICD can outperform COBYLA against a fix number of layers $p$, but executing FLICD on an ansatz with a smaller $p$ can achieve lower objective value on average than COBYLA on an ansatz with a bigger $p$. \Cref{fig:q=20_avg_convergence_p=1_COBYLA} shows that the averaged converged function value by FLICD with $p = 1$ is lower than that by COBYLA with $p = 5$. This means that a shallower circuit ansatz can outperform a deeper circuit ansatz, by solely adjusting the choice of optimizer. This is especially of importance to NISQ-era circuits where ansatz depth is limited by hardware noise. 


\textbf{FLICD can be more efficient or more accurate than ICD.} \Cref{fig:q=20_avg_convergence_p=1_ICD,fig:q=20_avg_convergence_p=3_ICD,fig:q=20_avg_convergence_p=5_ICD} show the results of ICD against FLICD across  different valus of $p$. When $p = 1$, ICD and FLICD have the same objective value on average regardless of parameter setting. However, when $p = 3$ or $5$ they perform differently. For low shot regime, ICD converges at a similar rate in terms of shot total as FLICD, however its converged objective value is higher than that of FLICD on average. For high shot regime, ICD is able to achieve similar objective value as FLICD, but its convergence rate is slower than that of FLICD.

\begin{figure}[p]
    \centering
    
    \begin{subfigure}[t]{0.48\linewidth}
        \centering
        \includegraphics[height=0.30\textheight, width=\linewidth, keepaspectratio]{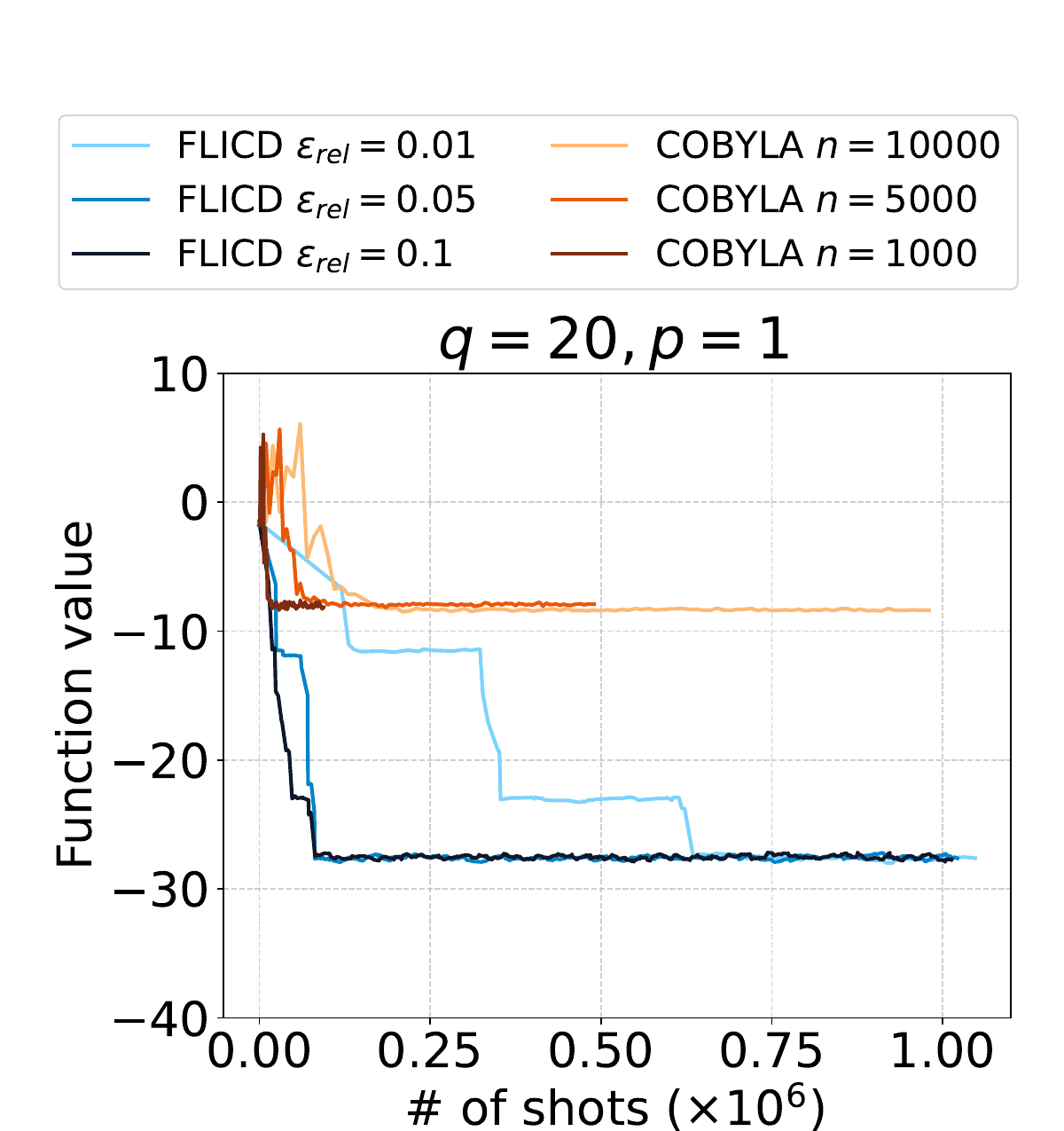}
        \phantomcaption
        \label{fig:q=20_avg_convergence_p=1_COBYLA}
    \end{subfigure}
    \hfill
    \begin{subfigure}[t]{0.47\linewidth}
        \centering
        \includegraphics[height=0.30\textheight, width=\linewidth, keepaspectratio]{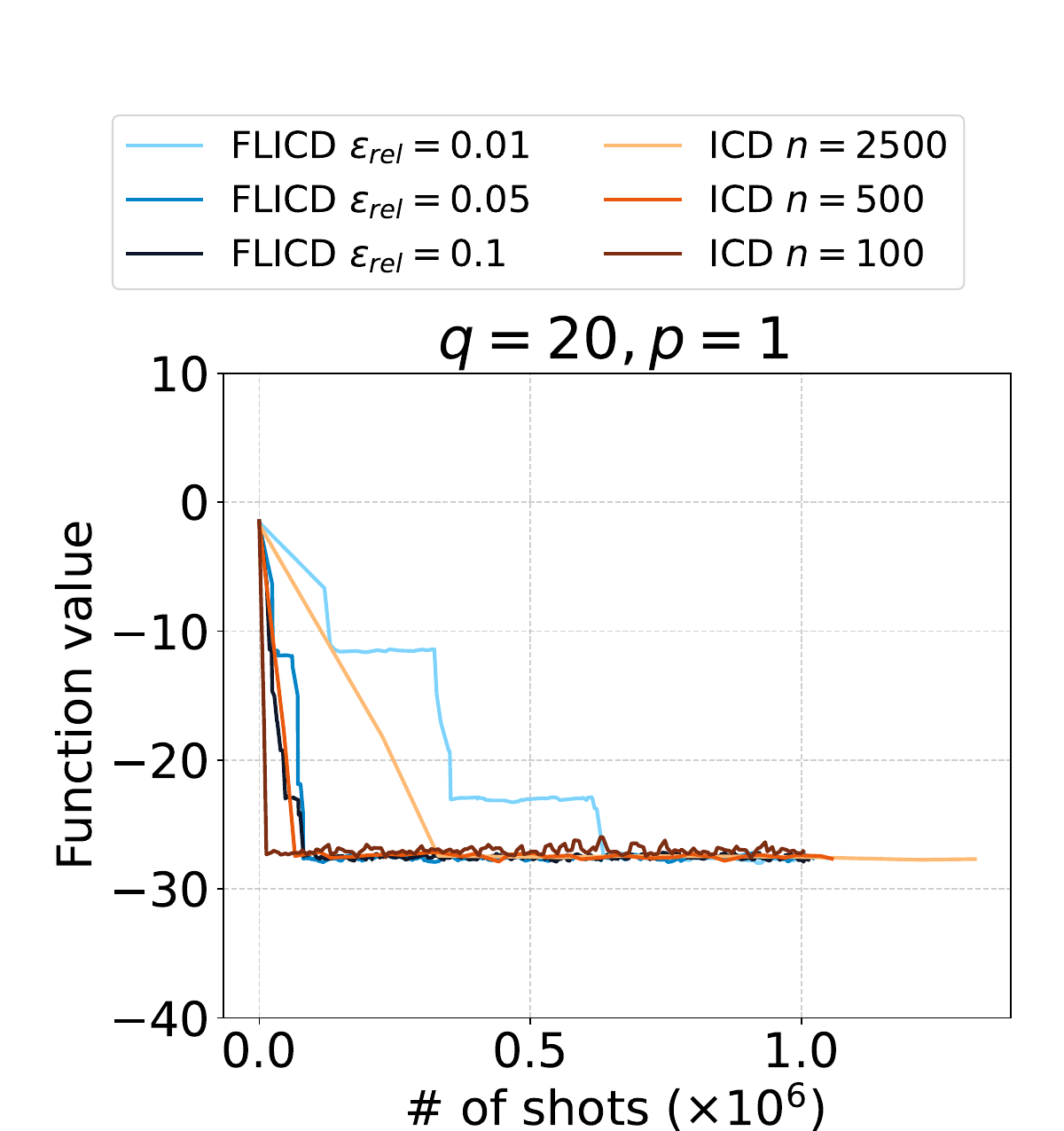}
        \phantomcaption
        \label{fig:q=20_avg_convergence_p=1_ICD}
    \end{subfigure}

    \vspace{0.4em}

    \begin{subfigure}[t]{0.48\linewidth}
        \centering
        \includegraphics[height=0.3\textheight, width=\linewidth, keepaspectratio]{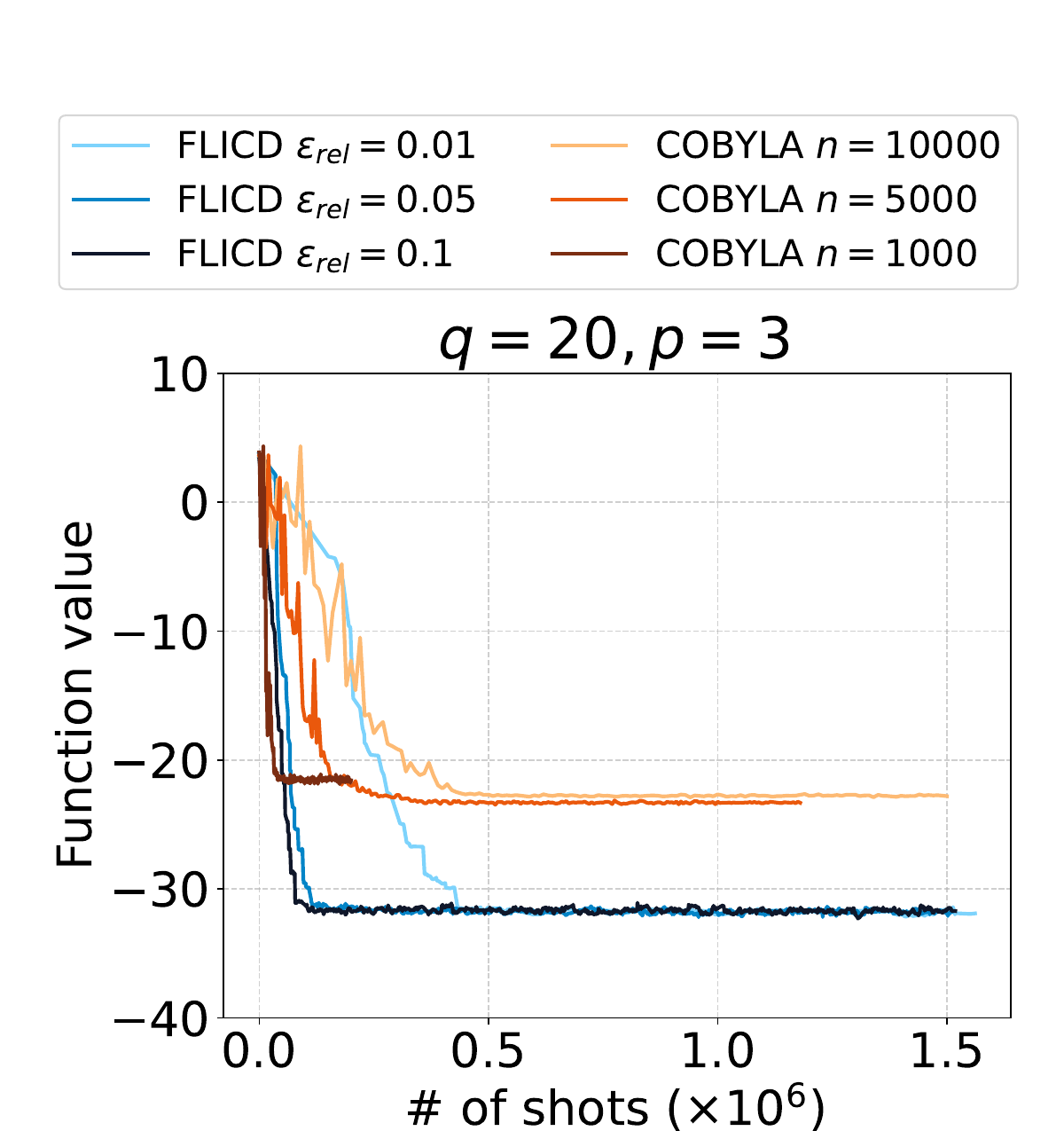}
        \phantomcaption
        \label{fig:q=20_avg_convergence_p=3_COBYLA}
    \end{subfigure}
    \hfill
    \begin{subfigure}[t]{0.48\linewidth}
        \centering
        \includegraphics[height=0.3\textheight, width=\linewidth, keepaspectratio]{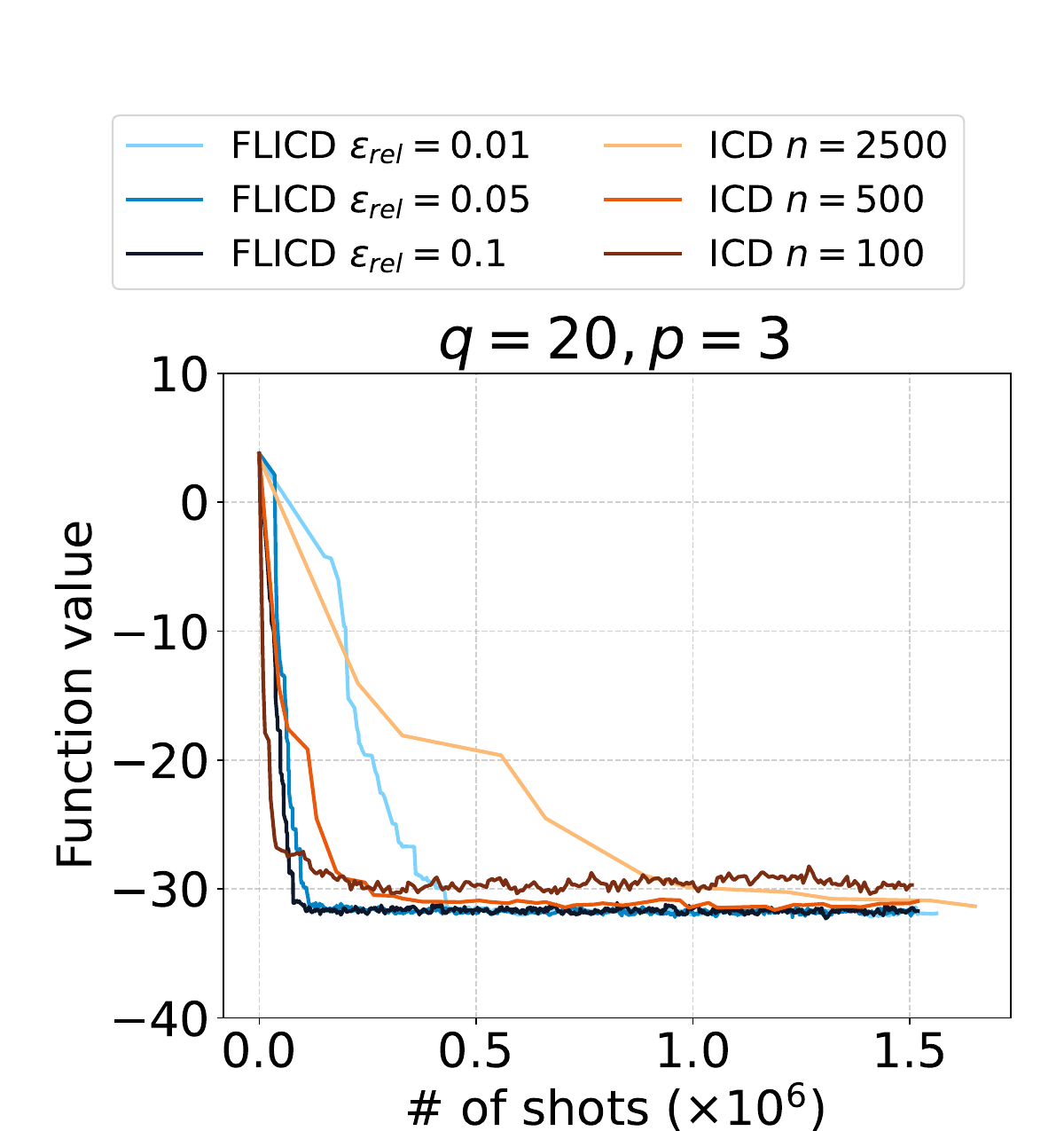}
        \phantomcaption
        \label{fig:q=20_avg_convergence_p=3_ICD}
    \end{subfigure}

    \vspace{0.4em}

    \begin{subfigure}[t]{0.48\linewidth}
        \centering
        \includegraphics[height=0.3\textheight, width=\linewidth, keepaspectratio]{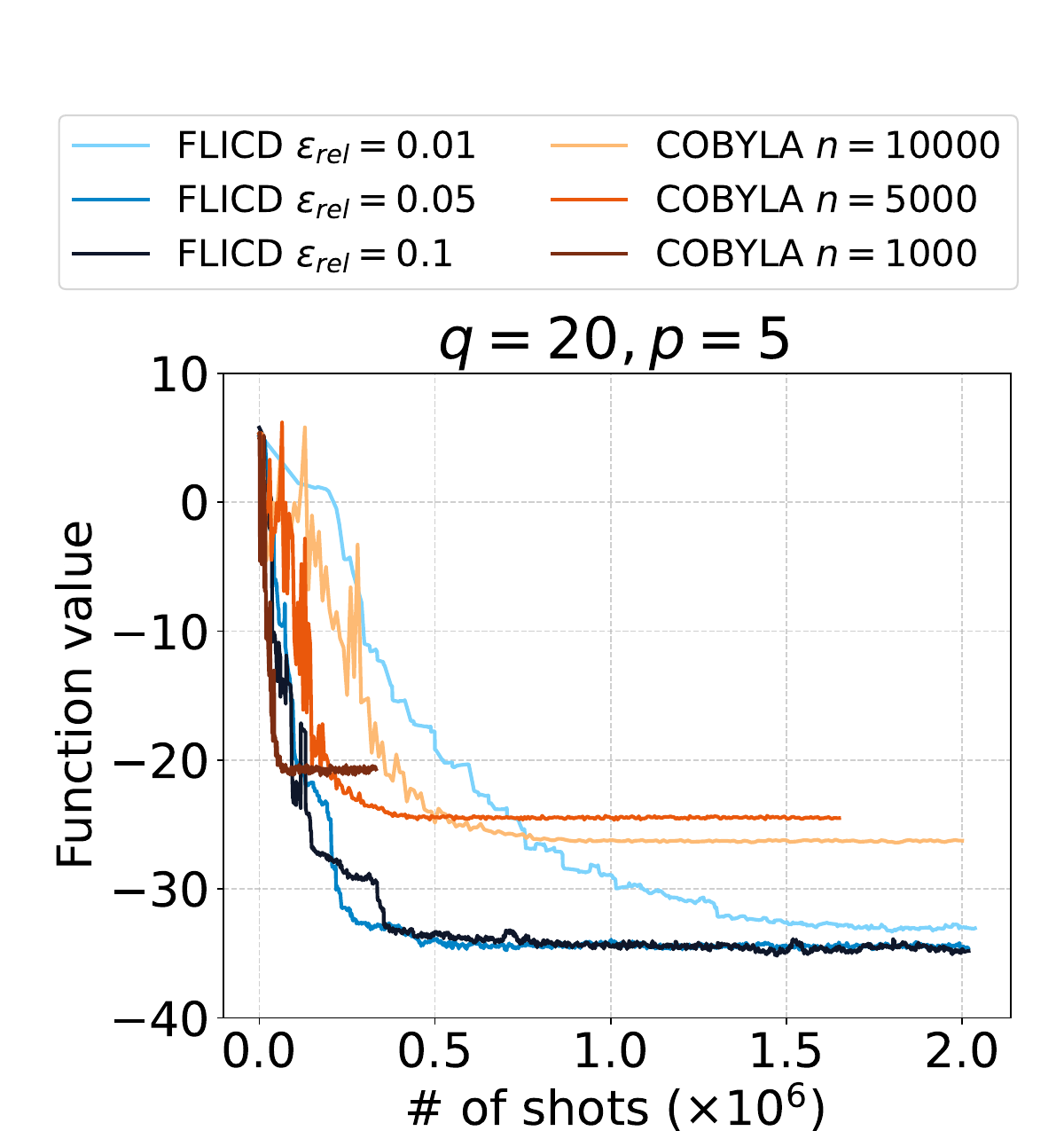}
        \phantomcaption
        \label{fig:q=20_avg_convergence_p=5_COBYLA}
    \end{subfigure}
    \hfill
    \begin{subfigure}[t]{0.48\linewidth}
        \centering
        \includegraphics[height=0.3\textheight, width=\linewidth, keepaspectratio]{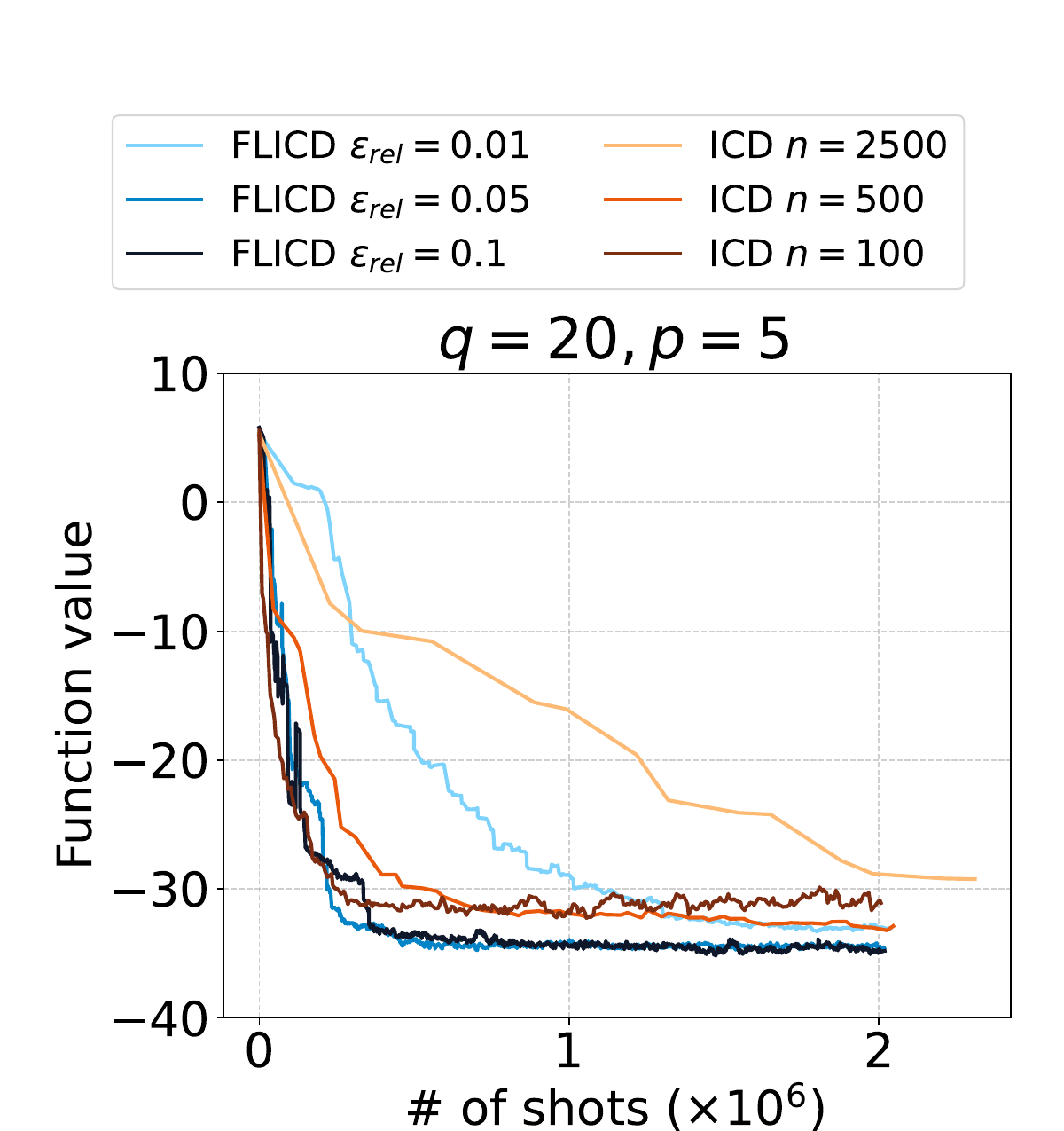}
        \phantomcaption
        \label{fig:q=20_avg_convergence_p=5_ICD}
    \end{subfigure}

    \caption{Average function value with respect to total shots across 5 runs for FLICD against COBYLA and ICD on simulator backend. Cost functions $f(\boldsymbol{\theta})$ generated from $p$-layer LC-QAOA ansatz of 20-vertex random regular degree-3 graph with $p \in \{1,2,5 \}$.}
    \label{fig:q=20_avg_convergence_combined}
\end{figure}

\subsubsection{FLICD on quantum hardware and at scale} \label{FLICD on quantum hardware and at scale}

\begin{figure}[htbp]
    \centering
    
    \begin{subfigure}[b]{0.47\textwidth}
        \centering
        \includegraphics[width=\textwidth]{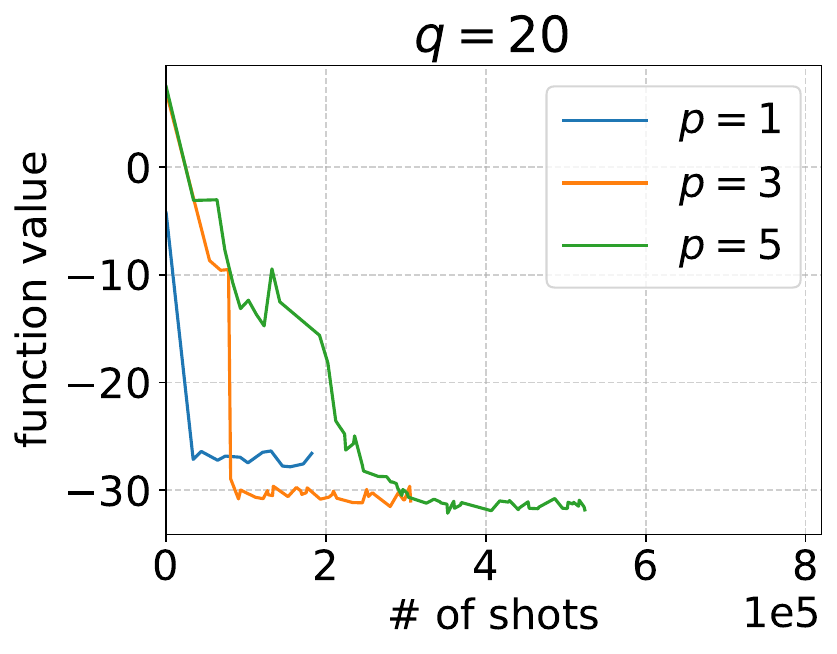}
        \caption{}
        \label{fig:q=20_sub-1}
    \end{subfigure}
    \hfill 
    \begin{subfigure}[b]{0.50\textwidth}
        \centering
        \includegraphics[width=\textwidth]{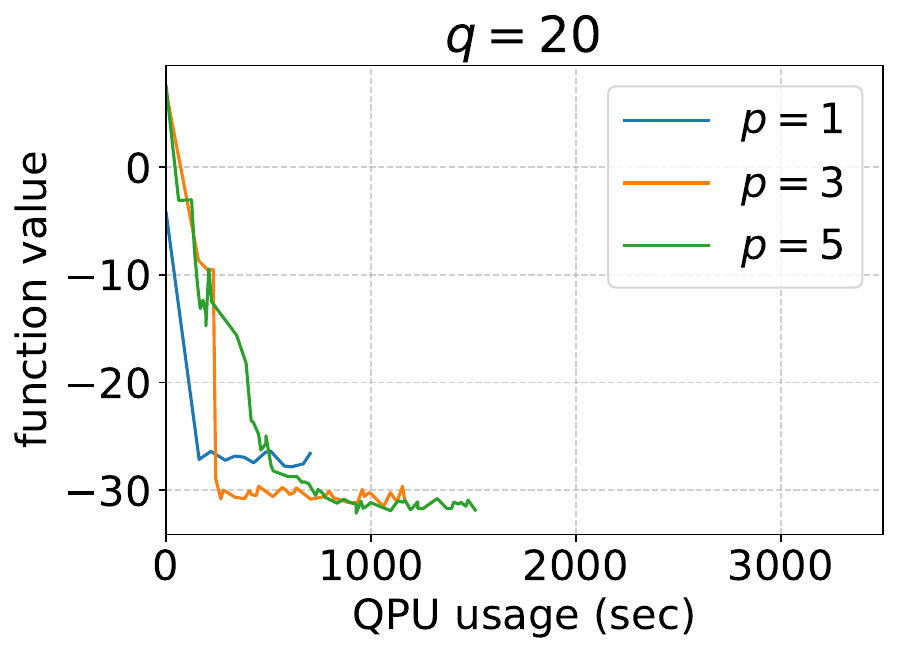}
        \caption{}
        \label{fig:q=20_sub-2}
    \end{subfigure}
    
    \vspace{0.4cm} 
    
    \begin{subfigure}[b]{0.47\textwidth}
        \centering
        \includegraphics[width=\textwidth]{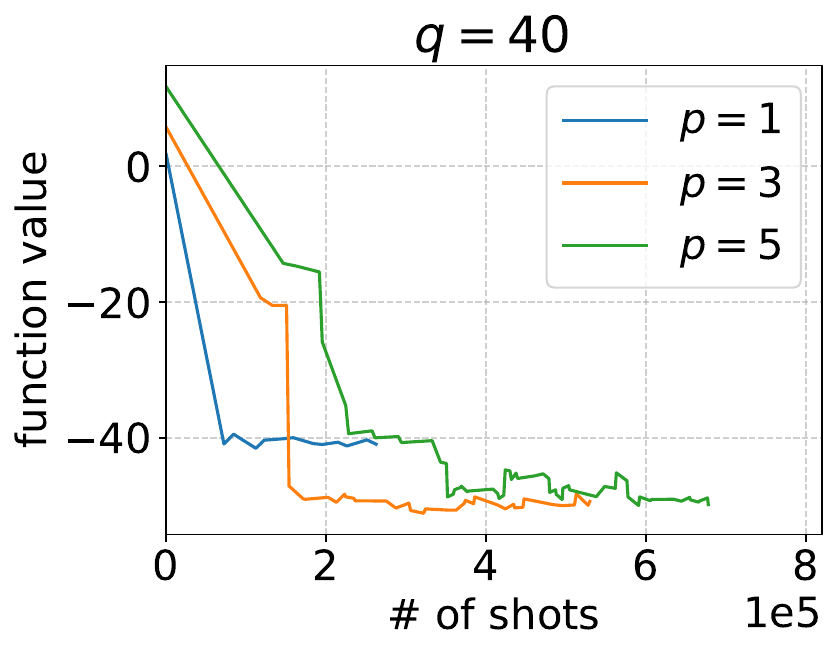}
        \caption{}
        \label{fig:q=40_sub-1}
    \end{subfigure}
    \hfill
    \begin{subfigure}[b]{0.50\textwidth}
        \centering
        \includegraphics[width=\textwidth]{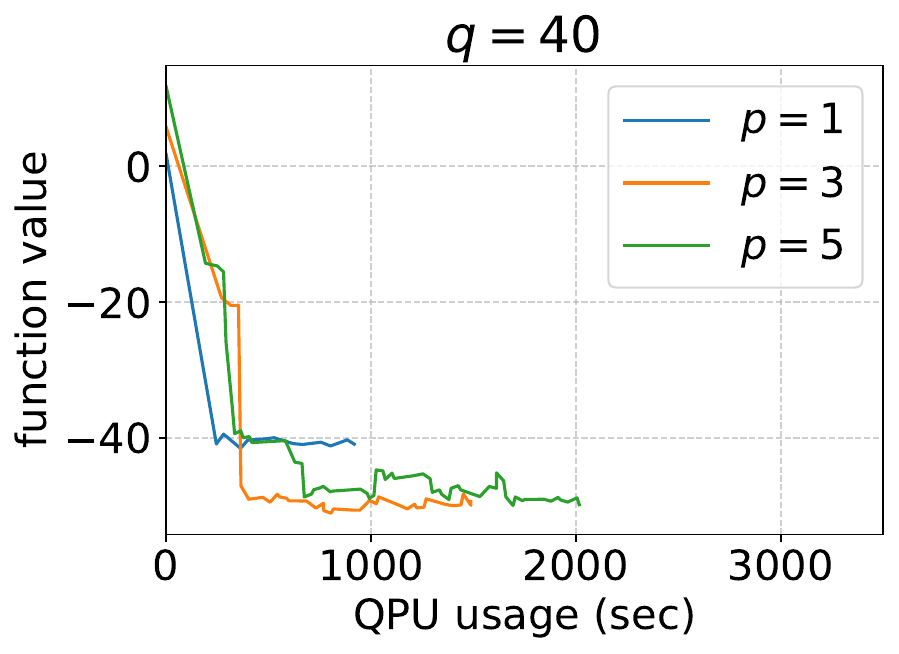}
        \caption{}
        \label{fig:q=40_sub-2}
    \end{subfigure}
    
    \vspace{0.4cm} 
    
    \begin{subfigure}[b]{0.47\textwidth}
        \centering
        \includegraphics[width=\textwidth]{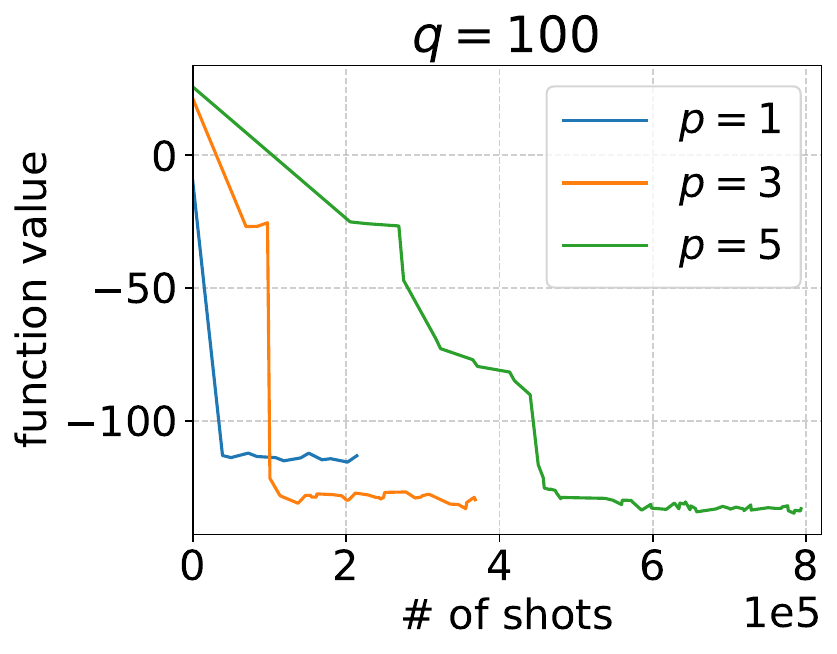}
        \caption{}
        \label{fig:q=100_sub-1}
    \end{subfigure}
    \hfill
    \begin{subfigure}[b]{0.50\textwidth}
        \centering
        \includegraphics[width=\textwidth]{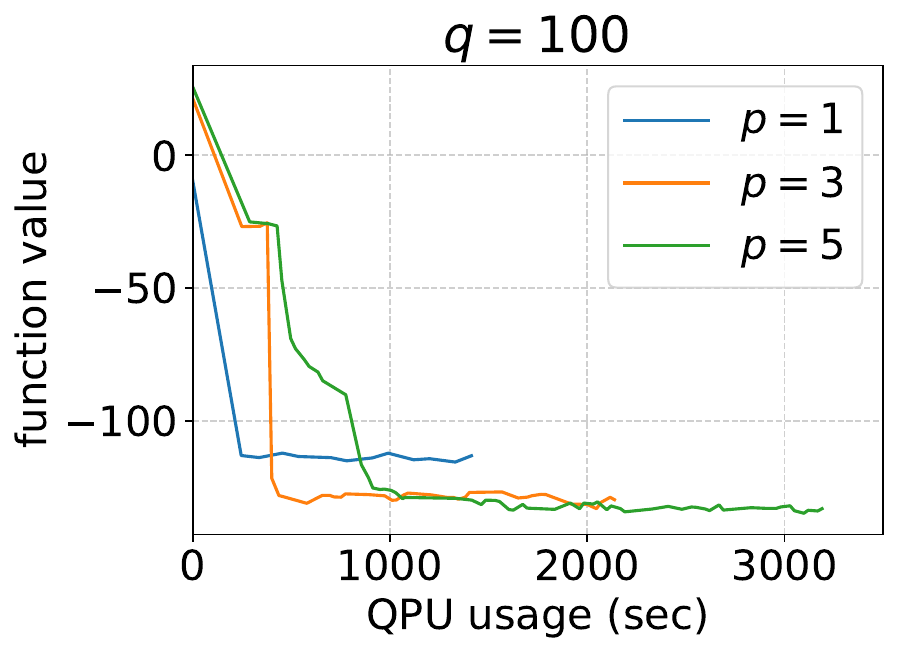}
        \caption{}
        \label{fig:q=100_sub-2}
    \end{subfigure}
    
    \caption{ FLICD function value with respect to total number of shots and total QPU usage on IBM quantum hardware. Cost function $f(\boldsymbol{\theta})$ generated from $p$-layer LC-QAOA ansatz with $p \in \{1,3,5 \}$ using random regular degree-3 graphs with weights in $\{1,2,5\}$ with probability distribution $\{0.4,0.4,0.2\}$. Experiments for $q = 20$ and $q =100$ performed on \textit{ibm\_marrakesh} and $q=40$ performed on \textit{ibm\_fez}.} 
    \label{fig:all_qubit_conv_hardware}
\end{figure}

\begin{figure}[htbp]
    \centering
    
    \begin{subfigure}[b]{0.32\textwidth}
        \centering
        \includegraphics[width=\textwidth]{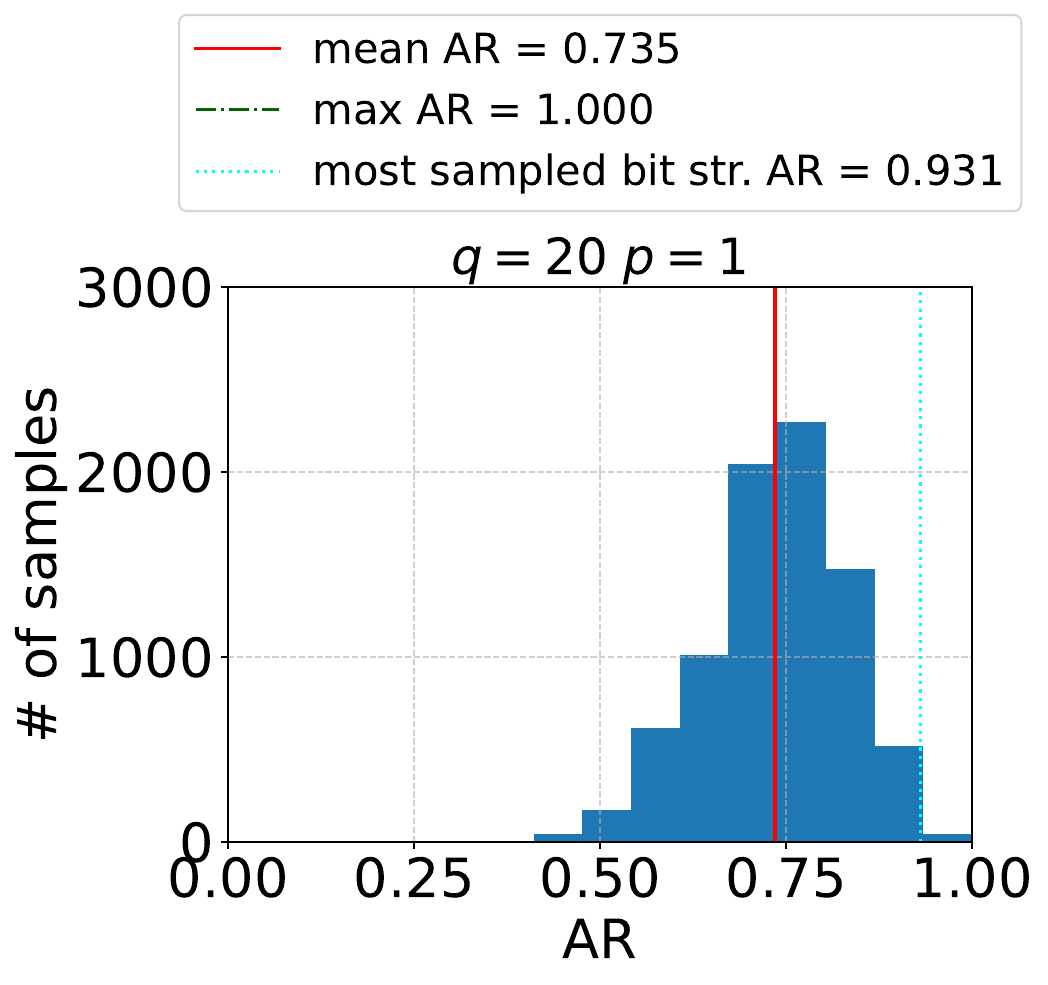}
        \caption{}
        \label{fig:hist_q20_p1}
    \end{subfigure}
    \hfill
    \begin{subfigure}[b]{0.32\textwidth}
        \centering
        \includegraphics[width=\textwidth]{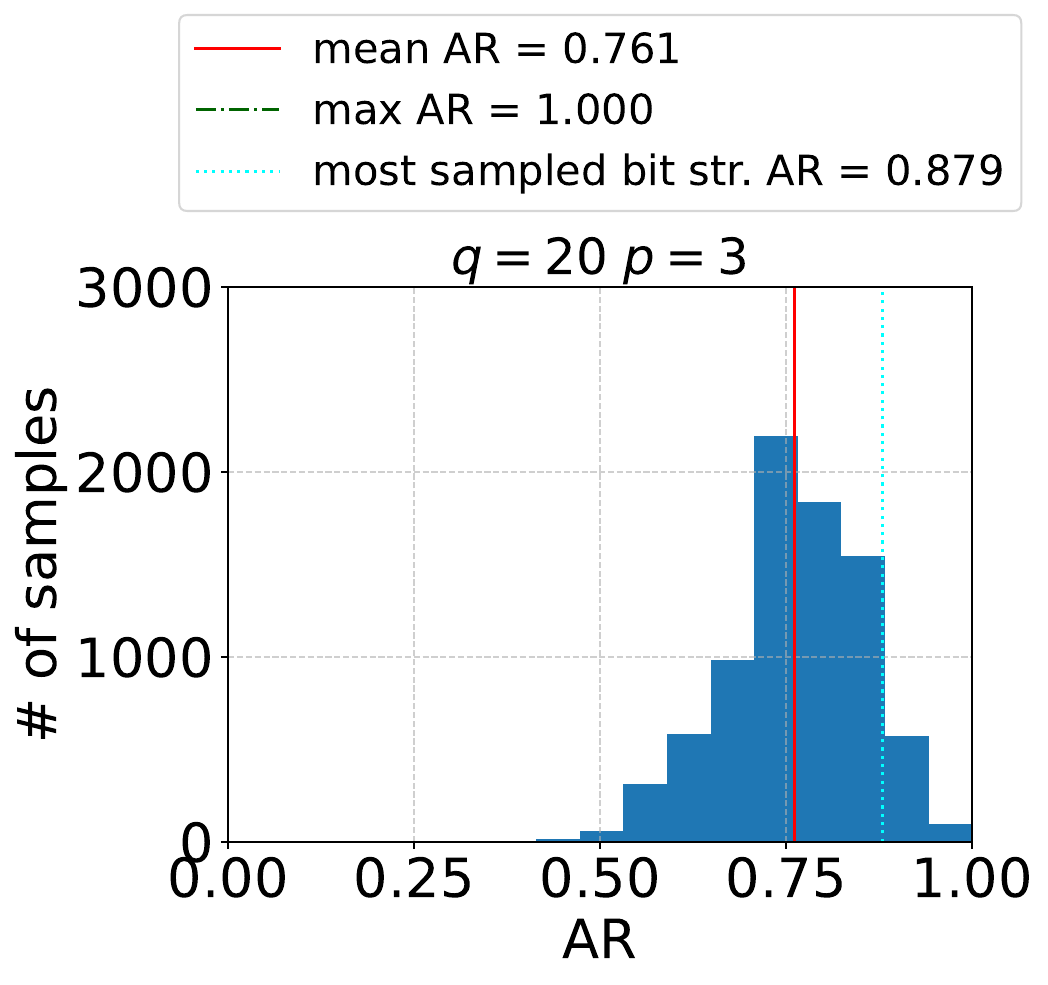}
         \caption{}
        \label{fig:hist_q20_p3}
    \end{subfigure}
    \hfill
    \begin{subfigure}[b]{0.32\textwidth}
        \centering
        \includegraphics[width=\textwidth]{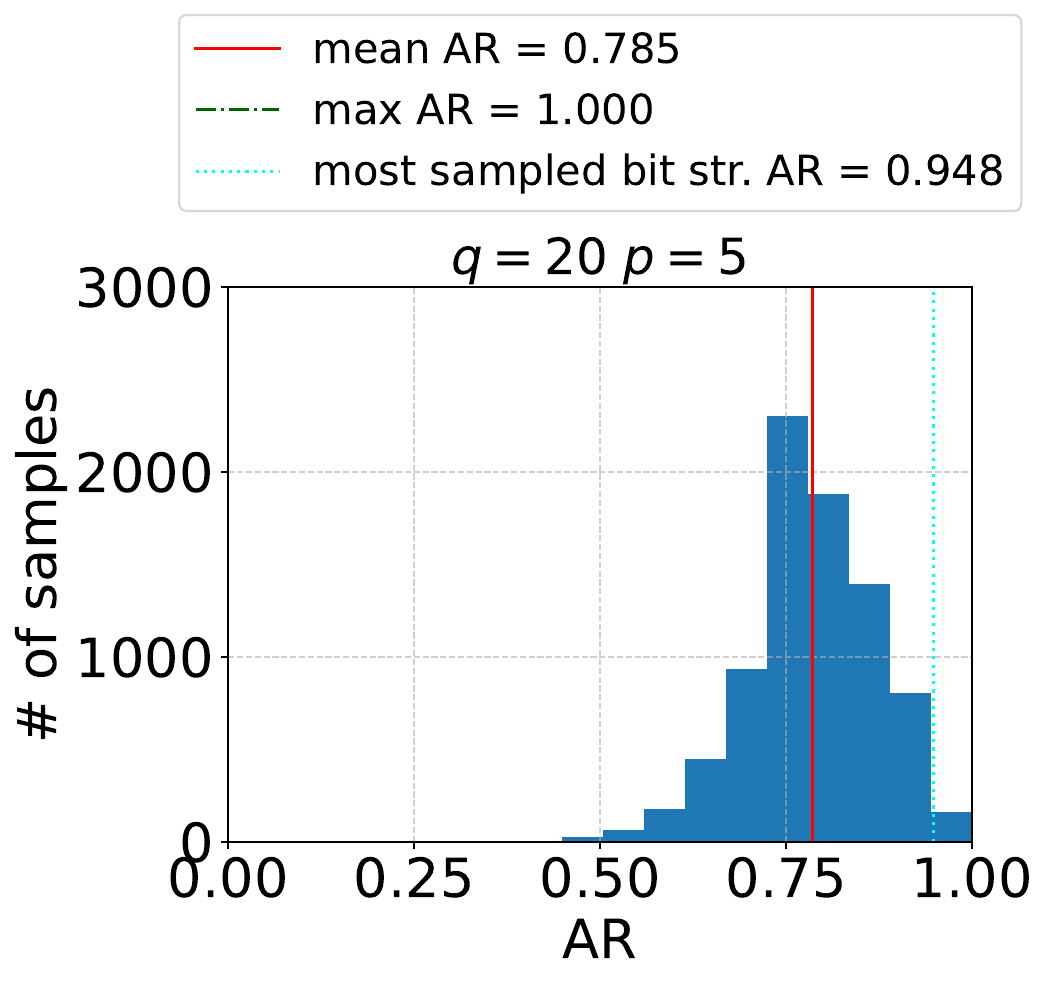}
         \caption{}
        \label{fig:hist_q20_p5}
    \end{subfigure}
    
    \vspace{0.4cm} 
    
    \begin{subfigure}[b]{0.32\textwidth}
        \centering
        \includegraphics[width=\textwidth]{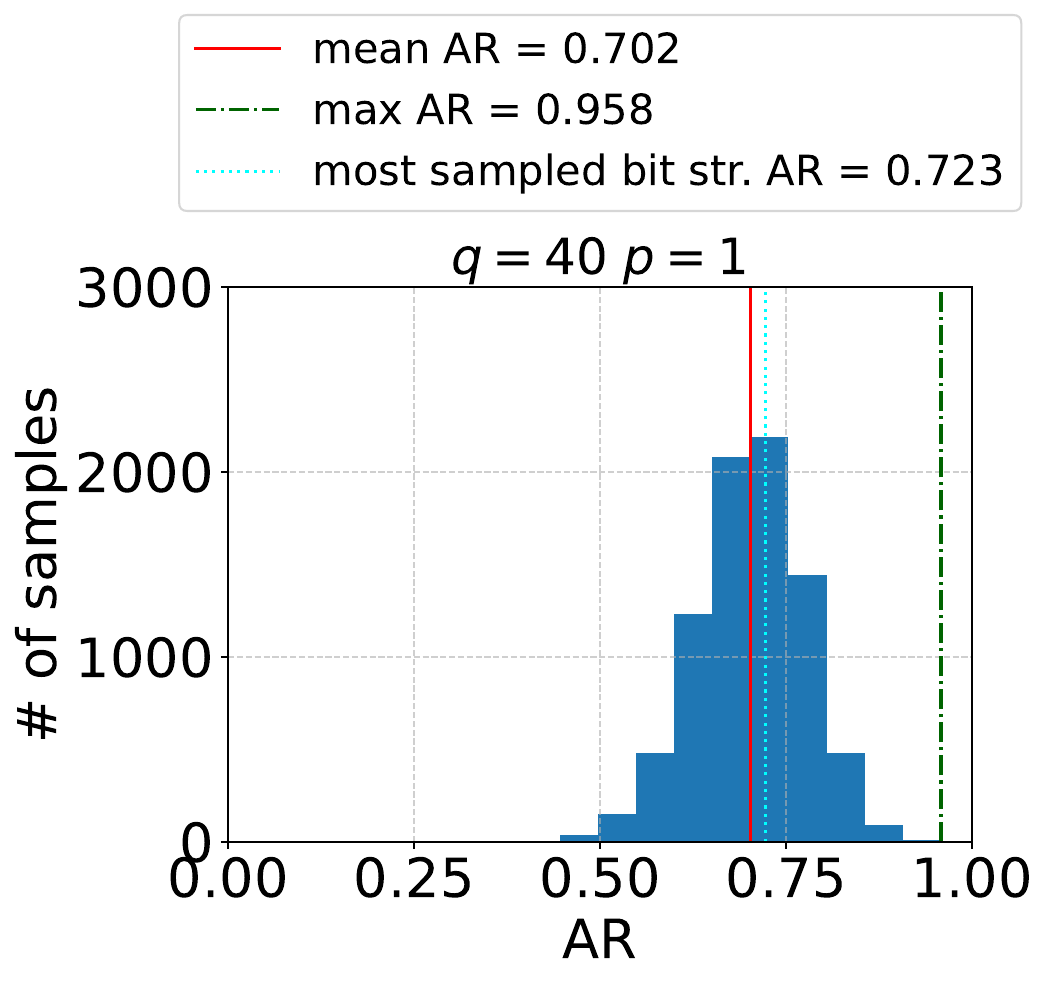}
         \caption{}
        \label{fig:hist_q40_p1}
    \end{subfigure}
    \hfill
    \begin{subfigure}[b]{0.32\textwidth}
        \centering
        \includegraphics[width=\textwidth]{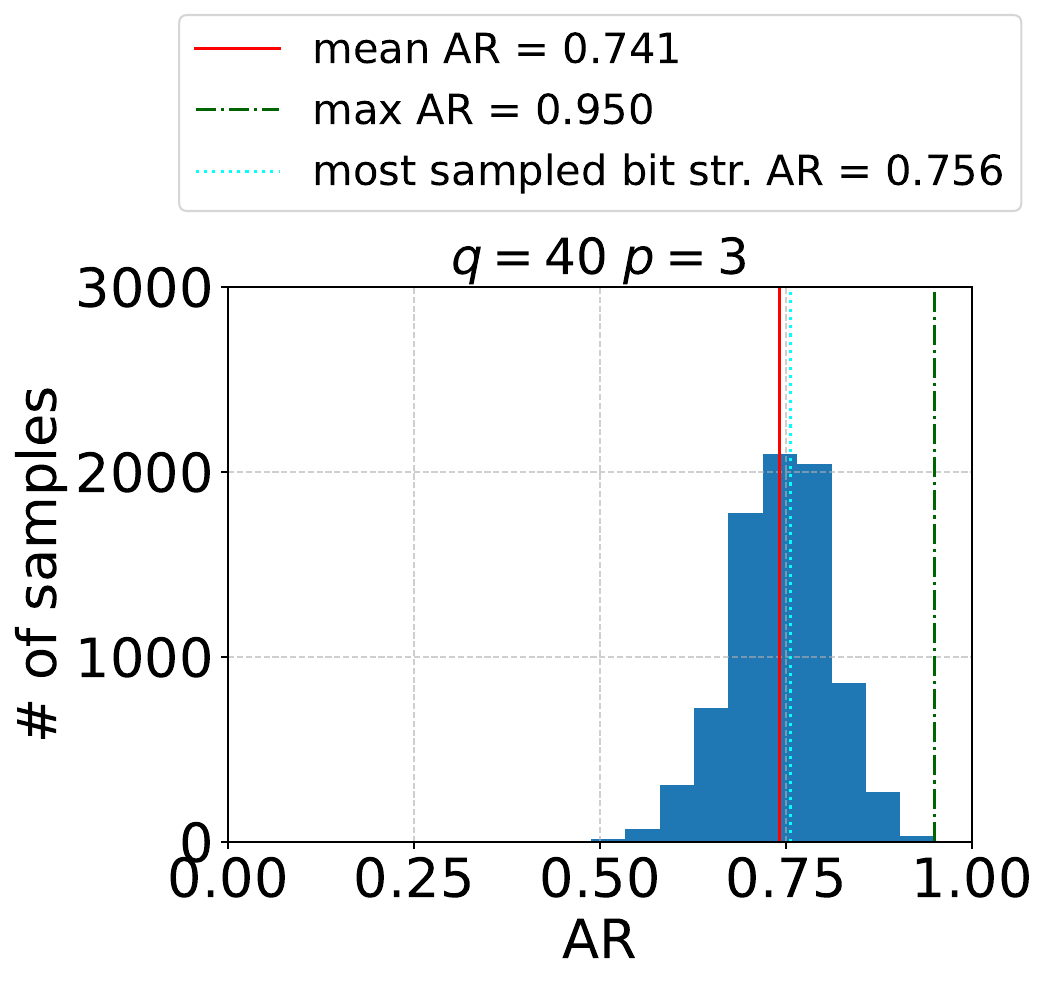}
         \caption{}
        \label{fig:hist_q40_p3}
    \end{subfigure}
    \hfill
    \begin{subfigure}[b]{0.32\textwidth}
        \centering
        \includegraphics[width=\textwidth]{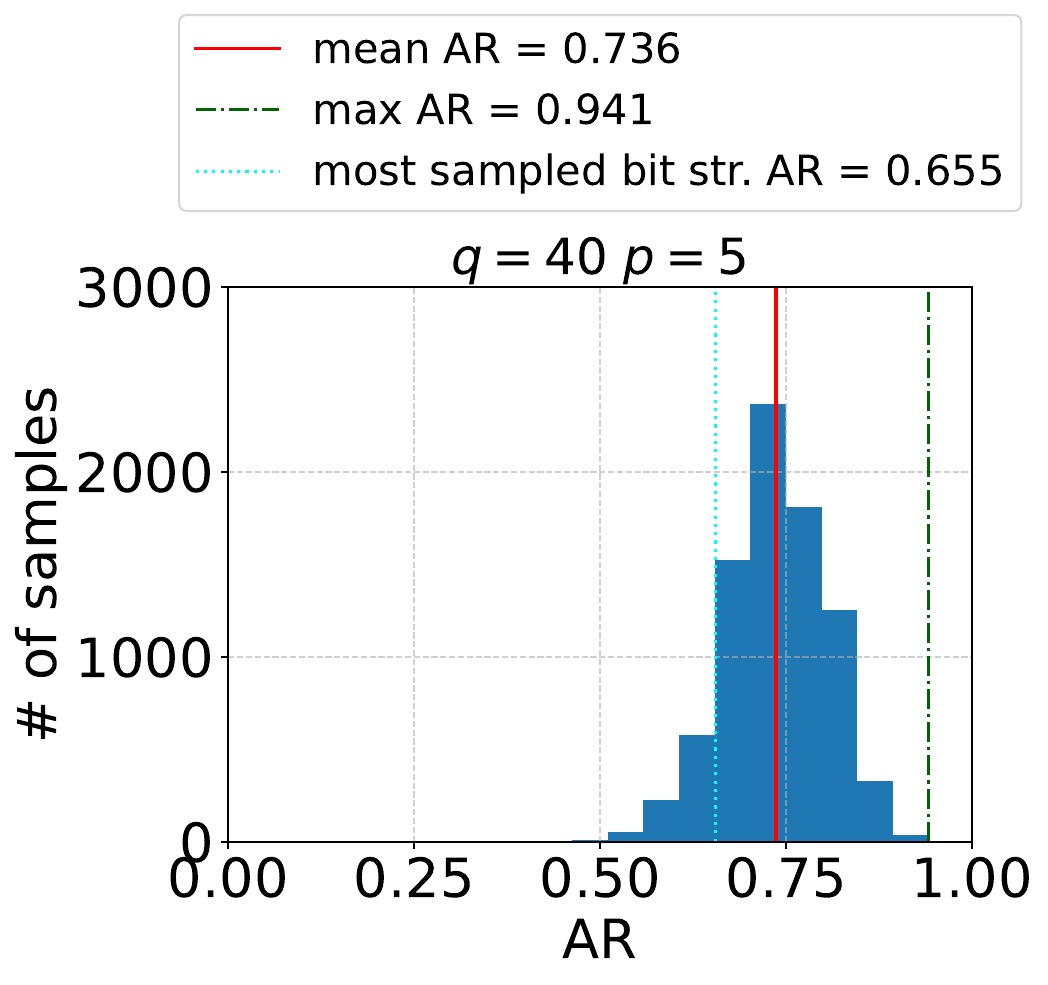}
         \caption{}
        \label{fig:hist_q40_p5}
    \end{subfigure}
    
    \vspace{0.4cm} 
    
    \begin{subfigure}[b]{0.32\textwidth}
        \centering
        \includegraphics[width=\textwidth]{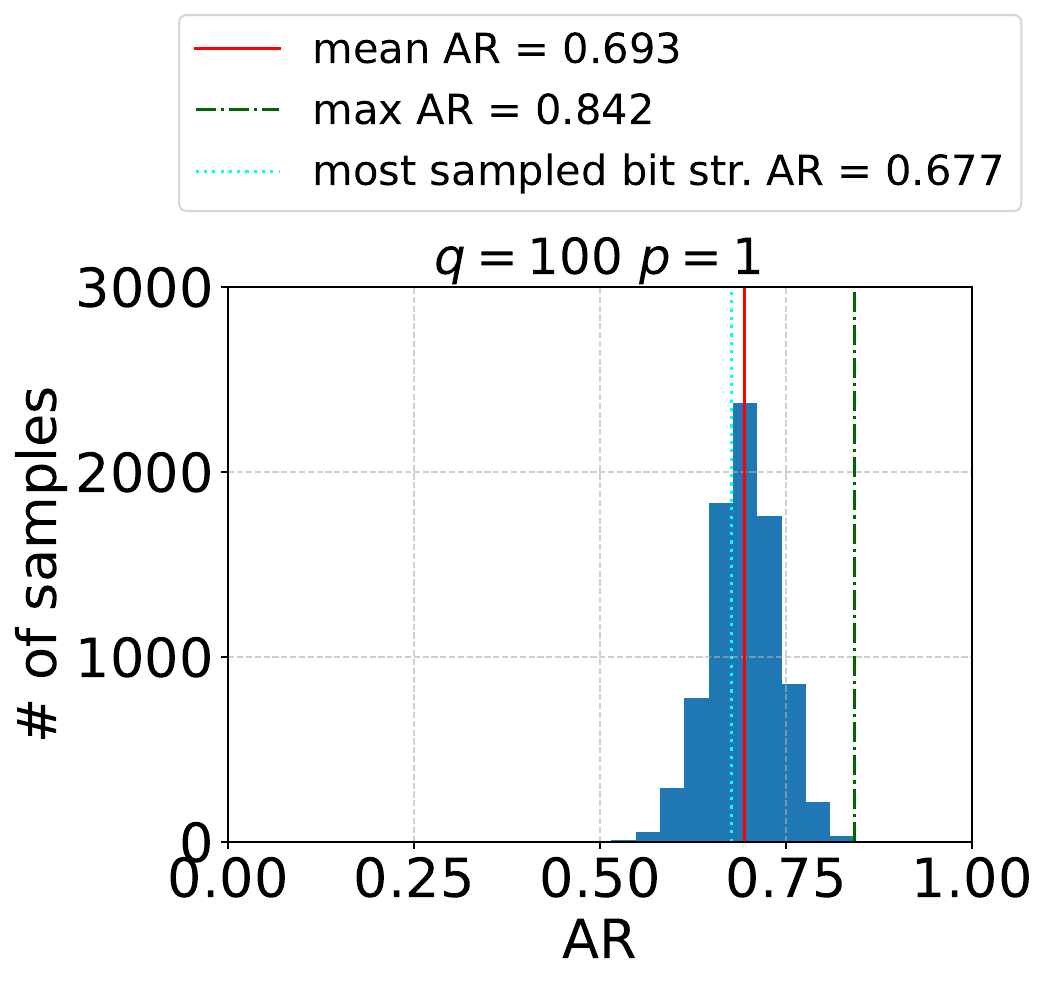}
         \caption{}
        \label{fig:hist_q100_p1}
    \end{subfigure}
    \hfill
    \begin{subfigure}[b]{0.32\textwidth}
        \centering
        \includegraphics[width=\textwidth]{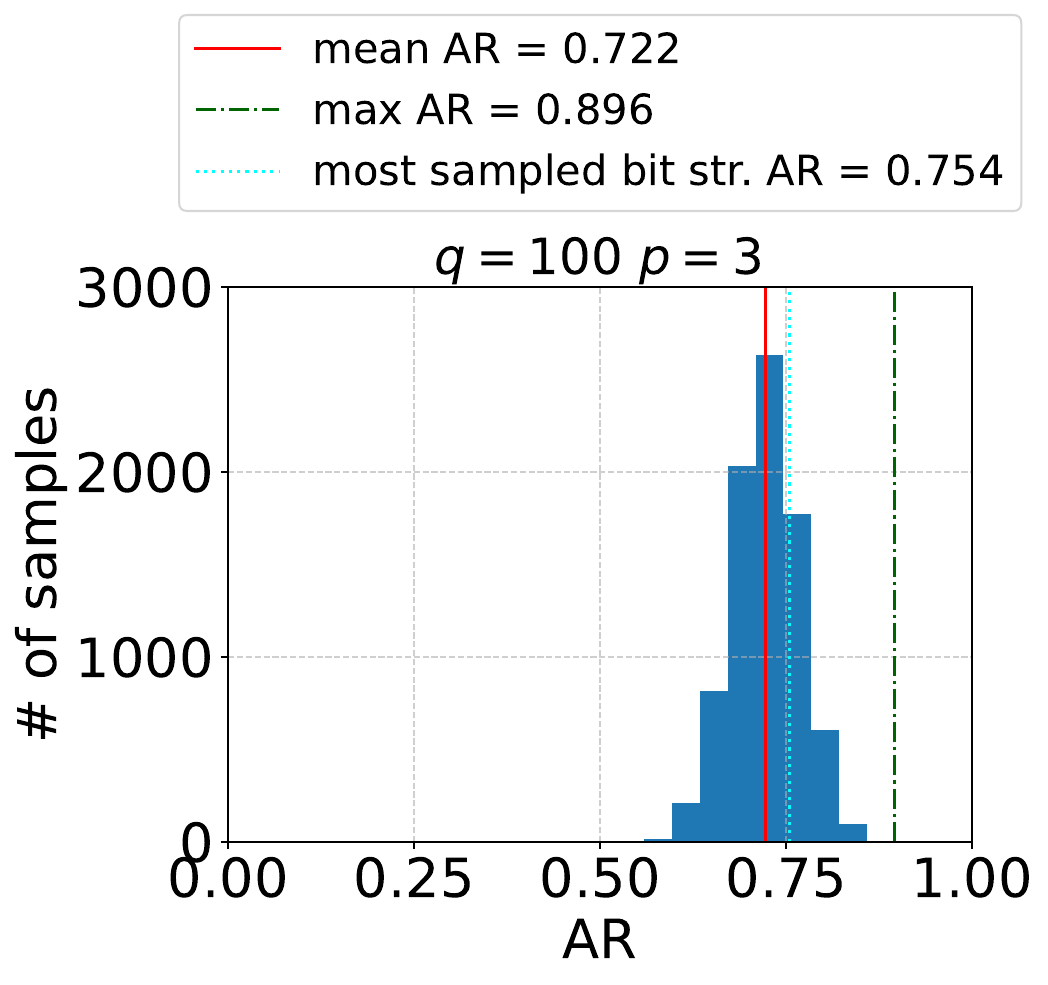}
         \caption{}
        \label{fig:hist_q100_p3}
    \end{subfigure}
    \hfill
    \begin{subfigure}[b]{0.32\textwidth}
        \centering
        \includegraphics[width=\textwidth]{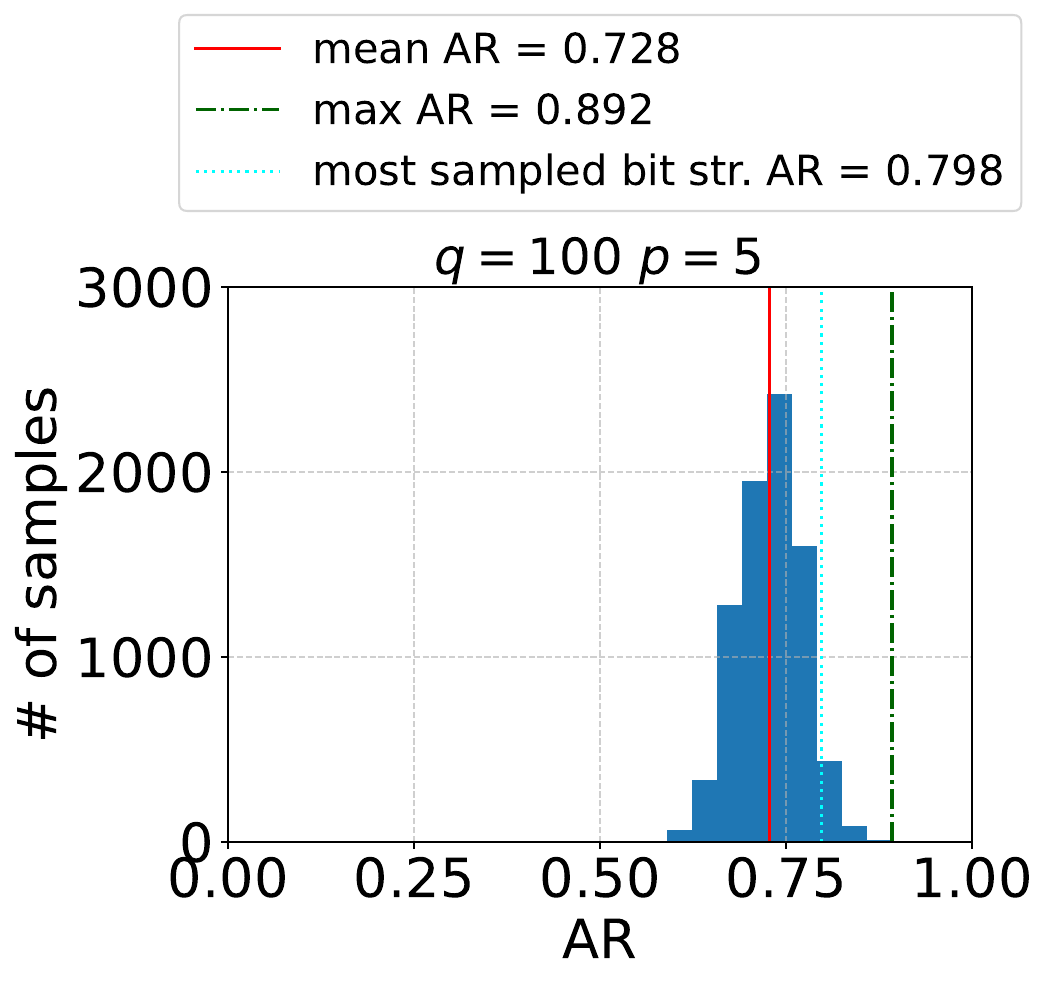}
        \caption{}
        \label{fig:hist_q100_p5}
    \end{subfigure}
    
    \caption{Approximation ratios (AR) of sampled bit strings at converged state across different qubit sizes $q \in \{20, 40, 100\}$ and ansatz layers $p \in \{1, 3, 5\}$ for experiments in Figure \ref{fig:all_qubit_conv_hardware}.}
    \label{fig:combined_histograms_grid}
\end{figure}

We now deploy FLICD on IBM quantum hardware and scale system size. Figure \ref{fig:all_qubit_conv_hardware} shows the convergences for all system sizes and Figure \ref{fig:combined_histograms_grid} shows how the converged state performs on the original graph. We summarize the setup for FLICD, stating only the hyper-parameters that are different from the experiments on quantum simulator in section \ref{subsubsec:Comparison of FLICD to  COBYLA and ICD}. 
\begin{itemize} 
    \item $\varepsilon_{rel}$: The relative error is chosen to be $\varepsilon_{rel} =  0.05 $.  
    \item $T$: Total layers updated is set to $T = 6p$, namely, each $\theta_j$ is updated $6$ times. 
\end{itemize}


\textbf{FLICD is deployable on quantum hardware.} The left and right columns of Figure~\ref{fig:all_qubit_conv_hardware} show the function value with respect to total shots and total QPU usage, respectively. The number of total shots and QPU usage as a function of system size $q$ and layers $p$ is non-trivial. For a fixed $q$, 
tripling $p$ 
only increases the total QPU usage by a factor less than $2$, 
while increasing $p$ by fivefold 
only doubles the QPU usage. 
These savings in QPU usage are primarily attributable to the algorithmic design and the adaptive frequency update technique. 
Since $S=p$, the number of adaptive frequency updates remains the same, regardless of $p$. As $p$ increases, the additional updates to parameters are fixed (potentially low) frequency updates, which require less resources than the adaptive frequency updates and full frequency updates.  

From the final objective values reached by FLICD, we observe that increasing $p$ yields consistently lower objective value with $q=20$ and $q = 100$. At $q=40$, FLICD reaches similar objective values for both $p=3$ and $p = 5$. 
This shows that FLICD may not always achieve a strictly better objective value as the number of layers increases, and may get stuck in local minima. Figure \ref{fig:combined_histograms_grid} shows the approximation ratios of the sampled bitstrings of the final state for the $20$, $40$, and $100$-qubit experiments, respectively. The average approximation ratio is improved with increasing $p$ for $q=20$ and $q=100$.

\section{Conclusions and discussions}
We have introduced Frequency-Learning Interpolation-based Coordinate Descent (FLICD) to minimize cost functions derived from expectation values of parameterized quantum circuits. Empirical results show that the Fourier coefficients of higher-order frequencies in the Fourier expansion of a uni-variate restriction function $f_j(\theta_j)$ are sparse in practice. Rather than 
manually selecting a frequency set, 
FLICD adaptively learns the key frequencies needed to reconstruct $f_j(\theta_j)$. 

We have benchmarked 
FLICD on Max-cut problems using random regular graphs and the LC-QAOA ansatz. In high-shot regimes, learning a smaller set of frequencies allowed FLICD to converge faster than the existing ICD method; in low-shot regimes, it achieved higher accuracy. FLICD also consistently outperformed COBYLA in accuracy across all shot regimes. On IBM quantum hardware, FLICD scaled effectively to learn important frequencies, with additional layers yielding lower 
objective values and higher mean approximation ratios at the 
system size of 100 qubits. Furthermore, because systematic noise primarily stretches function values rather than causing localized shifts, as hypothesized by \cite{jagerFastGradientfreeOptimization2025}, the locations of the restricted functions' minimizers remain largely unchanged. This allows our 
method to execute reliably on real hardware.


While our results demonstrate that FLICD scales effectively with system size, simulations and hardware experiments reveal that the LC-QAOA ansatz yields diminishing returns as the number of layers increases. Notably, the optimal objective value produced by FLICD for just a single layer of the linear chain ansatz is significantly lower than that of COBYLA, meaning additional layers beyond the first yield minimal relative gains. This trend is further supported by the approximation ratios—the histograms have only small relative improvement. Achieving results with an absolute higher approximation ratio requires improving not only the optimizer performance, but also the choice of ansatz. A slightly more expressive ansatz combined with FLICD may fair better at producing approximation ratios that are better in absolute terms.

\section*{Acknowledgements}
This work was supported by IBM through the IBM-RPI Future of Computing Research Collaboration. We thank Michael Sofka from FOCI RPI for his technical help in using the quantum hardware.

\printbibliography

@online{ibmquantum_estimatorv2_api,
	title = {{EstimatorV}2 (latest version) {\textbar} {IBM} Quantum Documentation},
	url = {https://quantum.cloud.ibm.com/docs/api/qiskit-ibm-runtime/estimator-v2},
	author = {{IBM Quantum}},
	urldate = {2026-11-02},
	date = {2026},
}

@online{ibmquantum_long_range_entanglement,
	title = {Long-range entanglement with dynamic circuits},
	url = {https://quantum.cloud.ibm.com/docs/en/tutorials/long-range-entanglement},
	author = {{IBM Quantum}},
	urldate = {2026-11-02},
	date = {2024},
}

@online{ibmquantum_fractionalgates,
	title = {Fractional gates},
	url = {https://quantum.cloud.ibm.com/docs/en/guides/fractional-gates},
	author = {{IBM Quantum}},
	date = {2026},
}

@misc{qiskit2024,
	title = {Quantum computing with Qiskit},
	doi = {10.48550/arXiv.2405.08810},
	author = {Javadi-Abhari, Ali and Treinish, Matthew and Krsulich, Kevin and Wood, Christopher J. and Lishman, Jake and Gacon, Julien and Martiel, Simon and Nation, Paul D. and Bishop, Lev S. and Cross, Andrew W. and Johnson, Blake R. and Gambetta, Jay M.},
	date = {2024},
	eprinttype = {arxiv},
	eprint = {2405.08810 [quant-ph]},
}

@article{bonet-monroigPerformanceComparisonOptimization2023,
	title = {Performance comparison of optimization methods on variational quantum algorithms},
	volume = {107},
	url = {https://link.aps.org/doi/10.1103/PhysRevA.107.032407},
	doi = {10.1103/PhysRevA.107.032407},
	pages = {032407},
	number = {3},
	journaltitle = {Physical Review A},
	shortjournal = {Phys. Rev. A},
	publisher = {American Physical Society},
	author = {Bonet-Monroig, Xavier and Wang, Hao and Vermetten, Diederick and Senjean, Bruno and Moussa, Charles and Bäck, Thomas and Dunjko, Vedran and O'Brien, Thomas E.},
	urldate = {2026-09-24},
	date = {2023-03-07},
}

@misc{cogburnHardwareAwareQAOAHoneypot2026a,
	title = {Hardware-Aware {QAOA} for Honeypot Traffic Partitioning on 100+ Qubit {IBM} Quantum Processors},
	url = {http://arxiv.org/abs/2606.09469},
	doi = {10.48550/arXiv.2606.09469},
	number = {{arXiv}:2606.09469},
	publisher = {{arXiv}},
	author = {Cogburn, Cameron V. and {DeCusatis}, Casimer and Spillane, Evan},
	urldate = {2026-09-24},
	date = {2026-06-08},
	eprinttype = {arxiv},
	eprint = {2606.09469 [quant-ph]},
	note = {version: 1},
}

@article{azfarImprovingQAOAFeasibility2026,
	title = {Improving {QAOA} feasibility and hardware performance for vehicle routing via linear chain and ramp schedules},
	rights = {2026 The Author(s)},
	issn = {2045-2322},
	url = {https://www.nature.com/articles/s41598-026-70696-y},
	doi = {10.1038/s41598-026-70696-y},
	journaltitle = {Scientific Reports},
	shortjournal = {Sci Rep},
	publisher = {Nature Publishing Group},
	author = {Azfar, Talha and Ke, Ruimin},
	urldate = {2026-09-24},
	date = {2026-09-10},
	langid = {english},
}

@article{anschuetzBarrenPlateausQuantum2022,
	title = {Beyond Barren Plateaus: Quantum Variational Algorithms Are Swamped With Traps},
	volume = {13},
	issn = {2041-1723},
	url = {http://arxiv.org/abs/2205.05786},
	doi = {10.1038/s41467-022-35364-5},
	shorttitle = {Beyond Barren Plateaus},
	pages = {7760},
	number = {1},
	journaltitle = {Nature Communications},
	shortjournal = {Nat Commun},
	author = {Anschuetz, Eric R. and Kiani, Bobak T.},
	urldate = {2026-09-24},
	date = {2022-12-15},
	eprinttype = {arxiv},
	eprint = {2205.05786 [quant-ph]},
}

@article{bittelTrainingVariationalQuantum2021,
	title = {Training Variational Quantum Algorithms Is {NP}-Hard},
	volume = {127},
	url = {https://link.aps.org/doi/10.1103/PhysRevLett.127.120502},
	doi = {10.1103/PhysRevLett.127.120502},
	pages = {120502},
	number = {12},
	journaltitle = {Physical Review Letters},
	shortjournal = {Phys. Rev. Lett.},
	publisher = {American Physical Society},
	author = {Bittel, Lennart and Kliesch, Martin},
	urldate = {2026-09-24},
	date = {2021-09-17},
}

@misc{pramanikOneCoordinateTime2026,
	title = {One Coordinate at a Time: Convergence Guarantees for Rotosolve in Variational Quantum Algorithms},
	url = {http://arxiv.org/abs/2604.25613},
	doi = {10.48550/arXiv.2604.25613},
	shorttitle = {One Coordinate at a Time},
	number = {{arXiv}:2604.25613},
	publisher = {{arXiv}},
	author = {Pramanik, Sayantan and Chandra, M. Girish},
	urldate = {2026-08-25},
	date = {2026-04-28},
	eprinttype = {arxiv},
	eprint = {2604.25613 [quant-ph]},
}

@article{pellow-jarmanEffectClassicalOptimizers2024,
	title = {The effect of classical optimizers and Ansatz depth on {QAOA} performance in noisy devices},
	volume = {14},
	rights = {2024 The Author(s)},
	issn = {2045-2322},
	url = {https://www.nature.com/articles/s41598-024-66625-6},
	doi = {10.1038/s41598-024-66625-6},
	pages = {16011},
	number = {1},
	journaltitle = {Scientific Reports},
	shortjournal = {Sci Rep},
	publisher = {Nature Publishing Group},
	author = {Pellow-Jarman, Aidan and {McFarthing}, Shane and Sinayskiy, Ilya and Park, Daniel K. and Pillay, Anban and Petruccione, Francesco},
	urldate = {2026-04-22},
	date = {2024-07-11},
	langid = {english},
}

@article{peruzzoVariationalEigenvalueSolver2014,
	title = {A variational eigenvalue solver on a photonic quantum processor},
	volume = {5},
	rights = {2014 The Author(s)},
	issn = {2041-1723},
	url = {https://www.nature.com/articles/ncomms5213},
	doi = {10.1038/ncomms5213},
	pages = {4213},
	number = {1},
	journaltitle = {Nature Communications},
	shortjournal = {Nat Commun},
	publisher = {Nature Publishing Group},
	author = {Peruzzo, Alberto and {McClean}, Jarrod and Shadbolt, Peter and Yung, Man-Hong and Zhou, Xiao-Qi and Love, Peter J. and Aspuru-Guzik, Alán and O’Brien, Jeremy L.},
	urldate = {2026-04-10},
	date = {2014-07-23},
	langid = {english},
}

@article{liEfficientRobustParameter2024,
	title = {Efficient and Robust Parameter Optimization of the Unitary Coupled-Cluster Ansatz},
	volume = {20},
	rights = {https://doi.org/10.15223/policy-029},
	issn = {1549-9618, 1549-9626},
	url = {https://pubs.acs.org/doi/10.1021/acs.jctc.4c00155},
	doi = {10.1021/acs.jctc.4c00155},
	pages = {3683--3696},
	number = {9},
	journaltitle = {Journal of Chemical Theory and Computation},
	shortjournal = {J. Chem. Theory Comput.},
	author = {Li, Weitang and Ge, Yufei and Zhang, Shi-Xin and Chen, Yu-Qin and Zhang, Shengyu},
	urldate = {2026-04-07},
	date = {2024-05-14},
	langid = {english},
}

@article{nakanishiSequentialMinimalOptimization2020,
	title = {Sequential minimal optimization for quantum-classical hybrid algorithms},
	volume = {2},
	issn = {2643-1564},
	url = {http://arxiv.org/abs/1903.12166},
	doi = {10.1103/PhysRevResearch.2.043158},
	pages = {043158},
	number = {4},
	journaltitle = {Physical Review Research},
	shortjournal = {Phys. Rev. Research},
	author = {Nakanishi, Ken M. and Fujii, Keisuke and Todo, Synge},
	urldate = {2026-04-06},
	date = {2020-10-29},
	eprinttype = {arxiv},
	eprint = {1903.12166 [quant-ph]},
}

@misc{parrishJacobiDiagonalizationAnderson2019,
	title = {A Jacobi Diagonalization and Anderson Acceleration Algorithm For Variational Quantum Algorithm Parameter Optimization},
	url = {http://arxiv.org/abs/1904.03206},
	doi = {10.48550/arXiv.1904.03206},
	number = {{arXiv}:1904.03206},
	publisher = {{arXiv}},
	author = {Parrish, Robert M. and Iosue, Joseph T. and Ozaeta, Asier and {McMahon}, Peter L.},
	urldate = {2026-04-06},
	date = {2019-04-05},
	eprinttype = {arxiv},
	eprint = {1904.03206 [quant-ph]},
}

@article{ostaszewskiStructureOptimizationParameterized2021,
	title = {Structure optimization for parameterized quantum circuits},
	volume = {5},
	issn = {2521-327X},
	url = {http://arxiv.org/abs/1905.09692},
	doi = {10.22331/q-2021-01-28-391},
	pages = {391},
	journaltitle = {Quantum},
	shortjournal = {Quantum},
	author = {Ostaszewski, Mateusz and Grant, Edward and Benedetti, Marcello},
	urldate = {2026-04-06},
	date = {2021-01-28},
	eprinttype = {arxiv},
	eprint = {1905.09692 [quant-ph]},
}

@article{jagerFastGradientfreeOptimization2025,
	title = {Fast gradient-free optimization of excitations in variational quantum eigensolvers},
	volume = {8},
	issn = {2399-3650},
	url = {http://arxiv.org/abs/2409.05939},
	doi = {10.1038/s42005-025-02375-9},
	pages = {418},
	number = {1},
	journaltitle = {Communications Physics},
	shortjournal = {Commun Phys},
	author = {Jäger, Jonas and Kaldenbach, Thierry Nicolas and Haas, Max and Schultheis, Erik},
	urldate = {2026-04-06},
	date = {2025-10-30},
	eprinttype = {arxiv},
	eprint = {2409.05939 [quant-ph]},
}

@misc{vidalCalculusParameterizedQuantum2018,
	title = {Calculus on parameterized quantum circuits},
	url = {http://arxiv.org/abs/1812.06323},
	doi = {10.48550/arXiv.1812.06323},
	number = {{arXiv}:1812.06323},
	publisher = {{arXiv}},
	author = {Vidal, Javier Gil and Theis, Dirk Oliver},
	urldate = {2026-04-06},
	date = {2018-12-31},
	eprinttype = {arxiv},
	eprint = {1812.06323 [quant-ph]},
}

@article{cookStochasticSamplingComputer1986,
	title = {Stochastic sampling in computer graphics},
	volume = {5},
	issn = {0730-0301, 1557-7368},
	url = {https://dl.acm.org/doi/10.1145/7529.8927},
	doi = {10.1145/7529.8927},
	pages = {51--72},
	number = {1},
	journaltitle = {{ACM} Transactions on Graphics},
	shortjournal = {{ACM} Trans. Graph.},
	author = {Cook, Robert L.},
	urldate = {2026-01-28},
	date = {1986-01},
	langid = {english},
}

@article{laiInterpolationbasedCoordinateDescent2026,
	title = {Interpolation-based coordinate descent method for parameterized quantum circuits},
	rights = {2026 The Author(s)},
	issn = {2399-3650},
	url = {https://www.nature.com/articles/s42005-025-02473-8},
	doi = {10.1038/s42005-025-02473-8},
	journaltitle = {Communications Physics},
	shortjournal = {Commun Phys},
	publisher = {Nature Publishing Group},
	author = {Lai, Zhijian and Hu, Jiang and Ko, Taehee and Wu, Jiayuan and An, Dong},
	urldate = {2026-01-21},
	date = {2026-01-07},
	langid = {english},
}

@misc{wangDepthIndependentLinearChain2025,
	title = {A Depth-Independent Linear Chain Ansatz for Large-Scale Quantum Approximate Optimization},
	url = {http://arxiv.org/abs/2509.17296},
	doi = {10.48550/arXiv.2509.17296},
	number = {{arXiv}:2509.17296},
	publisher = {{arXiv}},
	author = {Wang, Zixu and Mandell, Jack and Xu, Yangyang and Shi, Jian},
	urldate = {2025-12-26},
	date = {2025-09-22},
	eprinttype = {arxiv},
	eprint = {2509.17296 [quant-ph]},
}

@misc{farhiQuantumApproximateOptimization2014,
	title = {A Quantum Approximate Optimization Algorithm},
	url = {http://arxiv.org/abs/1411.4028},
	doi = {10.48550/arXiv.1411.4028},
	number = {{arXiv}:1411.4028},
	publisher = {{arXiv}},
	author = {Farhi, Edward and Goldstone, Jeffrey and Gutmann, Sam},
	urldate = {2025-09-30},
	date = {2014-11-14},
	eprinttype = {arxiv},
	eprint = {1411.4028 [quant-ph]},
}

\newpage

\section{Appendix}

\appendix
\renewcommand{\thesubsection}{\Alph{subsection}}
\counterwithin{figure}{subsection}

The Appendix section contains further details about theory, experiment setup, and results. In particular, Appendix \ref{sec:lemmas} contains the proof of Theorem \ref{theorem:MSE_optimality}. Appendix \ref{appendix:circuit_exection} discusses circuit execution on IBM hardware using Qiskit. Appendix \ref{appendix:Graph Transformations and Setup} contains details relating to Max-cut problem setup. Lastly, Appendix \ref{appendix:Comparison of algorithm performance over each run} provides further detailed results from experiments in section \ref{subsubsec:Comparison of FLICD to  COBYLA and ICD}. 

\subsection{Proofs of lemmas and theorem} \label{sec:lemmas}

We state and prove two lemmas that will be used to prove Theorem \ref{theorem:MSE_optimality}. 
\begin{lemma} \label{lemma:vandermonde}
Let $P$ and $r$ be positive integers with $P\geq 2r+1$. Define the $P$-th roots of unity $\omega_k = e^{2\pi i \frac{k}{P}}$ and construct the rectangular Vandermonde matrix with power $2r$ by   
\begin{align*}
   V := \begin{bmatrix}
        1 & \omega_0   & \omega_0 ^{2} & \cdots & \omega_0 ^{2r}\\
      1 & \omega_1   & \omega_1 ^{2} & \cdots & \omega_1 ^{2r}\\
        \vdots & \vdots & \ddots   & \vdots & \vdots   \\
         1 & \omega_{P-1}   & \omega_{P-1} ^{2} & \cdots & \omega_{P-1} ^{2r}\\  
    \end{bmatrix} \in \mathbb{C}^{P \times (2r+1)}.
\end{align*}
Then 
$V^\dagger V = PI \in \mathbb{C}^{(2r+1)\times (2r+1)}$. 
\end{lemma}
\begin{proof}
By $\omega_k^\dagger = e^{-2\pi i \frac{k}{P}}$, it is straightforward to have
\begin{align*}
    V^\dagger V =\begin{bmatrix}
    1 & 1 & \cdots & 1 \\
    \omega_0^{-1} & \omega_1^{-1} & \cdots & \omega_{P-1}^{-1} \\
    \omega_0^{-2} & \omega_1^{-2} & \cdots & \omega_{P-1}^{-2} \\
    \vdots & \vdots & \ddots & \vdots \\
    \omega_0^{-2r} & \omega_1^{-2r} & \cdots & \omega_{P-1}^{-2r}
\end{bmatrix} 
    \begin{bmatrix}
        1 & \omega_0   & \omega_0 ^{2} & \cdots & \omega_0 ^{2r}\\
      1 & \omega_1   & \omega_1 ^{2} & \cdots & \omega_1 ^{2r}\\
        \vdots & \vdots & \ddots   & \vdots & \vdots   \\
         1 & \omega_{P-1}   & \omega_{P-1} ^{2} & \cdots & \omega_{P-1} ^{2r}\\  
    \end{bmatrix}. 
\end{align*}
By looking at the entries of the above matrix component-wise:
\begin{align*}
    [V^\dagger V]_{jk} = \sum_{\ell = 0}^{P-1} \omega_\ell^{k-j}.
\end{align*}
If $j = k$, then $[V^\dagger V]_{jk} = \sum_{\ell = 0}^{P-1} \omega_\ell^{0} = P$. Now, take the case when $j \neq k$. Since $1\leq j,k\leq 2r+1$, it follows that $k - j \in \{ \pm 1,\pm 2,\dots, \pm 2r\}$. Hence $-P < -2r \leq k - j \leq 2r < P$. By applying Lemma 2 of \cite{laiInterpolationbasedCoordinateDescent2026}, it follows that $[V^\dagger V]_{jk} = 0$. It then follows that $V^\dagger V = PI$. 
\end{proof}

\begin{lemma} \label{lemma:trace_formula}
    Under Assumptions \ref{assump:noisy-est} and \ref{assump:eq-dist-freqs}, let $P$ be an integer such that $P \geq 2r+1$. Define $\bold{x}\in\mathbb{R}^P$ with $x_k = \frac{2\pi}{P}k, \forall\, k=0,1,\ldots,P-1$ and denote $ A_\bold{x} = A_{\bold{\Omega}, \bold{x}}$ by the definition in \eqref{eq:def-A-Omage-x}, then $\mathrm{tr}( (A_\bold{x}^{\dagger} A_\bold{x}) ^{-1}) = \frac{2(2r+1)}{P}. $ 
\end{lemma}
    
\begin{proof}
The proof follows from similar analysis from Appendix D of \cite{laiInterpolationbasedCoordinateDescent2026} of the square interpolation matrix. The difference is that the interpolation matrix is rectangular and some of the intermediate definitions need adjustment. For clarity, we give a brief analysis emphasizing the new definitions and dimensions to highlight the differences. Firstly, recall that the interpolation matrix in least-squares approximation is given by
\begin{align*}
    A_\bold{x} = \begin{bmatrix}
        1/\sqrt{2}&  \cos (\Omega_1x_0) & \sin (\Omega_1x_0) & \cdots &  \cos (\Omega_r x_0) & \sin (\Omega_r x_0) \\
        1/\sqrt{2}&  \cos (\Omega_1x_1) & \sin (\Omega_1x_1) & \cdots  & \cos (\Omega_r x_1)  & \sin (\Omega_r x_1)\\
        \vdots & \vdots & \vdots & \ddots & \vdots & \vdots  \\
        1/\sqrt{2}&  \cos (\Omega_1x_{P-1}) & \sin (\Omega_1x_{P-1}) & \cdots  & \cos (\Omega_r x_{P-1}) & \sin (\Omega_r x_{P-1})\\
    \end{bmatrix} \in \mathbb{R}^{P \times (2r+1)}.
\end{align*}
Define $\omega_k := e^{ix_k}= e^{2\pi i \frac{k}{P}} $. 
Now, define the interpolation matrix for the standard Fourier representation \eqref{c_rep},  
\begin{align*}
    A_\bold{x}^c := \begin{bmatrix}
        \omega_0 ^{-\Omega_r } & \cdots & \omega_0 ^{-\Omega_1 } &  1& \omega_0 ^{\Omega_1}&  \cdots & \omega_0 ^{\Omega_r}\\
        \omega_1 ^{-\Omega_r } & \cdots & \omega_1 ^{-\Omega_1 } &  1& \omega_1 ^{\Omega_1}&  \cdots & \omega_1 ^{\Omega_r}\\
        \vdots & \ddots & \vdots   & \vdots & \vdots & \ddots & \vdots  \\
        \omega_{P-1} ^{-\Omega_r } & \cdots & \omega_{P-1} ^{-\Omega_1 } &  1& \omega_{P-1} ^{\Omega_1}&  \cdots & \omega_{P-1} ^{\Omega_r}\\ 
    \end{bmatrix} \in \mathbb{C}^{P \times (2r+1)}.
\end{align*}
Then 
it holds that $A_\bold{x} = A_\bold{x}^c C$ by using the linear transformation \eqref{ab_to_c_coeff_change} where
\begin{align*}
    C:= \begin{bmatrix}
        & & & & \frac{1}{2}&\frac{i}{2} \\
         & & &\ddots  & & \\
          & \frac{1}{2}&\frac{i}{2} & & & \\
           \frac{1}{\sqrt{2}}& & & & & \\
            & \frac{1}{2}&-\frac{i}{2} & & & \\
             & & & \ddots & & \\
             & & & & \frac{1}{2}&-\frac{i}{2} \\
    \end{bmatrix} \in \mathbb{C}^{(2r+1)\times(2r+1)}
\end{align*}
satisfies $C^\dagger C = C C^\dagger = \frac{1}{2}I$. Next, we decompose $A_\bold{x}^c$. To do this, define
\begin{align*}
    D_x : = \text{Diag} (\omega_0 ^{-\Omega_r},\omega_1 ^{-\Omega_r},\dots,\omega_{P-1} ^{-\Omega_r}) \in \mathbb{C}^{P \times P}
\end{align*}
which satisfies $D_x^\dagger D_x = I$. Also define the rectangular Vandermonde matrix
\begin{align*}
   V_\bold{x} := \begin{bmatrix}
        1 & \omega_0 ^{\Omega_r - \Omega_{r-1} } & \cdots & \omega_0 ^{\Omega_r -\Omega_1 } & \omega_0 ^{\Omega_r}& \omega_0 ^{\Omega_r + \Omega_1}& \cdots & \omega_0 ^{2\Omega_r}\\
         1 & \omega_1 ^{\Omega_r - \Omega_{r-1} } & \cdots & \omega_1 ^{\Omega_r -\Omega_1 } & \omega_1 ^{\Omega_r}& \omega_1 ^{\Omega_r + \Omega_1}&  \cdots & \omega_1 ^{2\Omega_r}\\
        \vdots & \vdots & \ddots   & \vdots & \vdots & \vdots & \ddots & \vdots  \\
         1 & \omega_{P-1} ^{\Omega_r - \Omega_{r-1} } & \cdots & \omega_{P-1} ^{\Omega_r -\Omega_1 } & \omega_{P-1} ^{\Omega_r} & \omega_{P-1} ^{\Omega_r + \Omega_1}&  \cdots & \omega_{P-1} ^{2\Omega_r}    
    \end{bmatrix} \in \mathbb{C}^{P \times (2r+1)}.
\end{align*}
Then, $V_\bold{x}$ reduces down to the $P$-th root of unity Vandermonde matrix with power $2r$, since the frequencies are equidistant integers, by Assumption \ref{assump:eq-dist-freqs}. This can be seen by noting that $\Omega_r \pm \Omega_{j} = r\pm j$ for $j = 0,1,2,\dots,r$ with $\Omega_0 = 0$ for convenience of notation. Hence, by Lemma \ref{lemma:vandermonde}, $V_\bold{x}^\dagger V_\bold{x} = PI$. It can be shown that $A_\bold{x}^c =  D_\bold{x}V_\bold{x}$. Hence,
\begin{align*}
    A_\bold{x}^\dagger A_\bold{x} = (C^\dagger V_\bold{x}^\dagger D_\bold{x}^\dagger)( D_\bold{x} V_\bold{x} C) = C^\dagger V_\bold{x} \cdot I \cdot V_\bold{x} C = C^\dagger\cdot PI\cdot C = \frac{P}{2}I .
\end{align*}
Therefore, $(A_\bold{x}^\dagger A_\bold{x})^{-1} = \frac{2}{P}I$ and $\text{tr}((A_\bold{x}^\dagger A_\bold{x})^{-1}) = \frac{2(2r+1)}{P}.$
\end{proof}

\subsubsection{Proof of theorem \ref{theorem:MSE_optimality}}
\begin{proof}

In Lemma \ref{lemma:trace_formula}, we show that with the choice of $\bold{x}^*$ defined as $x_k^* = \frac{2\pi}{P}k, \forall\, k=0,1,\ldots,P-1$,  $\mathrm{tr}( (A_{\bold{x}^*}^\dagger A_{\bold{x}^*}) ^{-1}) = \frac{2(2r+1)}{P}$.  Hence, by \eqref{eq:E[SE] expression}, it follows that
\begin{align*}
    \mathbb{E}[\mathrm{SE}(\bold{\hat{z}_{\bold{x}^*}^*})] = \frac{2\sigma^2(2r+1)}{nP}.
\end{align*}
It suffices to show that for any other $\bold{x}$ set of sample points,
\begin{align*}
    \mathbb{E}[\mathrm{SE}(\bold{\hat{z}_x^*})] \geq \frac{2\sigma^2(2r+1)}{nP}.
\end{align*}
The proof method is an extension of the proof of Theorem 1 in \cite{laiInterpolationbasedCoordinateDescent2026}. Given the Vandermonde matrix $V_{\bold{x}} \in \mathbb{C}^{P \times (2r+1)}$, $V_{\bold{x}}^\dagger V_x = PI$. Hence, $\mathrm{tr}(V_{\bold{x}}^\dagger V_{\bold{x}} ) = (2r+1)P$. Also, $CC^\dagger  = \frac{1}{2}I$. Therefore, 
\begin{align*}
    \mathrm{tr}(A_{\bold{x}}^\dagger A_{\bold{x}}) = \mathrm{tr}(C^\dagger V_{\bold{x}}^\dagger D_{\bold{x}}^\dagger \cdot D_{\bold{x}}V_{\bold{x}}C) = \frac{1}{2}tr (V_{\bold{x}}^\dagger V_{\bold{x}}) = \frac{P(2r+1)}{2}
\end{align*}
and
\begin{align*}
    \mathbb{E}[\mathrm{SE}(\bold{\hat{z}_x^*})] = \frac{\sigma^2}{n} \mathrm{tr}( ( A_{\bold{x}}^\dagger A_{\bold{x}} )^{-1}) \geq  \frac{\sigma^2}{n} \cdot \frac{(2r+1)^2}{\mathrm{tr}(A_{\bold{x}}^\dagger A_{\bold{x}})} = \frac{\sigma^2(2r+1)^2}{\frac{nP(2r+1)}{2}} = \frac{2\sigma^2(2r+1)}{nP},
\end{align*}
which completes the proof.
\end{proof}

\subsection{Circuit execution} \label{appendix:circuit_exection}

\subsubsection{Setting shots}

Correctly setting the number of shots plays a central role in the design of FLICD. We describe how this is done in practice in Qiskit \cite{qiskit2024}. Suppose that the desired number of shots to estimate an observable expectation is set to $n$. The number of shots is not directly specified, but a \textit{precision} argument to EstimatorV2 \cite{ibmquantum_estimatorv2_api} is given. Setting $\textit{precision}= \frac{1}{\sqrt{n}}$ gives an allowed number of shots $n$. 

\subsubsection{Circuit transpilation}

\begin{figure}[htbp] 
    \centering
    \begin{subfigure}[c]{0.18\textwidth}
        \centering
        \includegraphics[width=\linewidth]{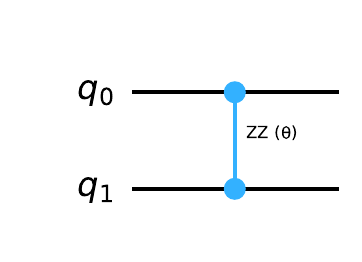}
        
    \end{subfigure}
    \hfill
    \begin{subfigure}[c]{0.74\textwidth}
        \centering
        \includegraphics[width=\linewidth]{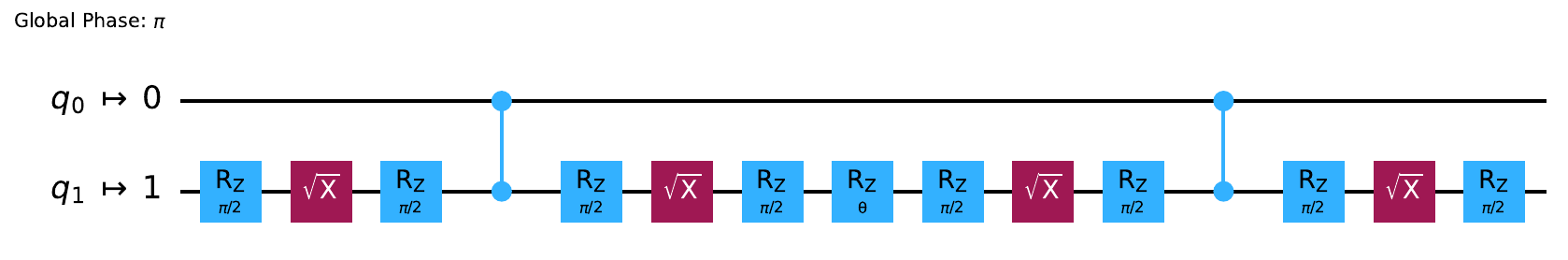}
    
    \end{subfigure}
    \caption{Logical RZZ($\theta$) gate  after transpilation on Heron processor with fractional gates (left) and without fractional gates (right).}
    \label{fig:frac_gate}
\end{figure}

When using IBM quantum hardware, the QPU backend determines how the circuit is transpiled into physical basis-native gate operations. The LC-QAOA ansatz used in experiments is composed of mostly two-qubit RZZ($\theta$) and single-qubit RX($\theta$) gates. The specific device backend will determine how these gates are transpiled. For Nighthawk r1 processors, the basis gates are from the set \{CZ, Id, RX, RZ, SX, X\}. Thus, RZZ and RX must be in general broken down into sequences of gates in the basis set when transpiling circuits onto Nighthawk r1 processors. On Heron r2 processors, the basis set includes an optional set of fractional gates \cite{ibmquantum_fractionalgates}, which are the RZZ and RX gates, that can be added to the existing set of basis gates. For arbitrary rotation angles $\theta \in [0,\frac{\pi} {2}]$, RZZ($\theta$) can be executed at the transpilation level. Similarly, RX($\theta$) can be executed at any angle $\theta$ at the transpilation level. Figure \ref{fig:frac_gate} shows the transpilation of an RZZ($\theta$) on an Heron r2 processor.  It has been shown that using fractional gates can dramatically reduce depth circuits of LC-QAOA circuits \cite{wangDepthIndependentLinearChain2025}.

When performing transpilation, \textit{optimization\_level} is set to 1. Above $\textit{optimization\_level} = 1$, we have witnessed that the transpiler can often give circuits with invalid gate angles for RZZ gates outside the $[0,\frac{\pi}{2}]$ interval, even if FoldRzzAngle is specified to the pass manager.

When mapping logical qubits to physical qubits, we choose the highest fidelity linear chain on the hardware topology. The layer fidelity string in backend properties can be used for this task, see \cite{ibmquantum_long_range_entanglement} for more information. 

\subsection{Graph setup} \label{appendix:Graph Transformations and Setup}

\subsubsection{Weights of graph instances} \label{subappendix:weights}

Experiments from sections \ref{sec:Empirical variance and norm experiments} and \ref{sec:Adaptive update experiments} both use the same 16-vertex graph in Figure \ref{fig:16 vertex graph}, though with different weight distributions. We state the weighted adjacency matrix for each instance. Given a graph $G=(V,E)$ with weight function $w:E \rightarrow \mathbb{R}$, the weighted adjacency matrix $A$ is given by  $A_{ij} = w_{ij}$ if $(i,j) \in E$ and $0$ otherwise. 
In section \ref{sec:Empirical variance and norm experiments}, the weighted adjacency matrix is 
\begin{align} \label{matrix:weights_1,2}
    \begin{bmatrix}
0 & 2 & 0 & 0 & 2 & 0 & 0 & 0 & 1 & 0 & 2 & 0 & 0 & 1 & 0 & 0 \\
2 & 0 & 1 & 0 & 0 & 2 & 0 & 0 & 0 & 0 & 1 & 0 & 2 & 0 & 0 & 0 \\
0 & 1 & 0 & 2 & 0 & 2 & 0 & 0 & 0 & 1 & 0 & 0 & 0 & 1 & 0 & 0 \\
0 & 0 & 2 & 0 & 2 & 0 & 0 & 1 & 0 & 2 & 2 & 0 & 0 & 0 & 0 & 0 \\
2 & 0 & 0 & 2 & 0 & 2 & 0 & 0 & 0 & 0 & 0 & 0 & 0 & 0 & 2 & 1 \\
0 & 2 & 2 & 0 & 2 & 0 & 2 & 0 & 0 & 0 & 0 & 1 & 0 & 0 & 0 & 0 \\
0 & 0 & 0 & 0 & 0 & 2 & 0 & 2 & 0 & 0 & 0 & 0 & 1 & 1 & 0 & 1 \\
0 & 0 & 0 & 1 & 0 & 0 & 2 & 0 & 1 & 0 & 0 & 0 & 1 & 0 & 0 & 2 \\
1 & 0 & 0 & 0 & 0 & 0 & 0 & 1 & 0 & 2 & 0 & 1 & 0 & 0 & 0 & 2 \\
0 & 0 & 1 & 2 & 0 & 0 & 0 & 0 & 2 & 0 & 1 & 0 & 0 & 0 & 2 & 0 \\
2 & 1 & 0 & 2 & 0 & 0 & 0 & 0 & 0 & 1 & 0 & 1 & 0 & 0 & 0 & 0 \\
0 & 0 & 0 & 0 & 0 & 1 & 0 & 0 & 1 & 0 & 1 & 0 & 1 & 0 & 1 & 0 \\
0 & 2 & 0 & 0 & 0 & 0 & 1 & 1 & 0 & 0 & 0 & 1 & 0 & 2 & 0 & 0 \\
1 & 0 & 1 & 0 & 0 & 0 & 1 & 0 & 0 & 0 & 0 & 0 & 2 & 0 & 2 & 0 \\
0 & 0 & 0 & 0 & 2 & 0 & 0 & 0 & 0 & 2 & 0 & 1 & 0 & 2 & 0 & 2 \\
0 & 0 & 0 & 0 & 1 & 0 & 1 & 2 & 2 & 0 & 0 & 0 & 0 & 0 & 2 & 0 
\end{bmatrix},
\end{align}
and in section \ref{sec:Adaptive update experiments} the weighted adjacency matrix is
\begin{align} \label{weights_1,2,4}
    \begin{bmatrix} 
0 & 4 & 0 & 0 & 1 & 0 & 0 & 0 & 4 & 0 & 4 & 0 & 0 & 1 & 0 & 0 \\
4 & 0 & 4 & 0 & 0 & 4 & 0 & 0 & 0 & 0 & 4 & 0 & 4 & 0 & 0 & 0 \\
0 & 4 & 0 & 1 & 0 & 1 & 0 & 0 & 0 & 4 & 0 & 0 & 0 & 2 & 0 & 0 \\
0 & 0 & 1 & 0 & 2 & 0 & 0 & 1 & 0 & 4 & 2 & 0 & 0 & 0 & 0 & 0 \\
1 & 0 & 0 & 2 & 0 & 2 & 0 & 0 & 0 & 0 & 0 & 0 & 0 & 0 & 4 & 2 \\
0 & 4 & 1 & 0 & 2 & 0 & 4 & 0 & 0 & 0 & 0 & 4 & 0 & 0 & 0 & 0 \\
0 & 0 & 0 & 0 & 0 & 4 & 0 & 4 & 0 & 0 & 0 & 0 & 1 & 1 & 0 & 1 \\
0 & 0 & 0 & 1 & 0 & 0 & 4 & 0 & 4 & 0 & 0 & 0 & 1 & 0 & 0 & 2 \\
4 & 0 & 0 & 0 & 0 & 0 & 0 & 4 & 0 & 2 & 0 & 4 & 0 & 0 & 0 & 1 \\
0 & 0 & 4 & 4 & 0 & 0 & 0 & 0 & 2 & 0 & 1 & 0 & 0 & 0 & 4 & 0 \\
4 & 4 & 0 & 2 & 0 & 0 & 0 & 0 & 0 & 1 & 0 & 1 & 0 & 0 & 0 & 0 \\
0 & 0 & 0 & 0 & 0 & 4 & 0 & 0 & 4 & 0 & 1 & 0 & 4 & 0 & 2 & 0 \\
0 & 4 & 0 & 0 & 0 & 0 & 1 & 1 & 0 & 0 & 0 & 4 & 0 & 2 & 0 & 0 \\
1 & 0 & 2 & 0 & 0 & 0 & 1 & 0 & 0 & 0 & 0 & 0 & 2 & 0 & 4 & 0 \\
0 & 0 & 0 & 0 & 4 & 0 & 0 & 0 & 0 & 4 & 0 & 2 & 0 & 4 & 0 & 4 \\
0 & 0 & 0 & 0 & 2 & 0 & 1 & 2 & 1 & 0 & 0 & 0 & 0 & 0 & 4 & 0 \\
\end{bmatrix}.
\end{align}
\subsection{Further results} \label{appendix:Comparison of algorithm performance over each run}

\subsubsection{Comparison of algorithm performance over each individual run }

\begin{figure}[htbp]
    \centering
    
    \begin{subfigure}[b]{0.32\textwidth}
        \centering
        \includegraphics[width=\textwidth]{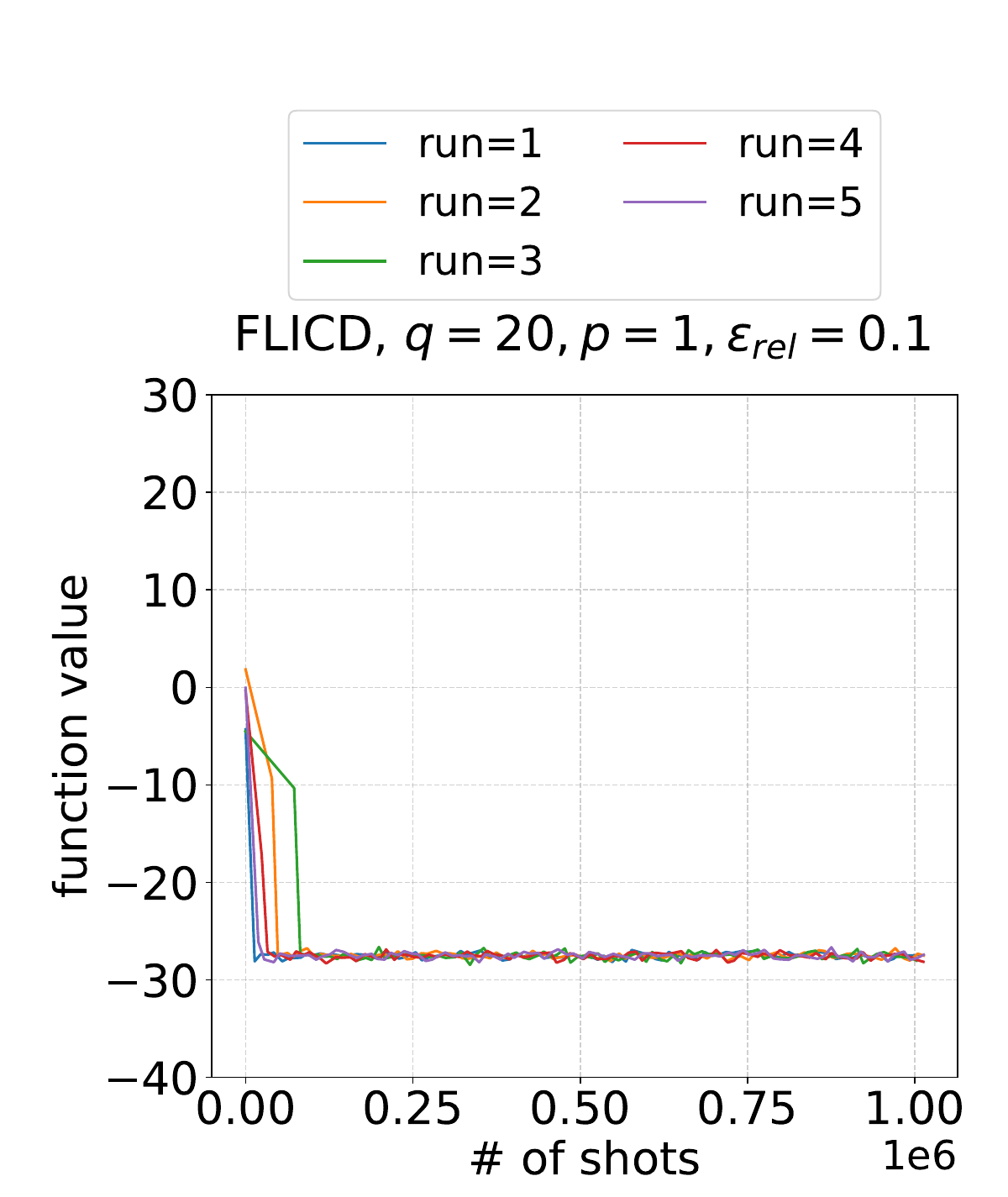}
        \caption{}
        \label{fig:sub1}
    \end{subfigure}
    \hfill 
    \begin{subfigure}[b]{0.32\textwidth}
        \centering
        \includegraphics[width=\textwidth]{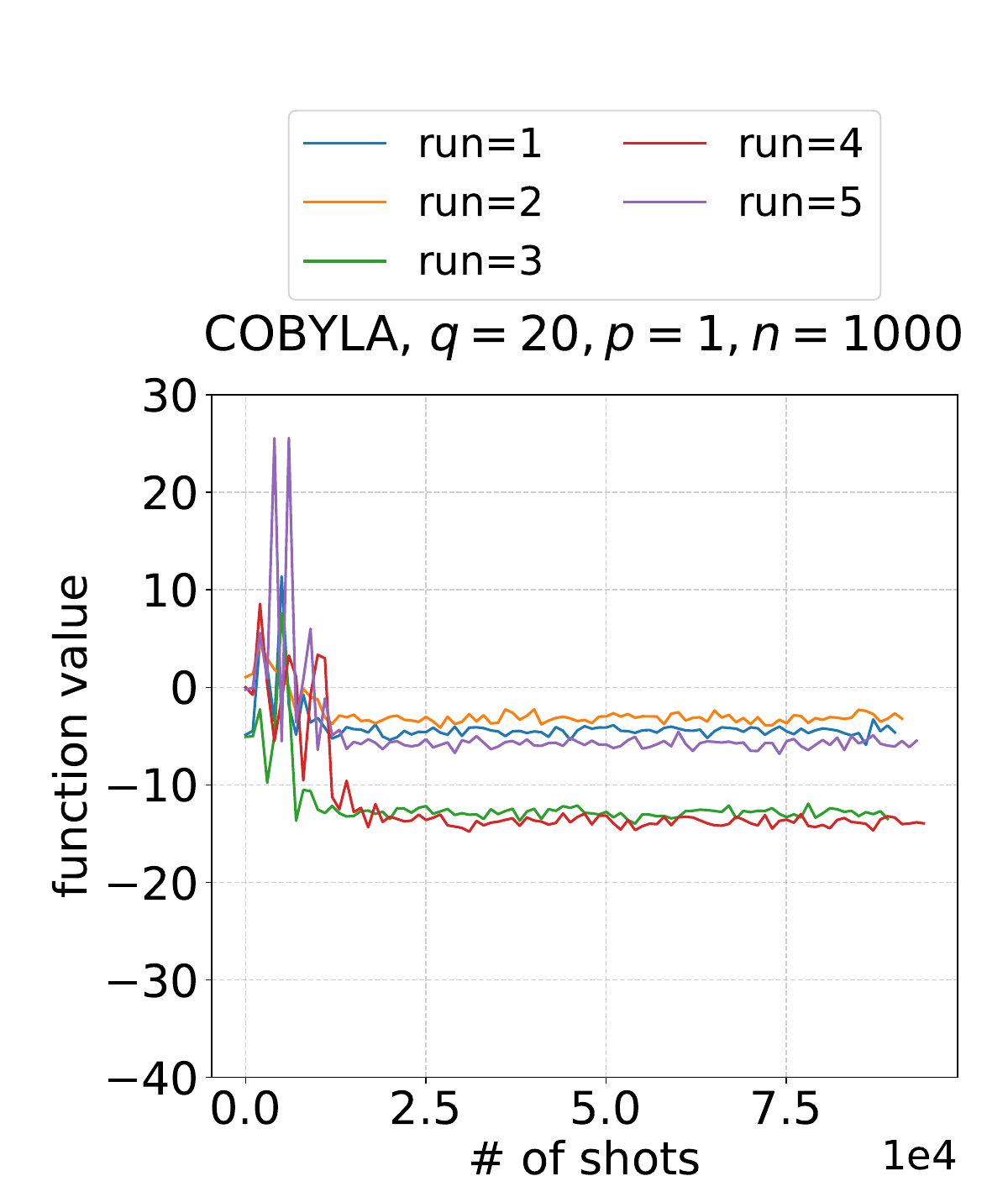}
        \caption{}
        \label{fig:sub2}
    \end{subfigure}
    \hfill 
    \begin{subfigure}[b]{0.32\textwidth}
        \centering
        \includegraphics[width=\textwidth]{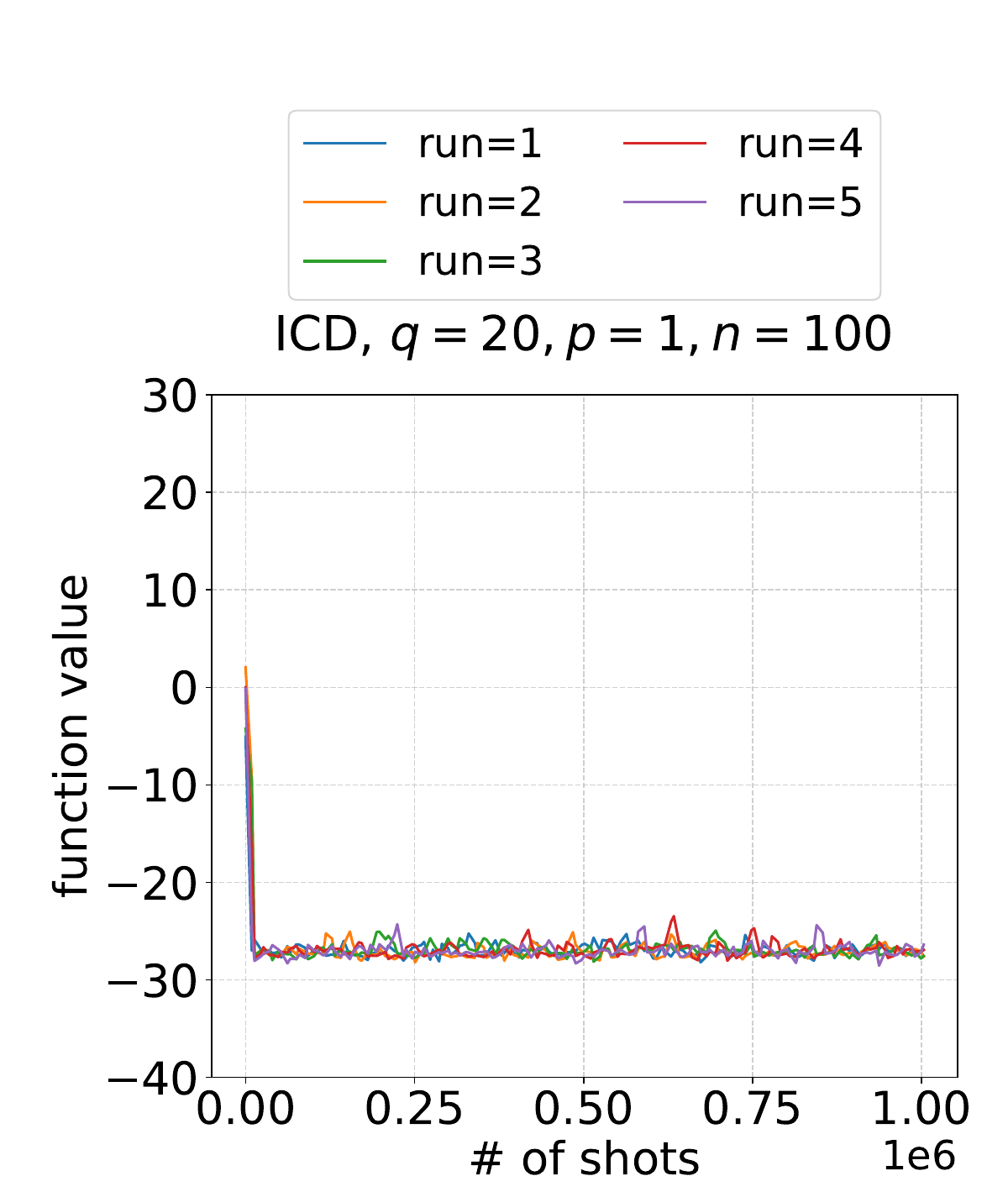}
        \caption{}
        \label{fig:top}
    \end{subfigure}

    \vspace{0.5cm} 

    \begin{subfigure}[b]{0.32\textwidth}
        \centering
        \includegraphics[width=\textwidth]{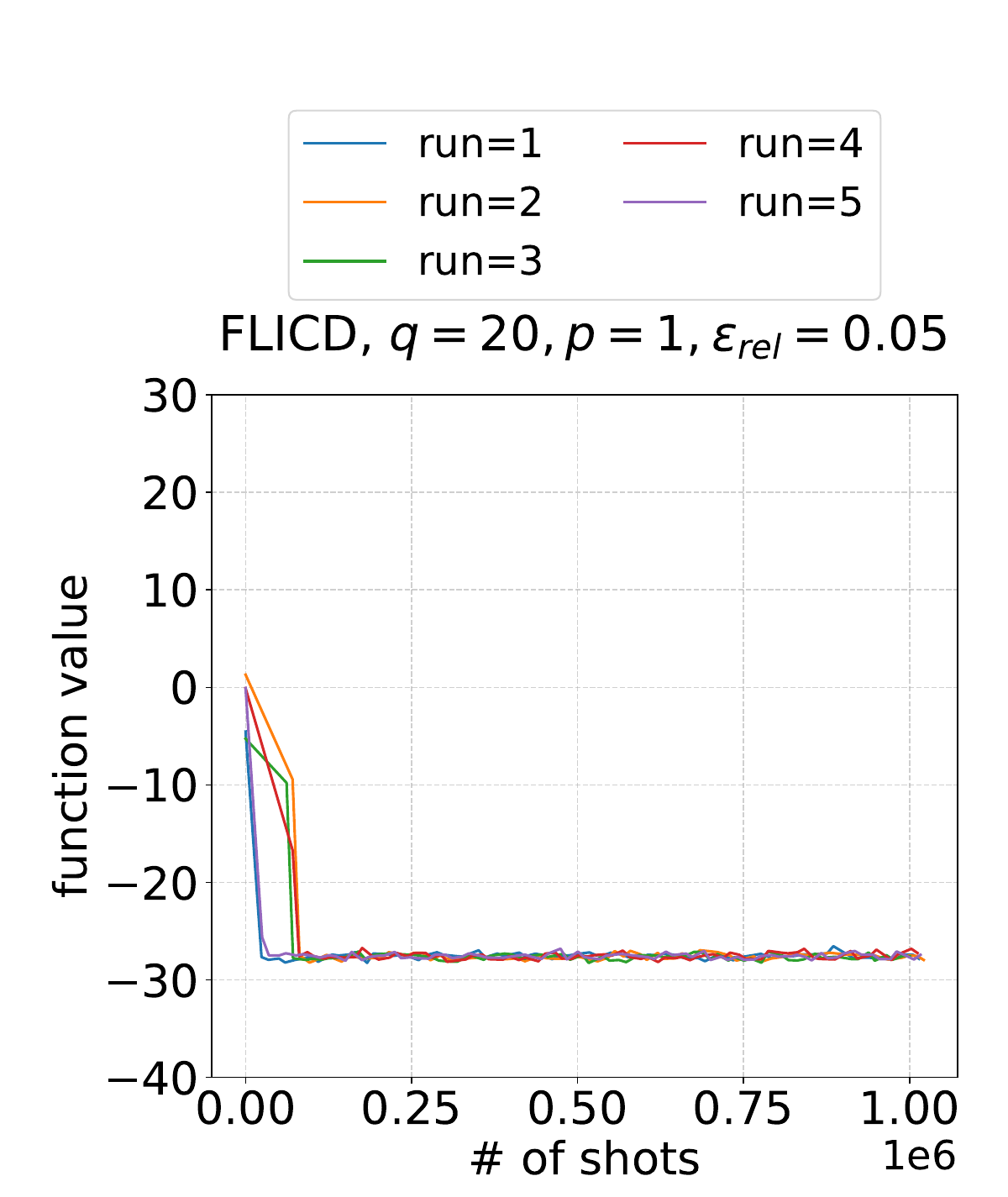}
        \caption{}
        \label{fig:sub3}
    \end{subfigure}
    \hfill
    \begin{subfigure}[b]{0.32\textwidth}
        \centering
        \includegraphics[width=\textwidth]{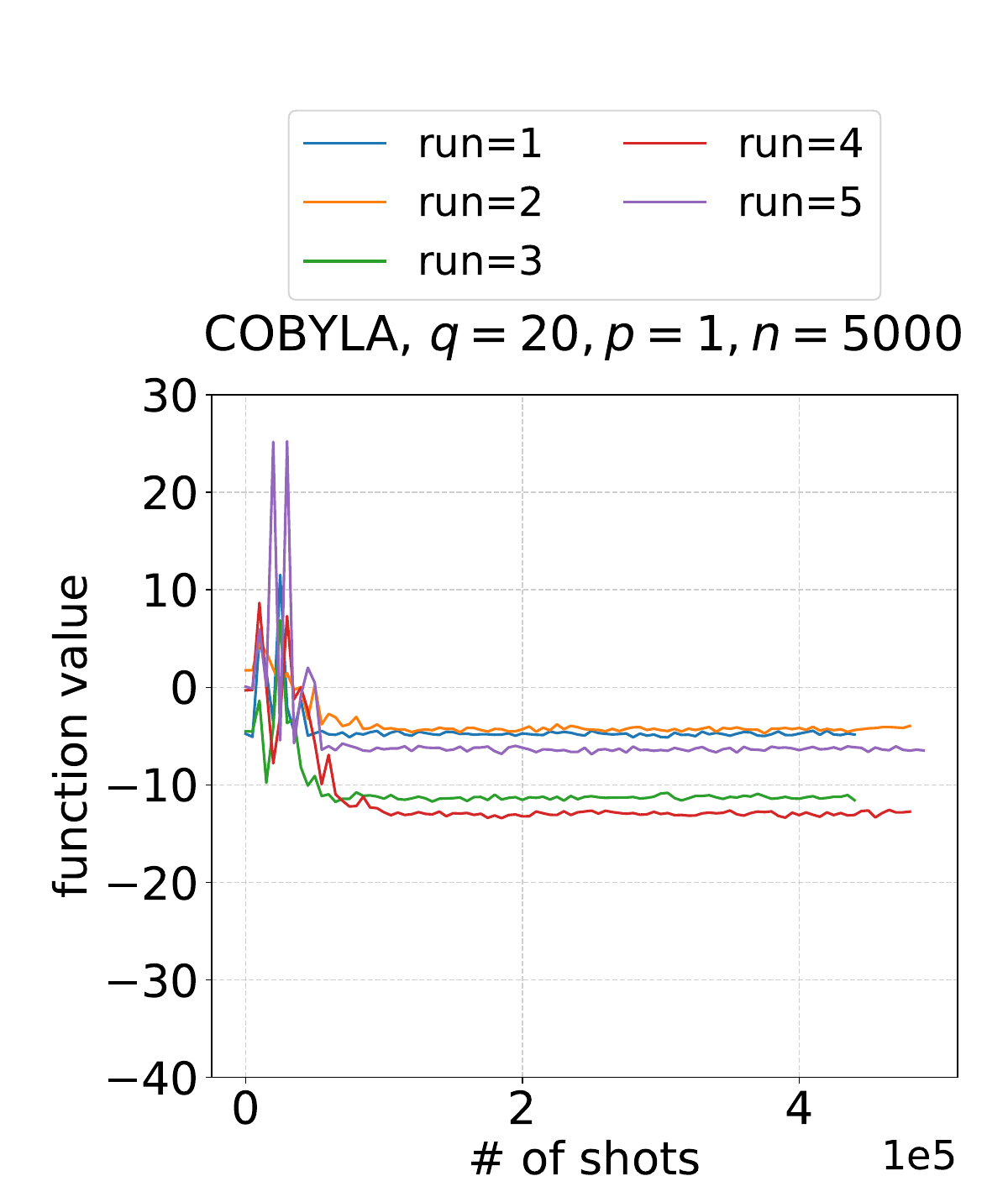}
        \caption{}
        \label{fig:sub4}
    \end{subfigure}
    \hfill 
    \begin{subfigure}[b]{0.32\textwidth}
        \centering
        \includegraphics[width=\textwidth]{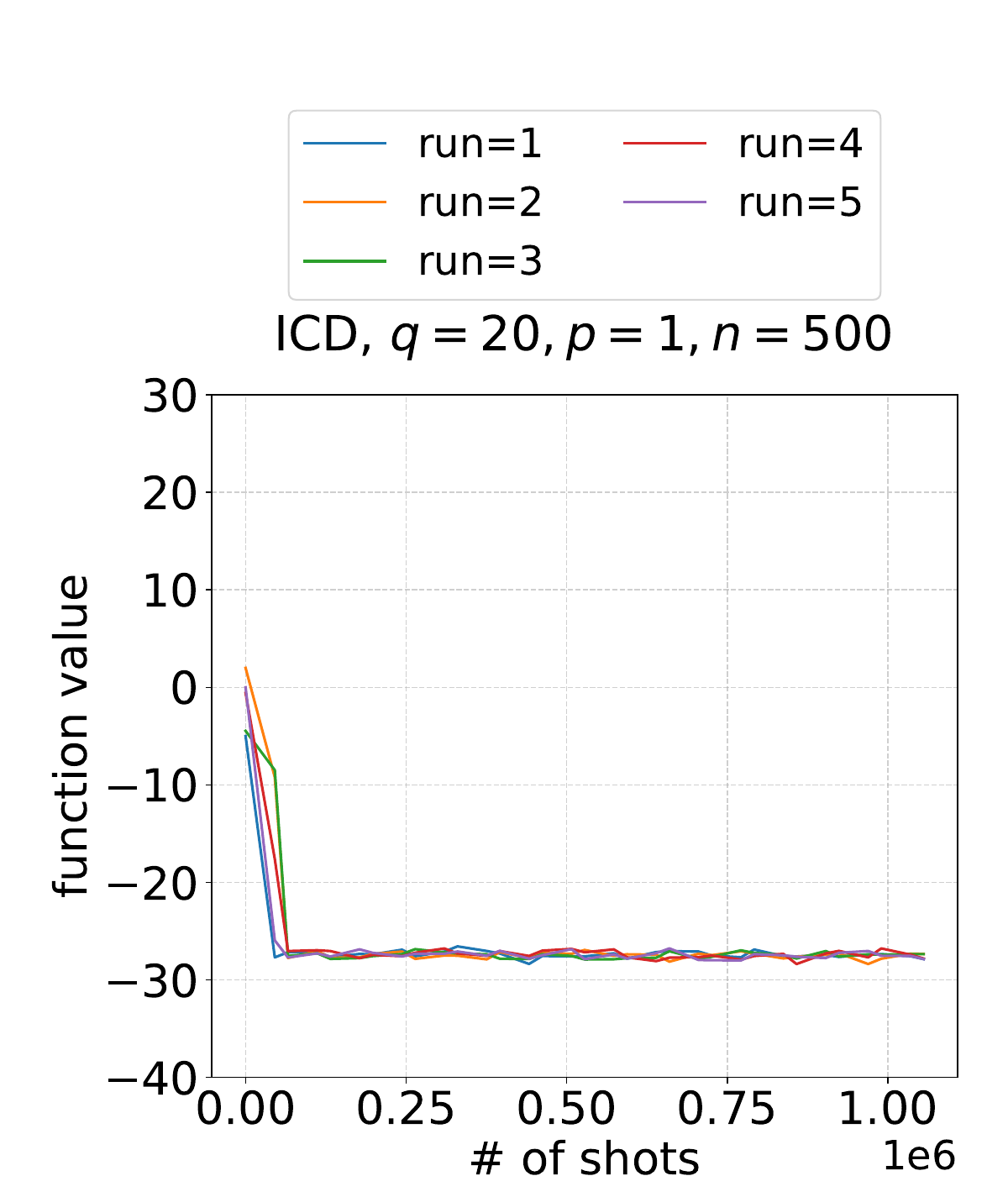}
        \caption{}
        \label{fig:top}
    \end{subfigure}

    \vspace{0.5cm} 

    \begin{subfigure}[b]{0.32\textwidth}
        \centering
        \includegraphics[width=\textwidth]{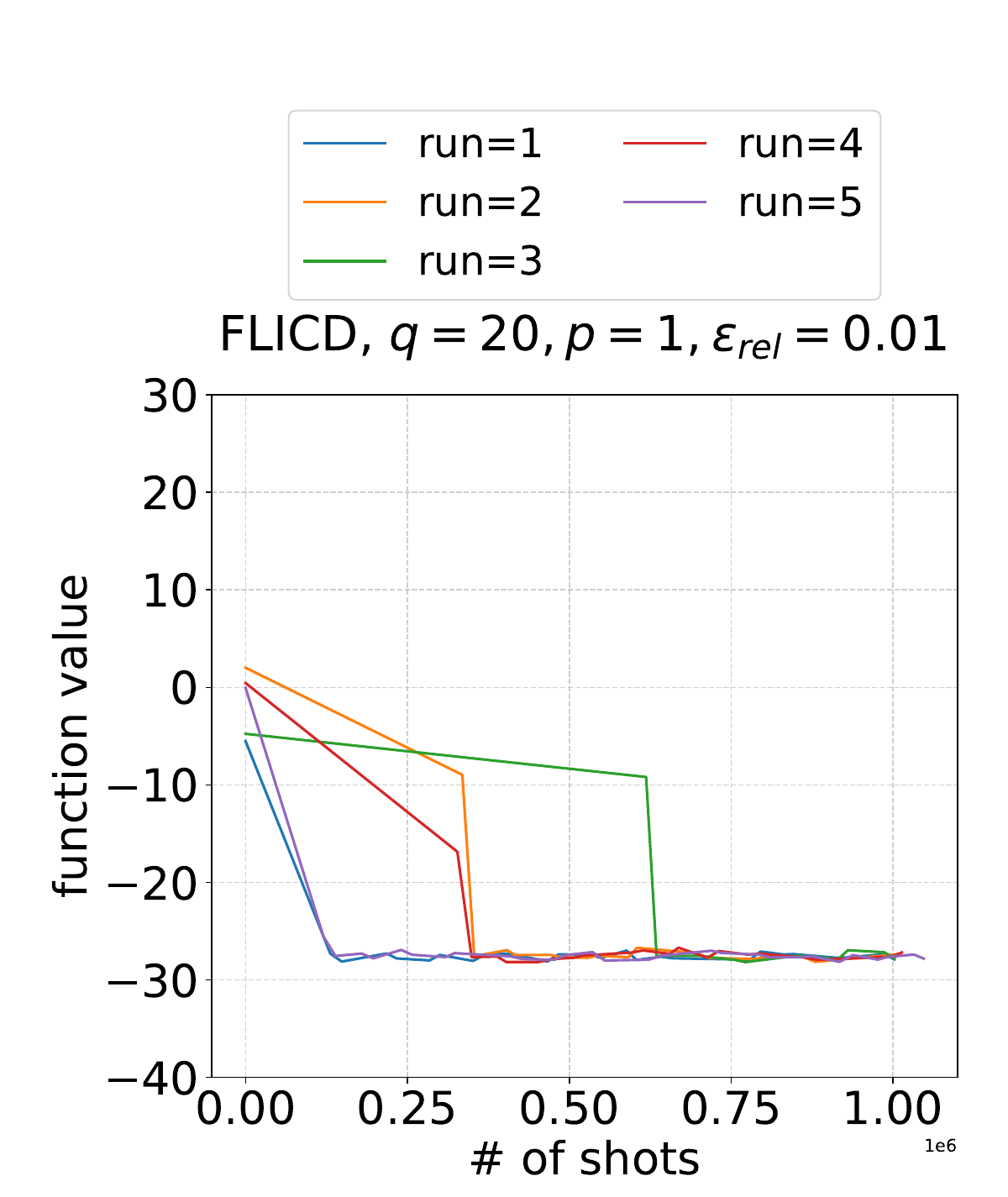}
        \caption{}
        \label{fig:sub5}
    \end{subfigure}
    \hfill
    \begin{subfigure}[b]{0.32\textwidth}
        \centering
        \includegraphics[width=\textwidth]{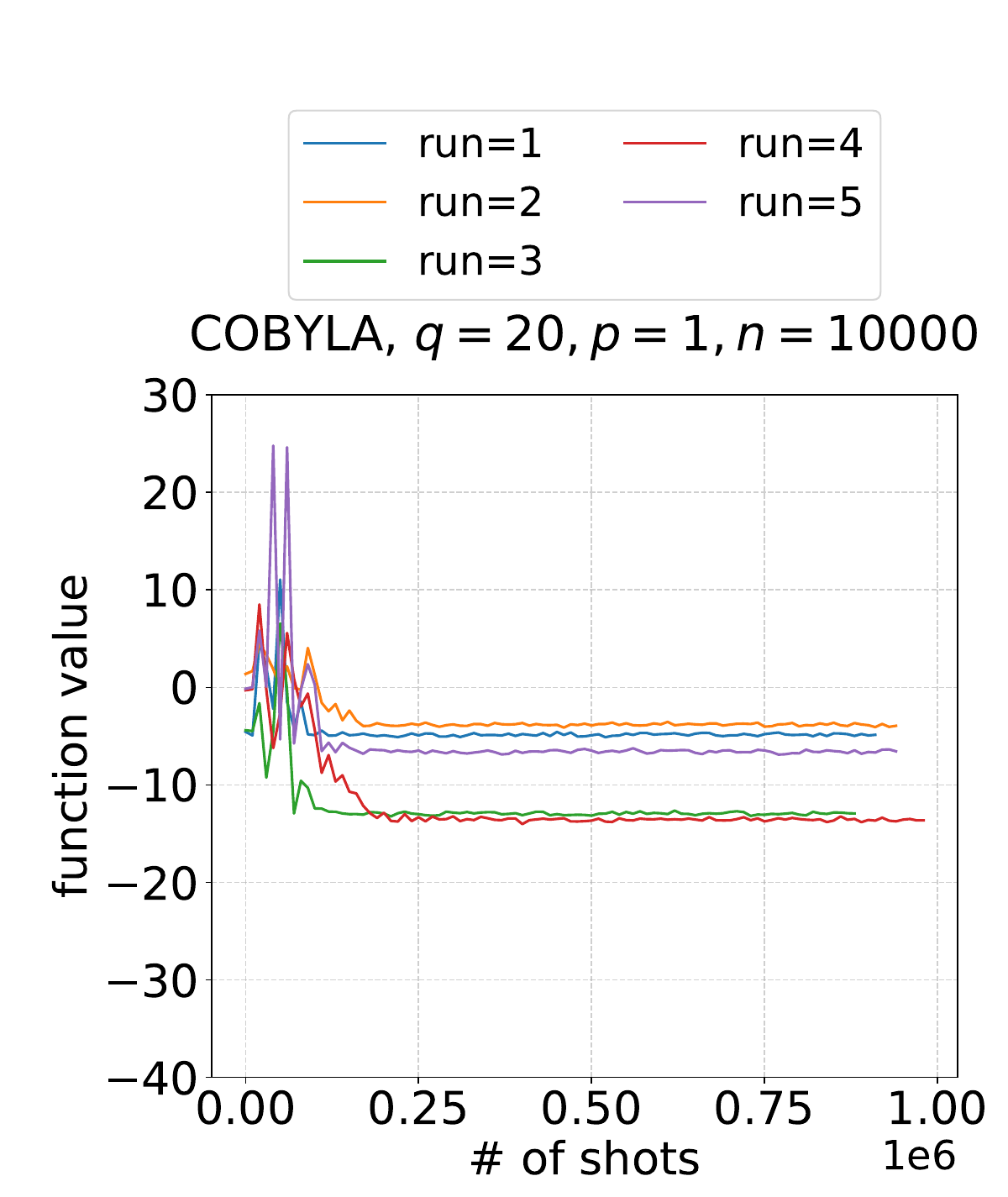}
        \caption{}
        \label{fig:sub6}
    \end{subfigure}
    \hfill 
    \begin{subfigure}[b]{0.32\textwidth}
        \centering
        \includegraphics[width=\textwidth]{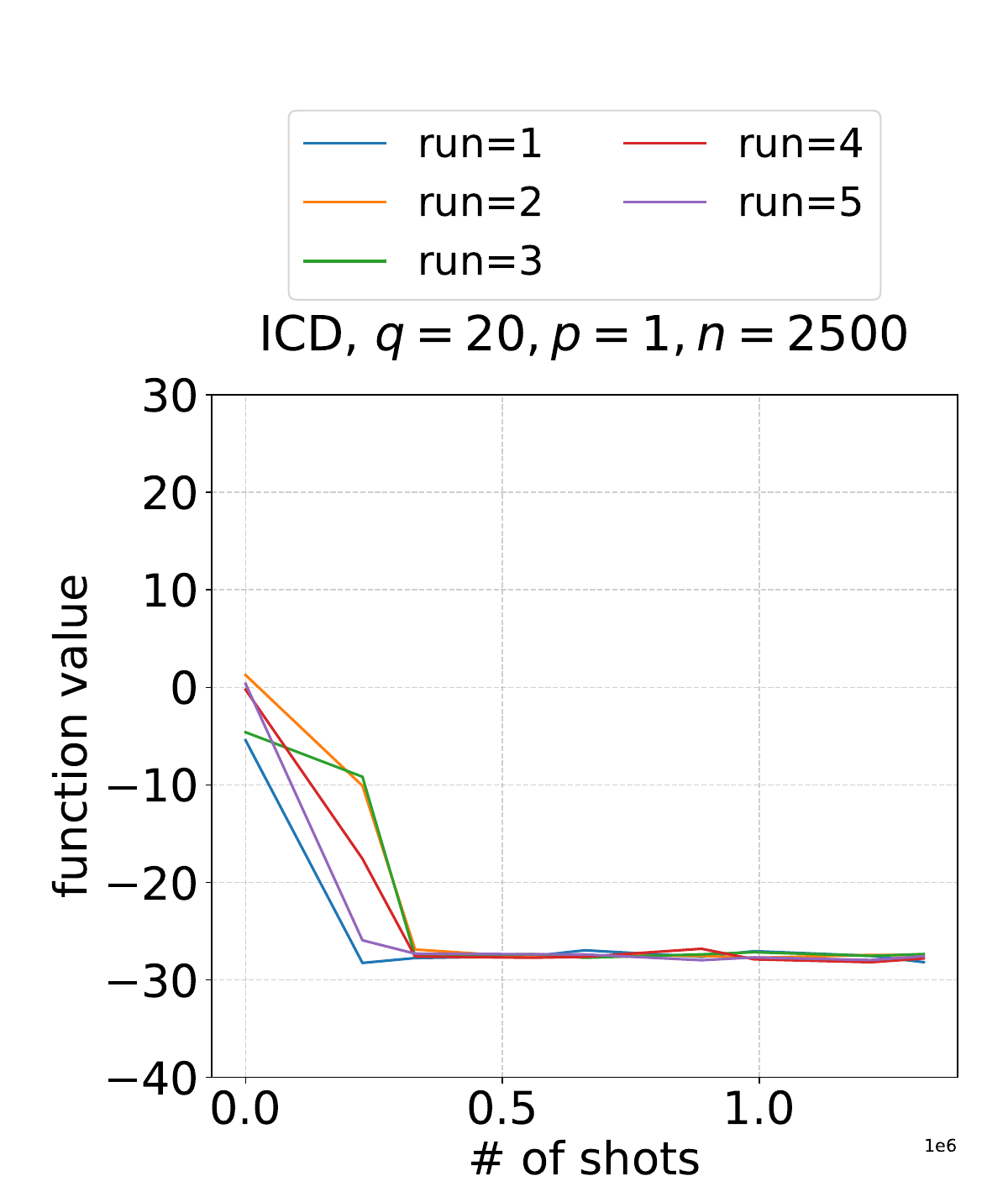}
        \caption{}
        \label{fig:top}
    \end{subfigure}

    \caption{Function value with respect to total shots across all 5 runs of FLICD (left), COBYLA (middle), and ICD (right) on simulator backend. The cost functions $f(\boldsymbol{\theta})$ generated from $1$-layer LC-QAOA ansatz of 20-vertex random regular degree-3 graph.}
    \label{fig:simulator_p=1}
\end{figure}

\begin{figure}[htbp]
    \centering
    
    \begin{subfigure}[b]{0.32\textwidth}
        \centering
        \includegraphics[width=\textwidth]{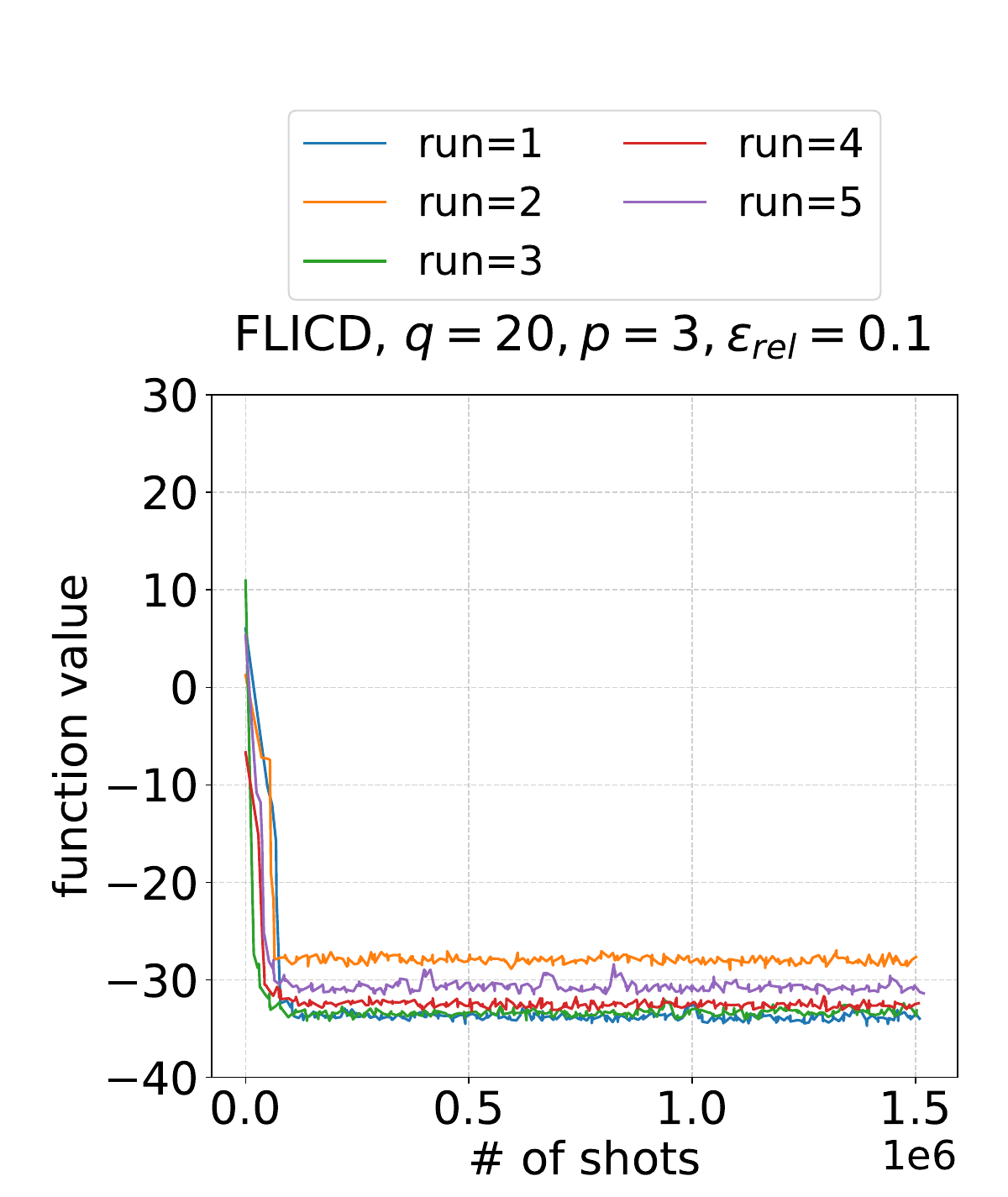}
        \caption{}
        \label{fig:sub1}
    \end{subfigure}
    \hfill 
    \begin{subfigure}[b]{0.32\textwidth}
        \centering
        \includegraphics[width=\textwidth]{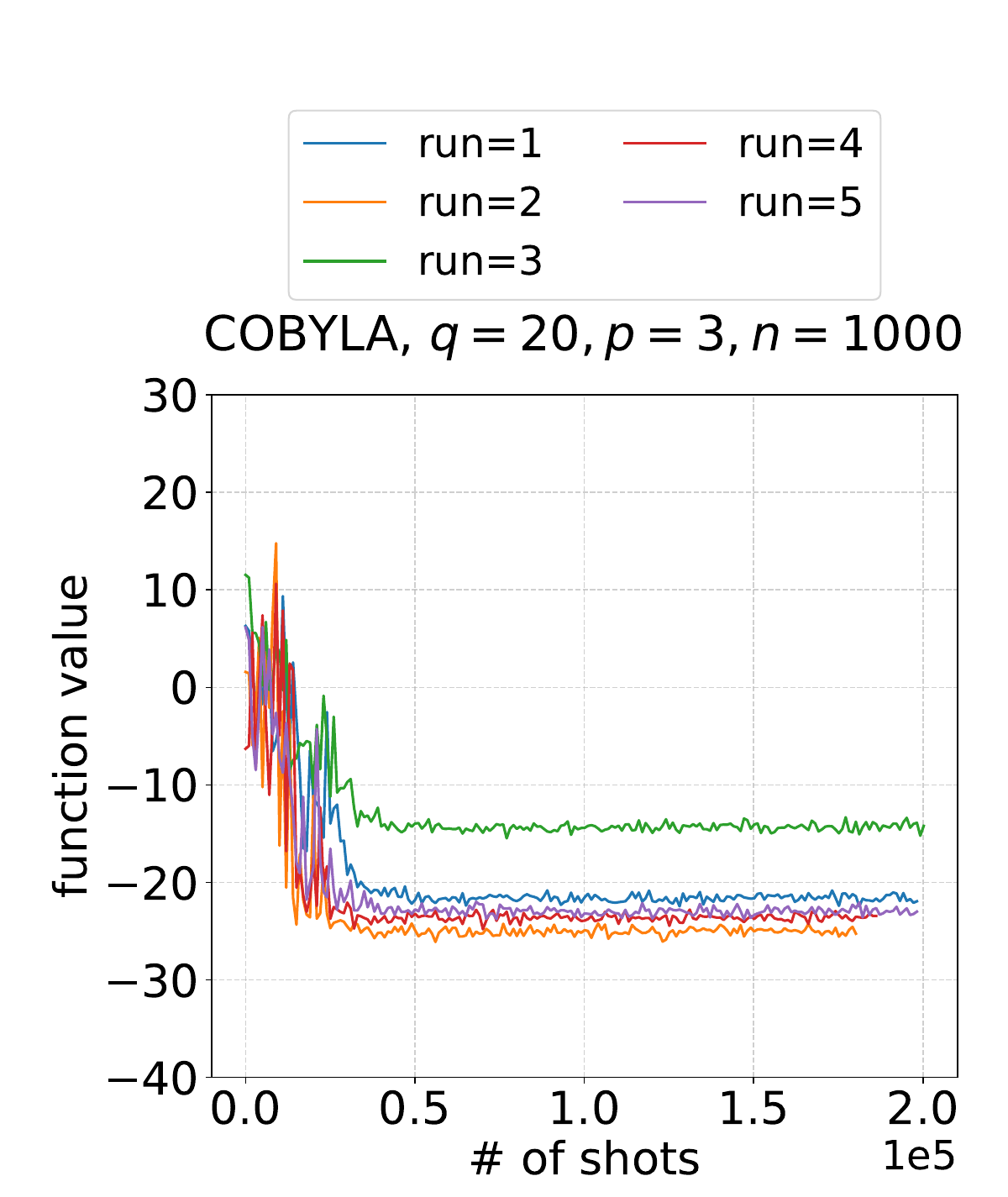}
        \caption{}
        \label{fig:sub2}
    \end{subfigure}
    \hfill
    \begin{subfigure}[b]{0.32\textwidth}
        \centering
        \includegraphics[width=\textwidth]{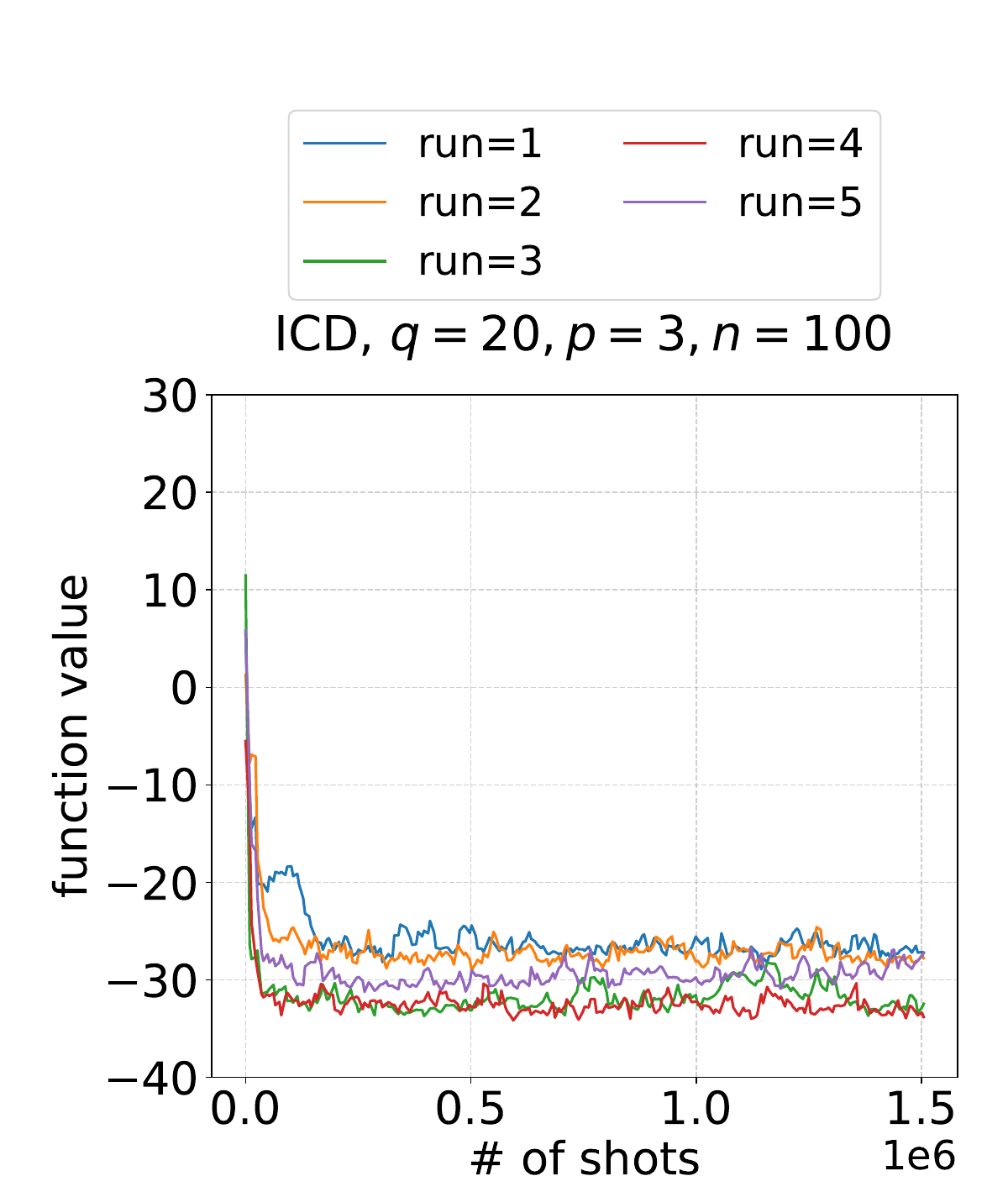}
        \caption{}
        \label{fig:top}
    \end{subfigure}

    \vspace{0.5cm} 

    \begin{subfigure}[b]{0.32\textwidth}
        \centering
        \includegraphics[width=\textwidth]{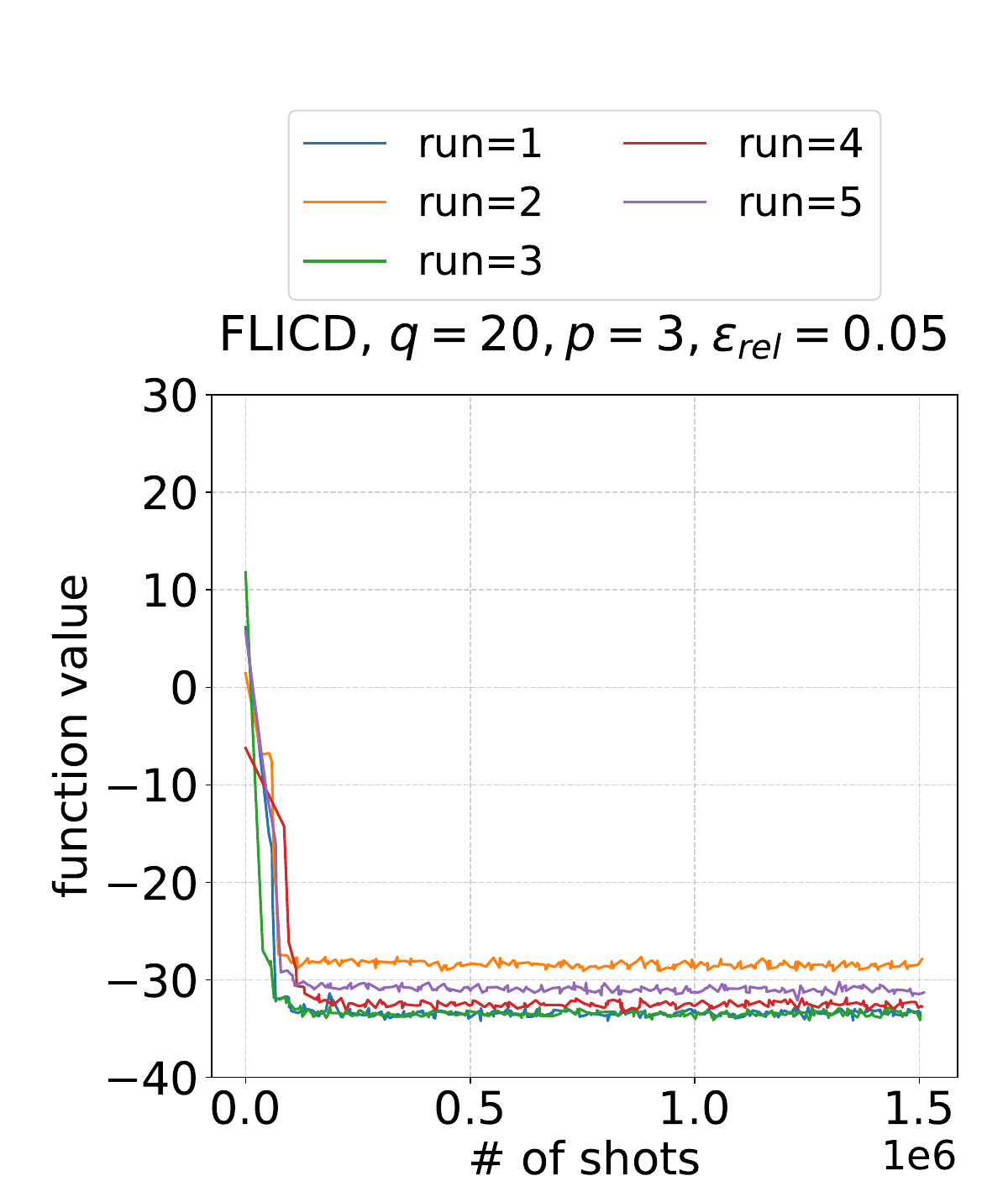}
        \caption{}
        \label{fig:sub3}
    \end{subfigure}
    \hfill
    \begin{subfigure}[b]{0.32\textwidth}
        \centering
        \includegraphics[width=\textwidth]{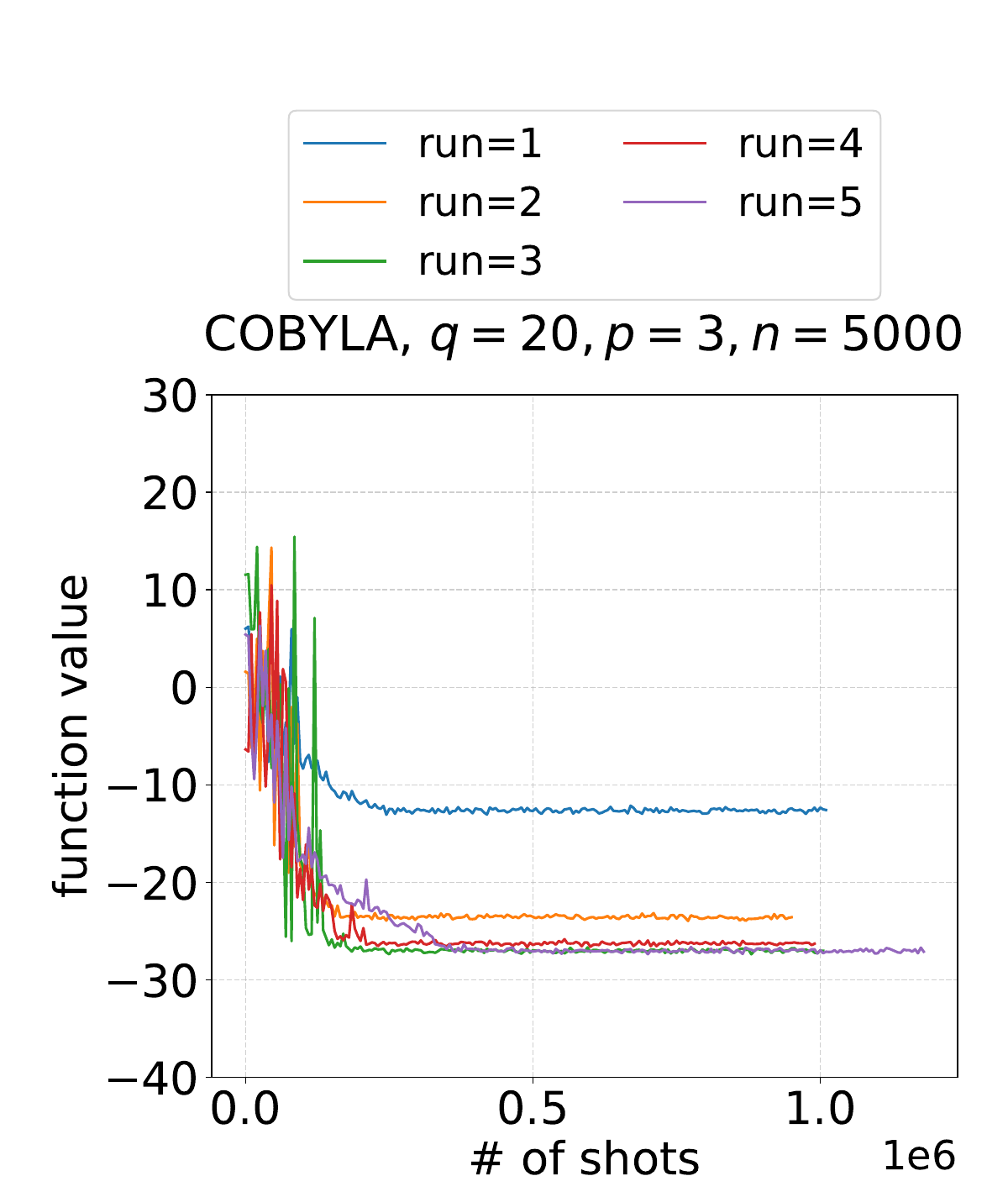}
        \caption{}
        \label{fig:sub4}
    \end{subfigure}
    \hfill
    \begin{subfigure}[b]{0.32\textwidth}
        \centering
        \includegraphics[width=\textwidth]{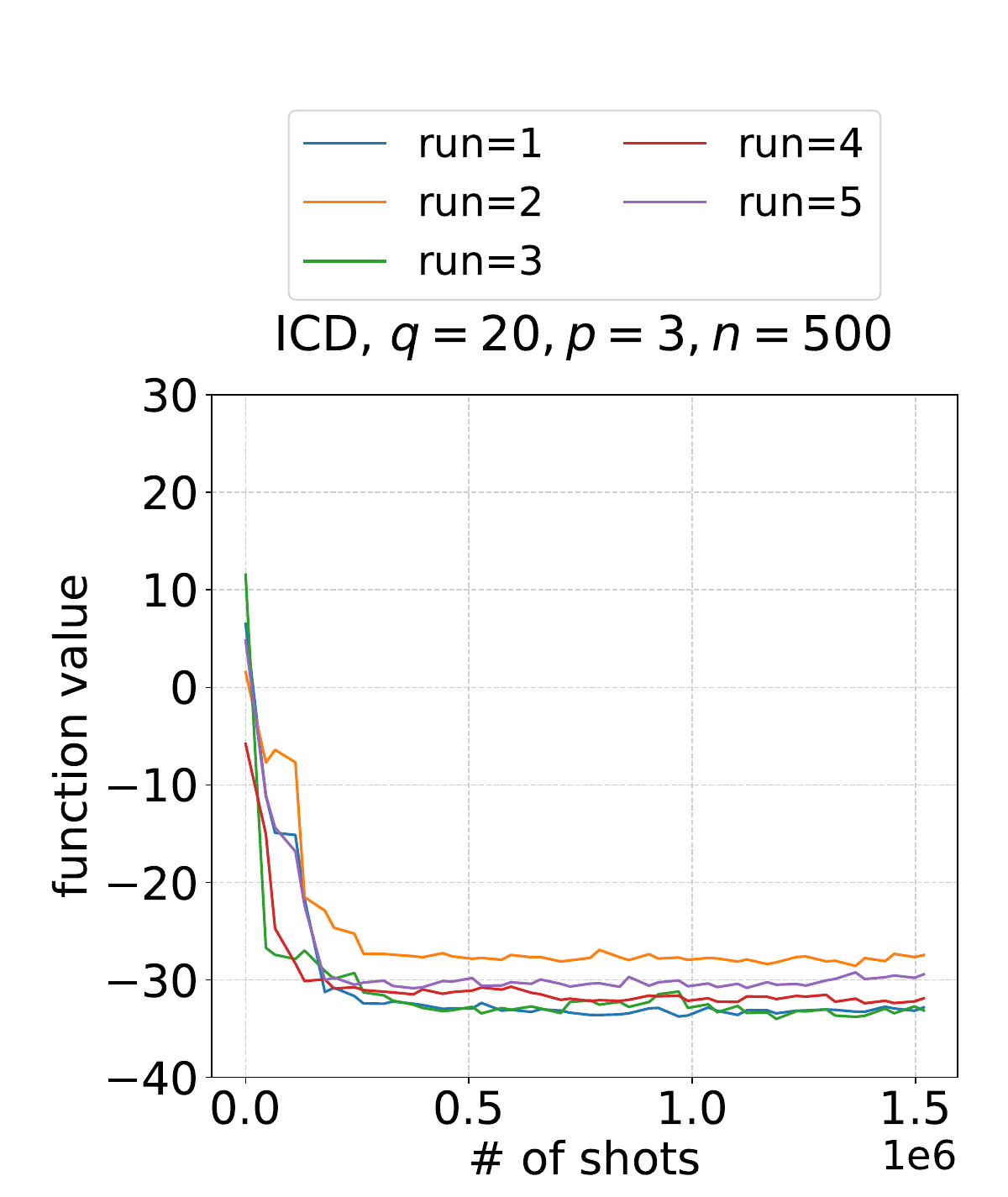}
        \caption{}
        \label{fig:top}
    \end{subfigure}

    \vspace{0.5cm} 

    \begin{subfigure}[b]{0.32\textwidth}
        \centering
        \includegraphics[width=\textwidth]{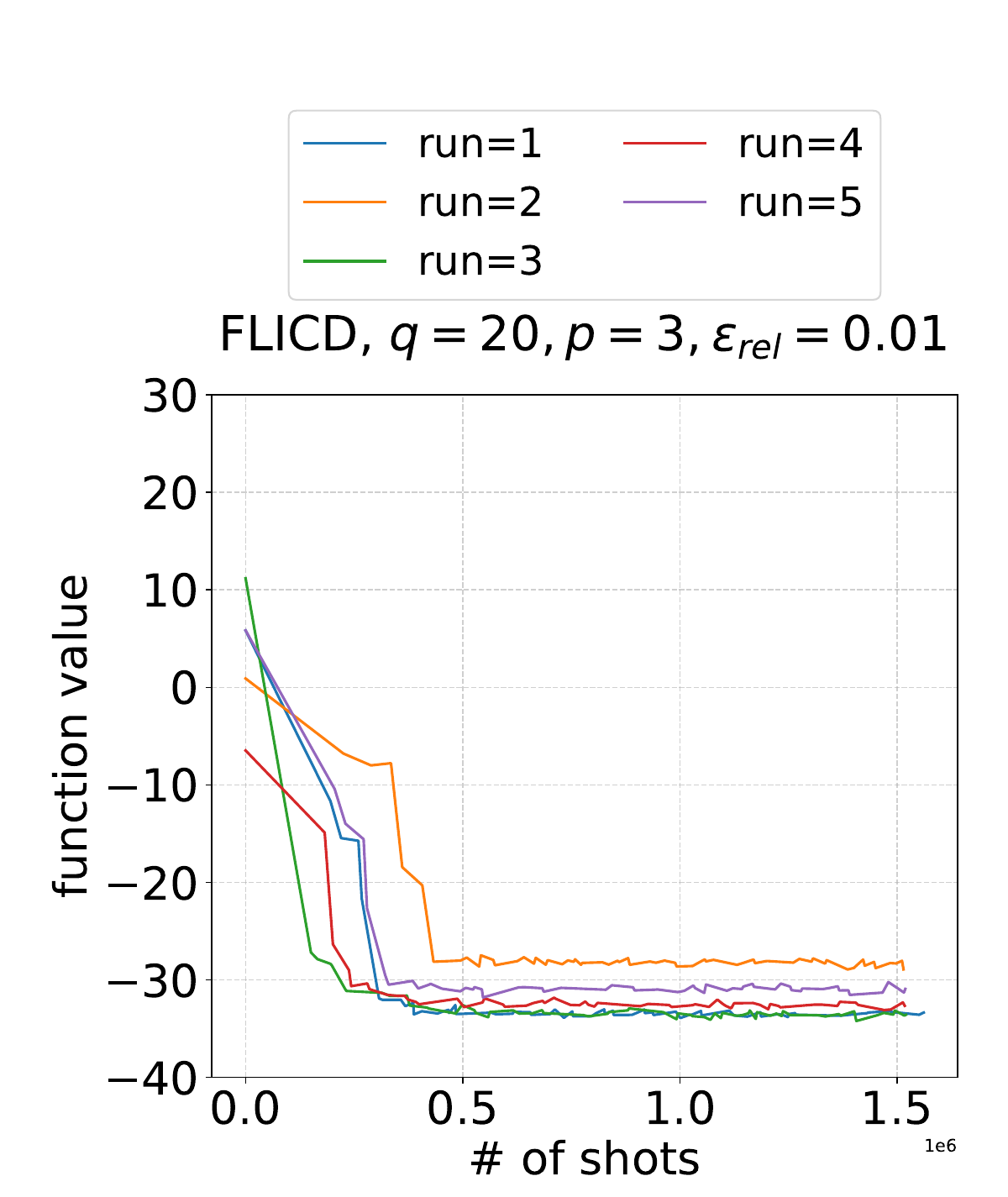}
        \caption{}
        \label{fig:sub5}
    \end{subfigure}
    \hfill
    \begin{subfigure}[b]{0.32\textwidth}
        \centering
        \includegraphics[width=\textwidth]{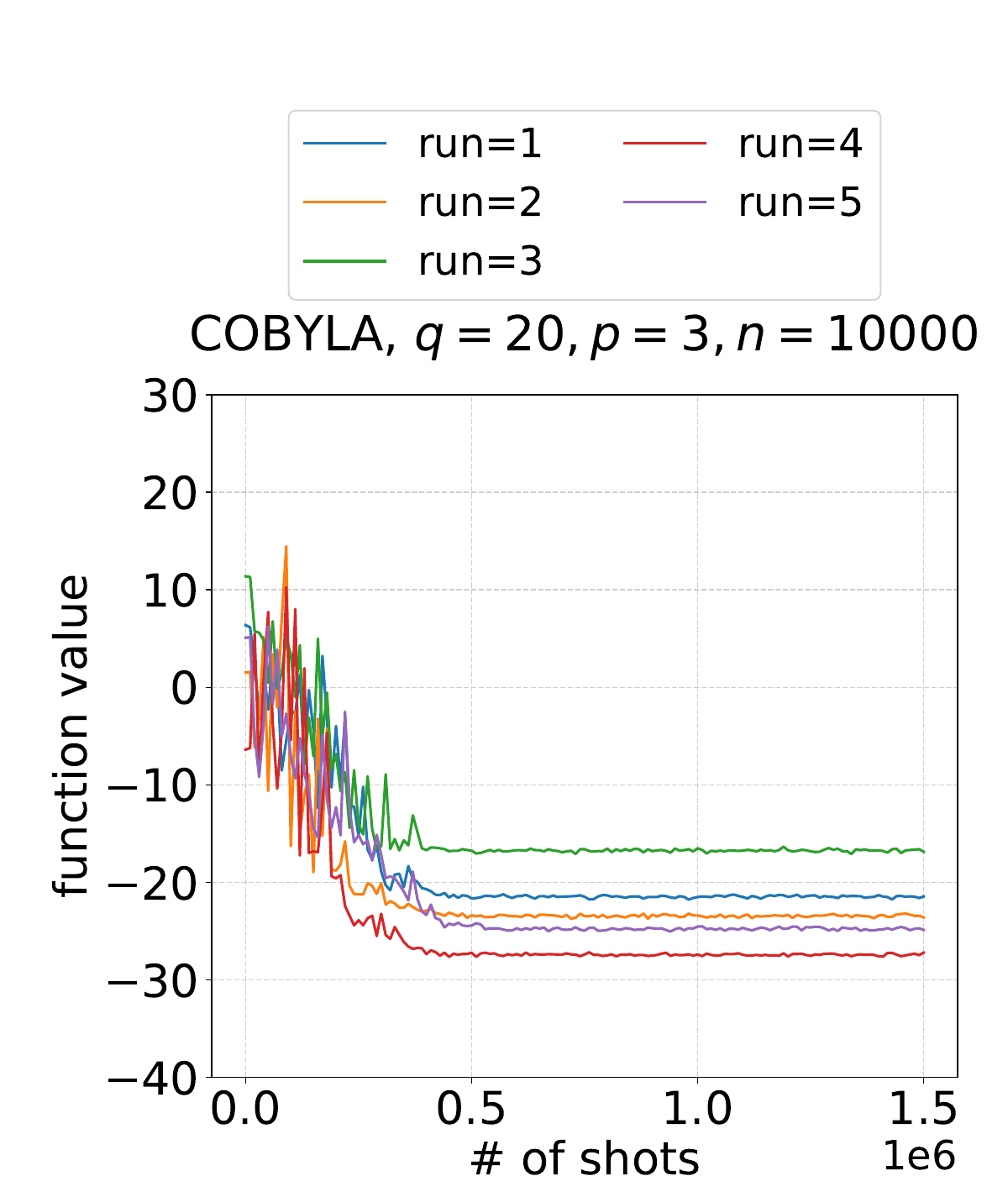}
        \caption{}
        \label{fig:sub6}
    \end{subfigure}
    \hfill
    \begin{subfigure}[b]{0.32\textwidth}
        \centering
        \includegraphics[width=\textwidth]{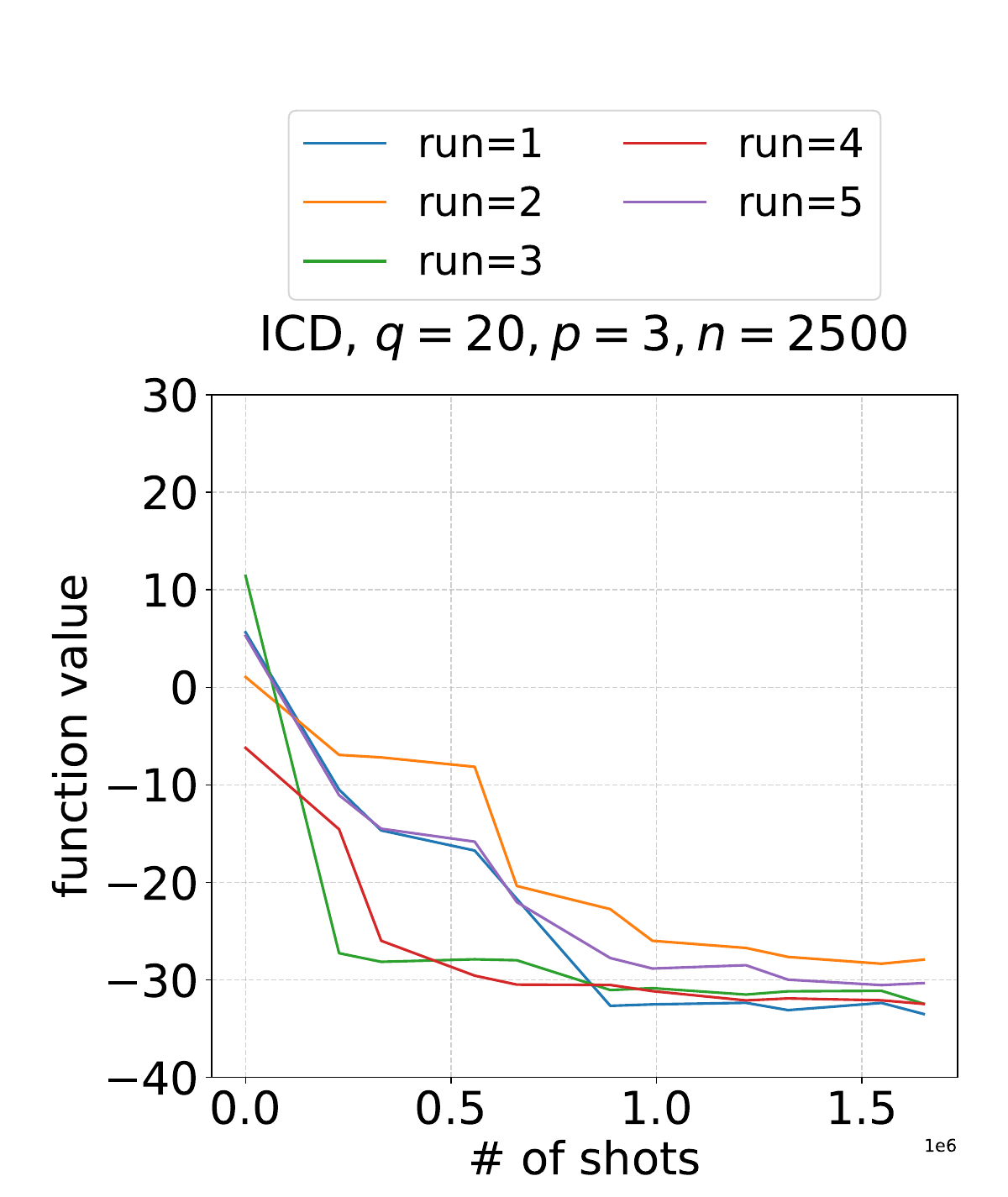}
        \caption{}
        \label{fig:top}
    \end{subfigure}

    \caption{Function value with respect to total shots across all 5 runs of FLICD (left), COBYLA (middle), and ICD (right) on simulator backend. Cost functions $f(\boldsymbol{\theta})$ generated from $3$-layer LC-QAOA ansatz of 20-vertex random regular degree-3 graph.}
    \label{fig:simulator_p=3}
\end{figure}

\begin{figure}[htbp]
    \centering
    
    \begin{subfigure}[b]{0.32\textwidth}
        \centering
        \includegraphics[width=\textwidth]{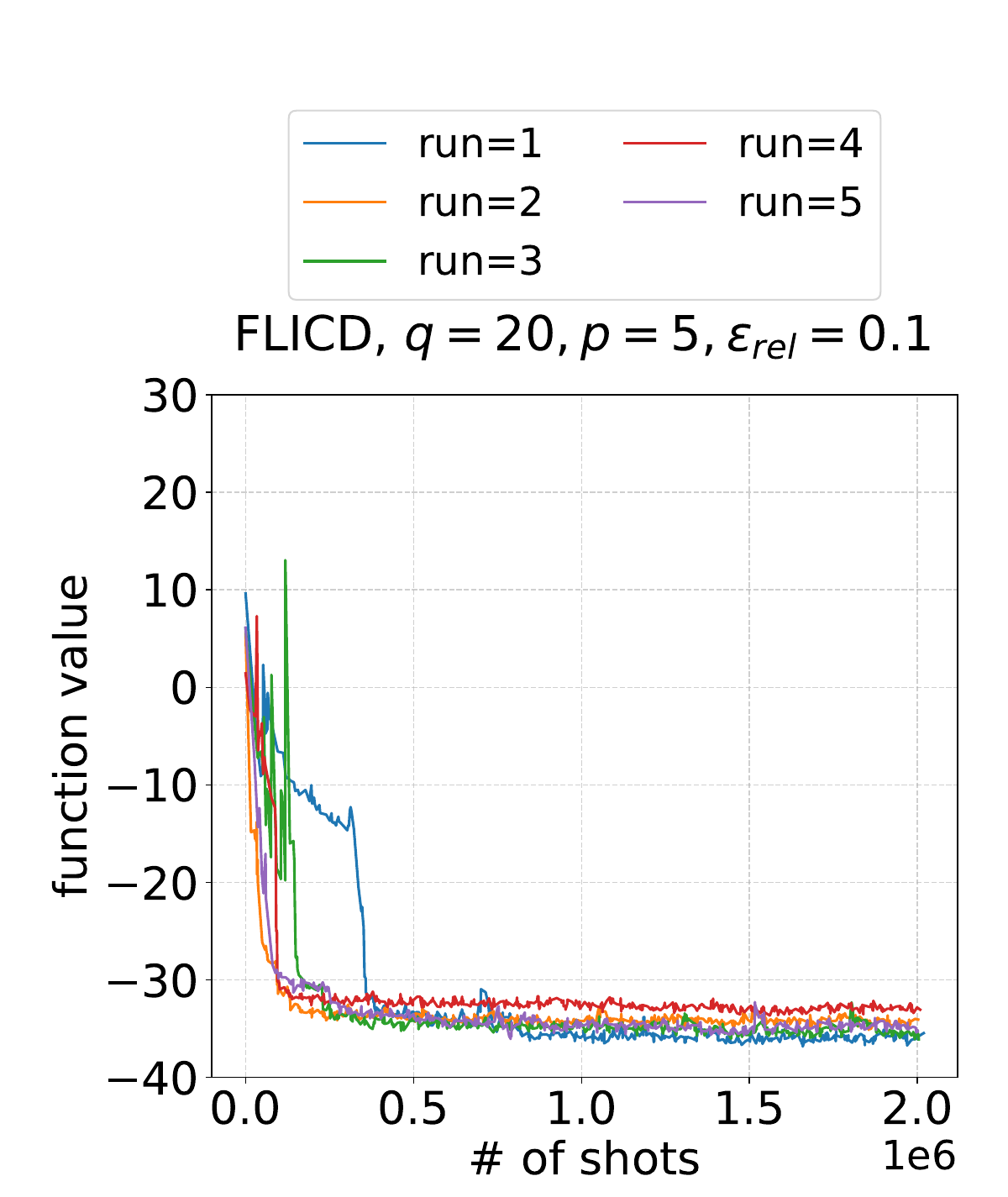}
        \caption{}
        \label{fig:sub1}
    \end{subfigure}
    \hfill 
    \begin{subfigure}[b]{0.32\textwidth}
        \centering
        \includegraphics[width=\textwidth]{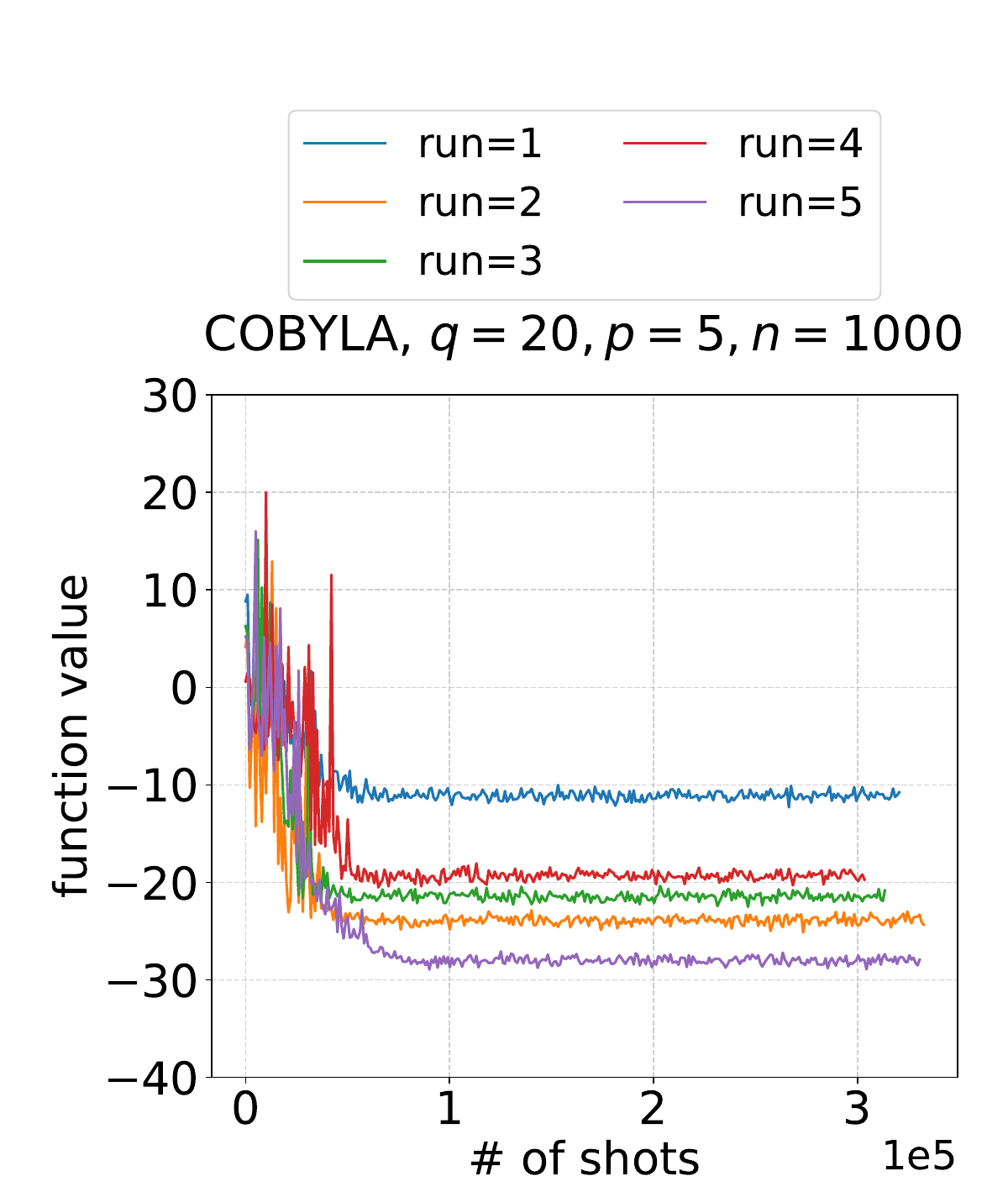}
        \caption{}
        \label{fig:sub2}
    \end{subfigure}
    \hfill
    \begin{subfigure}[b]{0.32\textwidth}
        \centering
        \includegraphics[width=\textwidth]{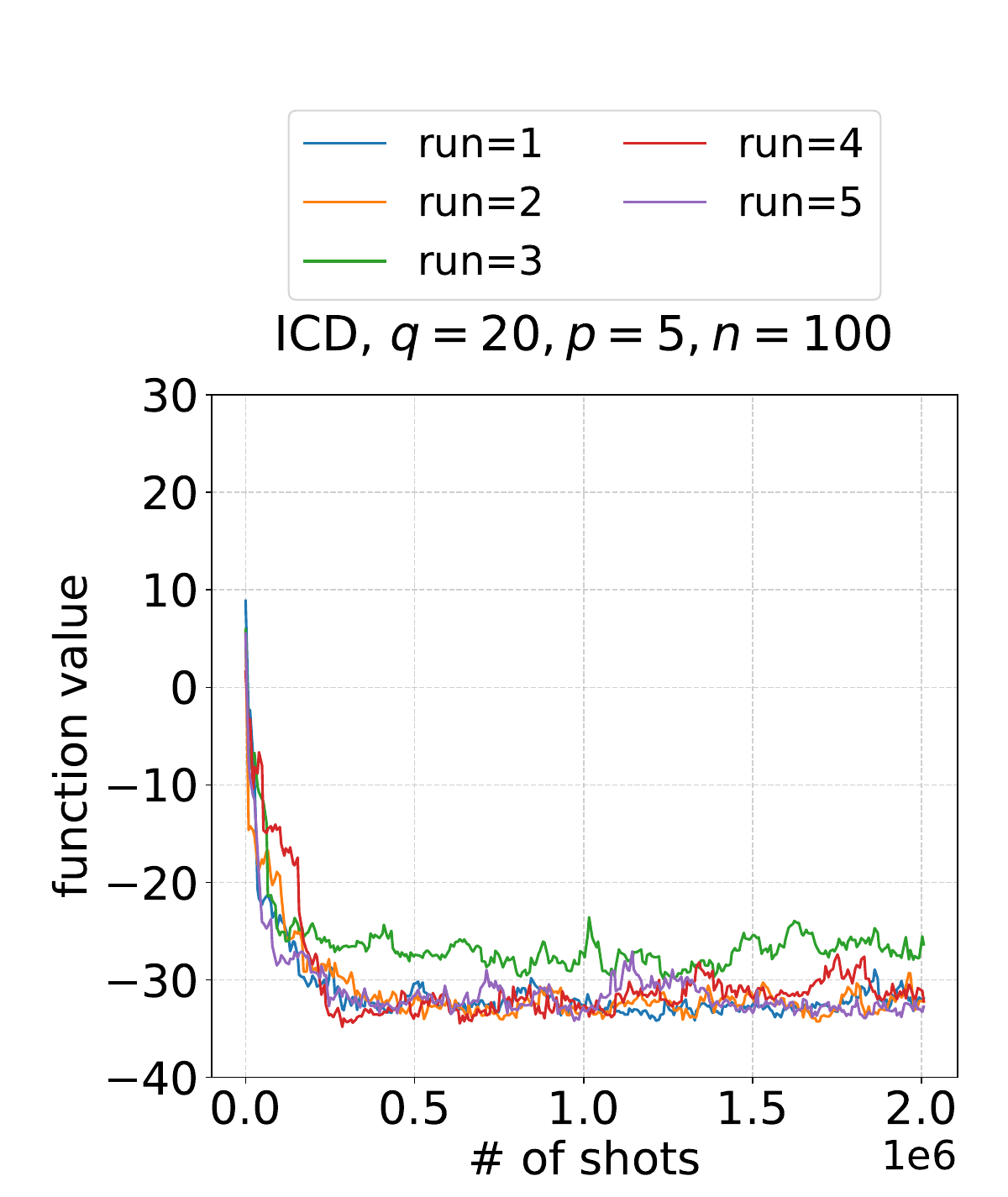}
        \caption{}
        \label{fig:top}
    \end{subfigure}

    \vspace{0.5cm} 

    \begin{subfigure}[b]{0.32\textwidth}
        \centering
        \includegraphics[width=\textwidth]{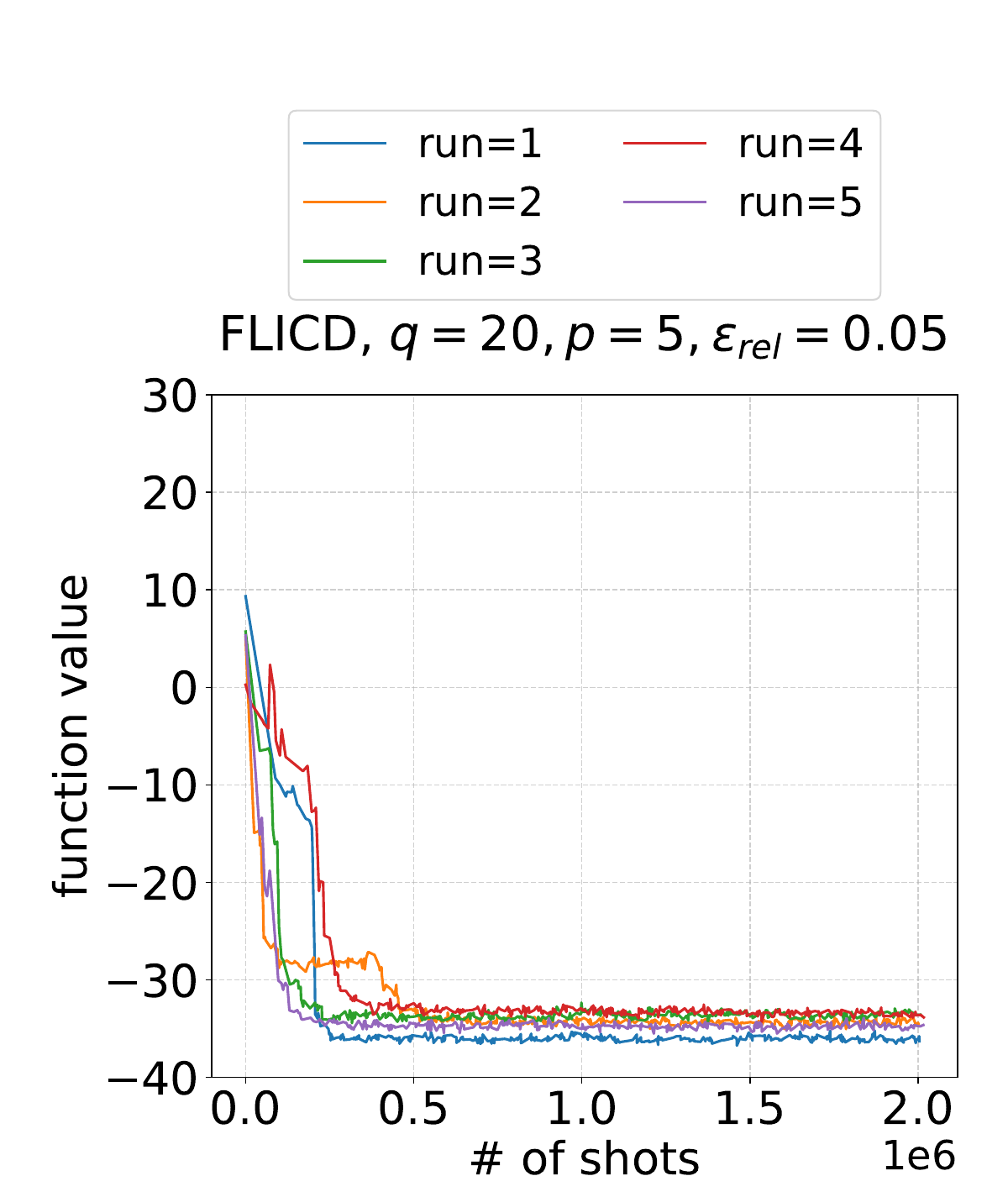}
        \caption{}
        \label{fig:sub3}
    \end{subfigure}
    \hfill
    \begin{subfigure}[b]{0.32\textwidth}
        \centering
        \includegraphics[width=\textwidth]{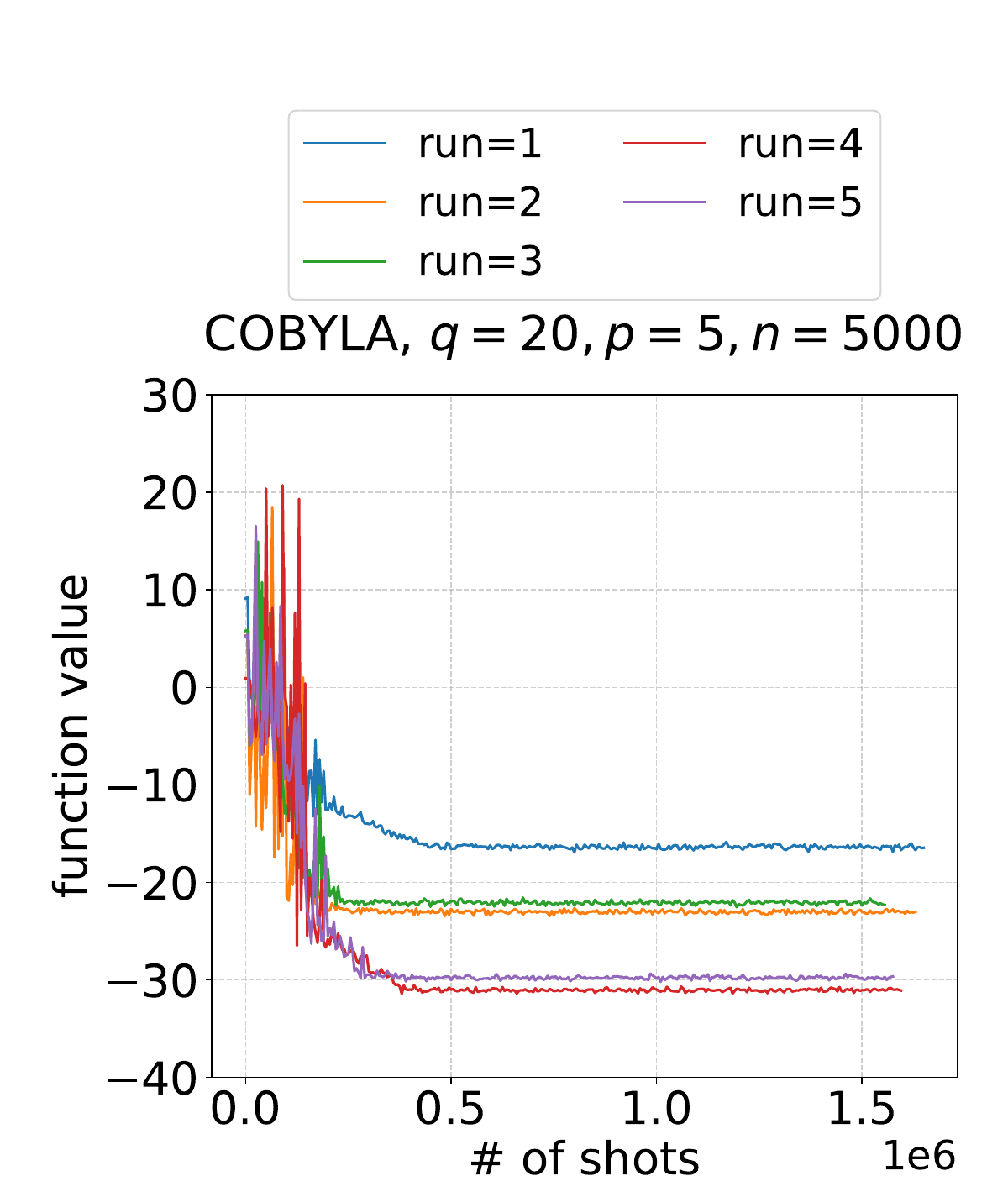}
        \caption{}
        \label{fig:sub4}
    \end{subfigure}
    \hfill
    \begin{subfigure}[b]{0.32\textwidth}
        \centering
        \includegraphics[width=\textwidth]{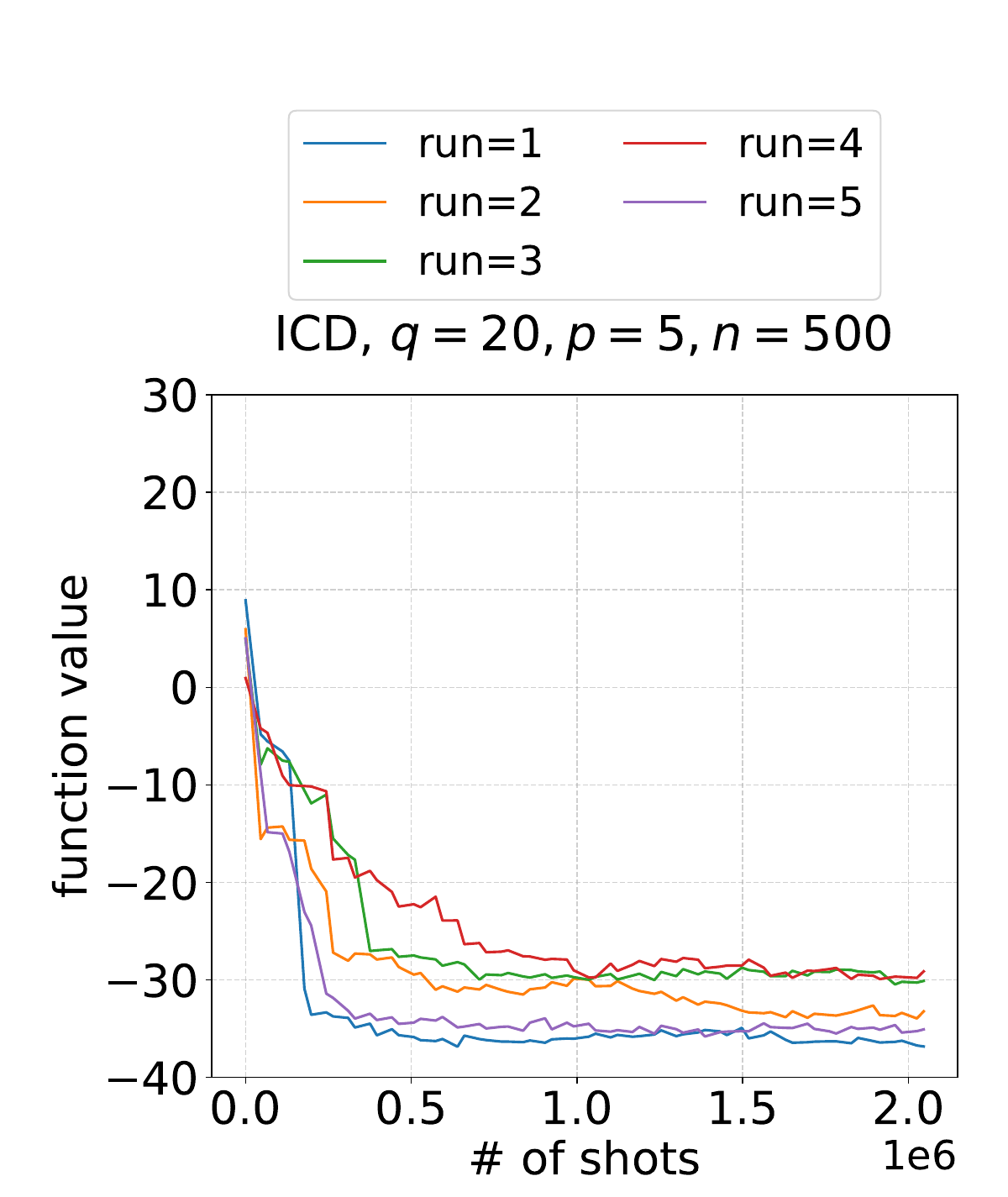}
        \caption{}
        \label{fig:top}
    \end{subfigure}
    \vspace{0.5cm} 

    \begin{subfigure}[b]{0.32\textwidth}
        \centering
        \includegraphics[width=\textwidth]{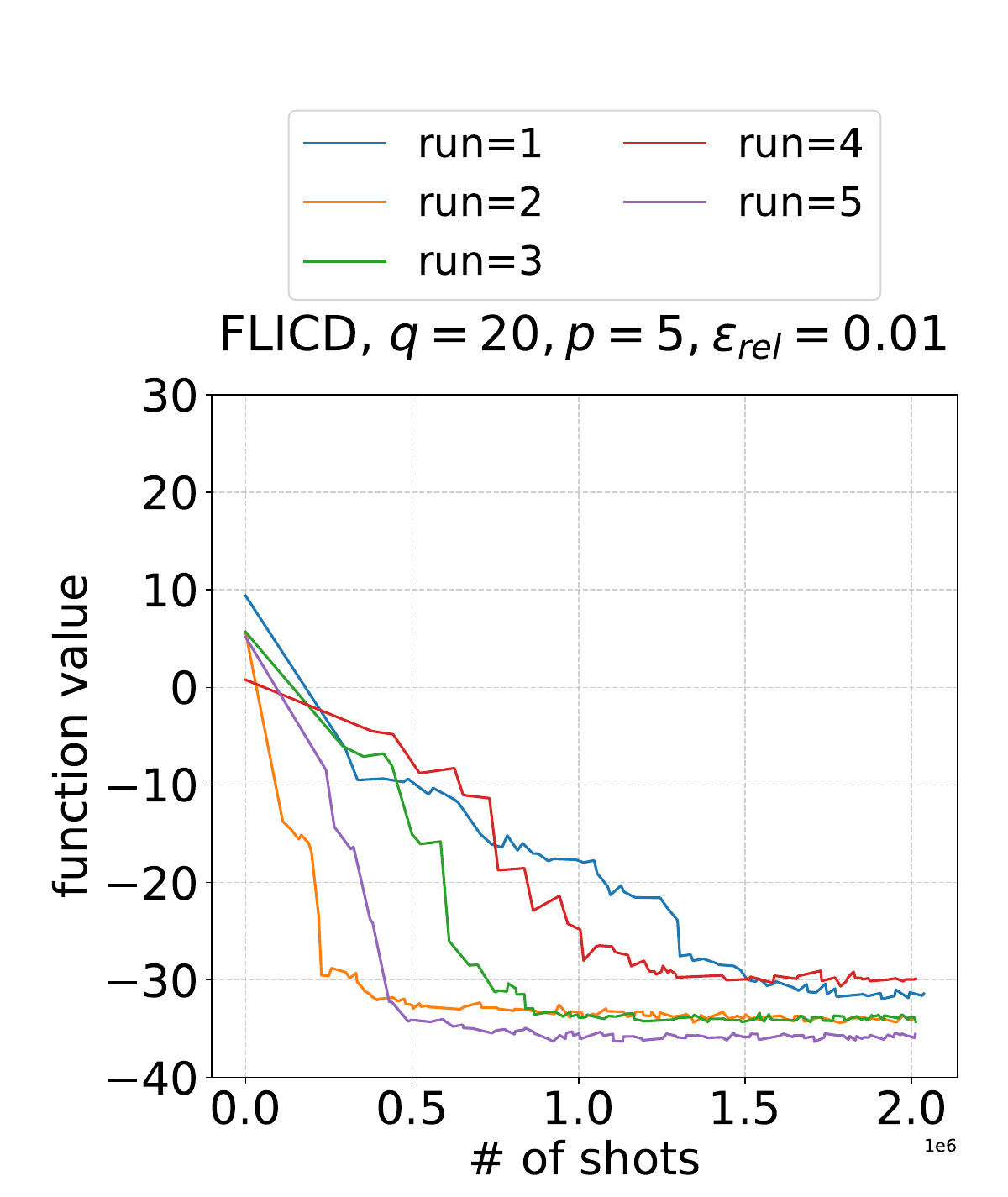}
        \caption{}
        \label{fig:sub5}
    \end{subfigure}
    \hfill
    \begin{subfigure}[b]{0.32\textwidth}
        \centering
        \includegraphics[width=\textwidth]{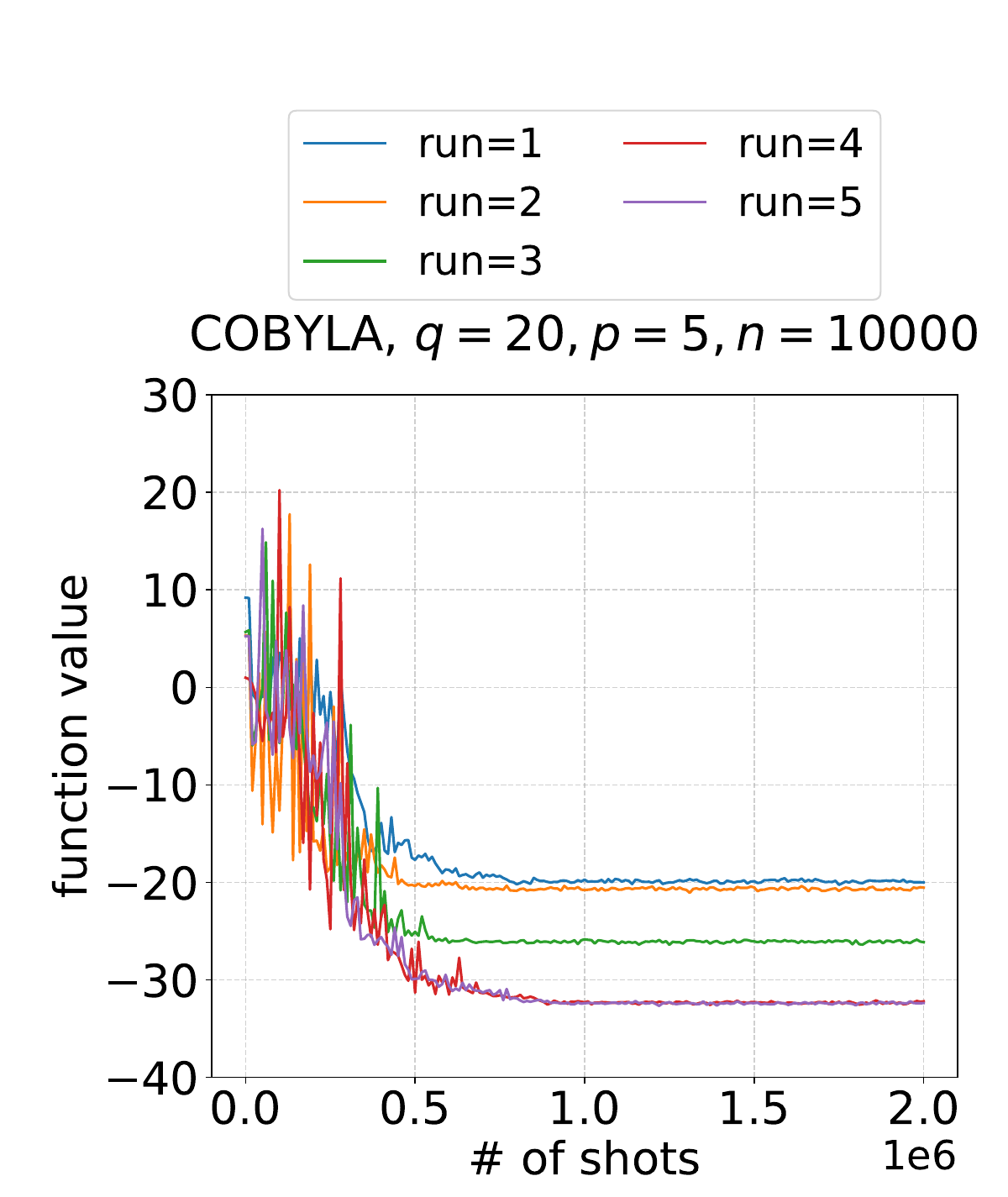}
        \caption{}
        \label{fig:sub6}
    \end{subfigure}
    \hfill
    \begin{subfigure}[b]{0.32\textwidth}
        \centering
        \includegraphics[width=\textwidth]{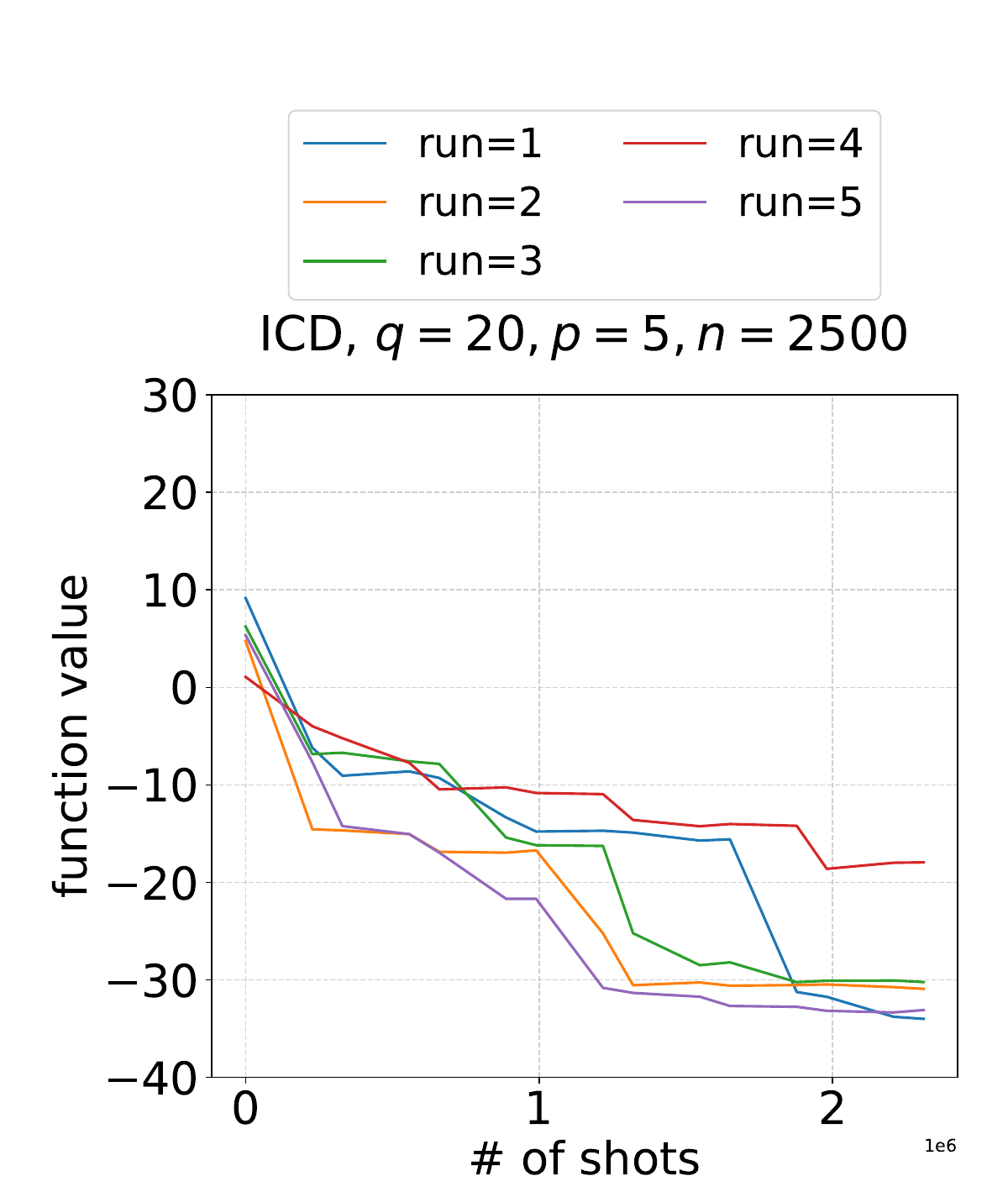}
        \caption{}
        \label{fig:top}
    \end{subfigure}
    \caption{Function value with respect to total shots across all 5 runs of FLICD (left), COBYLA (middle), and ICD (right) on simulator backend. Cost functions $f(\boldsymbol{\theta})$ generated from $5$-layer LC-QAOA ansatz of 20-vertex random regular degree-3 graph.}
    \label{fig:simulator_p=5}
\end{figure}

In section \ref{subsubsec:Comparison of FLICD to  COBYLA and ICD}, plots were given showing the performance of FLICD, COBYLA, and ICD on average across five different initial parameters. Here, we breakdown the performance of each algorithm across the five individual runs. Figures \ref{fig:simulator_p=1},~\ref{fig:simulator_p=3}, and \ref{fig:simulator_p=5} show the convergence of each algorithm across the three different choice of total layers $p \in \{1,3,5 \}$. 

\textbf{Almost all runs of FLICD outperform COBYLA, regardless of hyperparameter.} By comparing the first and second columns of Figures \ref{fig:simulator_p=1},~\ref{fig:simulator_p=3}, and \ref{fig:simulator_p=5}, the performance gap between FLICD and COBYLA can be accessed. Regardless of the hyperparameter choice $\varepsilon_{rel}$ for FLICD and shots $n$ for COBYLA, FLICD outperforms COBYLA from the same initial point, except run $4$ with $\varepsilon_{rel}  = 0.01$. Furthermore, when $p = 1$  or $3$, the FLICD run with the highest converged function value is still lower than the lowest converged function value of any of the COBYLA runs. This further demonstrates the struggles that an off-the-shelf optimizer may face with function landscapes containing many local minima.  

\textbf{FLICD is more stable than ICD.} The second and third columns of Figures \ref{fig:simulator_p=1},~\ref{fig:simulator_p=3}, and \ref{fig:simulator_p=5} reveal the convergence behavior of FLICD and ICD.  With $p=1$, both FLICD and ICD converge to the same objective value, regardless of the hyperparameter choice. The function value trajectories are characteristically different at higher $p$. For low shot ICD, the trajectory becomes more chaotic. It is likely that with a lower number of shots, the recovered restricted function at each iteration is not close enough to the true uni-variate function, leading to inefficient and inaccurate parameter update trajectories. For high shot ICD, the function trajectories are not near convergence or have higher variance in converged function value. This provides additional evidence as to why unnecessarily higher number of frequencies can be detrimental to algorithm performance.

\subsubsection{FLICD and analysis of inner adaptive frequency updates}

\begin{figure}[htbp]
    \centering
    
    \begin{subfigure}[b]{0.31\textwidth}
        \centering
        \includegraphics[width=\textwidth]{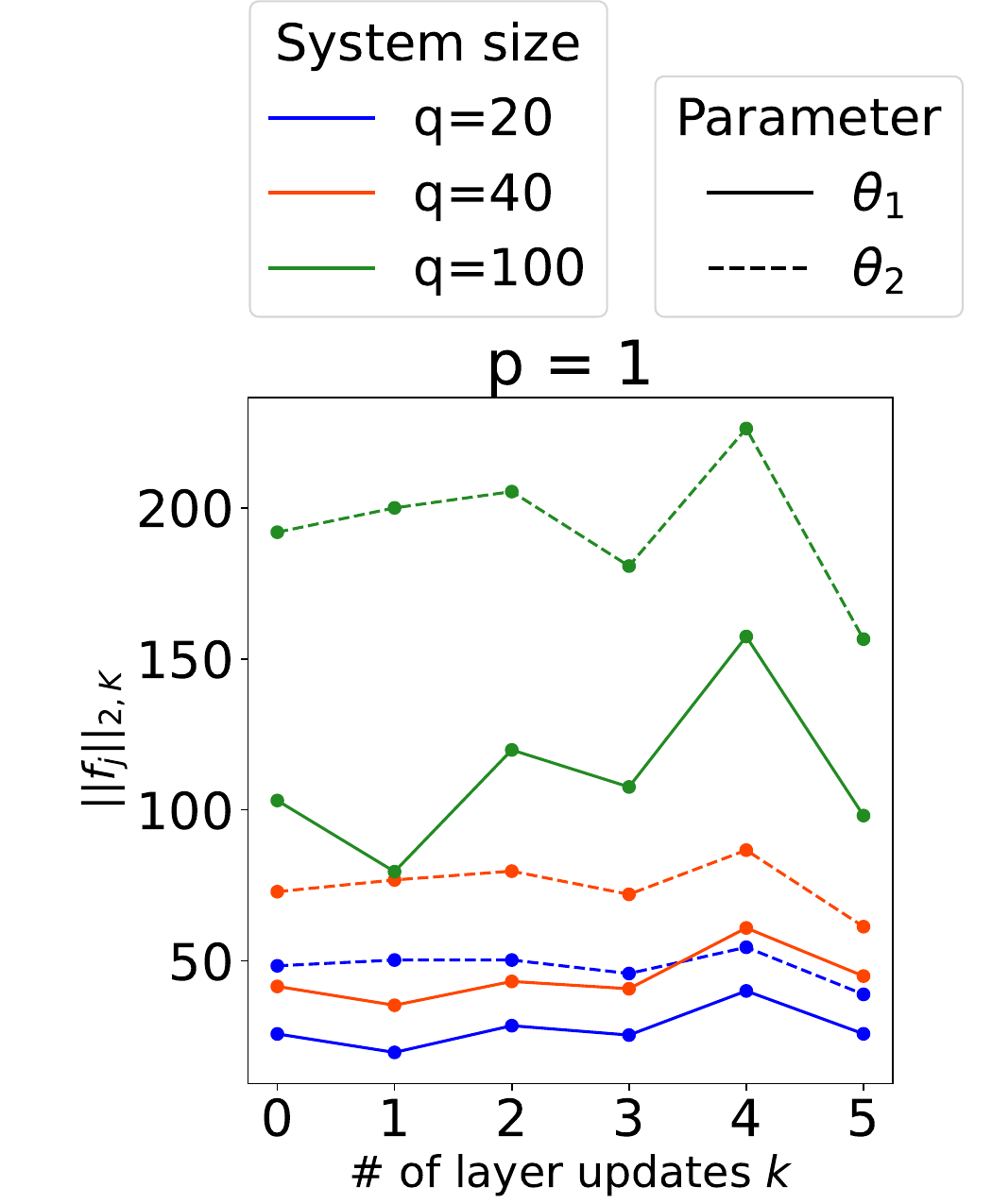}
        \caption{}
        \label{fig:sub1}
    \end{subfigure}
    \hfill
    \begin{subfigure}[b]{0.31\textwidth}
        \centering
        \includegraphics[width=\textwidth]{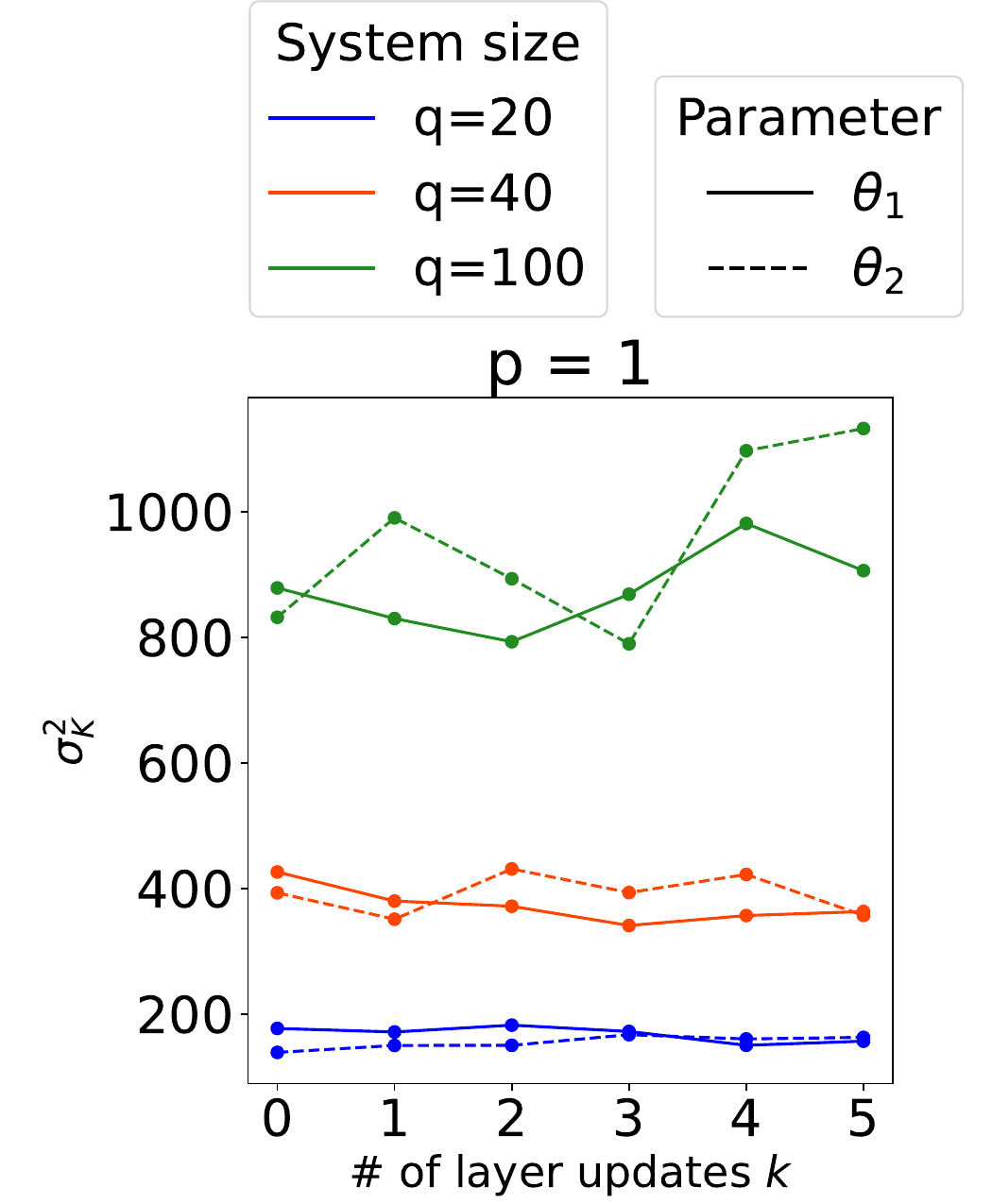}
        \caption{}
        \label{fig:sub2}
    \end{subfigure}
    \hfill
    \begin{subfigure}[b]{0.31\textwidth}
        \centering
        \includegraphics[width=\textwidth]{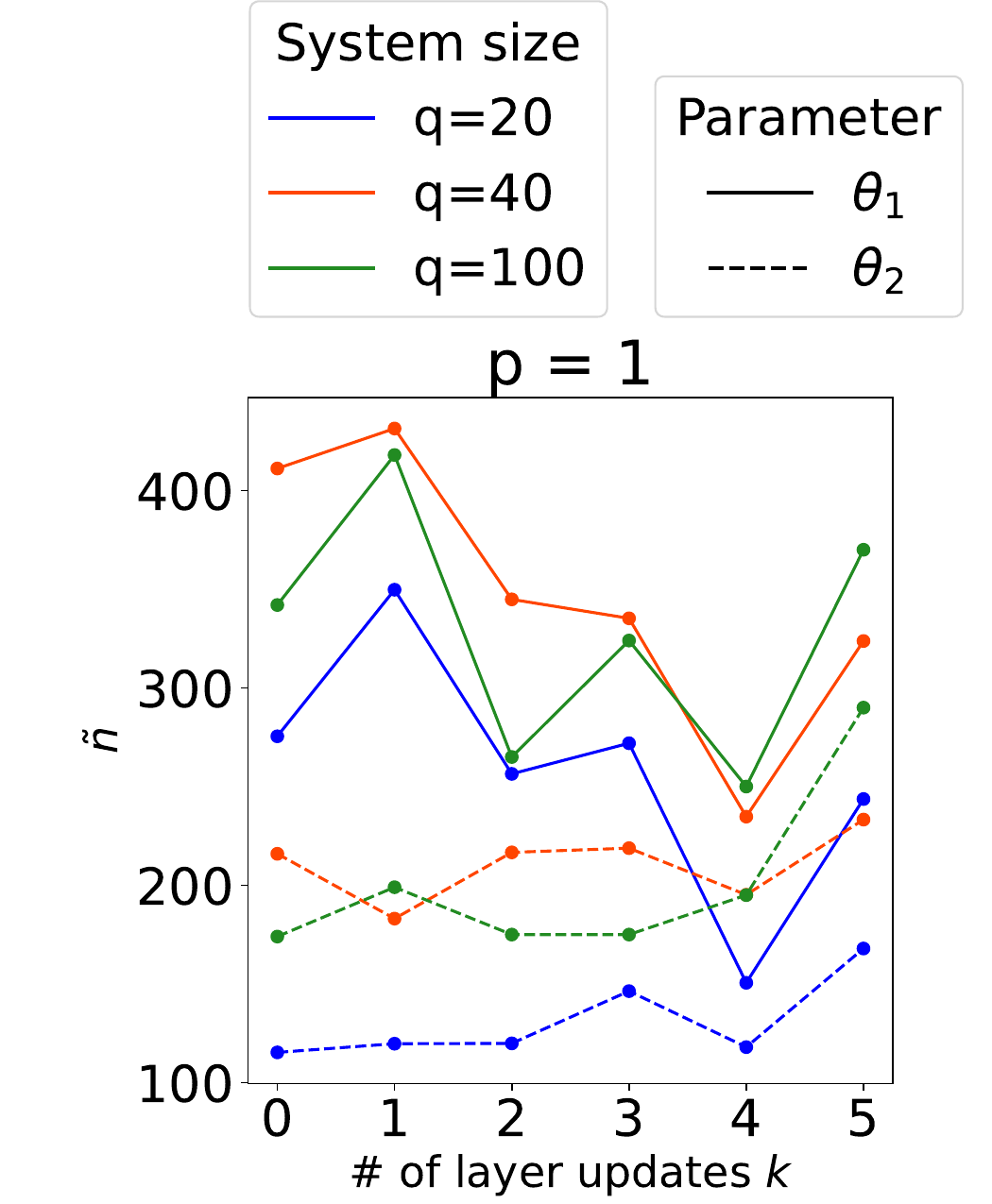}
        \caption{}
        \label{fig:sub3}
    \end{subfigure}

    \vspace{0.4cm} 

    \begin{subfigure}[b]{0.31\textwidth}
        \centering
        \includegraphics[width=\textwidth]{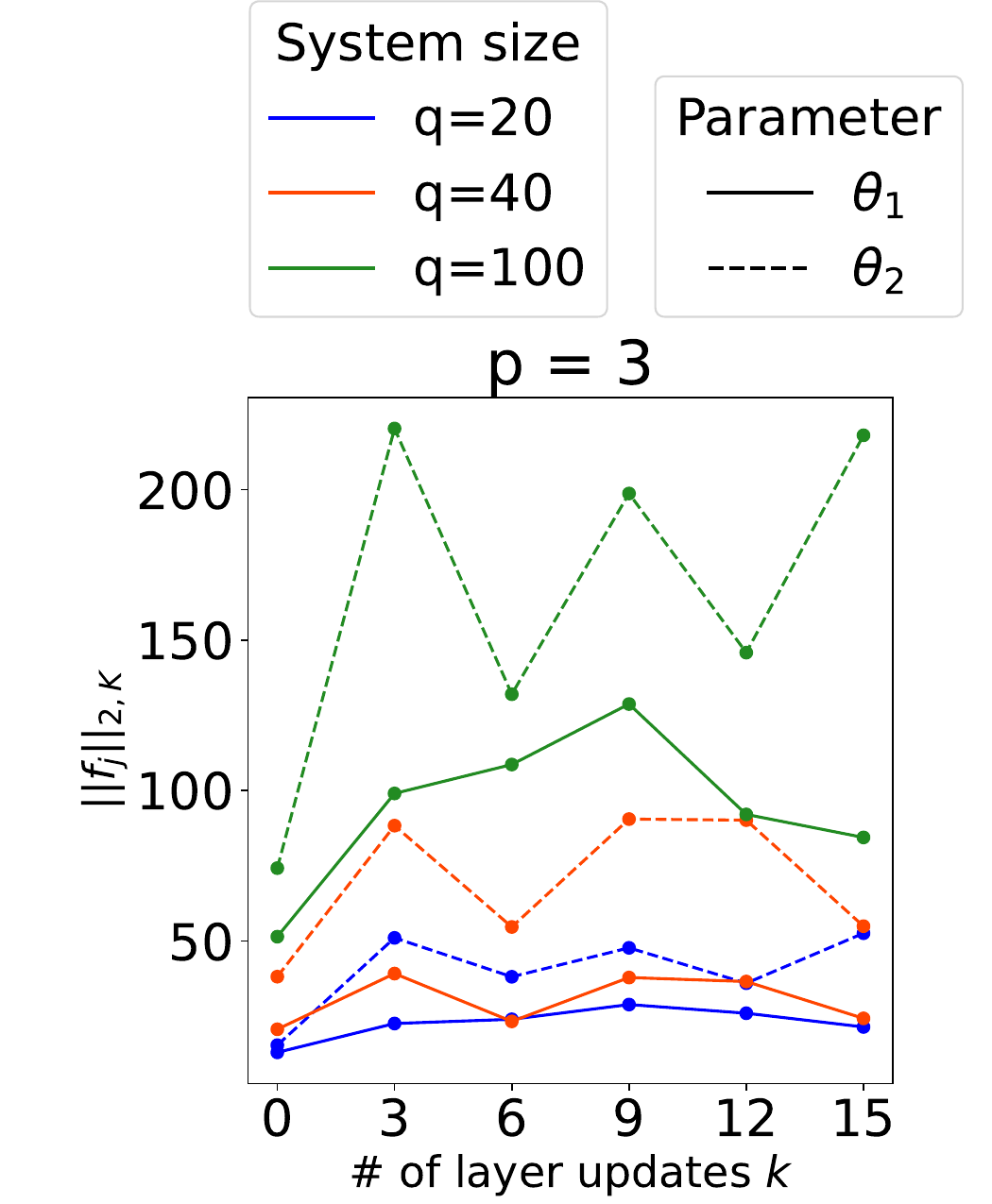}
        \caption{}
        \label{fig:sub4}
    \end{subfigure}
    \hfill
    \begin{subfigure}[b]{0.31\textwidth}
        \centering
        \includegraphics[width=\textwidth]{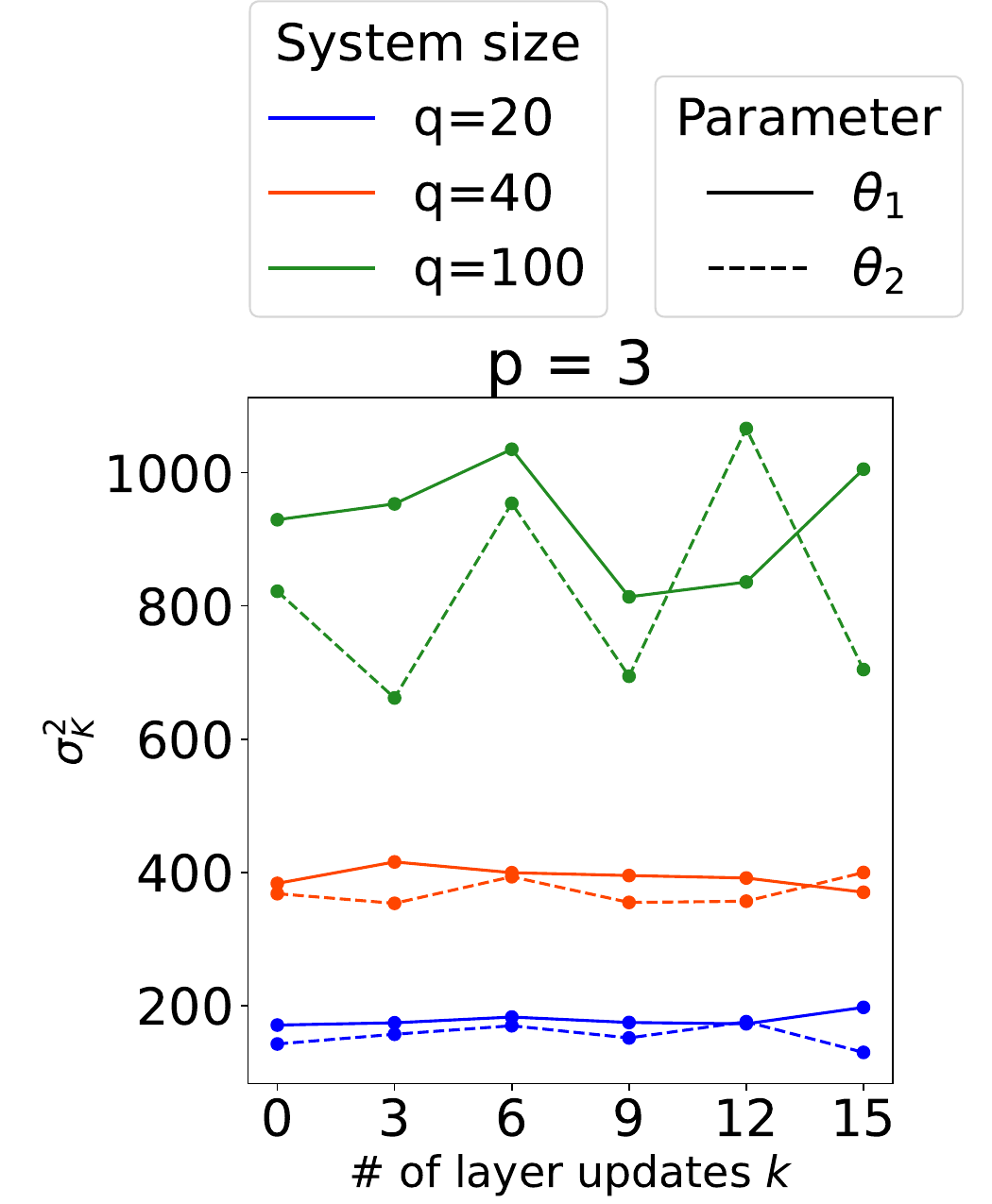}
        \caption{}
        \label{fig:sub5}
    \end{subfigure}
    \hfill
    \begin{subfigure}[b]{0.31\textwidth}
        \centering
        \includegraphics[width=\textwidth]{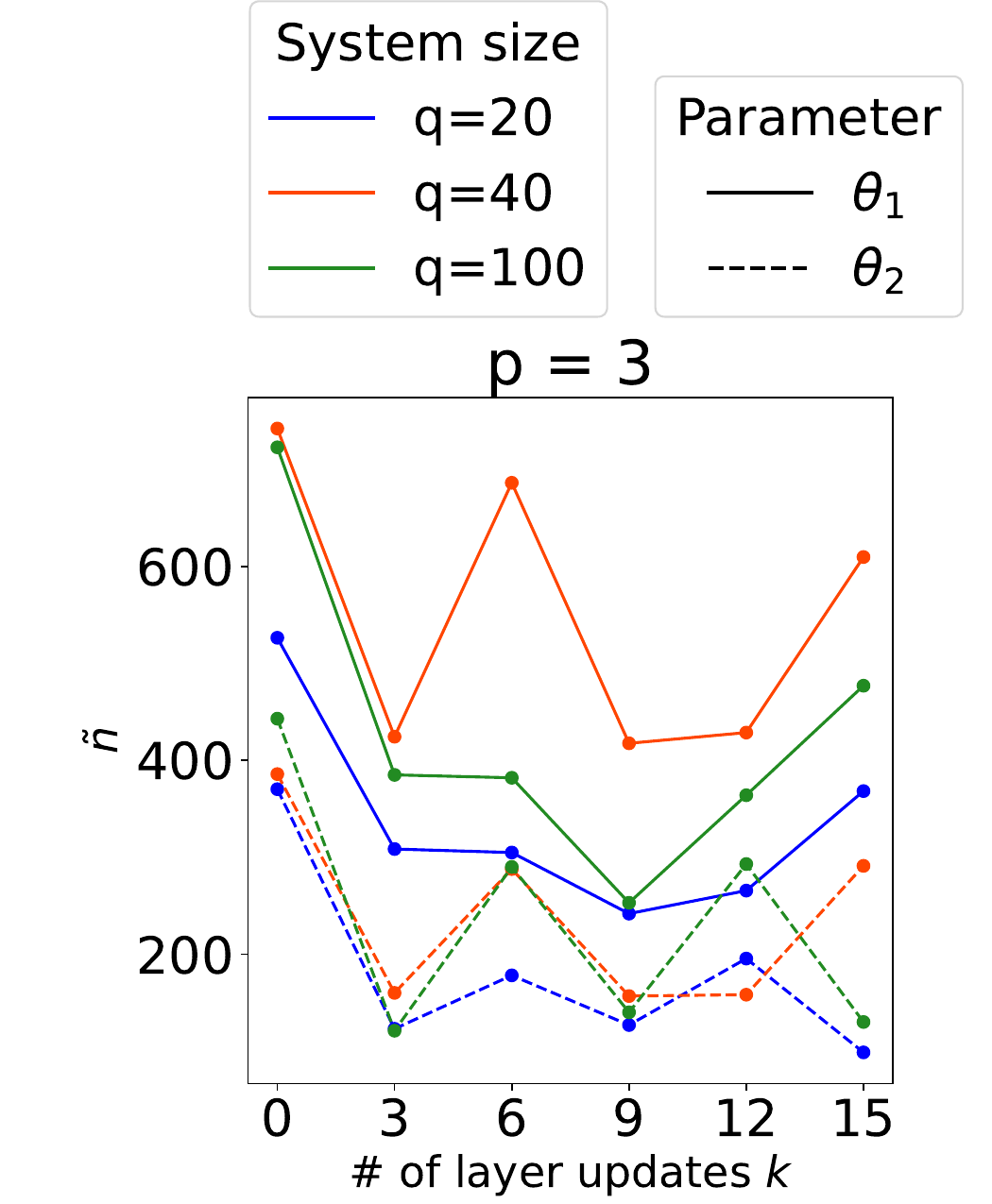}
        \caption{}
        \label{fig:sub6}
    \end{subfigure}

    \vspace{0.4cm} 

    \begin{subfigure}[b]{0.31\textwidth}
        \centering
        \includegraphics[width=\textwidth]{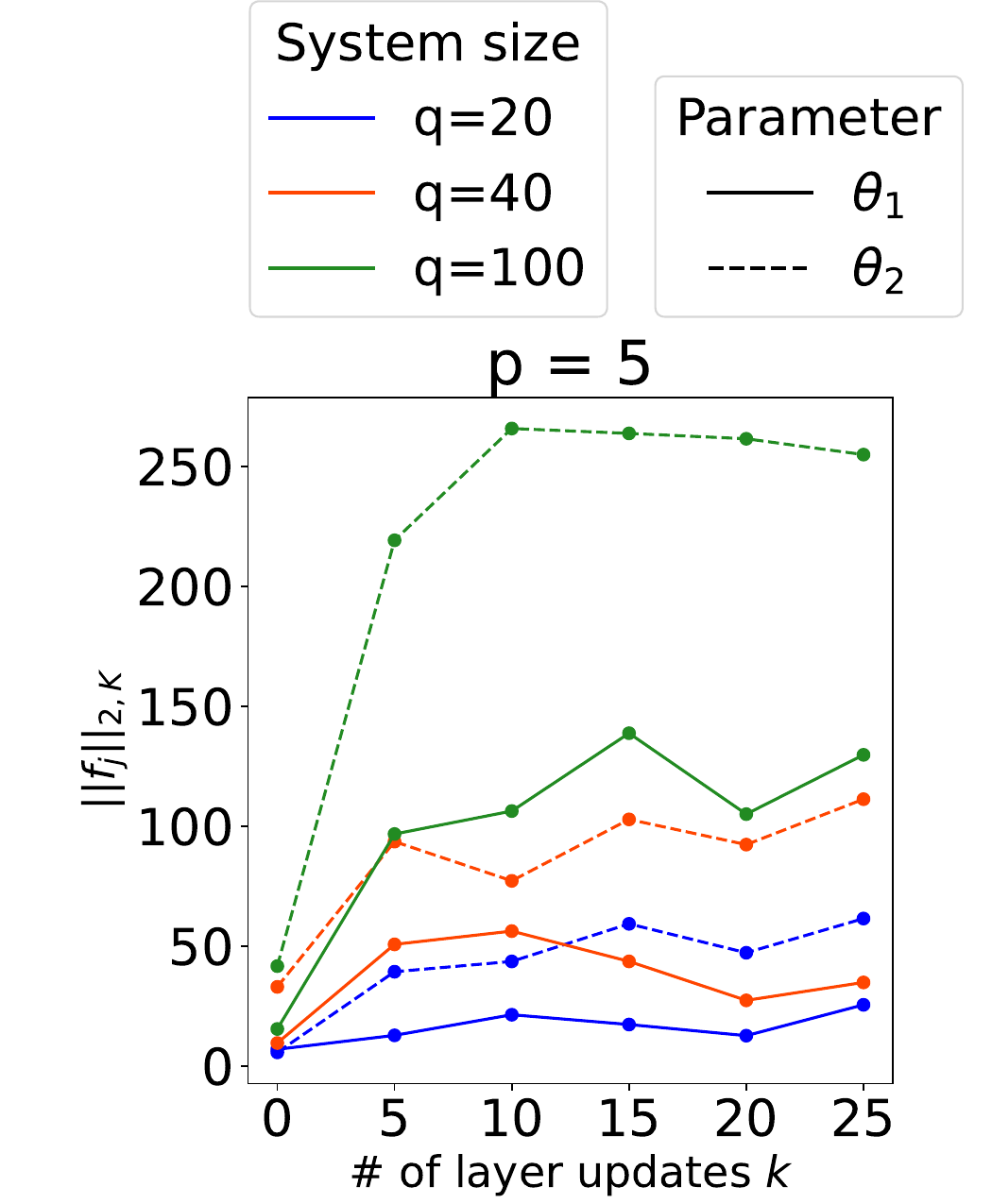}
        \caption{}
        \label{fig:sub7}
    \end{subfigure}
    \hfill
    \begin{subfigure}[b]{0.31\textwidth}
        \centering
        \includegraphics[width=\textwidth]{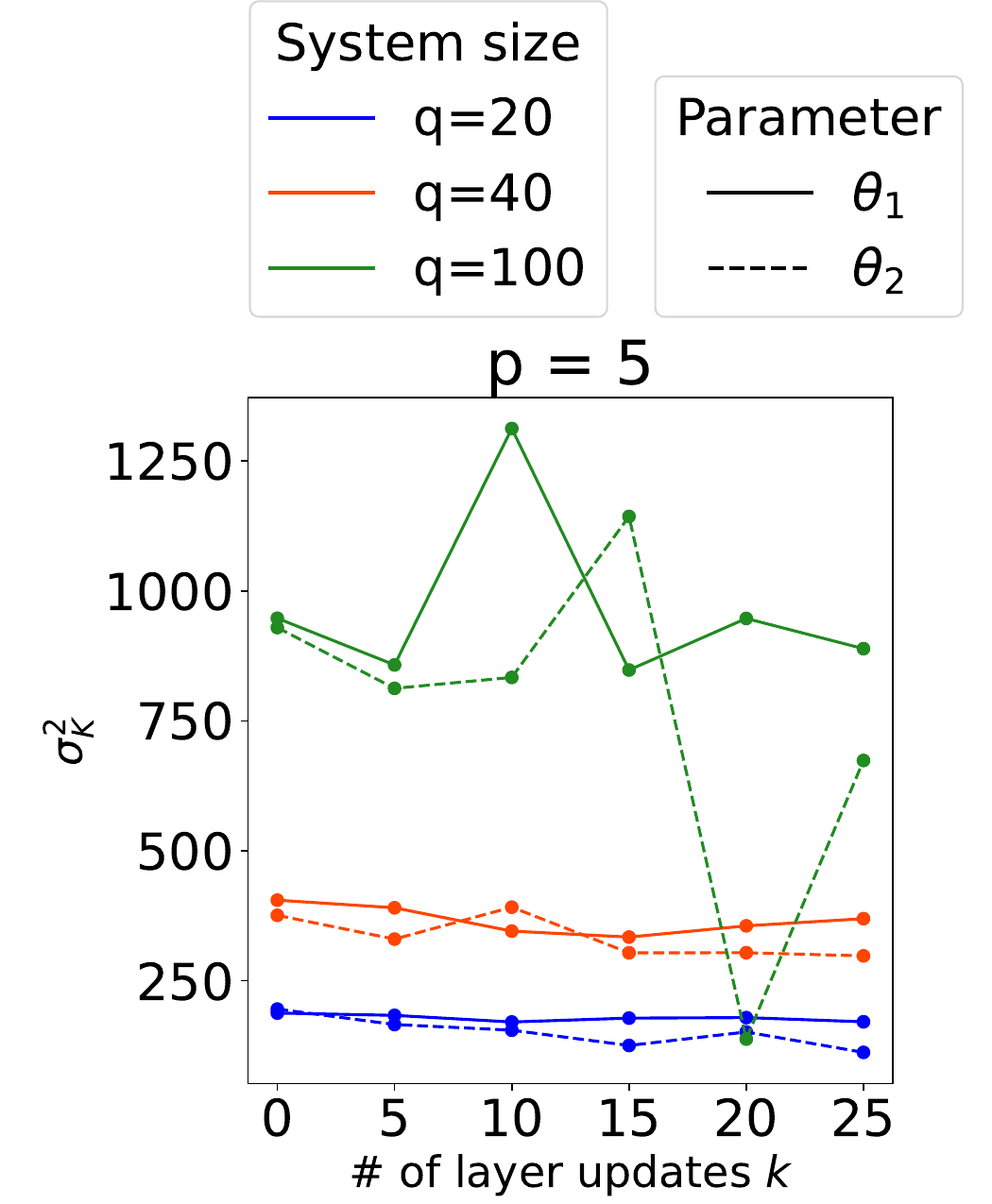}
        \caption{}
        \label{fig:sub8}
    \end{subfigure}
    \hfill
    \begin{subfigure}[b]{0.31\textwidth}
        \centering
        \includegraphics[width=\textwidth]{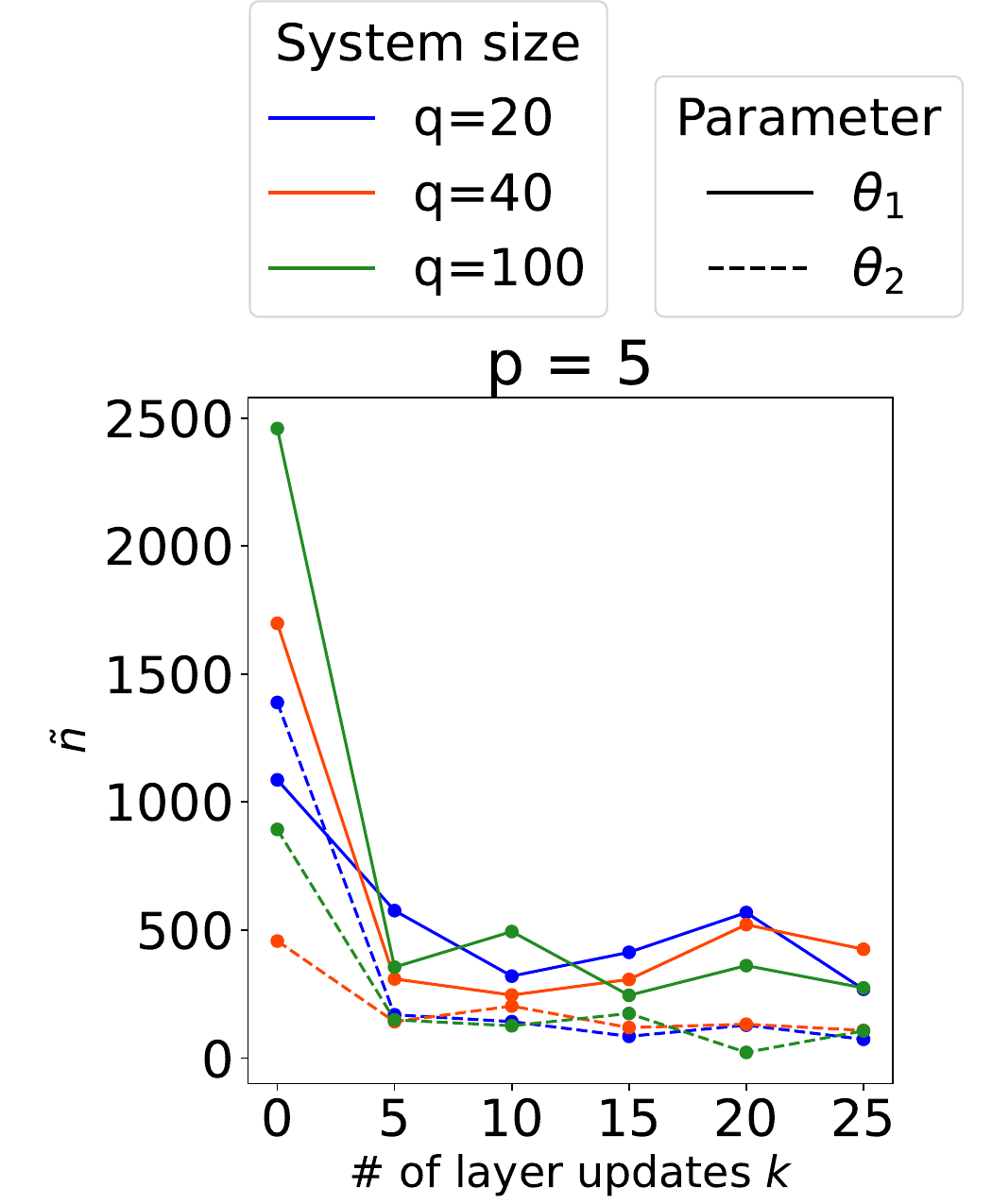}
        \caption{}
        \label{fig:sub9}
    \end{subfigure}

    \caption{Adaptive frequency update metrics during FLICD for experiments on IBM quantum hardware in Figure \ref{fig:all_qubit_conv_hardware}. The L2 norm (left) and function variance (middle) are used to compute the number of shots needed for interpolation (right) to achieve the desired relative error. 
    }
    \label{fig:overall_stats_hardware}
\end{figure}

\begin{figure}[p]
    \centering
    \begin{minipage}{6.5in}
        \centering
        \begin{subfigure}[b]{0.31\textwidth}
            \centering
            \includegraphics[width=\textwidth]{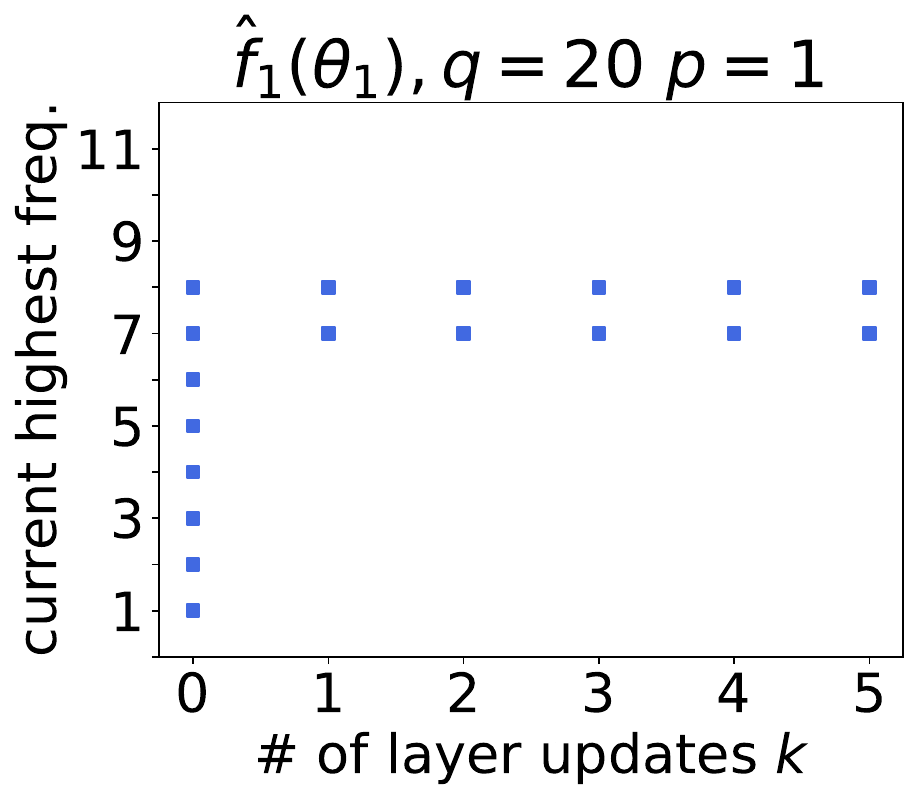}
            \caption{}
            \label{fig:l0_q20_p1}
        \end{subfigure}
        \hfill
        \begin{subfigure}[b]{0.31\textwidth}
            \centering
            \includegraphics[width=\textwidth]{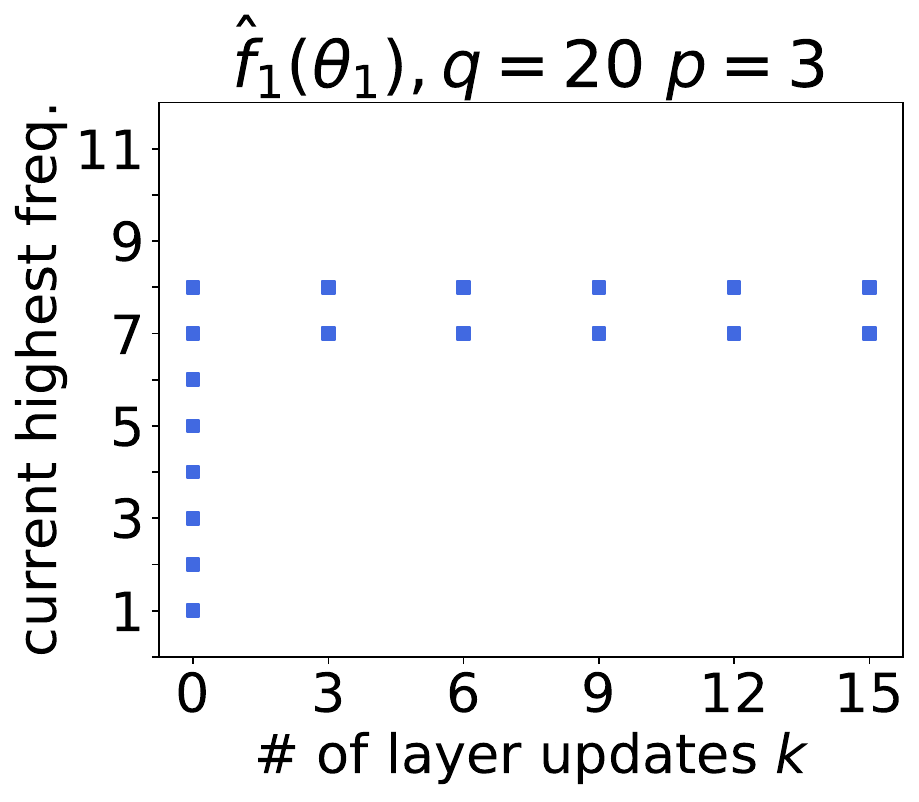}
            \caption{}
            \label{fig:l0_q20_p3}
        \end{subfigure}
        \hfill
        \begin{subfigure}[b]{0.31\textwidth}
            \centering
            \includegraphics[width=\textwidth]{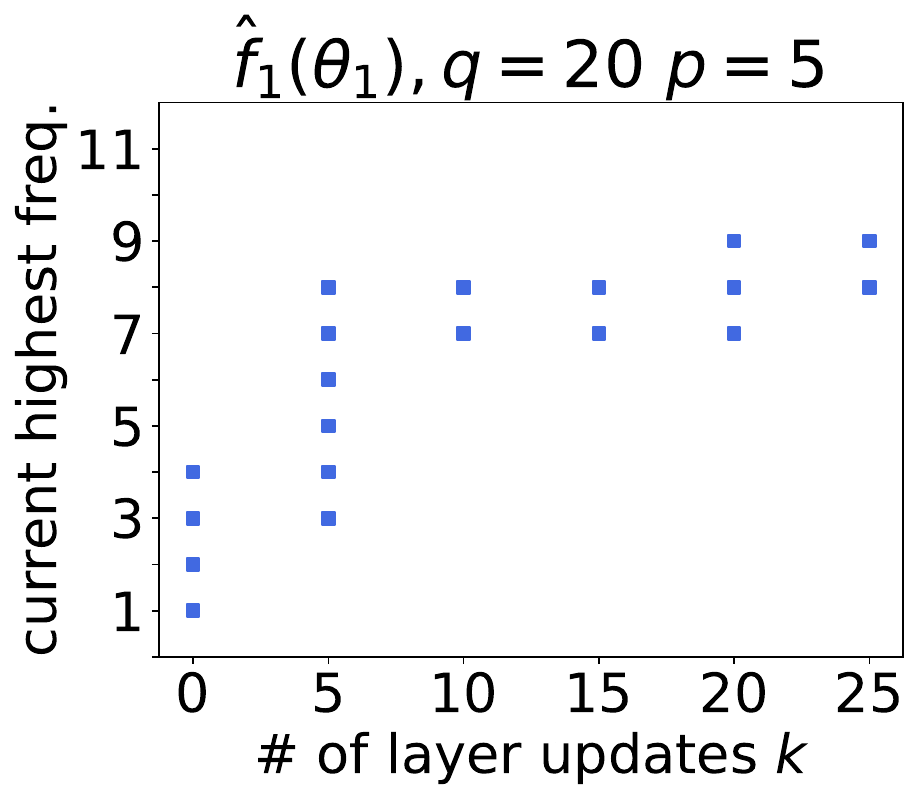}
            \caption{}
            \label{fig:l0_q20_p5}
        \end{subfigure}
        
        \vspace{0.2em}

        \begin{subfigure}[b]{0.31\textwidth}
            \centering
            \includegraphics[width=\textwidth]{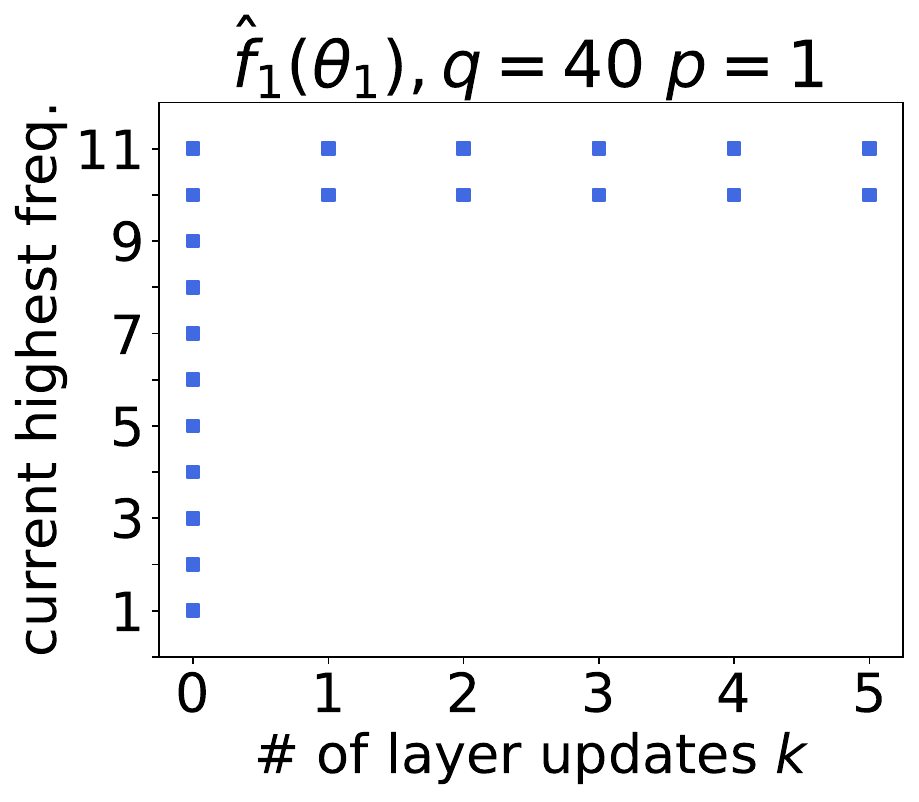}
            \caption{}
            \label{fig:l0_q40_p1}
        \end{subfigure}
        \hfill
        \begin{subfigure}[b]{0.31\textwidth}
            \centering
            \includegraphics[width=\textwidth]{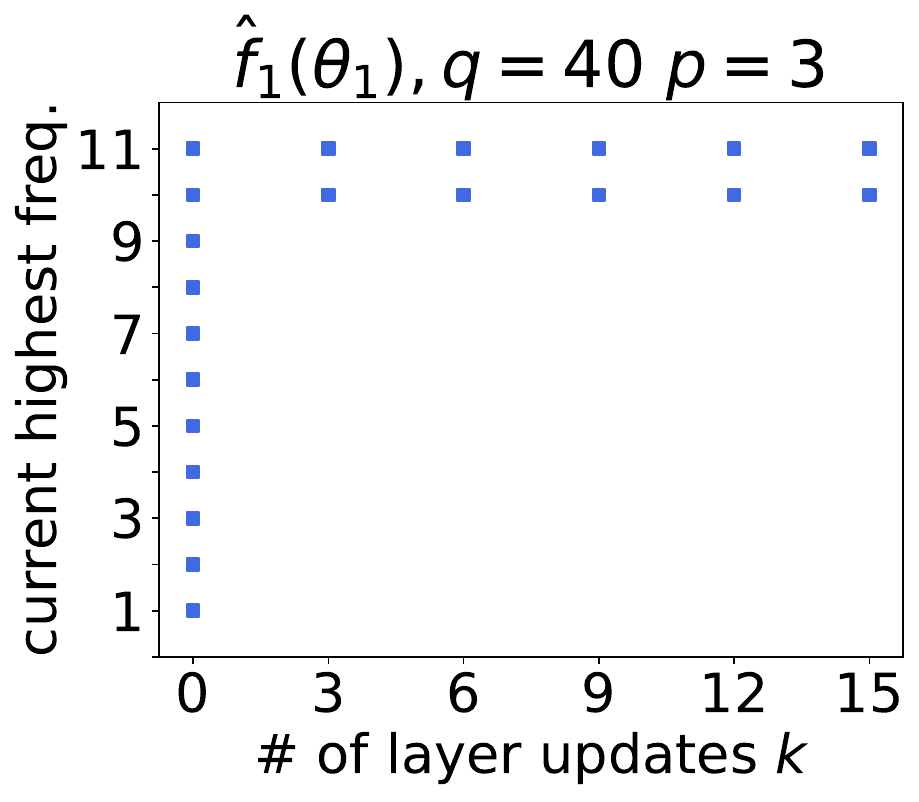}
            \caption{}
            \label{fig:l0_q40_p3}
        \end{subfigure}
        \hfill
        \begin{subfigure}[b]{0.31\textwidth}
            \centering
            \includegraphics[width=\textwidth]{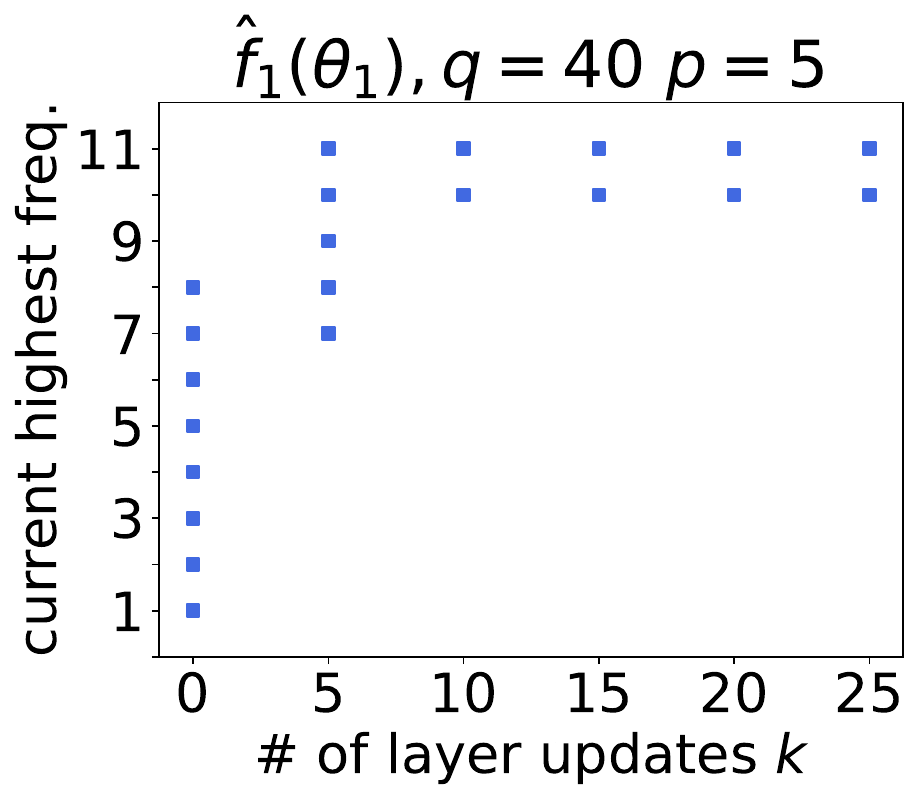}
            \caption{}
            \label{fig:l0_q40_p5}
        \end{subfigure}

        \vspace{0.2em}

        \begin{subfigure}[b]{0.31\textwidth}
            \centering
            \includegraphics[width=\textwidth]{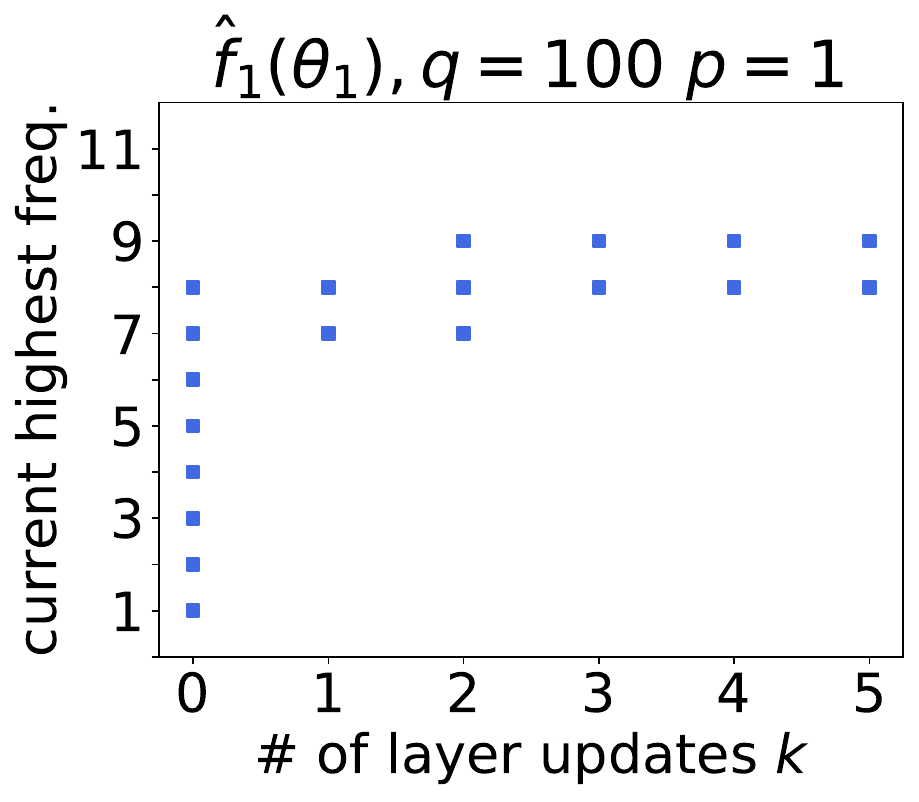}
            \caption{}
            \label{fig:l0_q100_p1}
        \end{subfigure}
        \hfill
        \begin{subfigure}[b]{0.31\textwidth}
            \centering
            \includegraphics[width=\textwidth]{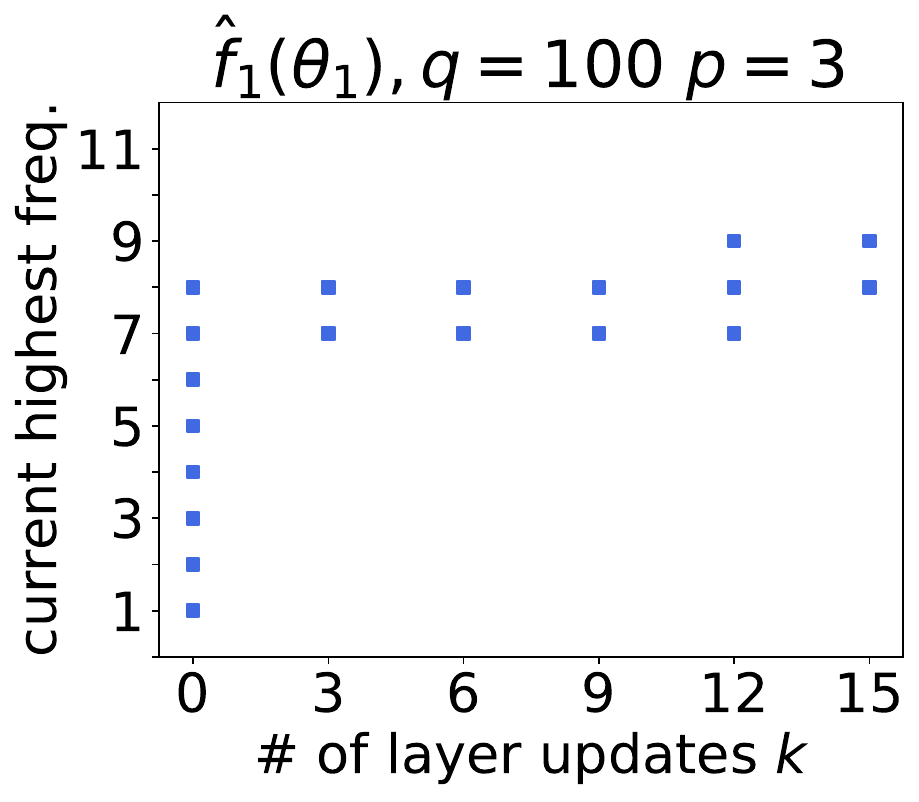}
            \caption{}
            \label{fig:l0_q100_p3}
        \end{subfigure}
        \hfill
        \begin{subfigure}[b]{0.31\textwidth}
            \centering
            \includegraphics[width=\textwidth]{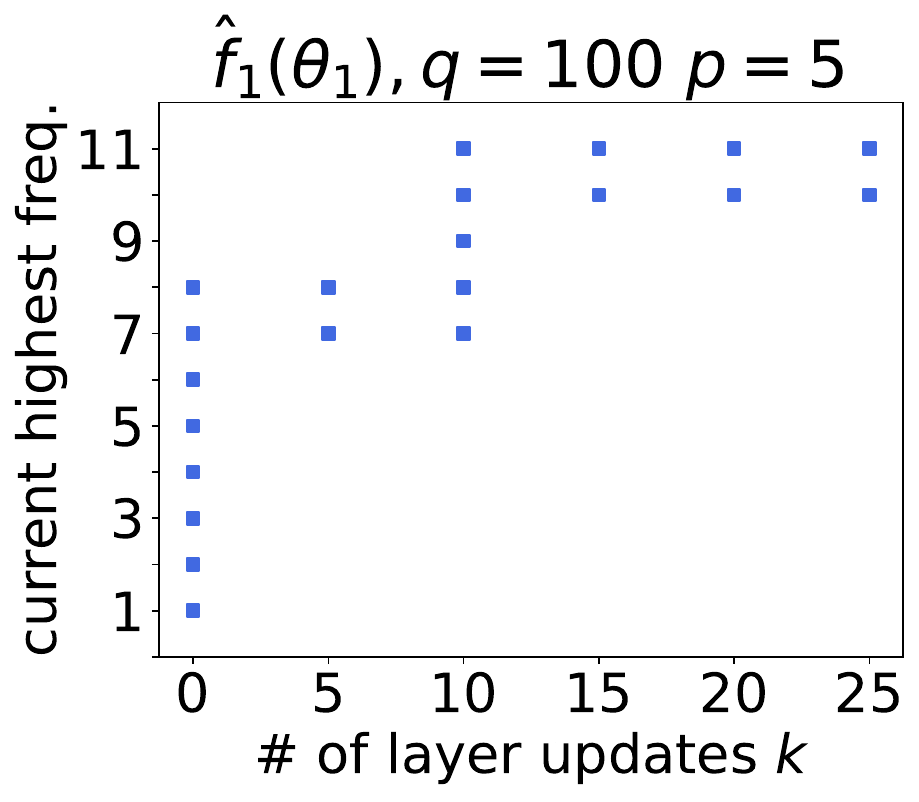}
            \caption{}
            \label{fig:l0_q100_p5}
        \end{subfigure}
    \end{minipage}
    
    \caption{The frequencies that are learned through the adaptive frequency updates on $\theta_1$, the cost layer, during FLICD for experiments on IBM quantum hardware in Figure \ref{fig:all_qubit_conv_hardware}.    }
    
    \label{fig:freqs_ell_0}
\end{figure}

\begin{figure}[p]
    \centering
    \begin{minipage}{6.5in}
        \centering
        \begin{subfigure}[b]{0.31\textwidth}
            \centering
            \includegraphics[width=\textwidth]{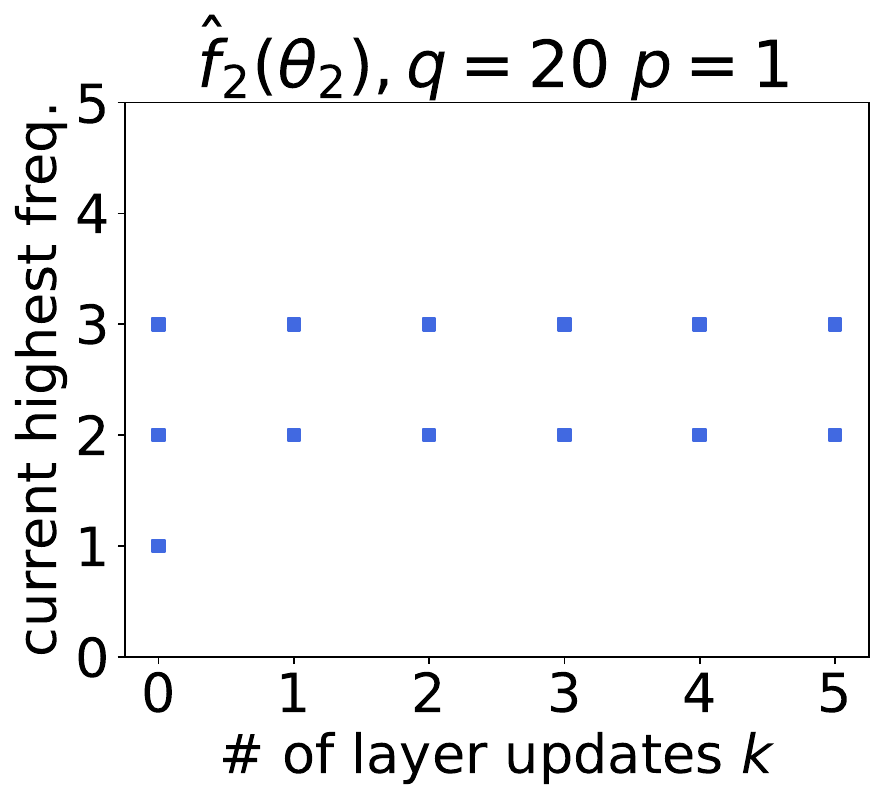}
            \caption{}
            \label{fig:l1_q20_p1}
        \end{subfigure}
        \hfill
        \begin{subfigure}[b]{0.31\textwidth}
            \centering
            \includegraphics[width=\textwidth]{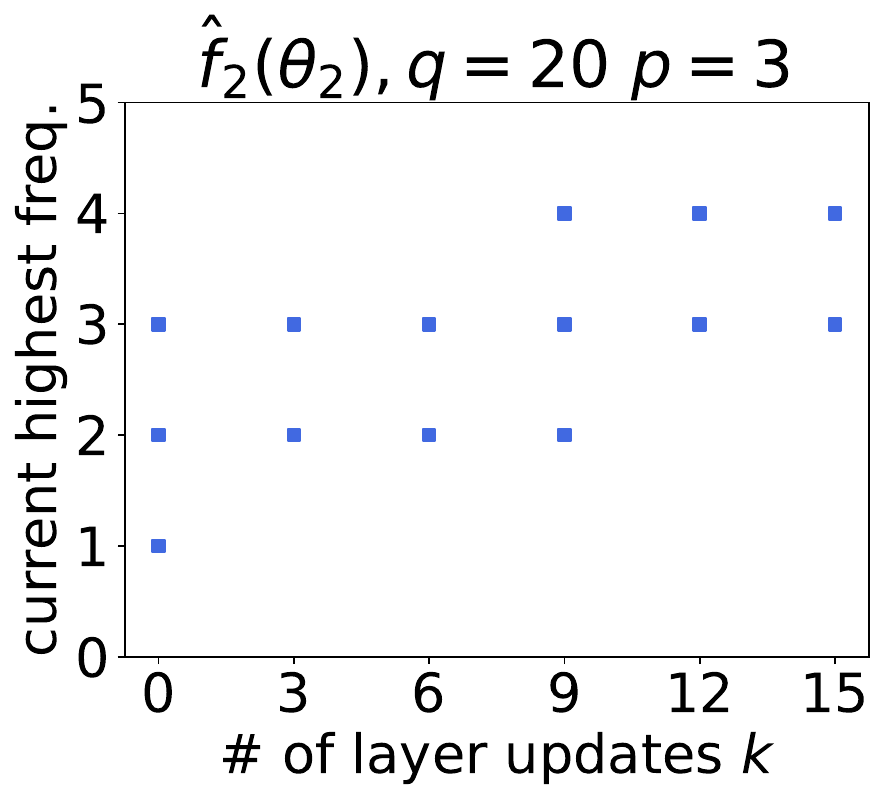}
           \caption{}
            \label{fig:l1_q20_p3}
        \end{subfigure}
        \hfill
        \begin{subfigure}[b]{0.31\textwidth}
            \centering
            \includegraphics[width=\textwidth]{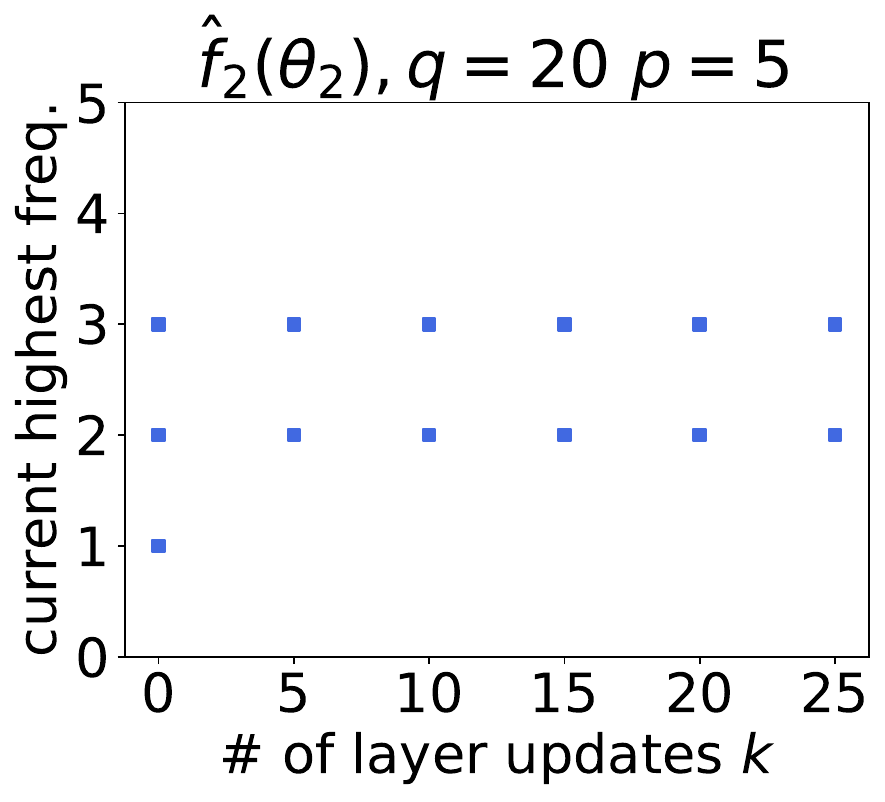}
            \caption{}
            \label{fig:l1_q20_p5}
        \end{subfigure}
        
        \vspace{0.2em}

        \begin{subfigure}[b]{0.31\textwidth}
            \centering
            \includegraphics[width=\textwidth]{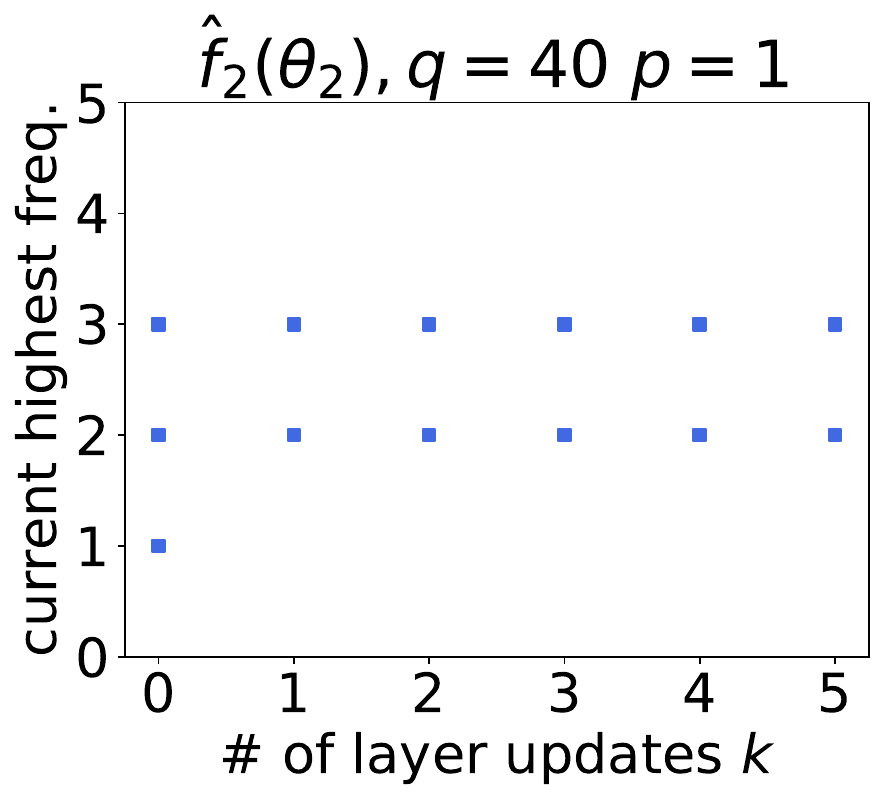}
            \caption{}
            \label{fig:l1_q40_p1}
        \end{subfigure}
        \hfill
        \begin{subfigure}[b]{0.31\textwidth}
            \centering
            \includegraphics[width=\textwidth]{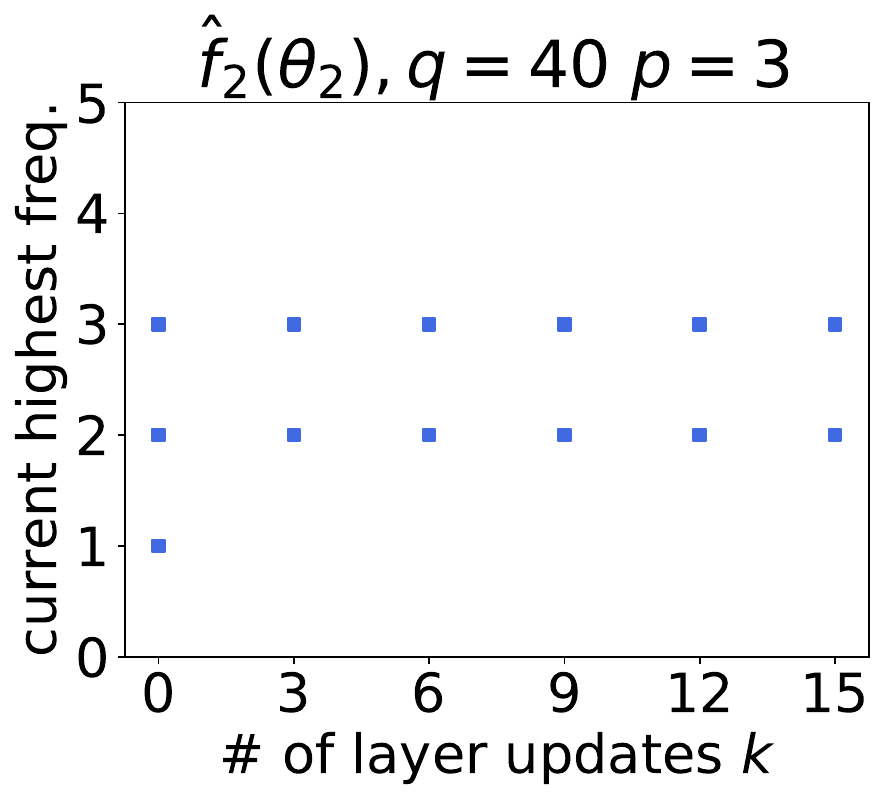}
            \caption{}
            \label{fig:l1_q40_p3}
        \end{subfigure}
        \hfill
        \begin{subfigure}[b]{0.31\textwidth}
            \centering
            \includegraphics[width=\textwidth]{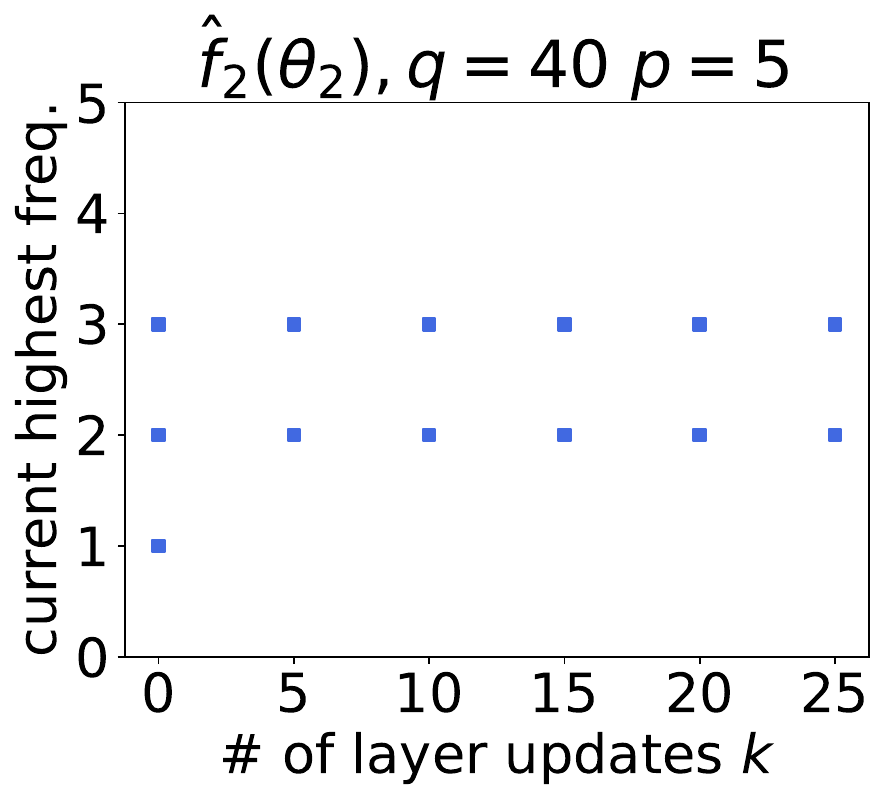}
            \caption{}
            \label{fig:l1_q40_p5}
        \end{subfigure}

        \vspace{0.2em}

        \begin{subfigure}[b]{0.31\textwidth}
            \centering
            \includegraphics[width=\textwidth]{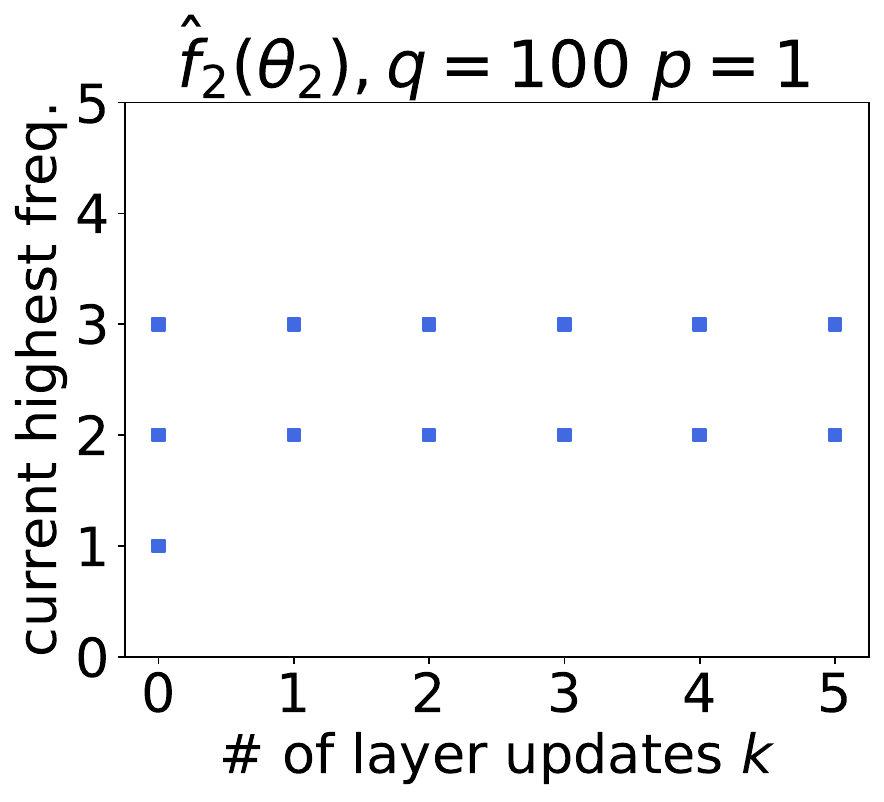}
            \caption{}
            \label{fig:l1_q100_p1}
        \end{subfigure}
        \hfill
        \begin{subfigure}[b]{0.31\textwidth}
            \centering
            \includegraphics[width=\textwidth]{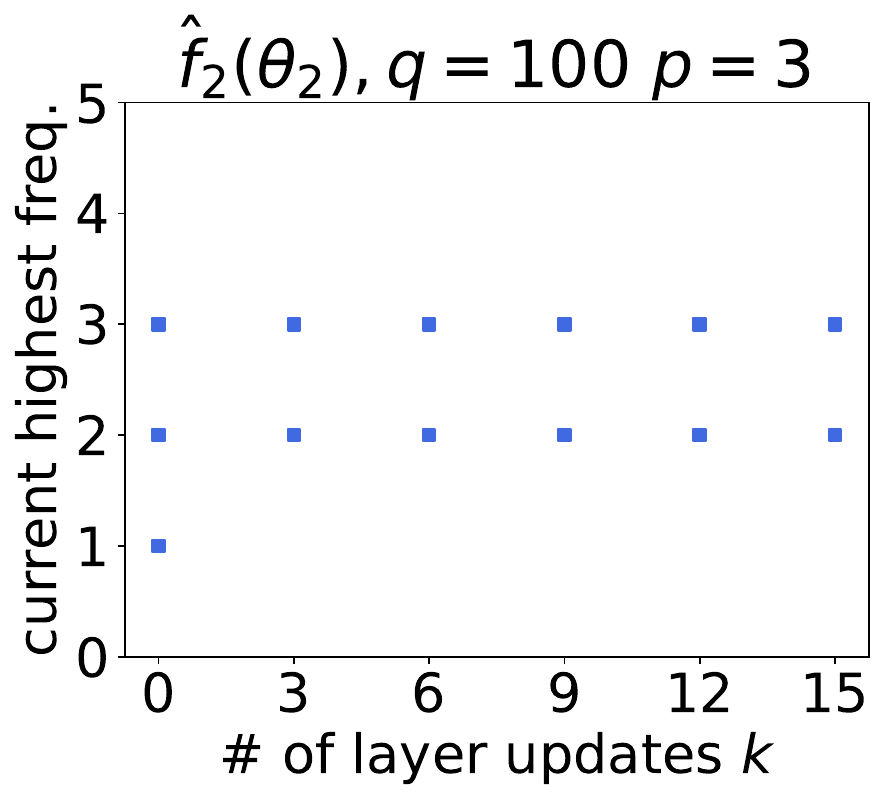}
            \caption{}
            \label{fig:l1_q100_p3}
        \end{subfigure}
        \hfill
        \begin{subfigure}[b]{0.31\textwidth}
            \centering
            \includegraphics[width=\textwidth]{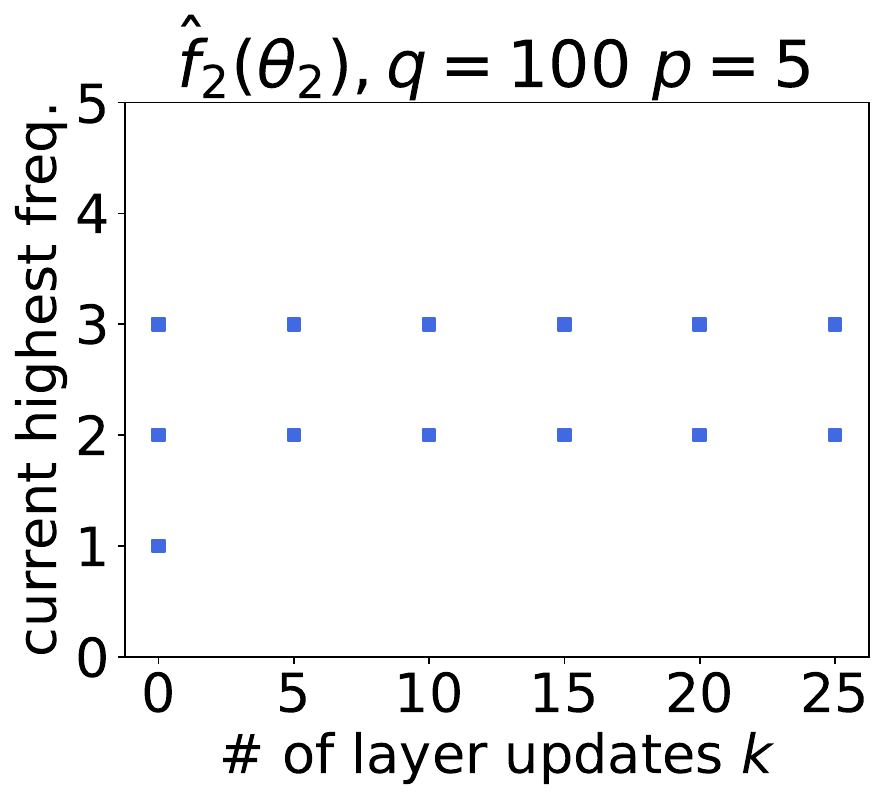}
            \caption{}
            \label{fig:l1_q100_p5}
        \end{subfigure}
    \end{minipage}
    
    \caption{The frequencies that are learned through the adaptive frequency updates on $\theta_2$, the mixing layer, during FLICD for experiments on IBM quantum hardware in Figure \ref{fig:all_qubit_conv_hardware}.   }
    \label{fig:freqs_ell_1}
\end{figure}

In section \ref{FLICD on quantum hardware and at scale}, the performance of FLICD on hardware and on larger problem sizes is given. Now, we reveal the inner working of FLICD across these experiments by showing the results of the adaptive frequency updates. Since $S = p$, by line $6$ of Algorithm \ref{alg:two}, the adaptive frequency update occurs when $k$ is a multiple of $p$. This $k$ is referenced in Figures \ref{fig:freqs_ell_0} and \ref{fig:freqs_ell_1}. Stated another way, the adaptive frequency update always occurs on updates to $\theta_1$ and $\theta_2$.  These two variables correspond to the frequency learning of the cost layer and mixing layer, respectively. Figure \ref{fig:overall_stats_hardware} shows the computed function norm, variance, and computed shots used for interpolation. Figures \ref{fig:freqs_ell_0} and \ref{fig:freqs_ell_1} correspond to the frequencies learned for the cost and mixing layer.

\textbf{Different layer positions require different number of shots for same accuracy.} By comparing the solid and dashed lines of Figure \ref{fig:overall_stats_hardware} across all cases of $q$ and $p$, we can compare the difference between the cost and mixing layer by function norm, variance, and computed shots for interpolation. In general, the cost layer has lower $L_2$-norm than the mixing layer, but the function variances are similar. It is clear that the mixing layer requires much fewer shots than the cost layer to acheive the same relative error. This may be of interest to ICD. Currently, one shot count is specified for interpolation, but allowing different $\theta_j$ to have different number of shots may lead to further cost savings.  


\textbf{Learned frequencies on the initial parameters may not be enough.} It is vital to execute the adaptive frequency update multiple times during FLICD, as opposed to solely at the beginning of the algorithm. For example, with $p = 5$ across all system sizes, FLICD added higher-order frequencies to interpolate the cost layer during subsequent adaptive frequency updates to achieve the same relative error bound. This motivates why a systematic method is needed to set a smaller set of frequencies, as initial information alone may be misleading about structure at further iterations.  

In most experiments, adaptive frequency updates do not learn significantly higher-order frequencies beyond the third execution. If no new frequencies were learned, this means a fixed frequency update could have been performed instead—at a cheaper shot and QPU cost. This observation implies that specifying the adaptive frequency update every $S=p$ layer updates may not be the most efficient. While our choice of $S$ may be slightly conservative, it nonetheless shows that the stopping condition is not overly sensitive to random fluctuations in the error metric about its expected value: FLICD can settle on frequency sets by the later iterations.   


\textbf{Learned frequencies can scale with system size or can be independent of system size.} The structure of the ansatz greatly affects which frequencies are learned as system size increases. The mixing layer has maximum frequency 3 even as system size goes to 100 qubits. However, for the cost layer, additional frequencies are learned to achieve the same desired relative error. This supports why setting frequencies manually for ICD based on small problem sizes may be tricky. Some frequency sets may be easy to guess at for large systems, while others may have contribution from higher-order frequencies that scales with system size in a way that is not easily predictable without experimentation.

\end{document}